\documentclass[reprint,showpacs,superscriptaddress,english,twocolumn,notitlepage,longbibliography,nofootinbib]{revtex4-1}
\usepackage[utf8]{inputenc}
\pdfoutput=1
\usepackage{times}
\usepackage{color}
\usepackage{array}
\usepackage{verbatim}
\usepackage{multirow}
\usepackage{amsmath}
\usepackage{amssymb}
\usepackage{dsfont}
\usepackage{graphicx}
\usepackage{esint}
\usepackage[unicode=true,
bookmarks=true,bookmarksnumbered=true,bookmarksopen=false,
 breaklinks=true,pdfborder={0 0 0},backref=false,colorlinks=true]
 {hyperref}
\usepackage{enumitem}
\usepackage{titlesec}
\usepackage{cleveref}
\usepackage{bbm}
\usepackage[normalem]{ulem}
\usepackage{listings}
\usepackage[thicklines]{cancel}
\usepackage{float}  %
\usepackage{subfigure}  %
\usepackage{tabularx}  %
\usepackage{makecell}  %
\usepackage{nicematrix} %

\hypersetup{
    linkcolor=red,          %
    citecolor=blue,        %
    filecolor=magenta,      %
    urlcolor=magenta           %
}

\usepackage{cleveref}
\crefname{appendix}{Sec.}{Secs.}
\crefname{equation}{Eq.}{Eqs.}
\crefname{figure}{Fig.}{Figs.}
\crefname{table}{Table}{Tables}
\crefname{section}{Sec.}{Secs.}

\renewcommand{\paragraph}[1]{\vspace{0.2cm}{\bf \textit{#1}}}
\def\ie{{\it i.e.},\ }
\def\eg{{\it e.g.},\ }

\newcommand{\SZD}[1]{\textcolor{red}{(SZD: #1)}}

\definecolor{Gray}{gray}{0.85}
\newcolumntype{a}{>{\columncolor{Gray}}c}

\usepackage{simpler-wick}
\allowdisplaybreaks[1] %

\newcommand{\mrm}{\mathrm}

\def\pare#1{\left( #1 \right)}
\def\brak#1{\left[#1\right]}

\def\bra#1{\langle #1 |}
\def\ket#1{| #1 \rangle}

\def\Ket#1{\left| #1 \right\rangle}
\def\inn#1{\langle #1 \rangle}

\def\Im{\mathrm{Im}}

\def\kk{ {\vec{k}} }

\begin{document}

\title{Anderson Critical Metal Phase in Trivial States Protected by Average Magnetic Crystalline Symmetry}

\author{Fa-Jie Wang}
\affiliation{International Center for Quantum Materials, School of Physics, Peking University, Beijing 100871, China}

\author{Zhen-Yu Xiao}
\affiliation{International Center for Quantum Materials, School of Physics, Peking University, Beijing 100871, China}

\author{Raquel Queiroz}
\affiliation{Department of Physics, Columbia University, New York, USA}

\author{B. Andrei Bernevig}
\affiliation{Department of Physics, Princeton University, Princeton, New Jersey 08544, USA}

\author{Ady Stern}
\affiliation{Department of Condensed Matter Physics, Weizmann Institute of Science, Rehovot 7610001, Israel}

\author{Zhi-Da Song}
\email{songzd@pku.edu.cn}
\affiliation{International Center for Quantum Materials, School of Physics, Peking University, Beijing 100871, China}
\affiliation{Hefei National Laboratory, Hefei 230088, China}
\affiliation{Collaborative Innovation Center of Quantum Matter, Beijing 100871, China}

\date{\today}

\begin{abstract}
Transitions between distinct obstructed atomic insulators (OAIs) protected by crystalline symmetries, where electrons form molecular orbitals centering away from the atom positions, must go through an intermediate metallic phase. In this work, we find that the intermediate metals will become a scale-invariant critical metal phase (CMP) under certain types of quenched disorder that respect the magnetic crystalline symmetries on average. We explicitly construct models respecting average $C_{2z}T$, $m$, and $C_{4z}T$ and show their scale-invariance under chemical potential disorder by the finite-size scaling method. Conventional theories, such as weak anti-localization and topological phase transition, cannot explain the underlying mechanism. A quantitative mapping between lattice and network models shows that the CMP can be understood through a semi-classical percolation problem. Ultimately, we systematically classify all the OAI transitions protected by (magnetic) groups $Pm$, $P2'$, $P4'$, and $P6'$ with and without spin-orbit coupling, most of which can support CMP.
\end{abstract}

\maketitle

\section{Introduction} 
The interplay between topology and the Anderson (de)localization has provided an understanding of the quantum Hall transition \cite{Pruisken1983,khmel1983quantization} and the classification  (not including crystalline symmetries) \cite{schnyder_classification_2008,kitaev2009periodic,Ryu2010tenfold} of topological insulators (TIs) \cite{Kane2005Z2,Bernevig2006BHZ,hasan2010review,qi2011review}. 
A remarkable result of this interplay is the delocalization in TIs protected by local symmetries \cite{schnyder_classification_2008,Ryu2010tenfold}: 
In the presence of disorder that does not induce a bulk phase transition, the TI surface states \cite{Lu2011localization,He2011localization,Chen2011localization,Wang2012localization} are guaranteed to be delocalized; In the bulk, a disorder-induced transition between trivial and topological insulators must go through a divergent localization length \cite{fulga_statistical_2014,morimoto_anderson_2015}. 
Similar behavior occurs also in topological phases that require crystalline symmetry to be preserved, despite the  breaking of that symmetry by disorders, as long as the symmetry is preserved on average. Examples include weak topological insulators \cite{Moore2012,ringel_strong_2012,fu_topology_2012,fulga_statistical_2014,wang2020disorder,Li2020Quadrupole,Su2019disorder,Araki2019disorder} and some topological crystalline insulators (TCIs) \cite{fu_topological_2011,mong_antiferromagnetic_2010,turner_entanglement_2010,Huse2011Inversion,Liu2014TCI,po_symmetry-based_2017,Slager2017Combinatorics,bradlyn_topological_2017,khalaf_symmetry_2017,song_quantitative_2018,song_topological_2019,cornfeld_classification_2019,shiozaki_classification_2022} (including higher-order states with hinge modes \cite{benalcazar_quantized_2017,schindler_higher-order_2018,PhysRevLett.126.206404,PhysRevB.103.085408, PhysRevB.106.L081410,song_d-2-dimensional_2017,langbehn_reflection-symmetric_2017}). 
For instance, inversion-symmetry-protected axion insulators \cite{Vanderbilt2009axion,li2010dynamical,Huse2011Inversion,Ashvin2012Axion} have hinge modes \cite{ZhangFan2013Axion}, and their phase transitions to trivial insulators must experience a delocalized diffusive metal phase if the disorder respects an average inversion symmetry \cite{song_delocalization_2021,Li2021AxionDisorder}. 
Here, as defined in Refs.~\cite{PhysRevB.89.155424,Ma_2023}, average symmetry is the symmetry of an ensemble comprising different disorder realizations on a local Hamiltonian. The average symmetry operation transforms an individual system into another with the same realization probability. Also, we require each system in the ensemble to be self-average.
Even though not mathematically proven, most TCIs with protected boundary states are believed to be stable against disorders respecting the crystalline symmetries on average.
This can be understood intuitively: Suppose the disorder potential slowly varies in real space. Then, during the transition, the disordered system can be divided into topological and trivial regions. Boundary states between the two types of regions must exist as promised by stable topology, giving rise to the delocalized phase transition.

In this work, we find that such delocalization behavior also generalizes to some topologically trivial states. 
There are two types of non-atomic states that are {\it not} conventional TIs or TCIs---the fragile topological insulators \cite{Po2018Fragile,Cano2018Topology,Slager2019Fragile,Alexandradinata2020Fragile} and the obstructed atomic insulators (OAIs) \cite{bradlyn_topological_2017,xu2021filling,Frank2021Noncompact,Nie2021Electrides}. 
The former has a Wannier obstruction that can be removed by adding trivial bands and was recently found to hold delocalized critical states \cite{Raquel2021APS}. 
However, the latter is completely trivial and can be wannierized to molecular orbitals with charge centers away from the atoms. 
Given OAIs' localized nature, it would be surprising if they can have delocalized states in the presence of disorder. 
In this work, we demonstrate that such delocalization does exist and is actually a common feature in transitions between magnetic OAIs.
Conventional scenarios for delocalization theories, such as weak anti-localization (requiring time-reversal symmetry) and topological phase transition (requiring stable boundary states) cannot explain the underlying mechanism behind the delocalized states that we find.

\section{Results}


\subsection{A \texorpdfstring{$C_{2z}T$}{C2T}-symmetric quantum network model}
\label{sec:main-network-model}

We first investigate several OAIs protected by the $C_{2z}T$ symmetry without time-reversal symmetry (TRS) \cite{po_faithful_2019,song_all_2019,ahn_failure_2019,Bradlyn2019Disconnected}.
Later, we will generalize the discussion to other magnetic point groups in \cref{sec:main-OAI-transition}.  
The OAIs belong to the Altland-Zirnbauer symmetry class A \cite{altland_nonstandard_1997}, where all states except the quantum Hall transition point were expected to be localized \cite{efetov_interaction_1980}. 
These OAIs are characterized by the $\mathbb{Z}_2$ Real Space Invariant (RSI) $\delta_w$, which is protected by $C_{2z}T$ \cite{song_twisted_2020,herzog2022hofstadter} and takes the value 
$\delta_w=1$ if the associated $C_{2z}T$ center $w$ is occupied by an odd number of Wannier functions and zero otherwise. 
When the $\delta_w$'s of a system have the same value at all the $C_{2z}T$ centers, they are equivalent to the second Stiefel-Whitney class $w_2$ \cite{ahn_failure_2019,fang_topological_2015}.

Two features of a $w_2=1$ insulator are worth emphasizing here: First, it can be regarded as a $C_{2z}T$-protected fragile insulator plus two additional trivializing bands. 
And second, to tune a $w_2=1$ insulator into a $w_2=0$ insulator, one has to first close the gap by creating pairs of Dirac points, which are locally stabilized by the $C_{2z}T$ symmetry,  then braid [\onlinecite{Wu2019Science},\onlinecite{ahn_failure_2019},\onlinecite{bouhon2020non}] these Dirac points with other Dirac points inside the valence bands, and only then reopen a gap by annihilating the Dirac points, as illustrated in Fig.~\ref{fig:main-Lattice-full}\textcolor{red}{$\mathrm{(b)}$}. As a consequence, when the transition is driven by the variation of one parameter, there is a gapless transition region rather than a transition point.
Using a finite-size scaling procedure, we find that this gapless region becomes a critical metal phase (CMP) when disorder is added, provided that the disorder respects $C_{2z}T$ on average.
Electronic states in CMP are delocalized and contribute to a scale-invariant conductance in the thermodynamic limit. 
We also find CMPs with other average symmetries and RSIs, suggesting CMP is a common feature of magnetic OAI transitions.

CMP has been numerically observed in systems with random fluxes \cite{xie_kosterlitz-thouless-type_1998,cerovski_critical_2001,xiong_metallic_2001} or random spin-orbit coupling combined with a magnetic field \cite{wang_band_2015}.  
Inspired by these works, we first argue the existence of $C_{2z}T$-stabilized CMP through a semi-classical percolation theory. We then relate the percolation theory to a quantum network model \cite{chalker_percolation_1988} and further map it to lattice models for the $C_{2z}T$-protected OAIs.

\begin{figure}[t]
	\centering
    \includegraphics[width=1\linewidth]{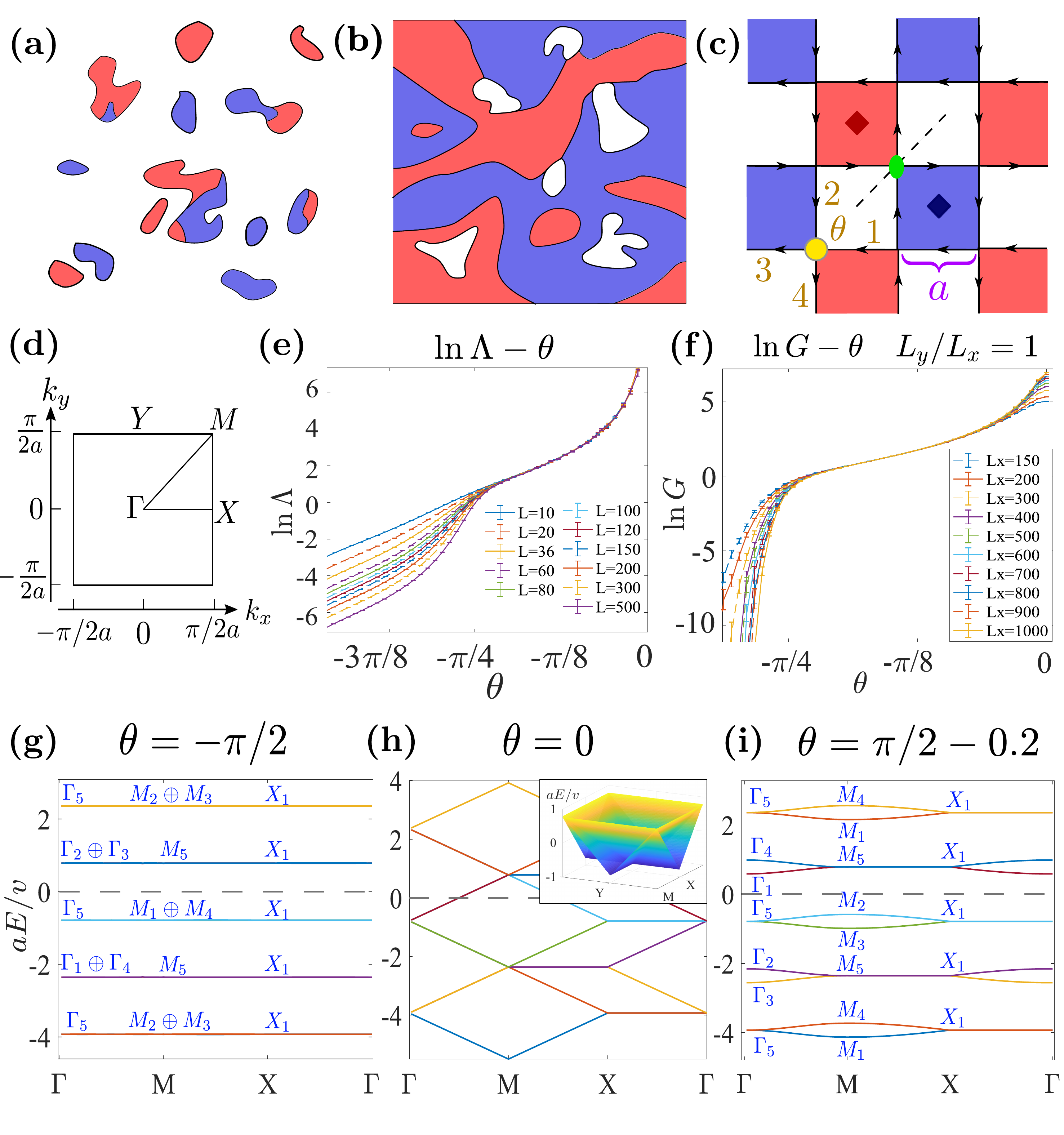}
    \caption[]{
    The network model $H_{N}$.   
    (a-b) Percolation systems with $p_{0}>1/2$ and $p_0<1/2$. 
    (c) The network model on the Manhattan lattice, where the dashed line, dark red and blue rhombuses, and green oval indicate the $M_{xy}$ mirror plane, $C_{4z}$ centers and $C_{2z}T$ center, respectively. 
    At every intersection (e.g. yellow circle), there is a scattering potential $\lambda$, and the scattering angle $\theta$ is determined by $\lambda=\theta v$ with $v$ being the velocity of chiral modes. 
    (d) Brillouin zone and high symmetry points. 
    (e) Normalized quasi-1D localization length $\Lambda$'s as functions of $\theta$ at different transversal system sizes $L$. The used longitudinal system size is $M=10^7$ and the precision ($\sigma_\Lambda/\Lambda$) has reached $1\%$. (For network model, `size' refers to the number of squares.) 
    $\Lambda$ only depends on $|\theta|$, hence only data with $\theta<0$ is shown. 
    (f) The mean conductances (over $10^3$ square-shaped samples) as the function of $\theta$ and the precision ($\sigma_G/G$) has reached $0.5\%$ in the delocalized phase. 
    (g-i) Band structures of the network model at $\theta=-\pi/2$, 0, $\pi/2-0.2$. 
    Blue capital letters indicate the associated irreps.
    The inset in $\mathrm{(h)}$ is the 3D plot of the dispersion of the middle two bands around the zero energy indicated by the dashed lines.
    }  
	\label{fig:main-Network}
\end{figure}

\begin{figure}[h]
	\centering
    \includegraphics[width=1\linewidth]{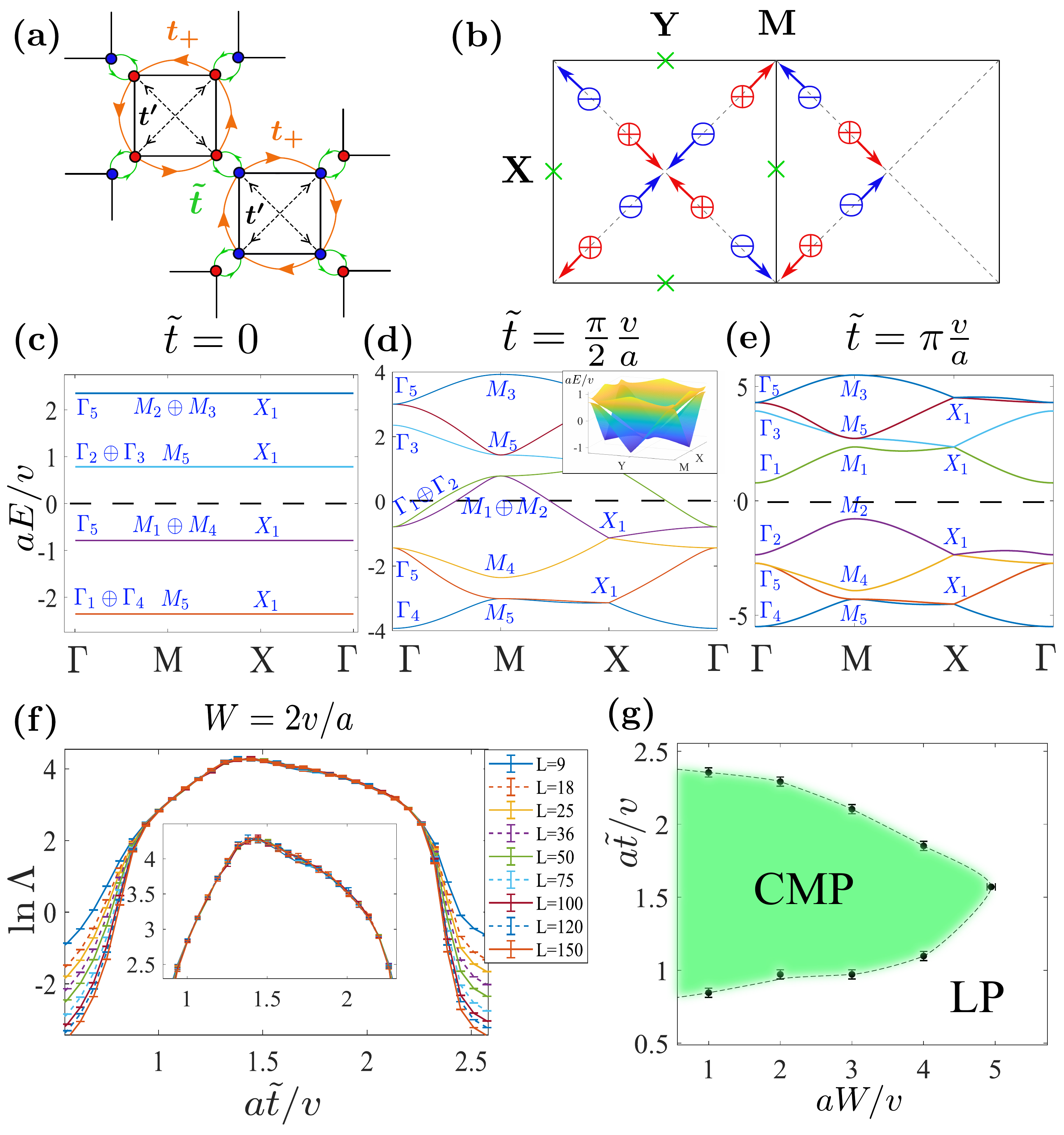}
    \caption[]{The lattice model $H_{8B}$. 
    (a) The unit cell and hoppings of $H_{8B}$.
    (b) The evolution of Dirac points between the fourth and fifth bands. 
    Red (blue) circles indicate Dirac points with positive (negative) chirality (defined in Sec.~$\color{red}\text{IV.D}$ of Ref.~\cite{sup}), and the green crosses correspond to Dirac points between the third and fourth bands. The blue and red arrows indicate the evolution directions of Dirac points when $\tilde{t}$ increases.
    (c-e) Band structures with $a\tilde{t}/v=0,\,\pi/2,\,\pi$ (correspond to $\theta=-\pi/2,\,0,\,\pi/2$, respectively).
    The blue capital letters indicate the irreps at high symmetry points. The dashed lines indicate zero energy.
    (f) The normalized localization length as functions of $\tilde{t}$ at various transversal sizes $L$ and fixed disorder strength $W=2v/a$ and Fermi level $E_F=0$.  
    The used longitudinal size is $M=10^7$ and the precision ($\sigma_{\Lambda}/\Lambda$) has reached $3\%$. (For lattice models, `size' refers to the number of unit cells.)
    (g) The phase diagram in $\tilde{t}-W$ plane for $E_F=0$, where the green region represents the CMP enclosed by the LP. 
    }  
	\label{fig:main-Lattice-full}
\end{figure}

To present a scenario that naturally leads to a CMP, we consider a system that is tessellated by three types of randomly sized and shaped insulating regions, whose Chern numbers are $0$, $1$ and $-1$, respectively, as shown in Fig.~\ref{fig:main-Network}\textcolor{red}{(a-b)}. 
Physically, the fluctuation of Chern numbers could arise from random fluxes. 
The area fraction of Chern number $C$ is $p_C$.
By definition $\sum_C p_C=1$. 
Since the operation $C_{2z}T$ reverses the sign of Chern numbers, an average $C_{2z}T$ symmetry means $p_1=p_{-1}$. 
Then there are two distinct phases. 
If $p_0>\frac12$, according to the classical percolation theory \cite{isichenko_percolation_1992}, the $C=0$ regions form an extensive cluster while the $C=\pm1$ regions form isolated islands and can be continuously shrunk to zero.
Thus, $p_0>\frac12$ should correspond to a localized phase (LP). 
If $p_0<\frac12$, it is instead the $C=0$ regions that can be shrunk to zero, and the system is equivalently tessellated by the $C=\pm1$ regions with the same area fraction $\frac12$. 
As in the quantum Hall transition \cite{chalker_percolation_1988}, the chiral edge states between the $C=\pm 1$ regions connect to an extensive cluster with a fractal dimension and contribute to a scale-invariant conductance in the thermodynamic limit. 
Thus, $p_0<\frac12$ should correspond to the CMP. 

We can simulate the above percolation problem with a quantum network model on the Manhattan lattice \cite{beamond2003quantum} (Fig.~\ref{fig:main-Network}\textcolor{red}{(c)}). 
The red and blue squares are Chern blocks with $C=1,-1$, respectively. 
The chiral edge modes between them and the trivial (white) regions form horizontal and vertical wires, and at each intersection, an electron can go straight or turn either left or right, depending on the type of intersection.
The scattering equation at one intersection reads 
\begin{align} \label{eq:main-scattering}
\begin{pmatrix}
\psi_3 \\ \psi_4
\end{pmatrix}
= \begin{pmatrix}
    \cos\theta &  -i\sin\theta \\
    -i\sin\theta & \cos\theta 
\end{pmatrix}
\begin{pmatrix}
    \psi_1 \\ \psi_2 
\end{pmatrix}\ ,
\end{align}
where $\psi_{3,4}$ and $\psi_{1,2}$ are the outgoing and incoming modes, respectively, and $\theta$ is the single parameter that determines the probability amplitudes of going straight ($\cos\theta$) and turning left or right ($-i\sin\theta$). 
The model in the clean limit has symmetries of the magnetic space group \href{https://www.cryst.ehu.es/cgi-bin/cryst/programs/magget_gen.pl}{$P_{C}4bm$} (\#100.177 in BNS setting) generated by $C_{2z}T$, $C_{4z}$, and $m_{xy}$ symmetries \cite{gallego_magnetic_2012}.
The symmetry elements of the generators are shown in Fig.~\ref{fig:main-Network}\textcolor{red}{(c)}. 
(Notice that $C_{2z}T$ centers do not coincide with $C_{4z}$ centers, and $C_{4z}^2 \cdot C_{2z}T$ is a magnetic translation, \ie a translation followed by time reversal. See Ref.~\cite{sup} for more details.)
One can see that the $C_{2z}T$ operation interchanges the $C=\pm1$ regions. 
Note that the Manhattan lattice is not the only way to simulate the percolation problem. A Kagome-like network also works, with a  localization behavior that is similar to the Manhattan lattice (See Sec.\textcolor{red}{II.A} in Ref.~\cite{sup}). 


Random sizes of the Chern blocks are simulated by the random propagation phases, or, equivalently, random vector potentials, along the bonds between intersections. 
When $\theta = \pm \frac{\pi}2$, the chiral modes form local current loops surrounding $C=\pm1$ regions, and the $C=0$ regions are effectively connected. 
When $\theta \to 0$, the chiral modes are almost decoupled wires, and $C=0$ regions are effectively separated. 
Thus, $\theta = \pm\frac{\pi}2$ and $\theta\to 0$ should correspond to the localized limits ($p_0=1$) and the CMP limit ($p_0=0$), respectively. 
According to the percolation argument, there will be a critical value $\theta_c$ below (above) which the system enters the critical (localized) phase. 

We use the transfer matrix techniques \cite{pichard_finite_1981} of quasi-1D systems, where longitudinal size $M$ is much larger than transversal size $L$, to calculate the quasi-1D localization length $\rho$ for finite $L$'s. 
More on this is summarized in \cref{subsec:method-Transfer-matrix}, and one can read Sec.\textcolor{red}{V} in Ref.~\cite{sup} for the full technical details. 
The normalized quasi-1D localization length $\Lambda = \rho/L$ is an indicator of the (de)localization: Divergent, finite, and vanishing $\Lambda$'s in the limit $L\to \infty$ indicate metallic, critical, and localized states, respectively. 
As shown in Fig.~\ref{fig:main-Network}$\textcolor{red}{\mathrm{(e)}}$, $\Lambda$ decreases with $L$ for $|\theta| > \theta_c \approx \frac{\pi}4$ and is almost independent of $L$ for $|\theta|<\theta_c$. 
Hence, $|\theta| \in (\theta_c,\pi/2]$ and $|\theta| \in (0,\theta_c)$ correspond to LP and CMP, respectively, which confirms the percolation argument.
Note that error bars in plots of this work represent the standard deviations (SD) of corresponding data points.
The conductance is also calculated and shown in \cref{fig:main-Network}\textcolor{red}{(f)}. 
The $\beta$-function $\beta=d\ln G/d\ln L$ derived from the conductance data vanishes in the thermodynamics limit above some critical conductance $G_{c}=2\sim3e^2/h$ (see \cref{subsec:method-beta-network} for details). The behavior of $\beta$-function further establishes the criticality of CMP and may suggest that the CMP-LP transition is similar to the Berezinskii–Kosterlitz–Thouless transition \cite{J-M-Kosterlitz_1973,Zhang_1994}.  

We can define the network model through its Hamiltonian $H_{N}$ rather than through the scattering matrix \cref{eq:main-scattering}. The Hamiltonian has only three parameters, velocity $v$ of the chiral modes, the $\delta$-potential $\lambda$ at each intersection, and the lattice constant $2a$. %
{\small
\begin{equation}
\label{eq:main-HN}
\begin{aligned}
    H_N&\!=\!\sum_{d,l} i (-1)^{l} v\!\int\!\mathrm{d}\xi \psi_{d,l}^{\dagger}(\xi)\partial_{\xi} \psi_{d,l}(\xi)
    \\&+\sum_{l l'} \lambda[\psi^{\dagger}_{\mrm{h},l}(l'a)\psi_{\mrm{v},l'}(la) + h.c.]
\end{aligned}
\end{equation}}%
The subscript $d=\mrm{v},\mrm{h}$ represents the vertical or horizontal orientations of the wires. 
$l=0,\pm1\cdots$ distinguishes different parallel wires.
$\xi$ is the coordinate inside one wire.
The operators $\psi_{\mrm{v},l}(\xi)$ and $\psi_{\mrm{h},l}(\xi)$ act at the real space positions $(x,y)\!=\!(la, \xi)$ and $(\xi,la)$, respectively. 
As established in Sec.~$\color{red}\mathrm{III.B}$ of Ref.~\cite{sup}, the scattering angle $\theta$ is determined by the Hamiltonian parameters as $\theta = \lambda/v$. 
The evolution of the band structure as $\theta$ changes from the localized limit $\theta=-\pi/2 $ to the other localized limit $\theta= \pi/2 $ is illustrated in Figs.~\ref{fig:main-Network}\textcolor{red}{(g-i)}.
At $\theta=0$, the network model decouples to vertical and horizontal chiral wires, and the corresponding dispersion becomes quasi-1D (\cref{fig:main-Network}\textcolor{red}{(h)}). 
The symbols appearing in the figure, $\Gamma_n$, $M_{n}$ ($n=1\cdots5$) and $X_1$, represent irreducible representations (irreps) of $P_{C}4bm$ and are defined in \cref{tab:main-Irreps}. 
It is worth mentioning that because the chiral modes have unbounded energies, this Hamiltonian has an infinite number of bands that are periodic in energy, \ie $E_{n}(\kk) = E_{n+8}(\kk)+2\pi v/a$. For example, the lowest branch (two connected bands) in \cref{fig:main-Network}\textcolor{red}{(g-i)} is identical to the highest branch. 
Tracing the evolution (detailed in Fig.~$\color{red}\mathrm{18}$ in Ref.~\cite{sup}), we find the phase transition process between these two LPs can be depicted by the irrep exchange at the zero energy (dashed lines in Fig.~\ref{fig:main-Network}\textcolor{red}{(g-i)})
\begin{equation} \label{eq:main-phase-transition}
    -\Gamma_1 - M_1  + \Gamma_2  + M_2  \ ,
\end{equation}
where a minus (plus) sign means an energy level with the associated irrep crosses the zero energy from below (above) to above (below) during the phase transition.
One may notice that the occupied $\Gamma_4$ state in \cref{fig:main-Network}\textcolor{red}{(g)} also changes into $\Gamma_3$ in \cref{fig:main-Network}\textcolor{red}{(i)}.
However, this change is realized by level exchanges between the lowest branch in the figures and the lower branches beyond the scope of the figures. 
Since $\Gamma_{3,4}$ do not cross the zero energy, they are not counted in the phase transition. 
We now relate the phase transition (\ref{eq:main-phase-transition}) to a change in the index $\delta_w$.

\subsection{Regularizing the network model to lattice models} 
In order to see the band topology, we need to regularize the network model to a lattice model. 
Our strategy is to use the local current loop states  (flat bands in Fig.~\ref{fig:main-Network}$\textcolor{red}{\mathrm{(g)}}$) in the localized limit $\theta=-\frac{\pi}2$ as a basis set and then truncate the basis according to their energies.
The minimal model that can reproduce the phase transition in \cref{eq:main-phase-transition} is constructed from the upper eight consecutive flat bands in \cref{fig:main-Network}\textcolor{red}{(g)} to  obtain the Hamiltonian $H_{8B}$ shown in Fig.~\ref{fig:main-Lattice-full}\textcolor{red}{$\mathrm{(a)}$}.
(See Sec.~$\color{red}\text{IV.C}$ in Ref.~\cite{sup} for more details.)
It has eight orbitals that are respectively located at the eight corners of the two squares in one unit cell, which correspond to the two Chern blocks in \cref{fig:main-Network}\textcolor{red}{(c)}. 
The explicit form of $H_{8B}$ can be expressed as 
{\small
\begin{equation}
\label{eq:main-H8B}
\begin{aligned}
H_{8B}&= \tilde{t} \sum_{\langle p,q \rangle}  c^{\dagger}_{p}c_q
    + t \sum_{\langle\!\langle p,q\rangle\!\rangle} e^{i\phi_{pq}} c^{\dagger}_{p}c_q
    + t^{\prime} \sum_{\langle\!\langle\!\langle p,q\rangle\!\rangle\!\rangle}  c^{\dagger}_{p} c_{q}\ ,
\end{aligned}
\end{equation}}%
where $p,q$ are the site indices, $\langle\cdot\rangle$, $\langle\!\langle \cdot\rangle\!\rangle$, and $\langle\!\langle\!\langle \cdot \rangle\!\rangle\!\rangle$ represent the green (nearest neighbor), orange (square edges), and dashed black (square diagonals) bonds in \cref{fig:main-Lattice-full}\textcolor{red}{(a)}, respectively.  
The hopping parameters are given by $\tilde{t} = (\theta + \frac{\pi}2) \frac{v}{a}$, $t = \frac{\sqrt2 \pi v}{4a}$, $t' = - \frac{\pi v}{4a}$. 
The phase factor $\phi_{pq}$ equals $\frac34 \pi$ ($-\frac34 \pi$) if the associated hopping is parallel (anti-parallel) to the orange arrows, which are clockwise and anticlockwise for squares formed by the blue and red sites, respectively.
For simplicity, we also denote the complex hopping $t e^{\pm i\frac34 \pi}$ as $t_{\pm}$ in the following.

We take the Fermi level to be at $E_F=0$ and focus on that energy. The Hamiltonian $H_{8B}$ reproduces the flat bands in the localized limit $\theta=-\frac{\pi}2$ when $\tilde{t}=0$, where blue and red squares are decoupled from each other (\cref{fig:main-Lattice-full}\textcolor{red}{(c)}).
The four flat valence bands are molecular orbitals at the square centers ($C_{4z}$ centers). 
Since they do not occupy the $C_{2z}T$ centers (squares corners), all the corresponding RSIs $\delta_w=0$ and the Stiefel-Whitney class $w_2=0$. 
As $\tilde{t}$ increases, $H_{8B}$ closes its gap, and when $\tilde{t}=\frac{\pi v}{2a}$, it reproduces the quasi-1D bands of the network in the decoupled wire limit $\theta=0$ except for small deviations (non-linear dispersions) as if the wires are weakly coupled. 
As a consequence of the $C_4$ symmetry, the Dirac points between the gapless bands occur at the same energy.
(See \cref{fig:main-Lattice-full}\textcolor{red}{(d)} and \cref{fig:main-Network}\textcolor{red}{(h)}.)
As $\tilde{t}$ continues to increase, $H_{8B}$ reopens a gap (\cref{fig:main-Lattice-full}\textcolor{red}{(e)}), and this gap continues to the limit $\tilde{t}\to \infty$. 
By tracing the evolution of the energy levels, one can verify that the irrep exchange at $E=0$ is indeed the same as \cref{eq:main-phase-transition}. 
In the limit $\tilde{t}\to \infty$, electrons form bonding states at the four $C_{2z}T$ centers (per cell), \ie green bonds in \cref{fig:main-Lattice-full}\textcolor{red}{(a)}.
Thus, the final state has $\delta_w=1$ at all $C_{2z}T$ centers, \ie $w_2=1$. 
As detailed in \cref{subsec:method-TQC}, the phase transition can be further confirmed through the machinery of Topological Quantum Chemistry \cite{bradlyn_topological_2017,elcoro_magnetic_2021}. 

As shown in Fig.~\ref{fig:main-Lattice-full}\textcolor{red}{$\mathrm{(d)}$}, (tilted) Dirac points between the fourth and fifth bands are created in the phase transition process.
The evolution of Dirac points is sketched in Fig.~\ref{fig:main-Lattice-full}\textcolor{red}{$\mathrm{(b)}$}, where the trajectory forms a closed path enclosing the underlying Dirac point at $X$ between the third and fourth bands [\onlinecite{Wu2019Science},\onlinecite{ahn_failure_2019},\onlinecite{bouhon2020non}]. 
A more detailed discussion is given in Sec.~$\color{red}\text{IV.D}$ of Ref.~\cite{sup}.

It is worth mentioning that the vector potential disorder we used in the network model is now mapped to a chemical potential disorder in $H_{8B}$ (plus two times weaker hopping disorders that will be omitted). 
We leave the mathematical analysis in Sec.~$\color{red}\text{IV.F}$ in Ref.~\cite{sup} and only present a heuristic argument here. 
The basis of $H_{8B}$ (\cref{fig:main-Lattice-full}\textcolor{red}{(a)}) can be thought as wave-packets of the chiral modes that simultaneously have position centers and momentum centers. 
The position centers, by construction, are located at the square corners. 
We denote their 1D momentum centers as $q_c$. 
Then a vector potential $A$ will shift a momentum center $q_c$ to $q_c+A$ and result in an energy shift $vA$.
Therefore, the resulting disorder potential in $H_{8B}$ should have a large on-site component. 
We also ignore the correlations among on-site random potentials for simplicity and efficiency.  In numerical calculations, we only use uncorrelated on-site disorder and choose the disorder potential equally distributed in $[-W/2,W/2]$. Test calculations with full projected disorder potentials and correlations show no qualitative difference (see Fig.~\textcolor{red}{30} in  Ref.~\cite{sup}).

We calculated the normalized localization length $\Lambda$ as a function of $\tilde{t}$ with $E_F=0$ and $W=2v/a$ (Fig.~\ref{fig:main-Lattice-full}\textcolor{red}{$\mathrm{(f)}$}).
The system is localized when $|\tilde{t}-\pi v/2a|>\Delta_c\approx0.6v/a$, corresponding to the two OAI limits, and becomes critical when $|\tilde{t}-\pi v/2a|<\Delta_c$. 
The criticality has been examined for large transversal sizes up to $L=500$ unit cells (2000 atoms).
We also calculate $\Lambda$ at other $\tilde{t}$'s and $W$'s.
From these data, we can determine a phase diagram shown in Fig.~\ref{fig:main-Lattice-full}\textcolor{red}{$\mathrm{(g)}$}, where the dashed line separates the CMP inside and the LP outside it.
(See Fig.~$\color{red}\text{8}$ in Ref.~\cite{sup} for details of large-scale examination and phase boundary determination.)
Since $H_{8B}$ does not have chiral or particle-hole symmetries, the choice $E_F= 0$ is not special in terms of symmetries. 
We have confirmed that CMP also exists when $E_F\neq 0$ as long as the OAI limits are intact (Fig.~$\color{red}\text{9}$ in Ref.~\cite{sup}). 

\begin{figure}[h]
	\centering
    \includegraphics[width=1\linewidth]{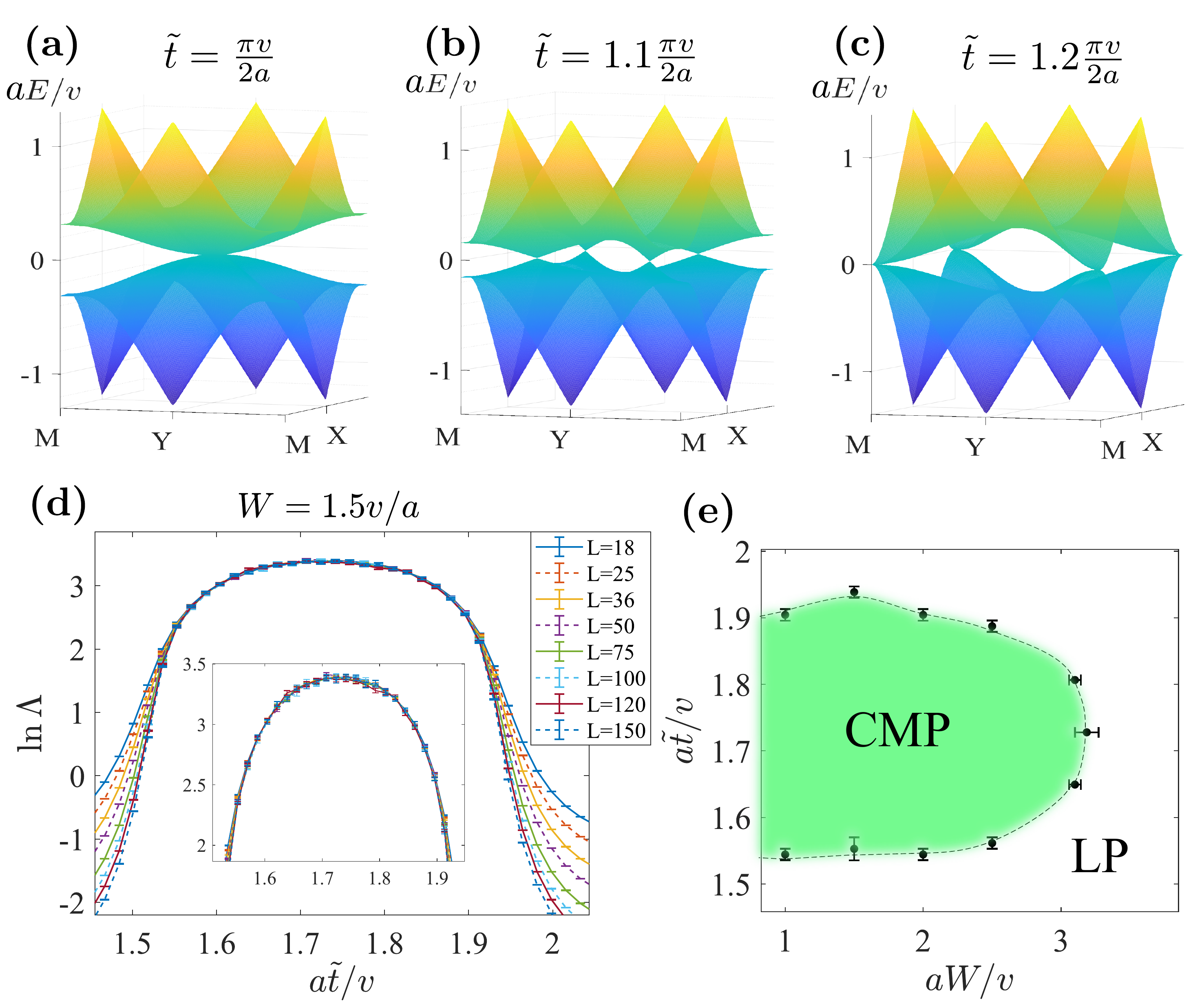}
    \caption[]{The lattice model $H_{8B}'$. 
    (a-c) Evolution of the middle two bands with $\tilde{t}$. 
    (d) Normalized localization length $\Lambda$ as functions of $\tilde{t}$ with $W=1.5v/a,\,E_F=0$ and various transversal sizes $L$. 
    The used longitudinal size is $M=10^7$ and the precision ($\sigma_{\Lambda}/\Lambda$) has reached $2\%$. 
     $\mathrm{(e)}$ The phase diagram in $\tilde{t}-W$ plane with $E_F=0$, where the green region represents the critical metal phase enclosed by the localized phase. 
     Note that $H_{8B}'$ in the clean limit has an additional chiral symmetry which fixes the Dirac points at zero energy. 
     However, the chiral symmetry is broken by the chemical potential disorder and is not responsible for CMP (see Sec.~$\color{red}\text{IV.E}$ and Fig.~$\color{red}\text{10}$ in Ref.~\cite{sup} for details).
    }
    \label{fig:main-Lattice-simplified}
\end{figure}

For the CMP in \cref{fig:main-Lattice-full}\textcolor{red}{(g)}, if we turn off the disorder, the resulting clean system has a finite density of states (DOS) around the zero energy, which may lead to a large localization length that may exceed the numerically accessible transversal size.
To rule out possible finite-size effects, we consider a lattice model $H_{8B}'$ that has the same crystalline symmetries and topology as $H_{8B}$ but vanishing DOS at the zero energy. 
Such a $H_{8B}'$ can be obtained from $H_{8B}$ by (i) removing diagonal hopping $t'$ and (ii) changing the edge hopping to $t_\pm = (\pm A i-1)t$, where $A$, chosen as 1.2 hereafter, is an extra parameter that controls the range of the critical phase.
(See Sec.~$\color{red}\text{IV.E}$ in Ref.~\cite{sup} for more details.) 
The hopping $\tilde{t}$ (green bond in \cref{fig:main-Lattice-full}\textcolor{red}{(a)}) at $C_{2z}T$ center remains unchanged. 
The $\tilde{t}=0$ and $\tilde{t}\to \infty$ limits still represent the two OAI limits with charge centers at the $C_{4z}$ and $C_{2z}T$ centers, respectively.
Hence, changing $\tilde{t}=0$ to $\tilde{t}=\infty$ will change $w_2=0$ to $w_2=1$ for the lower four bands as it did in $H_{8B}$.  
Figs.~\ref{fig:main-Lattice-simplified}\textcolor{red}{(a-c)} depict the evolution of the fourth and fifth bands, where the gap only closes at the four (untilted) Dirac points, resulting in a zero DOS at the zero energy. 

Parallel to the results of $H_{8B}$ shown in \cref{fig:main-Lattice-full}\textcolor{red}{(f)} and \textcolor{red}{(g)}, we show $\Lambda$ with $E_F=0$ and fixed $W$ and the phase diagram for $H_{8B}'$ in \cref{fig:main-Lattice-simplified}\textcolor{red}{(d)} and \textcolor{red}{(e)}, respectively, where no qualitative difference for the CMP is found. 
Thus, the potential finite-size effect of $H_{8B}$ due to large DOS is ruled out. 
Additionally, we did calculations with $E_F\neq0$ and obtained similar results as $E_F=0$ (Fig.~$\color{red}\text{10}$ in Ref.~\cite{sup}).

\begin{figure}[h]
	\centering
    \includegraphics[width=0.9\linewidth]{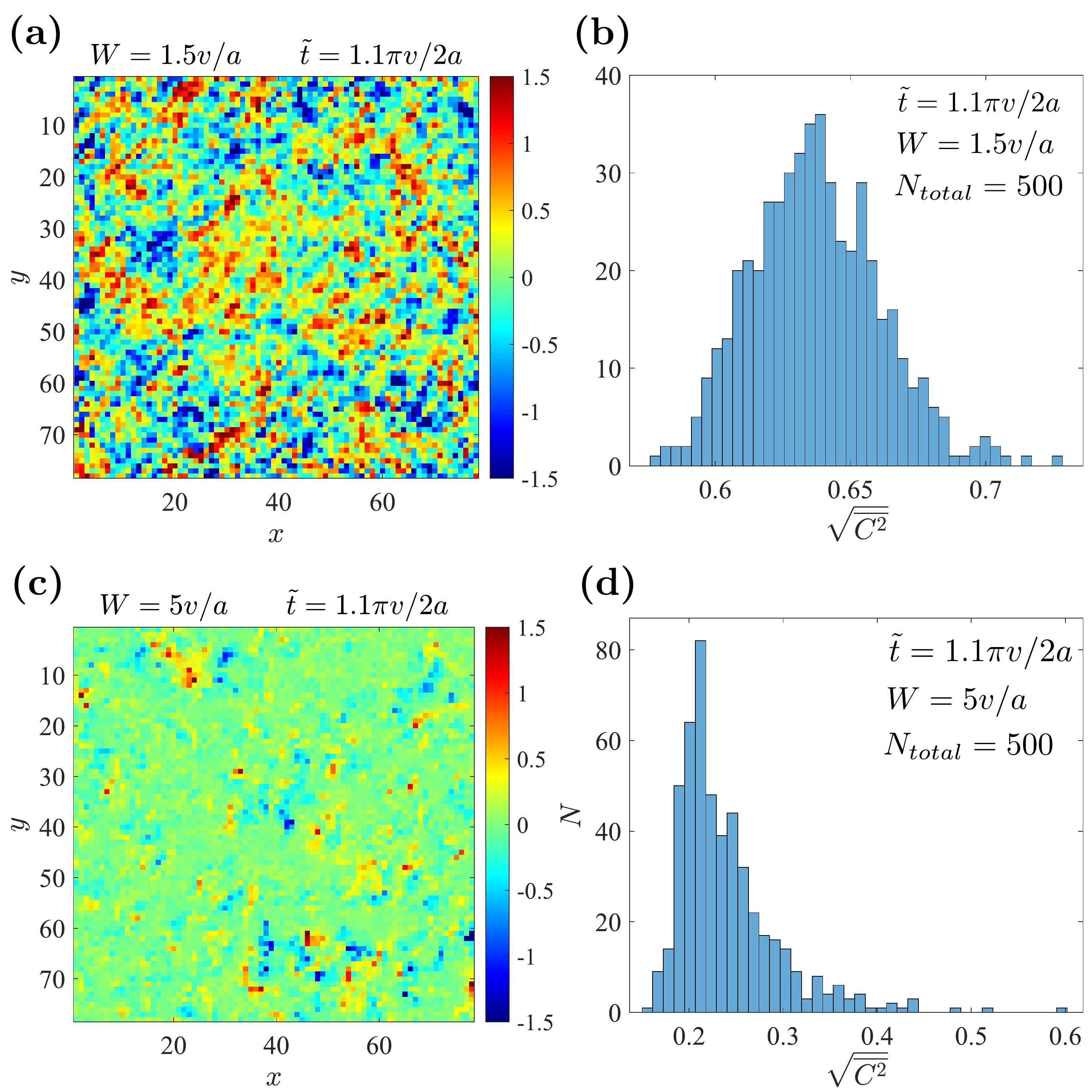}
    \caption[]{$\mathrm{(a)}$ Typical topography ($80\times80$ cells) of LCMs of $H_{8B}'$ in the CMP ($\tilde{t}=1.1 \pi v/2a$, $W=1.5v/a$). 
    $\mathrm{(b)}$ Distribution of $\sqrt{\inn{C^2}}$ in $500$ disorder configurations with the same parameters as (a). 
    (c) and (d) are the same plots as (a) and (b) but in the localized phase with $\tilde{t}=1.1 \pi v/2a$, $W=5v/a$. 
    }  
	\label{fig:main-LCM}
\end{figure}

\subsection{Local Chern markers}
In order to verify the percolation argument directly, we refer to a widely used local topological marker called local Chern marker (LCM) \cite{Prodan2010Entanglement,Bianco2011mapping,loring2011disordered}, which reformulates the Chern number locally  in real space without summing over the whole sample. The LCM of one unit cell is
\begin{equation}
    \label{eq:main-LCM}
        C(\vec{R})=\frac{4\pi}{A_c} \sum_{\alpha}\Im \left\langle \vec{R}\alpha\right\vert \hat{P} \hat{x} \hat{P} \hat{y} \hat{P} \left\vert \vec{R}\alpha\right\rangle 
\end{equation}
where $\vec{R}$ is the position of the unit cell, $\alpha$ indicates the orbitals inside one unit cell, $A_c$ is the area of one unit cell, and $\hat{P}$ is the projection operator of the occupied states.

Inside a macroscopic region where a gap is well preserved, the LCM converges to the quantized Chern number, whereas around gapless regions such as the boundaries, LCM may strongly fluctuate \cite{Bianco2011mapping}. 
In our models, due to the percolation argument, regions with a single Chern number $C$ ($=\pm1$) never become extended  because the fraction $p_C$, as required by the $C_{2z}T$, is always smaller than $\frac12$ for $p_0>0$.
Therefore, no macroscopic Chern block is expected for a general point in the CMP. 
Nevertheless, the microscopically inhomogeneous LCM can still reflect the local topological properties and can be applied to, for example, disordered systems near topological phase transitions \cite{Ul2020Kibbl-Zurek}. 
Fig.~\ref{fig:main-LCM}\textcolor{red}{(a)} and \textcolor{red}{(c)} show the topography of LCM of $H_{8B}'$ in the CMP and LP with given disorder configurations, respectively. 
In the CMP, the sample is dominated by randomly distributed positive and negative Chern cells, consistent with the percolation argument that $C=\pm1$ regions together are extended through the whole system. 
In the LP, the LCM almost vanishes everywhere. 
We also calculate the distribution of the second moment of LCM $\sqrt{ \inn{C^2}}$ over 500 disorder configurations in both the CMP and LP (Fig.~\ref{fig:main-LCM}\textcolor{red}{(b)} and \textcolor{red}{(d)}), where $\inn{\cdot}$ means averaging over all the cells. 
In the CMP (LP), $\sqrt{\inn{C^2}} > \frac12$ ($<\frac12$) for most configurations, \ie the regions with non-zero (zero) Chern numbers dominate.
These phenomena confirm our semi-classical percolation picture.

\subsection{OAI transitions in generic magnetic point groups} \label{sec:main-OAI-transition}

For a qualitative understanding of the CMP, we note that a local breaking of $C_{2z}T$ symmetry allows for a local gap of the Dirac nodes. This gap makes the region of the Dirac point carry a spread Berry curvature that integrates to $\pm \pi$, with the sign being determined by the chirality of the Dirac point, the sign of the gaping mass.
In the cases we considered here, all Dirac points occur at the same energy. Denoting the number of Dirac points by $2N$, there are $2^{2N}$ assignments of the signs of the gaping mass, out of which $(2N)!/(N!)^2$ lead to a total Chern number of zero. If the signs of the masses are uncorrelated, as would be expected for random disorder, then $p_0=\frac{(2N)!}{2^{2N}(N!)^2}\le \frac{1}{2}$, with the equality occurring only for $N=1$. 
In our case $N=2$ and $p_0 = \frac38$. 
Thus, for weak disorder, we expect a percolating network of edge states. 
If the masses are positively correlated, $p_0$ could be further suppressed. 
To be concrete, $p_0<\frac12$ even in the $N=1$ model if there is a higher possibility for the two masses to have the same sign. 
Under strong disorder, all local Chern insulators become trivial, and the system becomes localized. 

\begin{figure}[h]
	\centering
    \includegraphics[width=0.95\linewidth]{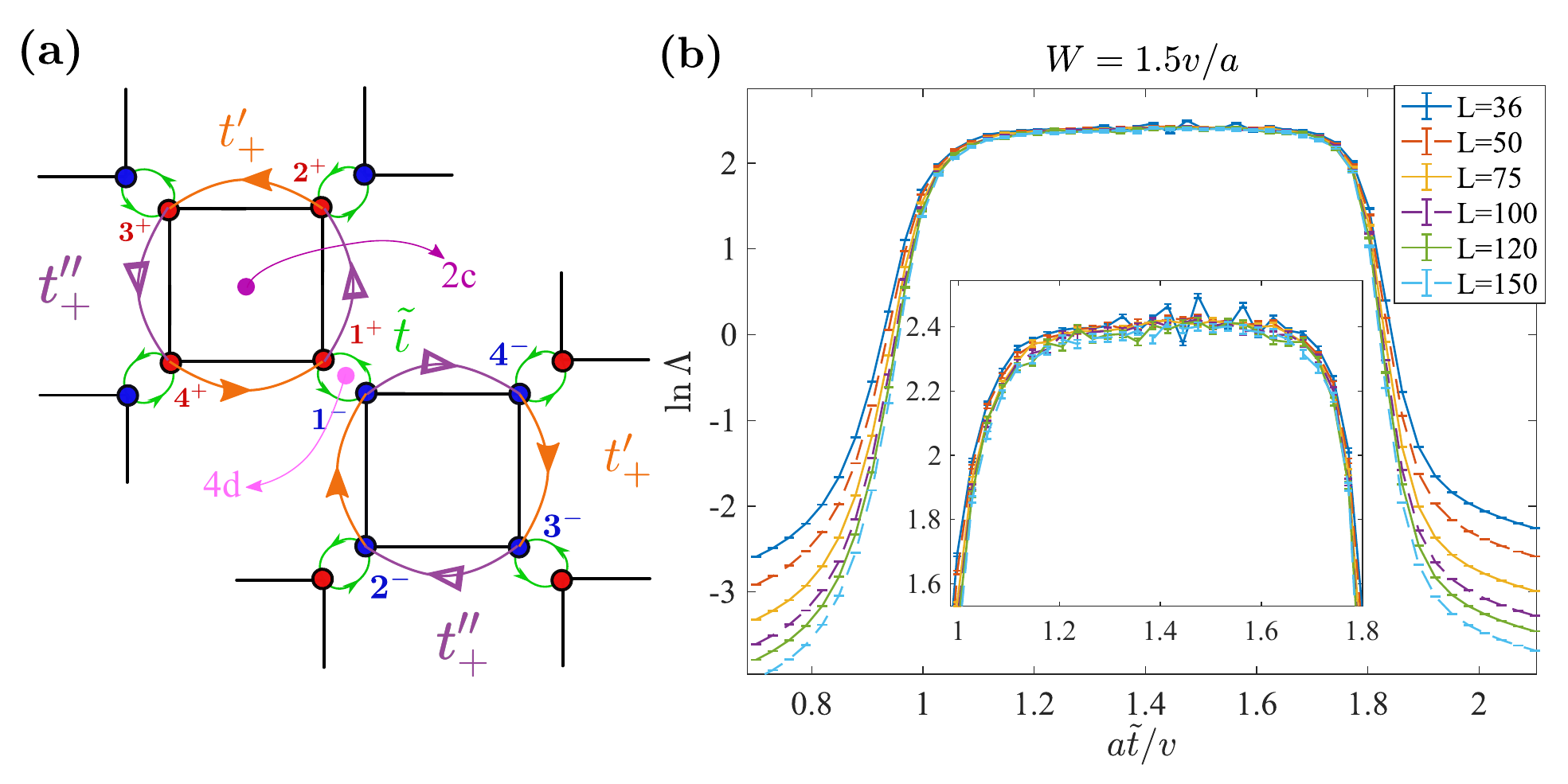}
    \caption[]{The lattice model $H_{8B}''$. (a) The unit cell and hoppings of $H_{8B}''$. (b) Normalized localization length $\Lambda$ as functions of $\tilde{t}$ with $A=1.2$, $q=0.625\exp(-0.15\pi i)$, $W=1.5v/a$, $E_F=0$ and various transversal sizes L. The used longitudinal size is $M = 10^7$ and the precision ($\sigma_{\Lambda}/\Lambda$) has reached $1.5\%$. }
    \label{fig:main-Lattice-simplified-mod}
\end{figure}

The above argument can be immediately generalized to generic OAI transitions beyond $C_{2z}T$. As a direct verification, we break $C_{2z}T$ in $H_{8B}'$ while keeping $m_{xy}$ and $C_{4z}T$ for the percolation mechanism and making two OAI limits at $a\tilde{t}/v=0$ and $\infty$ inequivalent. 
 The unit cell of the modified model $H_{8B}''$ is illustrated by \cref{fig:main-Lattice-simplified-mod}$\color{red}\mathrm{(a)}$, where the orange arrows indicate hopping $t_{+}'=(1+Ai)\frac{\pi v}{4a}$ and the purple arrows indicate $t_{+}''=qt_{+}'\,(q\in\mathbb{C})$. The magnetic space group of $H_{8B}''$ reduces to $P4'm'm$ ($\#99.165$ in BNS setting), and the Wannier centers of OAI limits at $\tilde{t}=0$ and $\infty$ are now located at the $C_{2z}$ centers $\mathrm{2c\,[2m'm']}$ and mirror planes $\mathrm{4d\,[m]}$, respectively. The band structure also goes through an evolution of four Dirac points when $\tilde{t}=0\rightarrow\infty$. \cref{fig:main-Lattice-simplified-mod}$\color{red}\mathrm{(b)}$ shows the localization length of $H_{8B}''$ with $A=1.2,\,q=0.625e^{-0.15\pi i}$. We can see that CMP indeed survives at $H_{8B}''$.

\begin{figure}[h]
	\centering
    \includegraphics[width=0.9\linewidth]{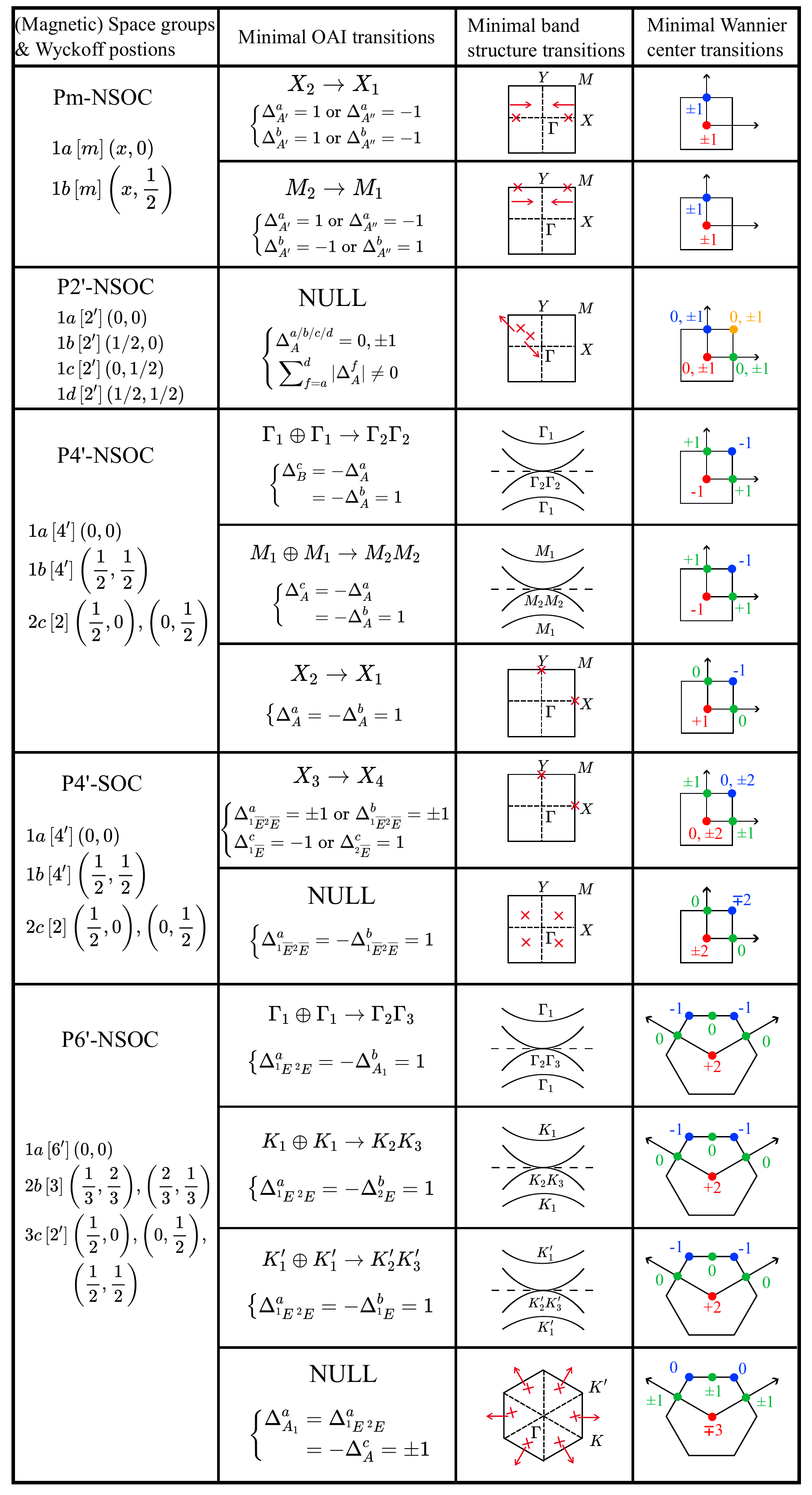}
    \caption[]{OAI transitions in (magnetic) space groups $Pm,\,P2',\,P4',\,P6'$. The first column contains symbols of the (magnetic) space groups, appearance of SOC, and the Wyckoff positions except for the general positions. The ``-NSOC'' indicates the absence of SOC in contrast to ``-SOC''. For symmetry groups having the same transition features with and without SOC, we only list the case without SOC. The square brackets $[\cdot]$ and parentheses $(\cdot)$ contain the site symmetries and coordinates of the Wyckoff positions, respectively. The second column contains the reciprocal and real space information of the minimal OAI transitions defined in \cref{sec:main-OAI-transition}. The transition with a symbol of momentum irrep exchange $R_n\!\rightarrow\!R_m$ induces band inversion(s) at high symmetry point $R$ (and its inequivalent symmetric partners), which replaces some occupied band(s) forming $R_n$ by band(s) with $R_m$. The transition with ``NULL" closes the band gap at general k-points. Equations inside ``$\{$" depict the minimal orbital occupation changes, where $\Delta_{Q}^f$ denotes the occupation change of irrep Q at each Wyckoff position $f$. The particle conservation is not explicitly stated, yet it is enforced in every transition by default.
    The third column illustrates the transitions in reciprocal space. A plot with a Brillouin zone is decorated by red crosses and arrows that indicate the positions and movements of Dirac points, respectively. A plot with four curves suggests that the band inversion will go through a gapless region where a quadratic touching from a 2-dim irrep dominates the physics near the Fermi level. The last column illustrates the possible minimal changes of Wannier centers during the transitions. The colored dots indicate the Wyckoff positions inside one cell, and the colored numbers show the possible occupation changes of each position.
    }  
	\label{fig:main-rsi-table}
\end{figure}

To further show the generality, in \cref{fig:main-rsi-table}, we enumerate all the minimal OAI transitions protected by magnetic point groups $Pm$, $P2'$, $P4'$, and $P6'$. The details of enumeration are summarized in Sec.~$\color{red}\mathrm{II.B}$ of Ref.~\cite{sup}. Here, for a given symmetry group, the minimal OAI transition is defined as the OAI transition with minimal band deformation and molecular orbital transition. The band deformation of a generic OAI transition can be viewed as a superposition of minimal band deformations that can be band inversions at high symmetry points or gap closures at generic k-points (see the third column of \cref{fig:main-rsi-table}).  Also, since OAI can be wannierized to molecular orbitals (each orbital forming a site symmetry irrep at some Wyckoff position), an OAI transition can be characterized by occupation changes of these orbitals. The minimal molecular orbital transition refers to the minimal occupation change that can realize the corresponding band deformation. For example, the third row of block ``P4'-NSOC" in \cref{fig:main-rsi-table} demonstrates a minimal OAI transition that replaces some occupied band forming irrep $X_2$ at $X$ point by a band with $X_1$ and moves an electron from irrep A at the Wyckoff position $\mathrm{1b}$ to irrep A at $\mathrm{1a}$. 

In terms of band deformation, these transitions can be divided into three categories: quadratic touching from a 2-dim irrep, Dirac points braiding, and immediate band gap closure-reopening. In the former two categories, the band structure is gapless for a finite parameter region in contrast to a single point in the last category. For the first category, if we add slow-varying disorders that mainly open a local gap with a nontrivial Chern number ($p_0<1/2$), a CMP is highly possible due to the percolation mechanism protected by the average symmetry. For the second category, as discussed above, the possibility of CMP is higher with more Dirac points and stronger positive Dirac mass correlations. For the third category, the transition can go through a critical point at most since the gapless band structure is necessary for delocalization. The number and correlations of Dirac points also influence the possibility of delocalizing the gapless point.

\section{Discussion}

For the first time, our work points out that transitions between {\it trivial} states (such as OAIs) without TRS can be critical under {\it simple} chemical potential disorders and, through a quantitative mapping, reveals that the criticality is due to a tricolor percolation mechanism of $C=0,\pm1$ regions. 
Since the chemical potential disorder is realistic and there are many topologically trivial magnetic materials with symmetries that forbid net (anomalous) Hall conductance  \cite{elcoro_magnetic_2021,xu_high-throughput_2020}, our work also has experimental relevance. 

We notice that CMPs in 2D class A systems have been observed in previous works \cite{onoda_localization_2007,xu_topologically_2012, Raquel2021APS, chen_effects_2019}.
Ref.~\cite{onoda_localization_2007} did not report a CMP, yet we find its Fig.~2(a) may suggest a CMP similar to the one found by Ref.~\cite{xu_topologically_2012}.
Ref.~\cite{xu_topologically_2012} reported a CMP in the Kane-Mele model ($\mathbb{Z}_2$ TI) in the presence of a weak Zeeman field and ascribed the criticality to the change of spin Chern number. 
Ref~\cite{chen_effects_2019} reported a CMP in the Bernevig-Hughes-Zhang  model ($\mathbb{Z}_2$ TI) in the presence of a random magnetic field and ascribed the criticality to two coupled quantum Hall transitions. 
Therefore, these CMPs exist in topological phase transitions with additional weak perturbation terms.
However, our CMP exists between two topologically trivial OAIs far from any topological state. 
The only difference between the two OAIs is the center and representation of Wanniers.
It is also worth mentioning that, to exclude possible finite-size effects, we have verified the criticality of CMP up to system sizes $L=2000$ for the network model and $L=500$ for the lattice model (see Sec.~\textcolor{red}{I} in Ref.~\cite{sup}), which are larger than the system sizes $L=24, 32, 128$ used in Refs.~\cite{onoda_localization_2007,xu_topologically_2012, chen_effects_2019}, respectively. 




\section{Methods}

\subsection{Topological quantum chemistry of the \texorpdfstring{$C_{2z}T$}{C2T} model}
\label{subsec:method-TQC}

The CMPs in the $C_{2z}T$ models arise between inequivalent OAIs. To depict the OAI transitions in the clean limit, we can use the tool of topological quantum chemistry, \ie analyzing the transition of occupied magnetic element band representations (MEBRs) \cite{elcoro_magnetic_2021}. In our models, all the OAIs can be wannierized to molecular orbitals centering at the $C_{4z}$ and $C_{2z}T$ centers (Wyckoff positions $\mathrm{2b}$ and $\mathrm{4c}$). These orbitals respect the site symmetries ($4m'm'$ and $2'm'm$ for $\mathrm{2b}$ and $\mathrm{4c}$, respectively) and hence can be characterized by the irreps of site symmetries. According to the topological quantum chemistry, these molecular orbitals will induce bands with MEBRs listed in \cref{tab:main-MEBRs}.  
One MEBR is the minimal group of bands formed by a type of molecular orbitals, and the left part of an MEBR notation indicates the site symmetry irrep of the orbitals, \eg $A_b\!\uparrow\!G$ indicates the MEBR formed by $s$ orbitals (trivial irrep $A$) at positions $2b$. Therefore, we can deduce the Wannier centers of OAIs from the occupied MEBRs, depicting the OAI transitions by MEBR transitions.

We start with the network model. Although a network model has infinite occupied bands, these bands form an infinite direct sum of MEBRs and can be viewed as a special kind of OAI. The band structure comprises disconnected branches, each containing two bands. 
One branch forms one of the following four MEBRs defined in \cref{tab:main-MEBRs}: $\mathrm{A_{b}\! \uparrow \! G,\, B_{b}\! \uparrow \! G,\,^{1}E_{b}\!\uparrow \! G,\, ^{2}E_{b}\! \uparrow \! G}$, result from effective $s$, $d_{x^2-y^2}+i d_{xy}$ (or equivalently $d_{x^2-y^2}- i d_{xy}$), $p_x-ip_y$, $p_x+ip_y$ orbitals, respectively. All these orbitals center at the Wyckoff position $\mathrm{2b}$. Since $2\mathrm{b}$ has multiplicity $2$ (two $C_{4z}$ centers per unit cell), each MEBR contains two bands. 
Also, notice that the band structure comprises infinite repeating units. Each unit contains eight bands, and one can generate the whole band structure by energy translations of a unit with step $2\pi v/a$. Hence, we can focus on the upper eight of the ten bands in \cref{fig:main-Network}{\color{red}(g-i)}. 
When $\theta\!<\!0 $, direct gaps separate different branches, and the lower four of the focused eight bands form a direct sum of two MEBRs: $\mathrm{^{2}E_{b}\! \uparrow \! G \oplus A_{b}\! \uparrow \! G}$. 
When $\theta\!=\!0$, the gaps close, and the band structure is indeed that of a ballistic 1D metal with linear dispersion relations. 
As $\theta$ increases across $0$, irreps $ \Gamma_1,\,M_1$ ($\Gamma_2,\,M_2$) go up (down) across the energy level. Similar irrep exchanges also happen above and below the repeating unit, \eg the lower two of the focused eight bands will exchange irreps with the lower bands that exceed the scope of \cref{fig:main-Network}{\color{red}(g-i)}.  
After the transition ($\theta\!>\!0$), the gaps reopen, and the MEBRs of the lower four of the focused eight bands change to $\mathrm{^{1}E_{b}\! \uparrow \! G \oplus B_{b}\! \uparrow \! G}$. Hence, $\theta=\pm\pi/2$ correspond to two inequivalent trivial phases, although the molecular orbitals of both phases center at the $C_{4z}T$ centers.

We now turn to the band evolution of $H_{8B}$ ($H_{8B}'$ is similar). 
See \cref{fig:main-Lattice-full}$\color{red}\text{(c-e)}$,  as $\tilde{t}$ increases from $0$ to $\pi v/a$, $\Gamma_1$ and $M_1$ rise across the Fermi level while $\Gamma_2$ and $M_2$ fall below the Fermi level. We encounter the same irreps exchange to the network model. 
Despite similarities near the Fermi level, the MEBR transition of $H_{8B}$ is different from the network model. The lower four bands of $H_{8B}$ change from $\mathrm{^{2}E_b\! \uparrow \!G \oplus A_b\! \uparrow\!G}$ to $\mathrm{A''_c\!\uparrow\! G}$ during the transition. The Wyckoff position of $\mathrm{A''_c\!\uparrow\! G}$ is $\mathrm{4c}$, which is the $C_{2z}T$ center rather than the $C_{4z}$ center (\cref{tab:main-MEBRs}). This difference in molecular orbital transition is unavoidable since the transition in the network model involves exchanges of representations between different repeating units, while $H_{8B}$ only has one unit. In the network model with a sufficient number of bands, the four bands below the Fermi level not only exchange representations with the bands above them but also with the bands (in another repeating unit) below them.  
However, $H_{8B}$ has no band below the lower four bands. Nevertheless, since the origin of this difference is well below the Fermi level, it should not affect the low-energy physics. Therefore, we can expect similar low-energy behaviors between $H_{8B}$ and the network model, which our numerical data confirms.

It is worth mentioning that the transition in $H_{8B}$ changes the position of the MEBRs from $C_{4z}$-centers (2b) to $C_{2z}T$-centers (4c). No $C_{2z}T$ center is occupied before the transition, and the RSI $\delta_{w}=0$. Given that there are four $C_{2z}T$ centers per cell and four occupied bands, every $C_{2z}T$ center is occupied by one electron after the transition, and the system has RSI $\delta_{w}=1$. Therefore, the second Stiefel-Whitney class $w_2$ \cite{ahn_failure_2019,fang_topological_2015} must change from 0 to 1. 
As we have discussed in the second paragraph of \cref{sec:main-network-model}, the transition process must involve braiding of the Dirac points. 
In addition, although the lower four bands form $\mathrm{A''_c\!\uparrow\! G}$ are always connected in our models, in general cases, $\mathrm{A''_c\!\uparrow\! G}$ can be decomposed into two fragile topological bands ($\Gamma_5,\,M_5,\,X_1$) and two trivial bands (forming MEBR $A_2\!\uparrow\!G$ with Wyckoff position $2a\,(\frac14\,\frac14,0),\,(\frac34\,\frac34,0)$, \ie the centers of white squares in \cref{fig:main-Network}$\color{red}\text{(c)}$), which is expected from $w_2=1$.

\begin{table*}[h]
\centering{
\begin{tabular*}{0.9\linewidth}{lccccc|lccccc|lc}
\hline
 & $\Gamma_1$ & $\Gamma_2$ & $\Gamma_3$ & $\Gamma_4$ & $\Gamma_5$ &  & $M_1$ & $M_2$ & $M_3$ & $M_4$ & $M_5$ &  & $X_1$ \\ \hline
\multicolumn{1}{l|}{$\{1\vert 0,0,0\}$} & 1 & 1 & 1 & 1 & 2 & \multicolumn{1}{l|}{$\{1\vert 0,0,0\}$} & 1 & 1 & 1 & 1 & 2 & \multicolumn{1}{l|}{$\{1\vert 0,0,0\}$} & 2 \\
\multicolumn{1}{l|}{$C_{2z} = \{2_{001}\vert - \frac12, \frac12, 0\}$} & 1 & 1 & 1 & 1 & -2 & \multicolumn{1}{l|}{$C_{2z} = \{2_{001}\vert - \frac12, \frac12, 0\}$} & $-1$ & $-1$ & $-1$ & $-1$ & 2 & \multicolumn{1}{l|}{$C_{2z} = \{2_{001}\vert - \frac12, \frac12, 0\}$} & 0 \\
\multicolumn{1}{l|}{$C_{4z}=\{4_{001}^+ \vert 0,\frac12,0\}$} & 1 & -1 & -1 & 1 & 0 & \multicolumn{1}{l|}{$C_{4z}=\{4_{001}^+ \vert 0,\frac12,0\}$} & $i$ & $-i$ & $-i$ & $i$ & 0 & \multicolumn{1}{l|}{$\{m_{100}\vert 0,\frac{1}{2},0\}$} & 0 \\
\multicolumn{1}{l|}{$M_{xy}=\{m_{1\bar{1}0}\vert 0,0,0\}$} & 1 & -1 & 1 & -1 & 0 & \multicolumn{1}{l|}{$M_{xy}=\{m_{1\bar{1}0}\vert 0,0,0\}$} & -1 & 1 & -1 & 1 & 0 & \multicolumn{1}{l|}{$\{m_{010}\vert\frac{1}{2},0,0\}$} & 0 \\ \hline
\end{tabular*}
}
\caption[]{Character table of irreps at high symmetry momenta in magnetic space group $P_C4bm$ ($\#100.177$ in BNS setting), taken from the  \href{https://www.cryst.ehu.es/cgi-bin/cryst/programs/corepresentations.pl}{COREPRESENTATIONS} program on the Bilbao Crystallographic Server \cite{elcoro_magnetic_2021}. 
Characters of the listed symmetry operations can uniquely determine the irreps
One should notice that we use a different convention of the origin point as the Bilbao Crystallographic Server. 
Our $C_{2z} = \{2_{001}\vert - \frac12, \frac12, 0\}$, $C_{4z}=\{4_{001}^+ \vert 0,\frac12,0\}$, $M_{xy}=\{m_{1\bar{1}0}\vert 0,0,0\}$, $\{m_{100}\vert 0,\frac{1}{2},0\}$, and  $\{m_{010}\vert\frac{1}{2},0,0\}$ correspond to $\{2_{001} | 0,0,0\}$, $\{4_{001}^+ | 0,0,0\}$, $\{m_{1\bar{1}0} | \frac12, - \frac12,0\}$, $\{m_{100}\vert \frac12,\frac{1}{2},0\}$, and $\{m_{010}\vert \frac12, - \frac{1}{2},0\}$ in the standard convention of the Bilbao Crystallographic Server, respectively. 
}
\label{tab:main-Irreps}
\end{table*}

\begin{table*}[h]
\centering{
\begin{tabular*}{0.6\linewidth}{l|c|c|c|c|c}
\hline
Wyckoff pos. & \multicolumn{4}{c|}{$\mathrm{2b}\, (\frac34 \frac14 0),\,(\frac{1}{4} \frac{3}{4} 0)$} & \begin{tabular}[c]{@{}c@{}}$\mathrm{4c} \,(0 0 0),\,(0 \frac12 0),$\\  $(\frac12 0 0), (\frac12 \frac12 0)$ \end{tabular} 
\\ \hline
Site sym.   &\multicolumn{4}{c|}{$4m^{\prime}m^{\prime},\,4$}         & $2^{\prime}m^{\prime}m,\,m$\\
\hline 
MEBR &\multicolumn{1}{c|}{$\mathrm{A_b\!\uparrow\!G}$} & \multicolumn{1}{c|}{$\mathrm{B_b\!\uparrow\!G}$} & \multicolumn{1}{c|}{$\mathrm{^{1}E_b\!\uparrow\!G} $} & $\mathrm{^{2}E_b\!\uparrow\!G} $ & $\mathrm{A^{\prime \prime}_c\!\uparrow\!G}$ \\ \hline
Orbital & \multicolumn{1}{c|}{$1$}&\multicolumn{1}{c|}{$ d_{x^2-y^2} + i d_{xy} $}& \multicolumn{1}{c|}{$p_x+i p_y$}  & $p_x-ip_y$ & $ p_y$ \\ \hline
Irreps at $\Gamma$ & $\Gamma_1\oplus\Gamma_4$ & $\Gamma_2\oplus\Gamma_3$ & $\Gamma_5$ & $\Gamma_5$ & $\Gamma_2\oplus\Gamma_4\oplus \Gamma_5$ \\
\hline
Irreps at $M$ & $M_5$ & $M_5$ & $M_2\oplus M_3$ & $M_1\oplus M_4$ & $M_2\oplus M_4\oplus M_5$ \\
\hline
Irreps at $X$ & $X_1$ & $X_1$ & $X_1$ & $X_1$ & $2 X_1$ \\
\hline
\end{tabular*}
}
\caption{MEBRs of $P_C4bm $ ($\#100.177$ in BNS setting) involved in this work, taken from the \href{https://www.cryst.ehu.es/cgi-bin/cryst/programs/mbandrep.pl}{MBANDREP} program on the Bilbao Crystallographic Server \cite{elcoro_magnetic_2021}. 
The real space orbital character of each MEBR is shown in the ``Orbital'' row.
For example, the MEBR $\mathrm{ B_b \uparrow G}$ can be generated by an $d_{x^2-y^2}+i d_{xy}$ type orbital at the first 2b position $(\frac34,\frac14,0)$. Note that $\mathrm{A^{\prime \prime}_c\!\uparrow\!G}$ is the only decomposable MEBR, although bands with this MEBR are always connected in this work.
One should notice that we use a different convention of the origin point as the Bilbao Crystallographic Server. 
}
\label{tab:main-MEBRs}
\end{table*}

\subsection{Quasi-1D localization length and transfer matrix method}
\label{subsec:method-Transfer-matrix}

A commonly used physical quantity in research of localization is the \emph{quasi-1D localization length} $\rho_{\mathrm{q-1D}}$. It is defined on a 2D/3D sample prepared in a quasi-1D shape, \eg a long, thin cylinder with $L_{\rm{axial}}\gg L_{\rm{radius}}$. 
$\rho_{\mathrm{q-1D}}$ reflects the decaying rate of eigenstates in the quasi-1D direction, \eg the axial direction of a long thin cylinder. Since any 1D system is localized under nonzero disorder strength, $\rho_{\mathrm{q-1D}}$ will always be finite except for a perfectly clean sample. 
Localization of the original 2D/3D system (in a 2D/3D shape) can be derived from a scaling analysis of the dimensionless quasi-1D localization length $\Lambda=\rho_{\mathrm{q-1D}}/L$, where $L$ is the transversal size of the quasi-1D sample. 
We denote the localization length of a normally shaped (scales of different directions are similar) and sufficiently large sample as $\rho$.
For a metallic system, $\rho$ in a (normally shaped and sufficiently large) sample is much larger than the sample size. Thus, $\rho_{\mathrm{q-1D}}(L)$ increases faster than $L$, \ie $\Lambda(L)\rightarrow\infty$ in the limit $L\rightarrow\infty$. 
For an insulating system, $\rho$ is finite in a (normally shaped and sufficiently large) sample. Hence, $\rho_{\mathrm{q-1D}}$ will converge to $\rho$ when $L\gg\rho$, \ie $\Lambda(L)\rightarrow 0$ as $L\rightarrow\infty$. 
In practice, we identify the region where $\Lambda(L)$ monotonically increases as the metallic phase and where $\Lambda(L)$ monotonically decreases as the localized phase. If a system contains both localized and extended phases in some parameter space, there will be a critical region (usually a boundary with measure zero) in the parameter space where $\Lambda(L)$ is independent of (sufficiently large) $L$. 

The transfer matrix method \cite{pichard_finite_1981} is a widely used numerical approach in calculating $\rho_{\mathrm{q-1D}}$. 
Although it has different formulae for (generic) network and lattice models, the basic ideas are the same. 
A quasi-1D sample is divided into layers with normals along the quasi-1D direction. 
The amplitudes of an energy-eigenstate on different layers are related by the Schrodinger equation. 
A ($2s\times2s$ shaped) transfer matrix $T_n$ generally transforms the amplitudes on the $(\rm{l\!-\!r_1)th\sim (l\!+\!r_2\!-\!1)th}$ layers to those on the $\rm{(l\!- \!r_1\!+\!1)th \sim ( l\!+\!r_2) th}$ layers. Here, $s\in \mathbb{N}^+$ is proportional to the number of degrees of freedom in one layer. And $r_1,\,r_2\!\in\!\mathbb{N}^+$ represent that only the $\rm{(l\!-\!r_1)th\sim(l\!+\!r_2)th}$ layers can hop/propagate (in one step) to the $\rm{l\thinspace th}$ layer. The values of $r_1$ and $r_2$ depend on the hopping of concrete models. Since the transfer matrix of a (general) network model is determined by the transmission matrix, the amplitudes in the $\mathrm{(l+1)th}$ layer depend only on the $\mathrm{lth}$ layer. Hence, $r_1\!=\!r_2\!=\!1$ for general network models (not only ours). For general lattice models, $r_1$ and $r_2$ can take arbitrary finite non-negative integer values. Nevertheless, in our models, $r_1\!=\!r_2\!=\!1$. 

To extract $\rho_1D$ from the transfer matrix, we can consider a consecutive product of transfer matrices $O_M\!=\!\prod_{n=1}^M T_n$.  
Because of the disorder, some elements in $T_n$ are random variables. According to the Oseledec theorem, the limit $P\!=\!\lim_{M\rightarrow\infty} (O_M^{\dagger}O_M)^{1/2M}$ exists and has eigenvalues $\{\exp(\nu_1),\,\exp(-\nu_1),...\exp(\nu_{s}),\,\exp(-\nu_{s})\}$ where $\nu_{i}\!\geq\!\nu_{i+1}\geq0,\,i=1,2...s$. These (positive) exponents are so-called \emph{Lyapunov exponents} (LEs). 
The definition of $P$ indicates that an eigenvector $\vec{\eta}_i$ of $P$ with eigenvalue $\exp(-\nu_i)$ satisfies $\Vert O_M\vec{\eta}_i
\Vert^2\!=\!\vec{\eta}_i^{\dagger}(O_M^{\dagger}O_M)\,\vec{\eta}_i\!=\!\vec{\eta}_i^{\dagger}[(O_M^{\dagger}O_M)^{1/2M}]^{2M}\,\vec{\eta}_i\approx\vec{\eta}_i^{\dagger}P^{2M}\,\vec{\eta}_i\!=\!\Vert\exp(-M\nu_i)\vec{\eta}_i\Vert^2$ for sufficiently large M. Therefore, the smallest LE $\nu_s$ determines the decaying rate of energy-eigenstates (along the quasi-1Dc direction) since any energy-eigenstate is a superposition of eigenvectors of $P$ with eigenvalues $\exp(-\nu_i)$ (the amplitudes cannot grow exponentially hence $\exp(\nu_i)$ is excluded). Because of this, one can define the quasi-1D localization length as the inverse of the smallest LE: $\rho_{\mathrm{q-1D}}=1/\nu_s$.

Due to the space limitation, there are three main aspects we cannot explain here: calculating conductance from the transfer matrix, the technique for numerical stability of LE, and the concrete formulae of the transfer matrices of our network and lattice models. Readers interested in these details can refer to Sec.~${\color{red}\mathrm{V}}$ of \cite{sup}.

\subsection{\texorpdfstring{$\beta$}{beta}-function of the CMP}
\label{subsec:method-beta-network}

\begin{figure}[h]
	\centering
    \includegraphics[width=0.8\linewidth]{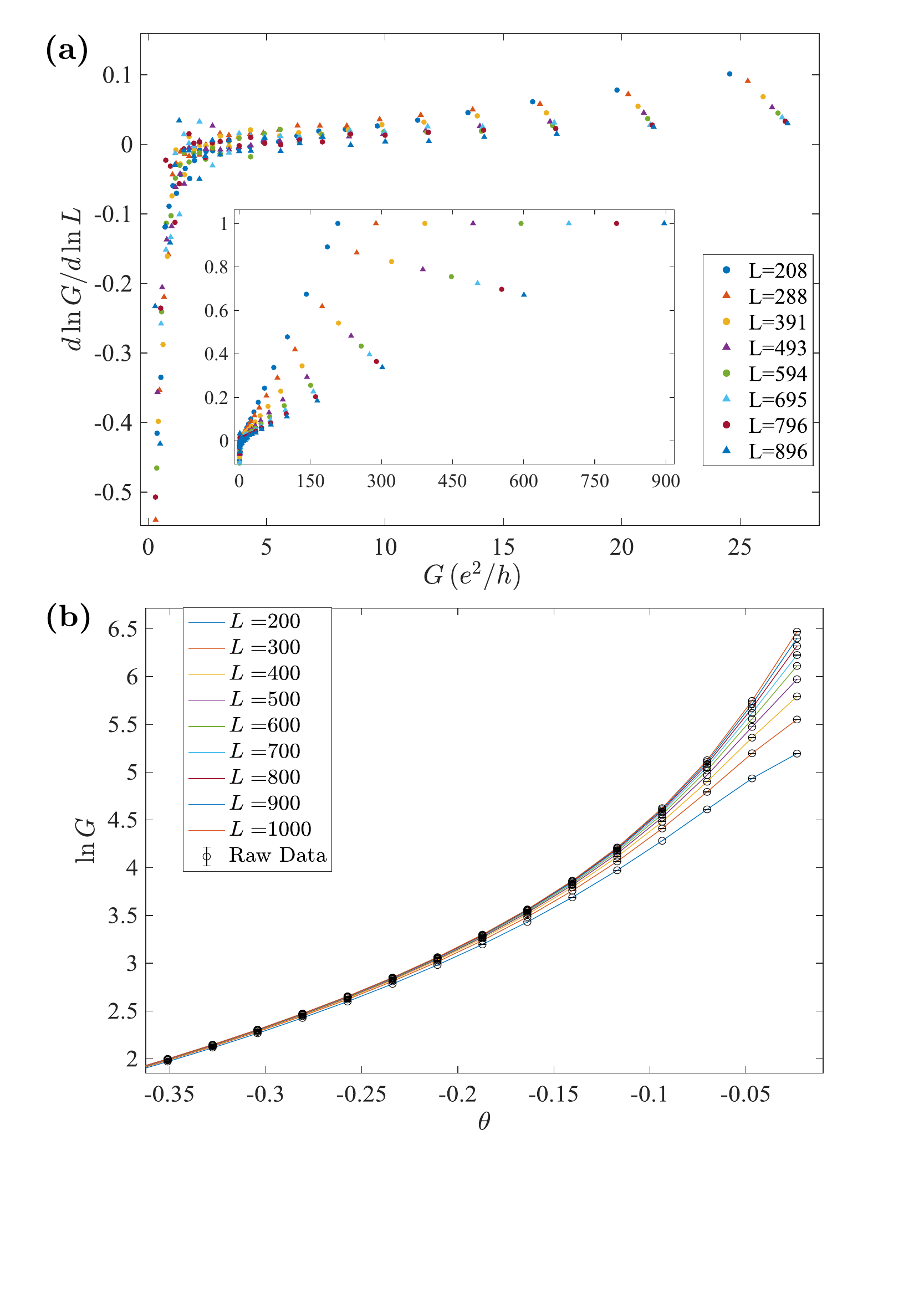}
    \caption[]{$\beta$-function and conductance fitting near the ballistic limit. (a) The $\beta$-function of the network model derived from finite differences of conductance data in \cref{fig:main-Network}${\color{red}\mathrm{(f)}}$. The legends inform the central sizes of the finite differences. The insect is the zoom out of the $\beta$-function for large $G$. (b) Conductance fitting according to the linear hypothesis of $\beta$-function for large $G$. Colored curves are fitting results while the circles are raw data. The goodness of fit has reached $0.23$ and will become higher as getting closer to the ballistic limit. ($1$ corresponds to a perfect fitting and $0.05$ is the frequently used threshold of accepting a fitting hypothesis.)  }
    \label{fig:main-beta-network}
\end{figure}

The $\beta$-function $\beta=d\ln G/d \ln L$ is an additional quantity to verify the criticality of the observed delocalized phases. \cref{fig:main-beta-network}${\color{red}\mathrm{(a)}}$ shows the $\beta$-function of the network model derived from the finite difference method (we have ignored the default conductance unit $e^2/h$). As we can see, all the data points fall into two parts corresponding to the delocalized and localized phases. A critical conductance $G_c=2\!\sim\!3$ divides these two phases. In the localized phase below $G_c$, data points collapse to one curve, demonstrating the Anderson localization. In the delocalized phases above $G_c$, the data points distribute around zero below $G\!\sim\!5$. And above $G\!\sim\!5$, data from different sizes deviate from each other and become significantly positive. These positive data points result from the finite size effect of the ballistic limit at $\theta=0$. To justify that the delocalized phase with $-\pi/4\lesssim\theta<0$ is indeed a CMP, we have to prove that $\beta(G>G_c,L\rightarrow\infty)\rightarrow 0^{+}$. 

 When $\theta=0$, the network degenerates to decoupled parallel chiral wires in four directions ($\pm \hat x,\pm \hat y$). Since there are $L$ channels in one direction, the ballistic conductance $G_{\max}(L)=L$. Hence, for a given system size $L$, the $\beta$-function should terminate at the point $(G\!=\!L,\beta\!=\!1)$, which is confirmed numerically (the inset of \cref{fig:main-beta-network}${\color{red}\mathrm{(a)}}$). Due to the finite size effects, the ballistic limit will induce a diffusive metal phase with $\theta\!\rightarrow\!0$ corresponding to the positive region of $\beta$-functions for large $G$. Further, the data suggest a linear hypothesis of $\beta$-function for large $G$, \ie $\beta(G,L)=G/L$ when $G\!\gg\!G_c$. To verify this hypothesis, we can view it as a differential equation and test its solution (as a hypothesis of conductance) on conductance data. To be concrete, $\beta(G,L)=G/L$ implies an ansatz of conductance $G(\theta,L)^{-1}\!=\!G(\theta,\infty)^{-1}\!+\!L^{-1}$ when $G\!\gg\!G_c$. Here, the fitting parameter $G(\theta,\infty)$ is the conductance in the thermodynamic limit for a given $\theta$. As illustrated in \cref{fig:main-beta-network}${\color{red}\mathrm{(b)}}$, this ansatz fits our conductance data very well when $G>5$. Although we have not understood the mechanism behind the linear $\beta$-function, we can conclude from the above results that $\beta(G\!>\!G_c,L\!\rightarrow\!\infty)\rightarrow 0^+$, proving the criticality of CMP.

\subsection{Localized OAI transition in class AI}

\begin{figure}[h]
	\centering
    \includegraphics[width=0.95\linewidth]{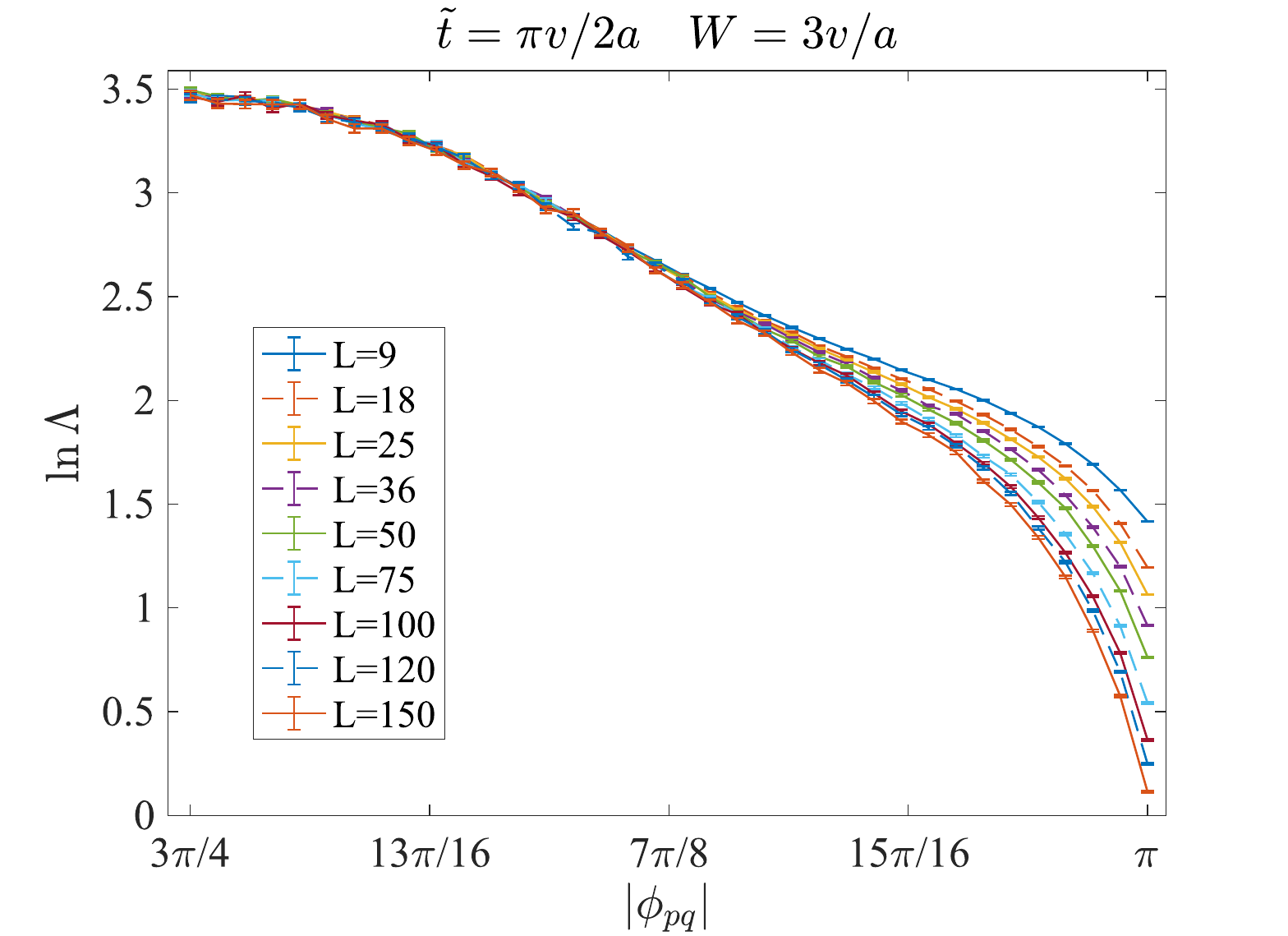}
    \caption[]{Normalized localization length $\Lambda$ as functions of $|\phi_{pq}|$ with $\tilde{t}=\frac{\pi v}{2a}$,$W=3v/a$, $E_F=0$, and various transversal sizes $L$. The used longitudinal size is $M=10^7$ and the precision ($\sigma_\Lambda/\Lambda$) has reached $2\%$. The point with $|\phi_{pq}|=3\pi/4$ corresponds to the middle of \cref{fig:main-Lattice-full}\textcolor{red}{(f)}. The point with $|\phi_{pq}|=\pi$ recovers TRS and is localized under any finite on-site disorders. }
    \label{fig:main-TRSOAI}
\end{figure}

One should not conclude from the above results that CMP exists in any transition between inequivalent 2D OAIs. Due to the localized nature of OAI, it is commonly believed that disorders should localize general transitions between inequivalent 2D OAIs. 
We can obtain a localized OAI transition by modifying our lattice model $H_{8B}$. Recall \cref{eq:main-H8B}, $H_{8B}$ is described by four real parameters $t$, $t'$, $\tilde{t}$, and $\phi_{pq}$. The only parameter breaking TRS is $|\phi_{pq}|=3\pi/4$. If we take $|\phi_{pq}|=0$ or $\pi$, the system will respect TRS even if on-site disorders are present. Now, the previous $C_{2z}T$ centers are promoted to $C_{2z}$ centers. 
According to Ref.~\cite{song_twisted_2020}, in the presence of TRS, the Wyckoff position with site symmetry $2$ has a $\mathbb{Z}$-valued RSI $\delta=m_{+}-m_{-}$ where $m_{+}$ ($m_{-}$) is the occupation number of orbital even (odd) under $C_{2z}$. Hence, when $|\phi_{pq}|=0$ or $\pi$, tuning $\tilde{t}$ from $0$ ($\delta=0$) to $+\infty$ ($\delta=-1$) still drives a transition between inequivalent OAIs. On the other hand, due to the weak localization effect in the presence of spinless TRS, the system must be localized regardless of $t$, $t'$, and $\tilde{t}$. In \cref{fig:main-TRSOAI}, we take $\tilde{t}=\frac{\pi v}{2a}$, corresponding to the middle of the CMP in \cref{fig:main-Lattice-full}$\color{red}\mathrm{(f)}$, and tune $|\phi_{pq}|$ from $3\pi/4$ to $\pi$. 
We can see a CMP-LP transition in \cref{fig:main-TRSOAI}. For all the $|\phi_{pq}|$ scanned, the band structure in the clean limit is gapless near the Fermi level $E_F=0$, \ie the clean system stays at the metallic intermediate state of the OAI transition but will be localized by disorders. 
Although the LP shrinks with weaker disorder strength $W$, $|\phi_{pq}|=\pi$ is always localized. 

\begin{acknowledgments}
We are grateful to  Roni Ilan, Ryuichi Shindou, Chen Wang,  Chui-Zhen Chen, and Jian Li for helpful discussions. 
Z.-D. S. and F.-J. W. were supported by
National Natural Science Foundation of China (General Program No.\ 12274005), 
Innovation Program for Quantum Science and Technology (No.\ 2021ZD0302403), 
National Key Research and Development Program of China (No.\ 2021YFA1401903).
B. A. B. was supported by the European Research Council (ERC) under the European Union’s Horizon 2020 research and innovation program (grant agreement No. 101020833), the ONR Grant No. N00014-20-1-2303, the Schmidt Fund for Innovative Research, Simons Investigator Grant No. 404513, the Gordon and Betty Moore Foundation through the EPiQS Initiative, Grant GBMF11070 and Grant No. GBMF8685 towards the Princeton theory program. 
Further support was provided by the NSF-MRSEC Grant No. DMR-2011750, BSF Israel US foundation Grant No. 2018226, the Princeton Global Network Funds. 
R. Q. is supported by the Simons Foundation award 990414, the U.S. Department of Energy (DOE) under award DE-SC0019443, and the NSF MRSEC  DMR-2011738
\end{acknowledgments}

\onecolumngrid
\clearpage
\appendix{\bf{Appendix}}
\footnotetext{We use {\color{red}red characters} to indicate figures, equations, and sections in this supplementary material, while {\color{green}green characters} indicate those of the main text. }

\tableofcontents
\clearpage 
\section{Supplementary numerical data of localization behavior}
\label{sec:numerical-localization-length}

\subsection{Network model on the Manhattan lattice}
\label{subsec:numerical-network}

\begin{figure}[h]
	\centering
    \includegraphics[width=.75\linewidth]{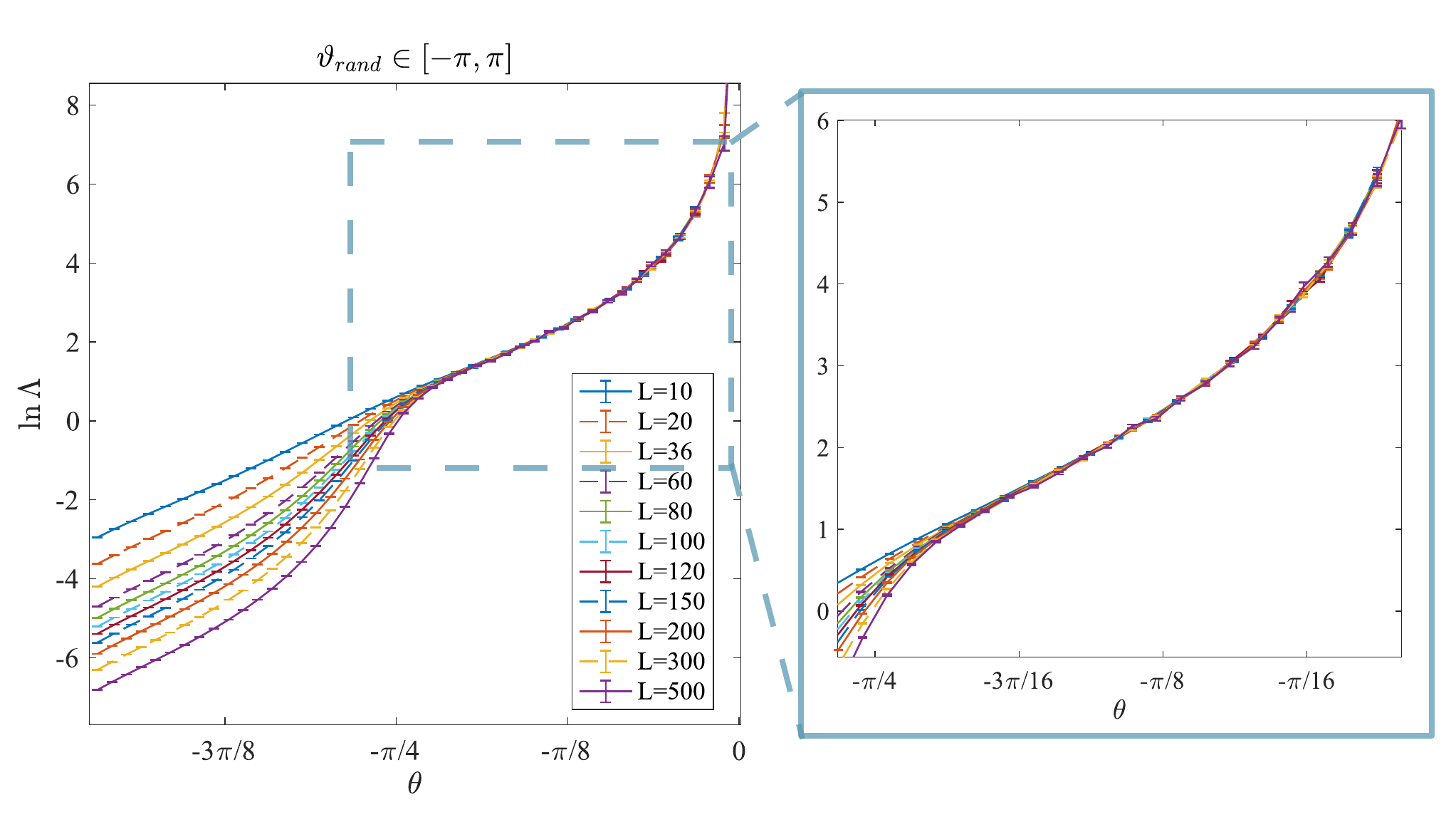}
    \caption[]{Localization length $\ln \Lambda$ versus scattering angle $\theta$ with different transversal sizes $L$ in the network model defined in \cref{network-decoupled-chiral-wires} \& (\ref{network-delta-scattering-potential}). The quasi-1D direction (longitudinal direction) is indicated in \cref{transfer-network-a} and the longitudinal size $M=10^7$. The data precision $\sigma_{\Lambda}/\Lambda$ reaches 1\% for $\theta<-\pi/8$, 2.5\% for $-\pi/8<\theta<-\pi/16$ and 5\% for other data points. $\sigma_{\Lambda}$ is the unbiased estimation of error mentioned in \cref{subsec:basic-localization-length}. The right side is the magnified critical region.}
    \label{Lambda-theta--L-network}
\end{figure}

In this subsection, we exhibit the numerical results of the network model defined in \cref{network-decoupled-chiral-wires} \& (\ref{network-delta-scattering-potential}) with random phases $\vartheta_{rand}$ on every chiral edges. Using the transfer matrices defined in \cref{concrete-transfer-matrix-network-1} \& (\ref{concrete-transfer-matrix-network-2}), we can calculate the quasi-1D localization length and conductance.
Fig.~\ref{Lambda-theta--L-network} shows the normalized quasi-1D localization length $\Lambda=\rho_{\mathrm{q-1D}}/L$ as a function of scattering angle $\theta\in[-\pi/2,0]$ with longitudinal size $M=10^7$ and different transversal sizes $L$. The plot with $\theta\in[0,\pi/2]$ is merely a mirror reflection of it. The critical region is magnified on the right side of Fig.~\ref{Lambda-theta--L-network}. 
As we can see, $\Lambda$ is independent of $L$ when $|\theta|\lesssim \pi/4$, \ie when the absolute value of transmission amplitude is larger than reflection amplitude. 
This is indeed what percolation argument predicts: The system will become critical once $(p_{1}+p_{-1})/p_{0}>1/2$. 

\begin{figure}[h]
	\centering
    \includegraphics[width=.75\linewidth]{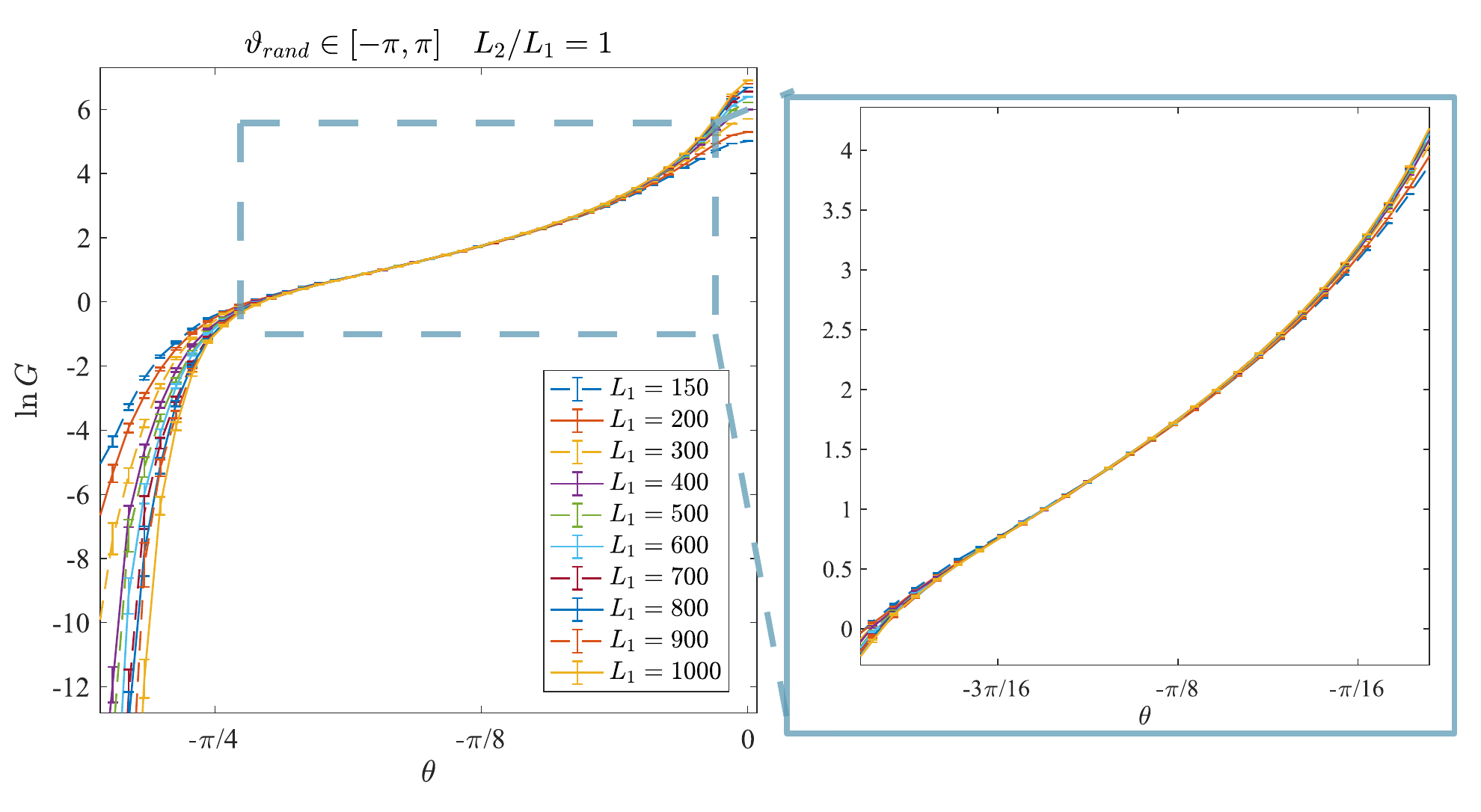}
    \caption[]{Conductance $\ln G$ versus scattering angle $\theta$ with different transversal sizes $L_2$ in the network model (\cref{network-decoupled-chiral-wires} \& (\ref{network-delta-scattering-potential})). The sample is normal shaped whose longitudinal length $L_1=L_2$. And the longitudinal direction is the same as the quasi-1D direction of \cref{transfer-network-a}. Each data points is averaged over $10^3$ samples and the precision $\sigma_{G}/G$ reaches 0.5\% near the critical region. $\sigma_{G}$ is the unbiased estimation of error of the conductance average. And the critical region is magnified and shown on the right side.}
    \label{G-theta--L-network}
\end{figure}

\cref{G-theta--L-network} shows the results of conductance $G$ versus $\theta$ with different transversal sizes $L_2$ and longitudinal sizes $L_1=L_2$ (so that conductance = conductivity). Here, we use $L_1,\,L_2$ instead of $M,\,L$ to emphasis that $G$ is calculated on normal shaped samples rather than quasi-1D samples. The plot with $\theta>0$ is a mirror reflection of \cref{G-theta--L-network}. We can see that $G$ is independent of $L_2$ when $|\theta|\lesssim \pi/4$, which is in agreement with \cref{Lambda-theta--L-network}. It seems that there is an extended phase when $\theta\rightarrow0$. However, the network model belongs to the class A and no extended phase is expected. Also, in contrast with the localized phase, \cref{G-theta--L-network} shows that $d \ln G/d L\rightarrow 0$ as $L\rightarrow\infty$ in the `extended phase'. Hence, we attribute this `extended phase' to the finite size effects. 

\begin{figure}[h]
	\centering
    \includegraphics[width=.75\linewidth]{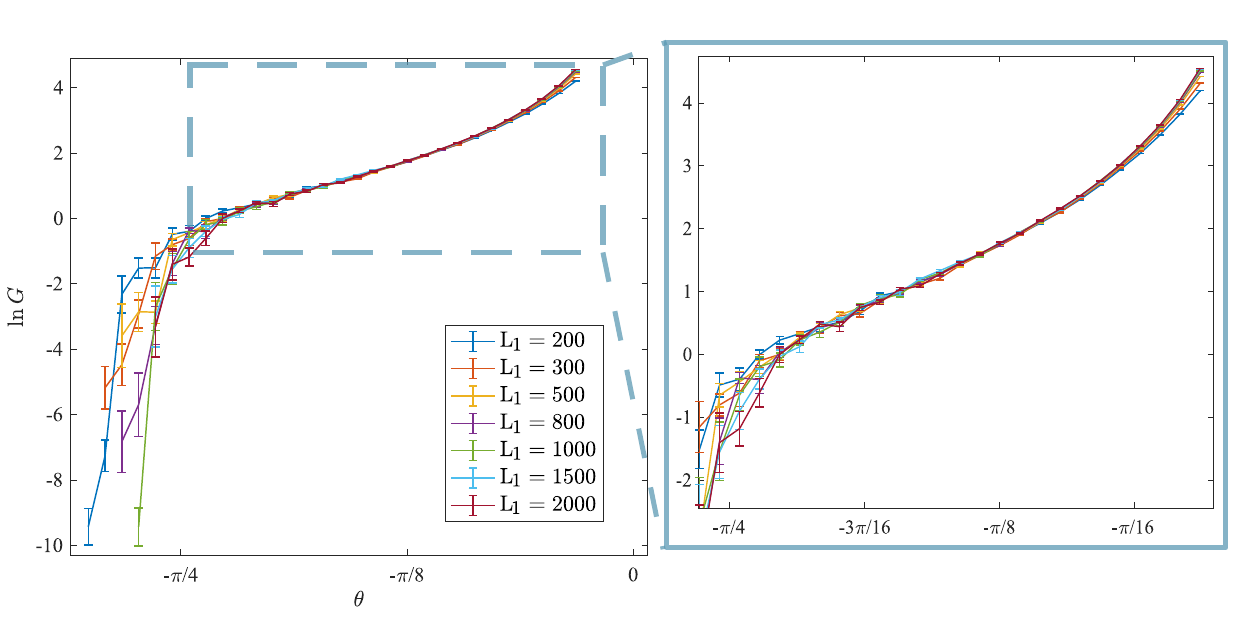}
    \caption[]{The same to \cref{G-theta--L-network} but with larger transversal sizes and $10$ samples. The precision reaches $5\%$ near the critical region.}
    \label{G-theta--L-network-larger}
\end{figure}

\subsection{Eight-band lattice model \texorpdfstring{$H_{8B}$}{H8B} }
\label{subsec:numerical-lattice-full}

\begin{figure}[h]
	\centering
	\subfigure[~Phase diagram of $H_{8B}$ with $E_F=0$ in the parameter space spanned by $W,\tilde{t}$]{\label{phase-diagram-full}
    \includegraphics[width=.6\linewidth]{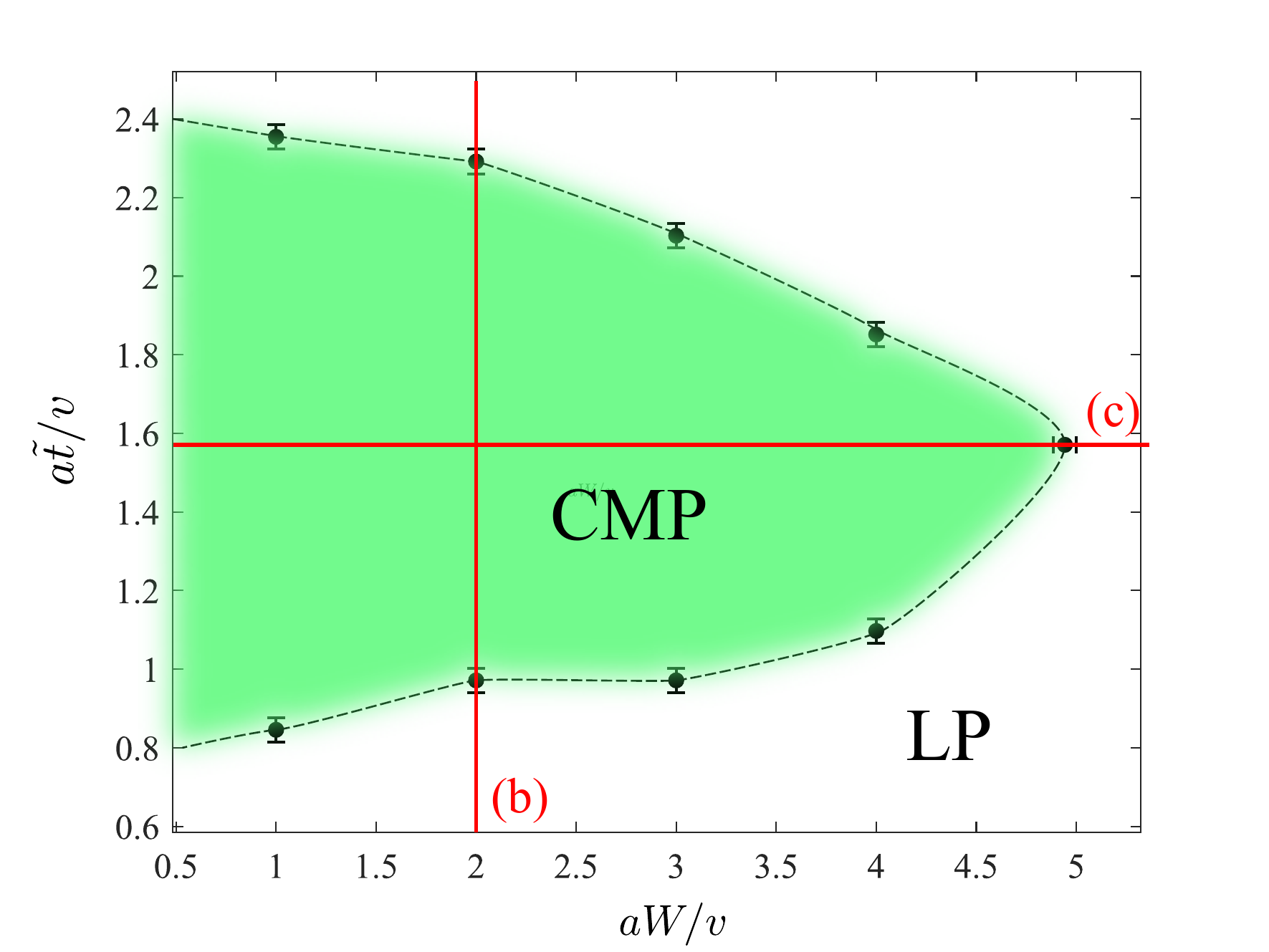}}
    \\
	\subfigure[~$\ln\Lambda$ versus $\tilde{t}$ with $W=2v/a$]{\label{Lambda-ga-W2-full}
	\includegraphics[width=.43\linewidth]{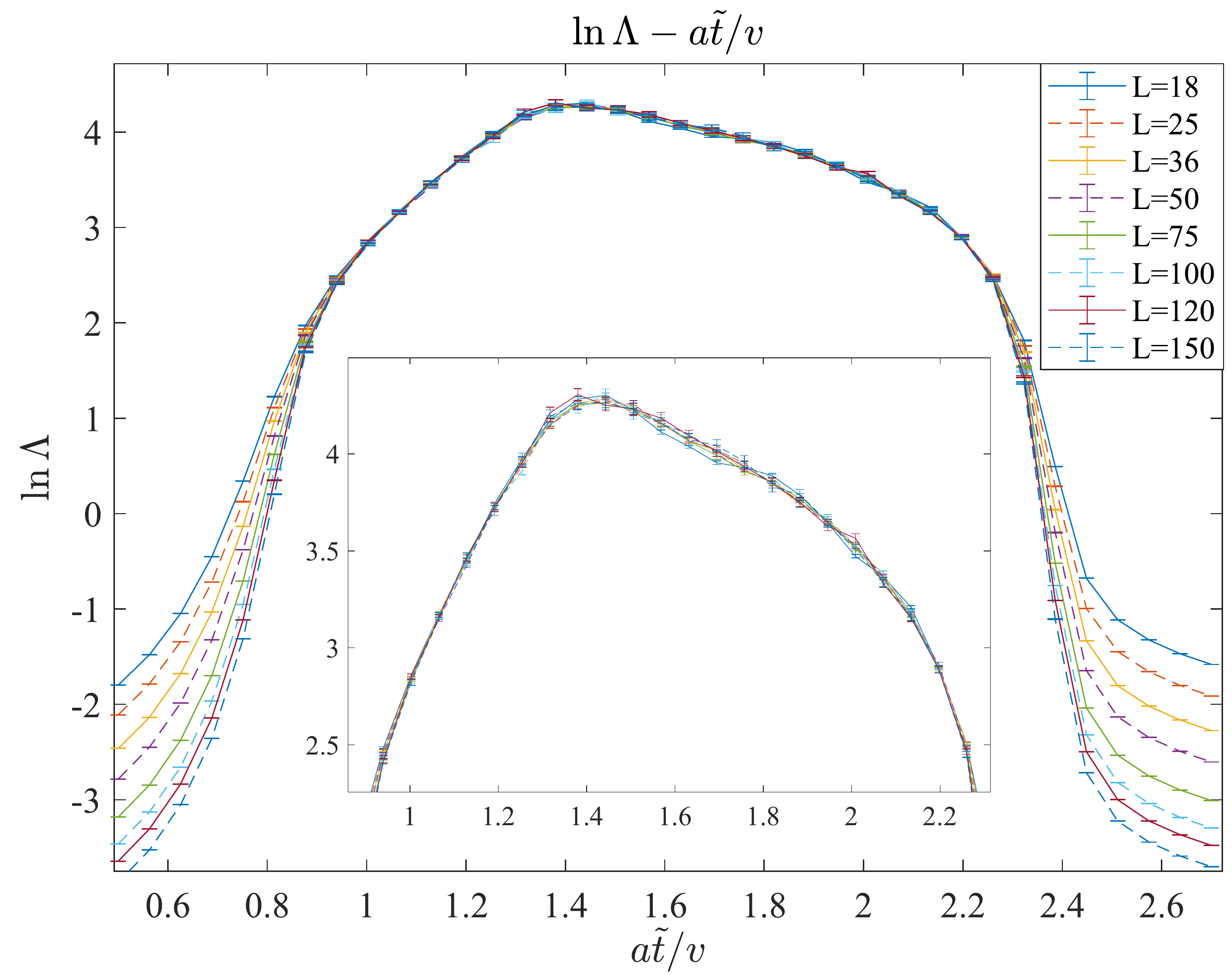}}
    \subfigure[~$\ln\Lambda$ versus $W$ with $\tilde{t}=\pi v/2a$]{\label{Lambda-W-ga0.5pi-full}
	\includegraphics[width=.44\linewidth]{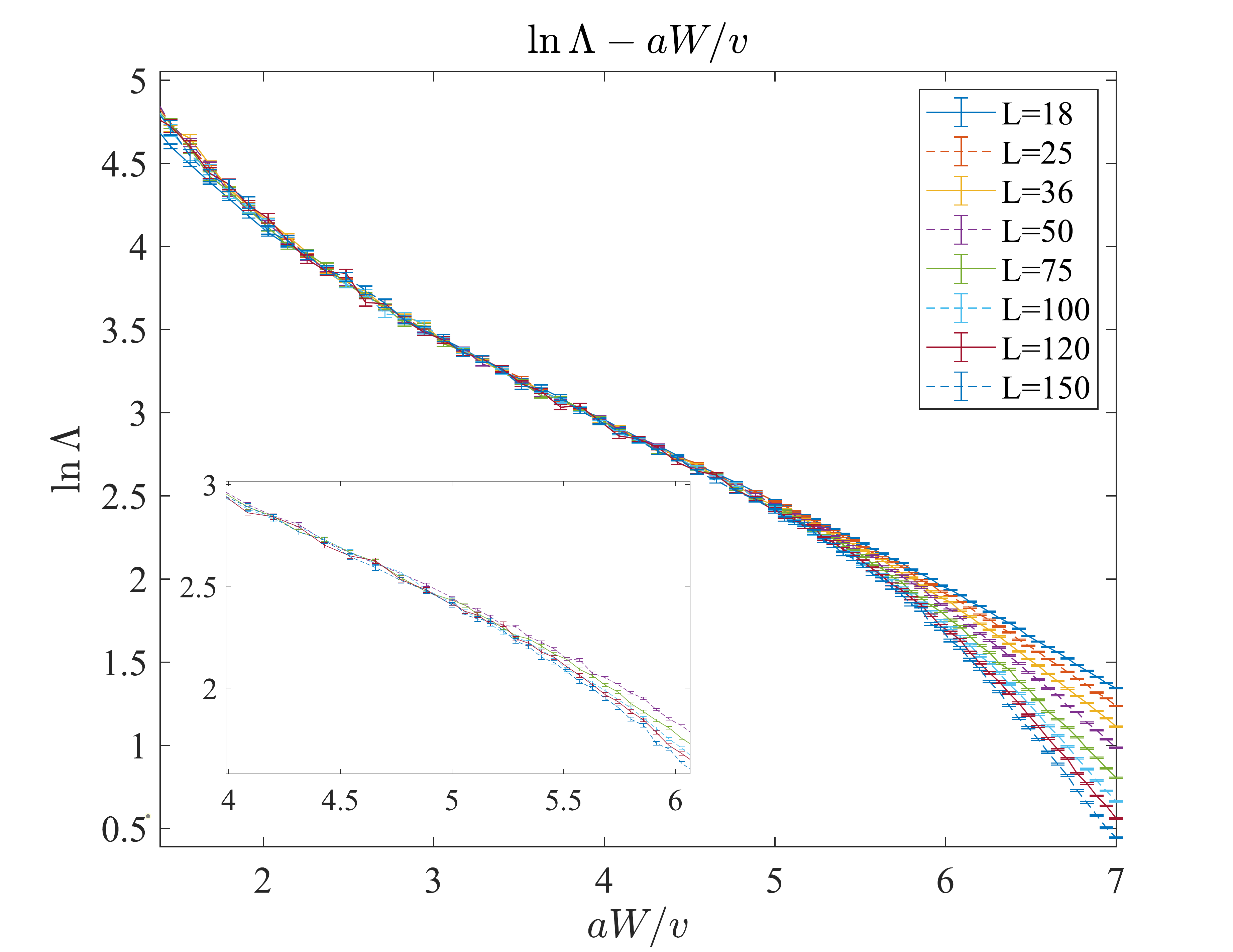}}
    \caption[]{Phase diagram and localization lengths of lattice model $H_{8B}$ (\cref{TB-H-real}) with $E_F=0$. (a) The phase diagram in $\tilde{t}-W$ plane. The phase boundary is determined by the p-value of weighted linear regression of $1/\Lambda-L$ for given $\tilde{t}, W$ and $L\geq50$. The regression weight of a data point with variance $\sigma^2$ is proportional to $1/\sigma^2$. We regard the system as localized if the p-value of fitted slope is less than $0.05$, \ie the positive fitted slope is significant at the 5\% significance level. Otherwise, we regarded it as critical. (See \cref{subsec:numerical-lattice-full} for more details) The green region in (a) indicates the critical metal phase enclosed by the localized phase. And the red lines in (a) represent the parameters used in (b) and (c), respectively. (b) Localization length as the function of $\tilde{t}$ with $W=2v/a$ and various transversal sizes $L$. The longitudinal size $M=10^7$ and the data precision ($\sigma_{\Lambda}/\Lambda$)  reaches 3\%. The inset in (b) is a zoomed plot of the critical region. (c) Localization length as the function of $W$ with $\tilde{t}=\pi v/2a$, where the inset magnifies the transition zone between the critical and localized regions (we ignore the curves of $L=18\sim36$ in the inset to avoid the finite size effects). The longitudinal size and data precision are the same as (b).}
    \label{ll-full}
\end{figure}

We then proceed to the lattice model $H_{8B}$ defined in \cref{TB-H-real}. Fig.~\ref{phase-diagram-full} depicts the phase diagram in parameter space $(W,\tilde{t})$ at $E_F=0$, where the green region represents the critical metal phase (CMP) and the white region corresponds to localized phase (LP). 
In LP, the normalized localization length $\Lambda$ decreases as the transversal size $L$ increases, while in CMP, $\Lambda$ remains invariant as $L$ increases. Hence, the scale dependence of $\Lambda(L)$ can be used for determining the phase boundary. 
Actually, the phase boundary is determined by the p-value of t-statistic of weighted linear regression $1/\Lambda=b_0+b_1 L$ for given $\tilde{t}$ and $W$. (We offer a heuristic description of t-test and p-value in later paragraphs. Readers can refer to Ref.~\cite{t-test} for more details.) We denote the fitted slope of some data set as $\hat{b}_1$ and the true slope as $b_1$. There should be $\hat{b}_1>0$ and $\hat{b}_1\approx0$ inside LP and CMP, respectively. Since there is no extended state (inside which $\hat{b}_1<0$), we can use $|\hat{b}_1|$ as a criterion.
More concretely, we regard the system as localized when the p-value of $|\hat{b}_1|$ is smaller than the commonly used threshold $0.05$. 
That means, if the true slope $b_1=0$, we have less than $5\%$ probability to obtain a fitted slope whose absolute value $>|\hat{b}_1|$. Therefore, we should no longer view the system as critical ($b_1=0$) and regard it as localized instead. In statistical terminologies, scale dependence of $\Lambda$ is significant at the 5\% significance level, so that we reject the null hypothesis. 
The system is regarded as critical when the p-value is greater than $0.05$.

In this paragraph, we justify the choice of weighted linear regression $1/\Lambda=b_0+b_1 L$. First, the scale dependence of $\Lambda$ (also $1/\Lambda$) is insignificant near the phase boundary. Hence, no matter what the relation is, we can always take a linear approximation and use linear regression to investigate the scale dependence. Second, estimation techniques of linear regression, \eg p-value used here, need independently normally distributed data. We derive $\Lambda(L)$ from the smallest Lyapunov exponent $\nu_s=L/\Lambda(L)$, and $\nu_s$ is obtained from Eq.~(\ref{true-mean-LE}) so that only $1/\Lambda\!\propto\!\nu_s$ is normally distributed. Thus, we choose $1/\Lambda$ rather than $\Lambda$ to do the linear regression. Third, the variance of $1/\Lambda$ varies with $L$. Therefore, we should use the least square method with weights $\{1/\sigma_i^2\vert i=1,2,\cdots,n\}$, where $\{\sigma_i^2\}$ is the variance of corresponding data points $\{1/\Lambda(L_i)\}$. Last, we take the data with $L\geq 50$ to avoid the finite size effects, which will result in an underestimation of CMP in our case.

In the following paragraphs, we present a brief introduction of t-test based on the p-value. Recall that the unknown true values of regression slope and intercept are $b_1$ and $b_0$, respectively. The fitted slope and intercept from some data set $\{1/\Lambda(L_i)\vert i\in[1,\,n]\}$ are $\hat{b}_1$ and $\hat{b}_0$, respectively. As mentioned in the last paragraph, each data point satisfies $1/\Lambda(L_i)=1/\tilde{\Lambda}(L_i)+\delta(1/\Lambda(L_i))$, where $\tilde{\Lambda}(L_i)$ is the unknown true value of $\Lambda(L_i)$ and $\delta(1/\Lambda(L_i))$ is a normally distributed random variable with standard error $\sigma_i$. The unbiased estimation of $\sigma_i$ has been described in \cref{subsec:basic-localization-length}. Each of the transversal sizes $\{L_i\}$ certainly has no error, nevertheless we can take an average of them $\bar{L}=\frac{1}{n}\sum_{i=1}^n L_i$.  Then, we point out without proof \cite{strutz2011data} that the unbiased estimation of the variance of $b_1$ is given by
\begin{equation}
    \hat{\sigma}_{b_1}^2=\frac{\frac{1}{n-2}\sum_{i=1}^n(\frac{1}{\Lambda(L_i)}-\hat{b}_0-\hat{b}_1 L_i)^2/\sigma_i^2}{\sum_{i=1}^n (L_i-\bar{L})^2/\sigma_i^2 }\ . 
\end{equation}
The factor $\frac{1}{n-2}$ represents that the statistical degrees of freedom is $n-2$. Then, the following random variable satisfies the Student's t-distribution with $n-2$ degrees of freedom (denoted as $t_{n-2}$):
\begin{equation}
\begin{aligned}
    \Delta_1=\frac{\hat{b}_1-b_1}{\hat{\sigma}_{b_1}}\sim t_{n-2} \ . 
\end{aligned}
\end{equation}
We omit the concrete formula of $t_{n-2}$. We merely point out that, similar to the normal distribution, the curve of probability density of Student's t-distribution is symmetric and bell-shaped around zero but with heavier tails. As the number of degrees of freedom $(n\!-\!2)\rightarrow \infty$, $t_{n-2}$ approaches the normal distribution with mean $0$ and variance $1$. Now, we can define the p-value as the probability of event $|\Delta_1|>C\ge0$ and denote it as $P(C)$. Since $P(C)$ decreases as $C$ increases, it is unlikely to obtain a large $\hat{b}_1/\hat{\sigma}_{b_1}$ when $b_1=0$. If the regression results in a large $\hat{b}_1/\hat{\sigma}_{b_1}$, we can reject the hypothesis that $b_1=0$ ($\Lambda$ is independent of $L$ and the system is critical). As a consequence, we should accept that the system is localized. In our case, we choose $0.05$ as the threshold, \ie the system is regarded as localized when $P(|\hat{b}_1/\hat{\sigma}_b|)\leq5\%$ and otherwise regarded as critical. The above procedure is called a t-test (based on p-value) with threshold $0.05$, where $t$ represents the Student's t-distribution.

The numerical results of $H_{8B}$ are shown in \cref{ll-full}. Fig.~\ref{phase-diagram-full} illustrate the phase diagram of $H_{8B}$ in the space of $W-\tilde{t}$ ($E_F=0$). The phase boundary close to the clean limit ($W<v/a$) is hard to determine since the localization lengths have large relative errors. Hence, we focus on $W>v/a$. From Fig.~\ref{phase-diagram-full} and Fig.~\ref{bands-lattice}, we can see that CMP roughly arises inside the metallic region in the clean limit (about $1<a\tilde{t}/v<2.5$).
And the bandwidth (of all bands) of $H_{8B}$ in the metallic region is about $8v/a$, which is comparable to the maximal disorder strength reached by CMP ($5v/a$). 
Fig.~\ref{Lambda-ga-W2-full} and \ref{Lambda-W-ga0.5pi-full} show $\Lambda$ as functions of $\tilde{t}$ and $W$, respectively. The parameters used in these plots are indicated by the red lines in Fig,~\ref{phase-diagram-full}. As we can see, $\Lambda$ is indeed independent of $L$ in CMP. During the growth of $\tilde{t}$, the transition LP-CMP-LP (\cref{Lambda-ga-W2-full}) is in accord with the network model (\cref{Lambda-theta--L-network} and its mirror reflection for $\theta>0$). We also calculate $\ln\Lambda$ vs. $\tilde{t}$ in LP, and the results are shown in \cref{full-LP}. As expected, $\Lambda$ always decreases as $L$ increases.

As explained in \cref{subsec:basic-localization-length}, we need further validations of the CMP in $H_{8B}$, since the localization length is fairly large in CMP: $\rho_{\mathrm{q-1D}}= e^{3\sim4}L\approx20\sim50L$. 
To rule out this problem, we take two points in CMP and calculate $\Lambda(L)$ up to $L_{\max}=500$. 
The results are shown in \cref{Lambda-L-ga0.5pi-full} \& \ref{Lambda-L-ga2-full}, and no significant scale dependence of $\Lambda$ is found. As explained in \cref{subsec:basic-localization-length}, we have two more evidences that validate the CMP in $H_{8B}$: CMP does not shrink as $L$ increases (\cref{Lambda-L-ga2-full}) and the local Chern markers highly fluctuate in CMP (\cref{Cmarker-fullon-site-a}). In addition, to make sure that the choice $E_F = 0$ is not special, we calculate $\Lambda$ versus $\tilde{t}$ with $E_F=-\pi v/10a$ (\cref{full-EF_no_zero}), which still shows the existence of CMP.

\subsection{Simplified eight-band lattice model \texorpdfstring{$H_{8B}'$}{H8Bp} }
\label{subsec:numerical-lattice-simplified}

\begin{figure}[h]
	\centering
	\subfigure[~Phase diagram of $H_{8B}'$ with $E_F=0,\,A=1.2$ in the parameter space spanned by $W,\tilde{t}$]{\label{phase-diagram-simplify}
    \includegraphics[width=.7\linewidth]{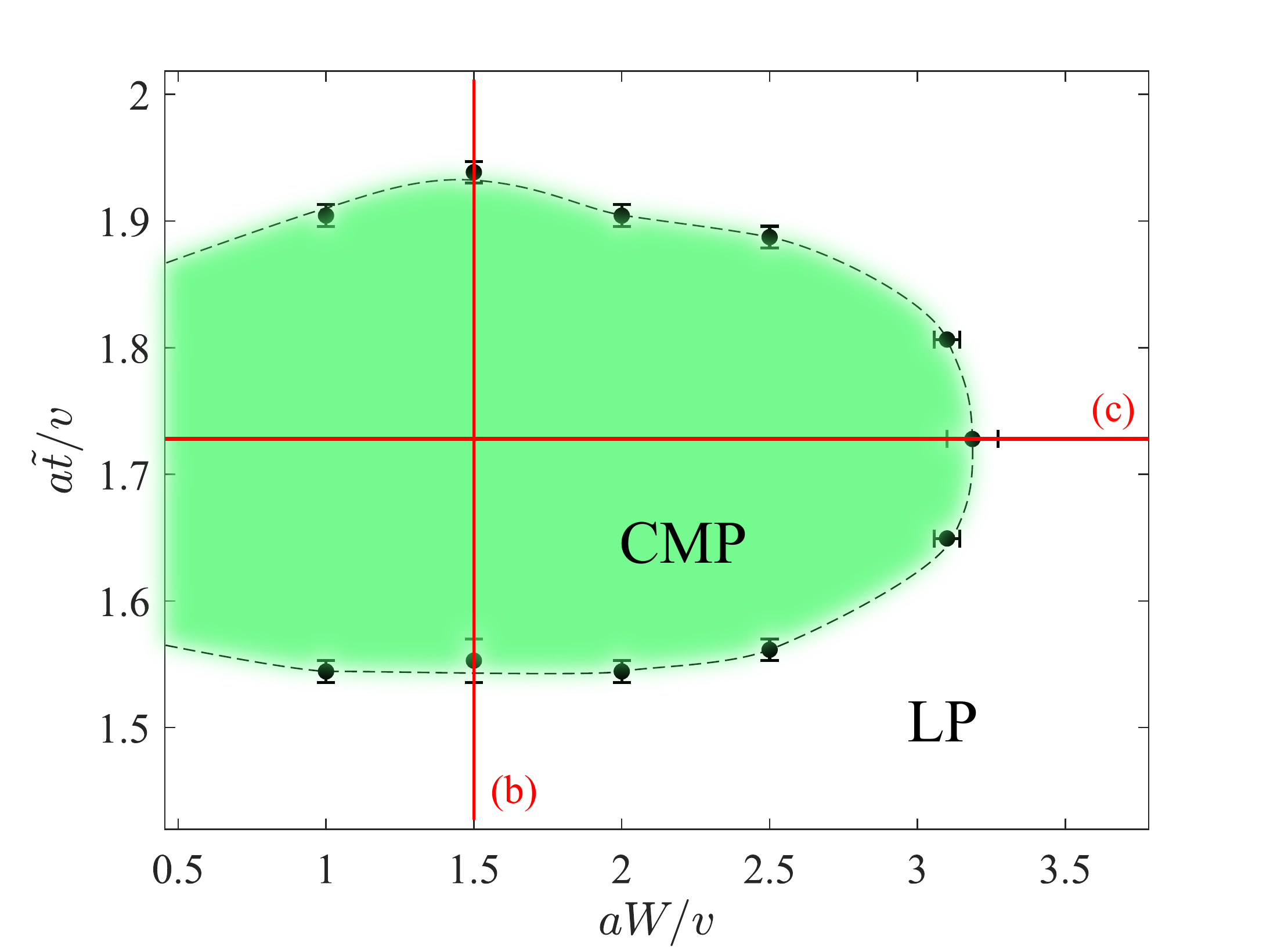}}
    \\
	\subfigure[~$\ln\Lambda$ versus $\tilde{t}$ with $W=1.5v/a$]{\label{Lambda-ga-W1.5-simplify}
	\includegraphics[width=.47\linewidth]{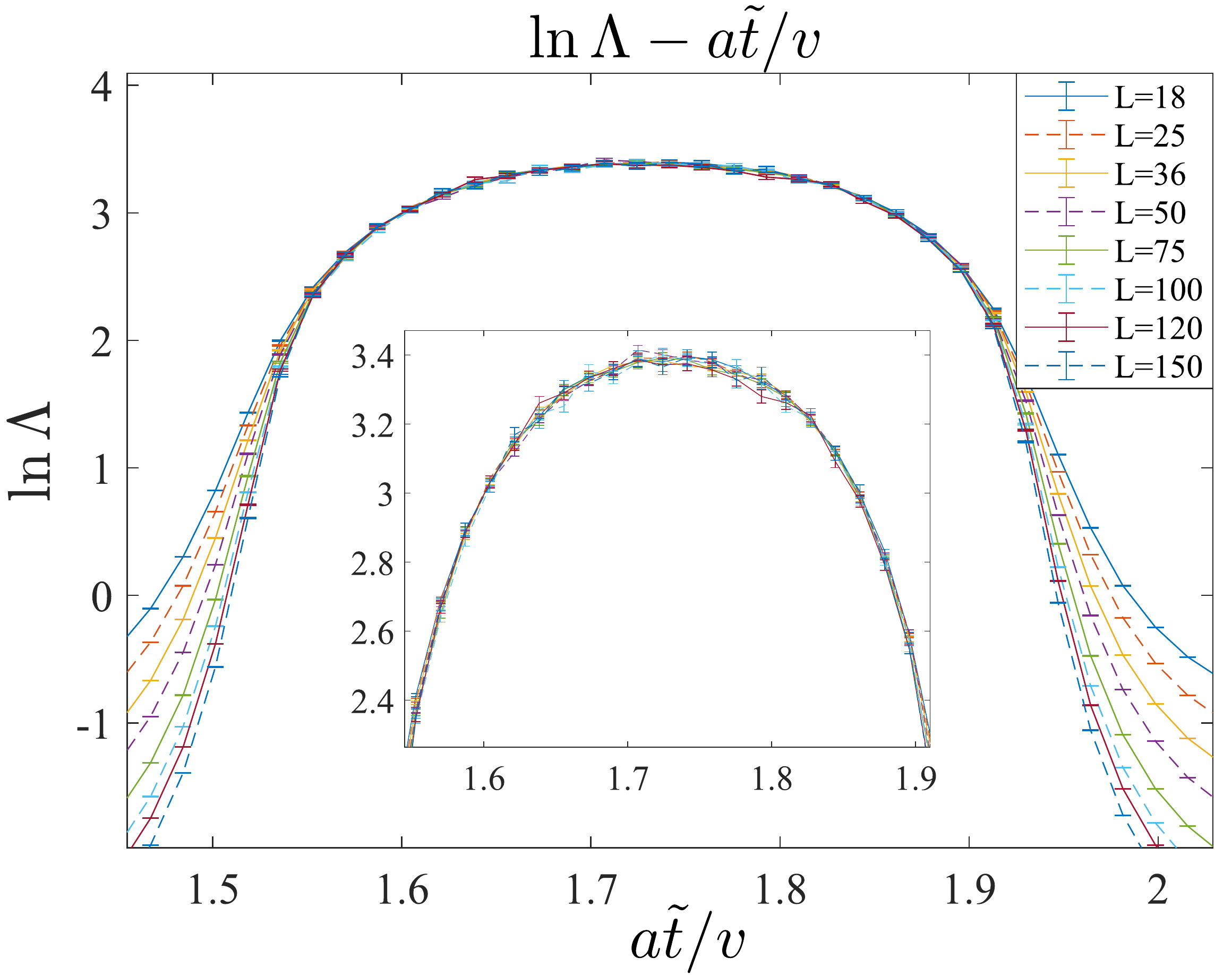}}
    \subfigure[~$\ln\Lambda$ versus $W$ with $\tilde{t}=1.1\pi v/2a$]{\label{Lambda-W-ga0.55pi-simplify}
	\includegraphics[width=.48\linewidth]{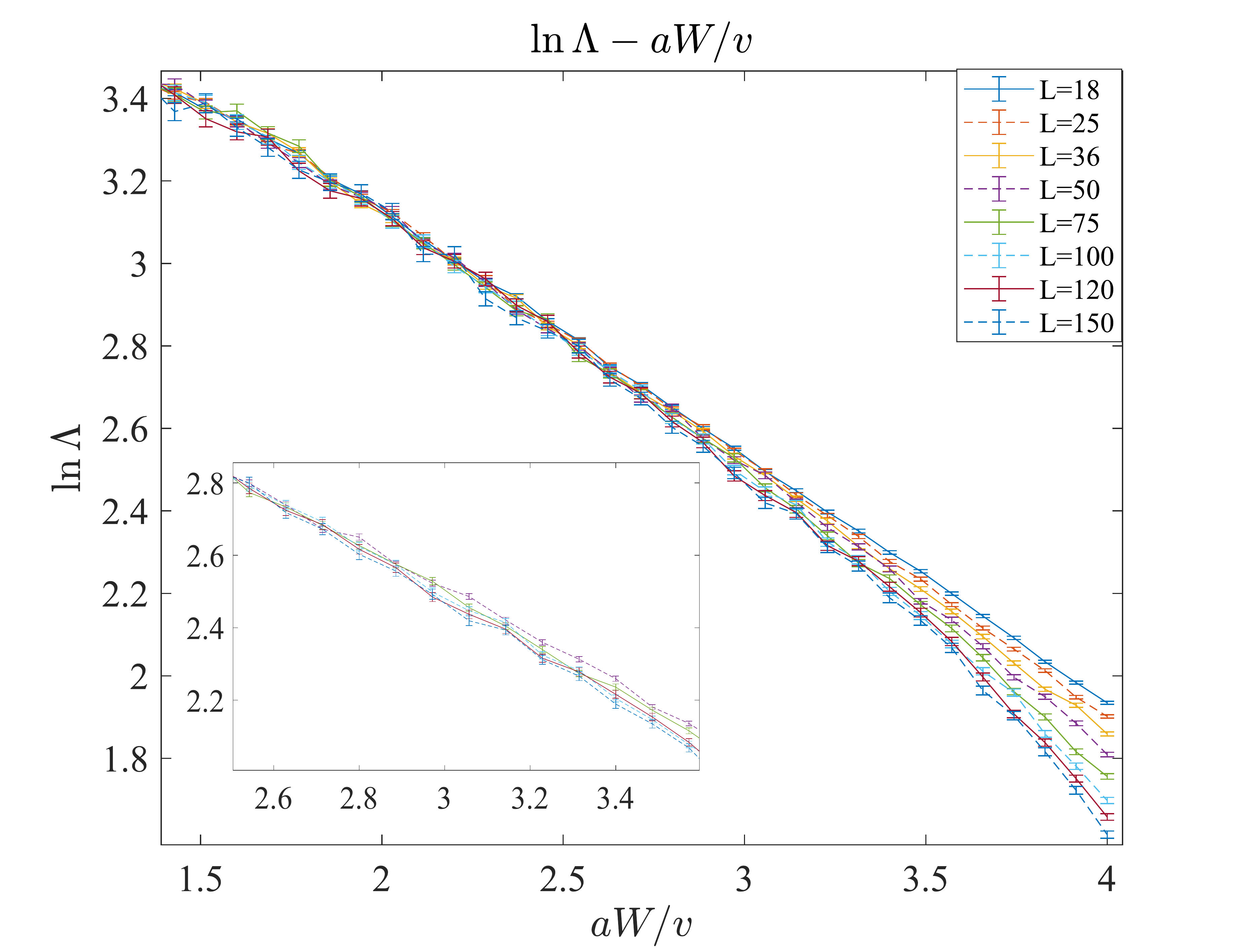}}
    \caption[]{Phase diagram and localization length of $H_{8B}'$ (\cref{simplified-H-reciprocal}) with $E_F=0$ and $A=1.2$. (a) The phase diagram in $\tilde{t}-W$ plane. The phase boundary is determined in the same way as Fig.~\ref{ll-full}. The green region and red lines also have the same meanings as that of Fig.~\ref{ll-full}. (b) Localization length as the function of $\tilde{t}$ with $W=1.5v/a$ and various transversal sizes. The longitudinal size $M=10^7$ and the data precision ($\sigma_{\Lambda}/\Lambda$) reaches 2\%. The inset in (b) is a zoomed plot of the critical region. (c) Localization length as the function of $W$ with $\tilde{t}=1.1\pi v/2a$, where the inset magnifies the transition zone between the critical and localized regions (we ignore the curves of $L=18\sim36$ in the inset to avoid the finite size effects). The longitudinal size and data precision are the same as (b).}
    \label{ll-simplify}
\end{figure}

\begin{figure}[h]
	\centering
    \includegraphics[width=.7\linewidth]{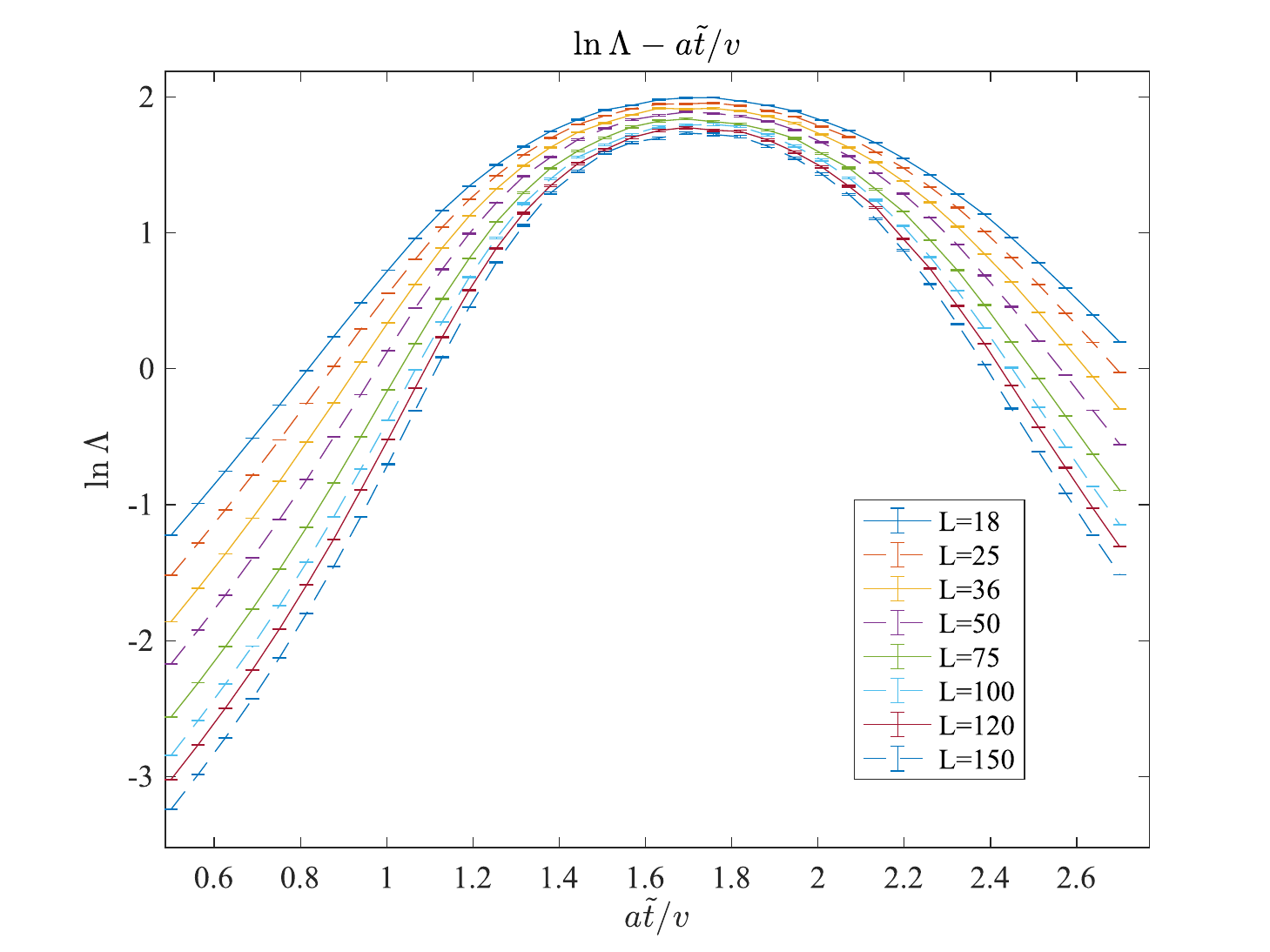}
    \caption[]{Localization length $\Lambda$ as the function of $\tilde{t}$ of $H_{8B}$ (\cref{TB-H-real}) with $W=6v/a$ (inside the localized phase) and $E_F=0$. The longitudinal size $M=10^7$ and the data precision ($\sigma_{\Lambda}/\Lambda$) reaches $1\%$. As expected for the localized phase, $\Lambda$ decreases as the transversal size $L$ increases. }
    \label{full-LP}
\end{figure}

\begin{figure}[h]
	\centering
    \includegraphics[width=.7\linewidth]{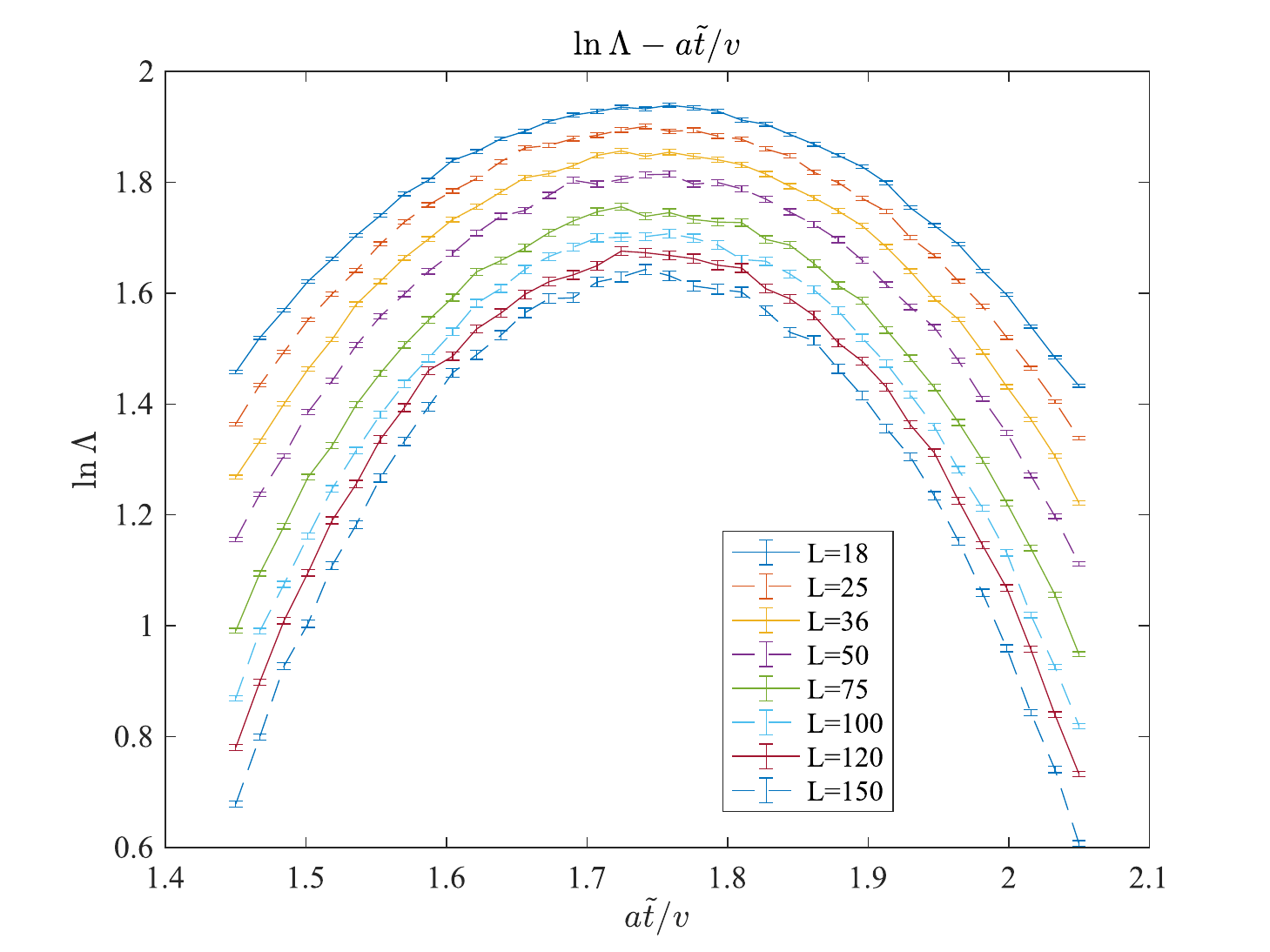}
    \caption[]{Localization length $\Lambda$ as the function of $\tilde{t}$ of $H_{8B}'$ (\cref{simplified-H-reciprocal}) with $A=1.2,\,W=4v/a$ (inside the localized phase) and $E_F=0$. The longitudinal size $M=10^7$ and the data precision ($\sigma_{\Lambda}/\Lambda$) reaches $0.9\%$. As expected for the localized phase, $\Lambda$ decreases as the transversal size $L$ increases. }
    \label{simp-LP}
\end{figure}

The results of $H_{8B}'$ (\cref{simplified-H-reciprocal}) are shown in Fig.~\ref{ll-simplify} and \cref{simp-LP}. Fig.~\ref{phase-diagram-simplify} shows the phase diagram in $\tilde{t}-W$ plane ($E_F=0$), and the phase boundary is determined in the same way as $H_{8B}$. Fig.~\ref{Lambda-ga-W1.5-simplify} \& \ref{Lambda-W-ga0.55pi-simplify} show the localization length along paths indicated by the red lines in Fig.~\ref{phase-diagram-simplify}. The same to $H_{8B}$, the transition pattern along the $\tilde{t}$ axis is LP-CMP-LP (\cref{Lambda-ga-W1.5-simplify}). \cref{simp-LP} shows $\ln \Lambda-\tilde{t}$ in LP, and, as expected, $\Lambda$ decreases as $L$ increases.

Combining Fig.~\ref{phase-diagram-simplify} with Fig.~\ref{bands-simplified-TB}, we see that CMP can arises inside the semi-metal region ($2a\tilde{t}/\pi v\in[1,A=1.2]$) where the four Dirac cones in \cref{bands-simplified-TB} exist. And CMP of $H_{8B}'$ persists to a disorder strength ($\approx 3v/a$) that is comparable to the bandwidth of the two bands in \cref{bands-simplified-TB} ($\approx 3v/a$). Numerical calculations on other $A$ show that CMP centers at $2a\tilde{t}/\pi v=\sqrt{A}$ and extends most to around $\pi v/2a\!\sim \!A\pi v/2a$.
It is trivial that CMP cannot arise away from the semi-metal region, since the band structure is gapped. But it is nontrivial that, for moderate disorder strength $W<1.5v/a$, CMP can approximately fill up (in terms of $\tilde{t}$) the entire semi-metal region ($a\tilde{t}/v=\pi/2\!\sim \!0.6\pi$). Therefore, the existence of the four Dirac cones in \cref{bands-simplified-TB}, \ie the transition region between the two OAI's, must have a tight connection with the critical metal phase. Also, considering that we offer a quantitative mapping from the network model to $H_{8B}'$ in \cref{sec:map-network-lattice}, there must be a connection between the OAI transition and the conductive percolation state.  

Again, the phase boundary close to the clean limit ($W<v/a$) is hard to determine. 
Besides the large error in $\Lambda$ as in $H_{8B}$, here we have another problem when $W$ is small: fluctuation in $DOS_F$ (the density of states at the Fermi level). 
As we know, the Bloch wave vector is discrete for a finite system. 
In our case, although the longitudinal size $M$ can reach $10^7$, the transversal size $L$ will not exceed the order of $10^2$. 
See \cref{bands-simplified-TB}, due to the coarse quantization of the transversal momentum component, only a (clean and $E_F=0$) system with fine-tuned $\tilde{t}\,\&\,L$ can `see' the Dirac cones; otherwise, the system will become a gapped insulator. 
Since $\tilde{t}$ affects the position of the Dirac cones, when the system is clean enough, $\Lambda$ will fluctuate as $\tilde{t}$ varies in the semi-metal region. Note that $DOS_F$ is also sensitive to $\tilde{t}$ and $L$ when $E_F\neq0$, hence the fluctuation of $\Lambda$ also exist in these cases. The weaker $W$ is, the larger $L$ is needed to erase the fluctuation. This makes it expensive to investigate the phase boundary near the clean limit. In practice, we ignore the data with $L<50$ for the determination of phase boundary to avoid the finite size effect. 

\begin{figure}[h]
	\centering
	\subfigure[~ $\tilde{t}=\pi v/2a,\,W=2v/a$ in $H_{8B}$]{\label{Lambda-L-ga0.5pi-full}
	\includegraphics[width=.48\linewidth]{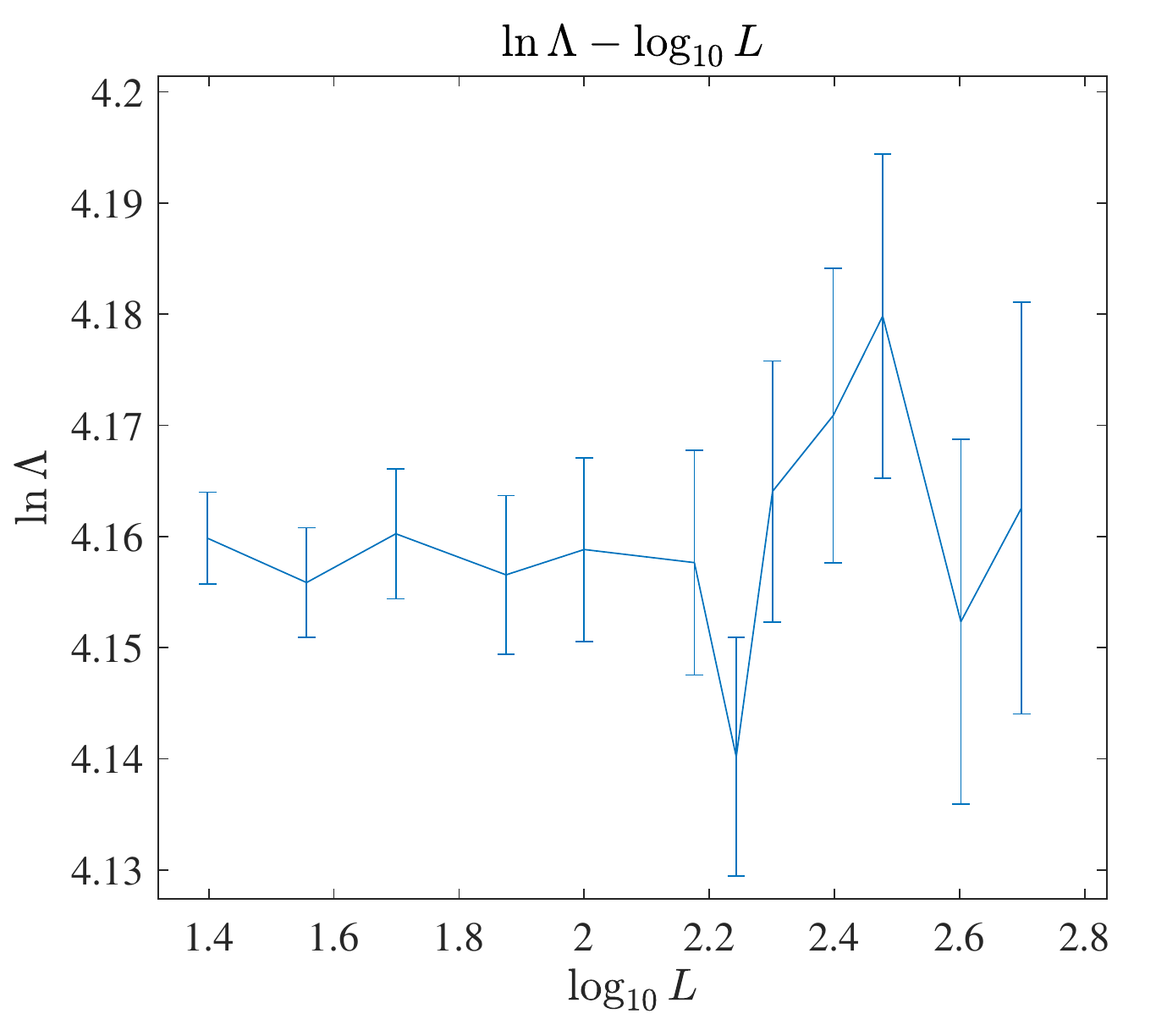}}
    \subfigure[~ $\tilde{t}=2v/a,\,W=2v/a$ in $H_{8B}$]{\label{Lambda-L-ga2-full}
	\includegraphics[width=.48\linewidth]{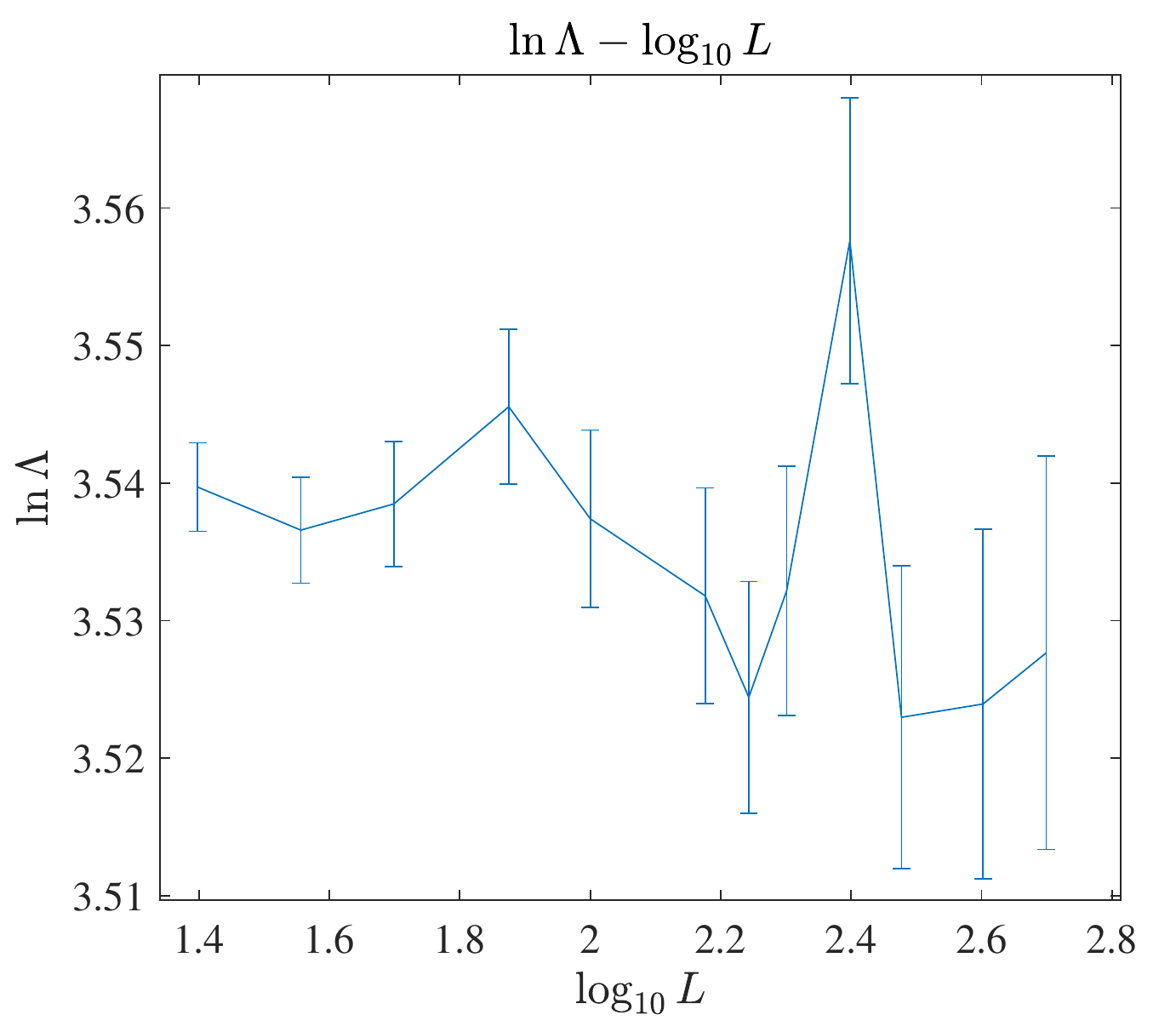}}\\
	\subfigure[~$A=1.2,\,\tilde{t}=1.1\pi v/2a,\,W=1.5v/a$ in $H_{8B}'$]{\label{Lambda-L-ga0.55pi-simp}
	\includegraphics[width=.48\linewidth]{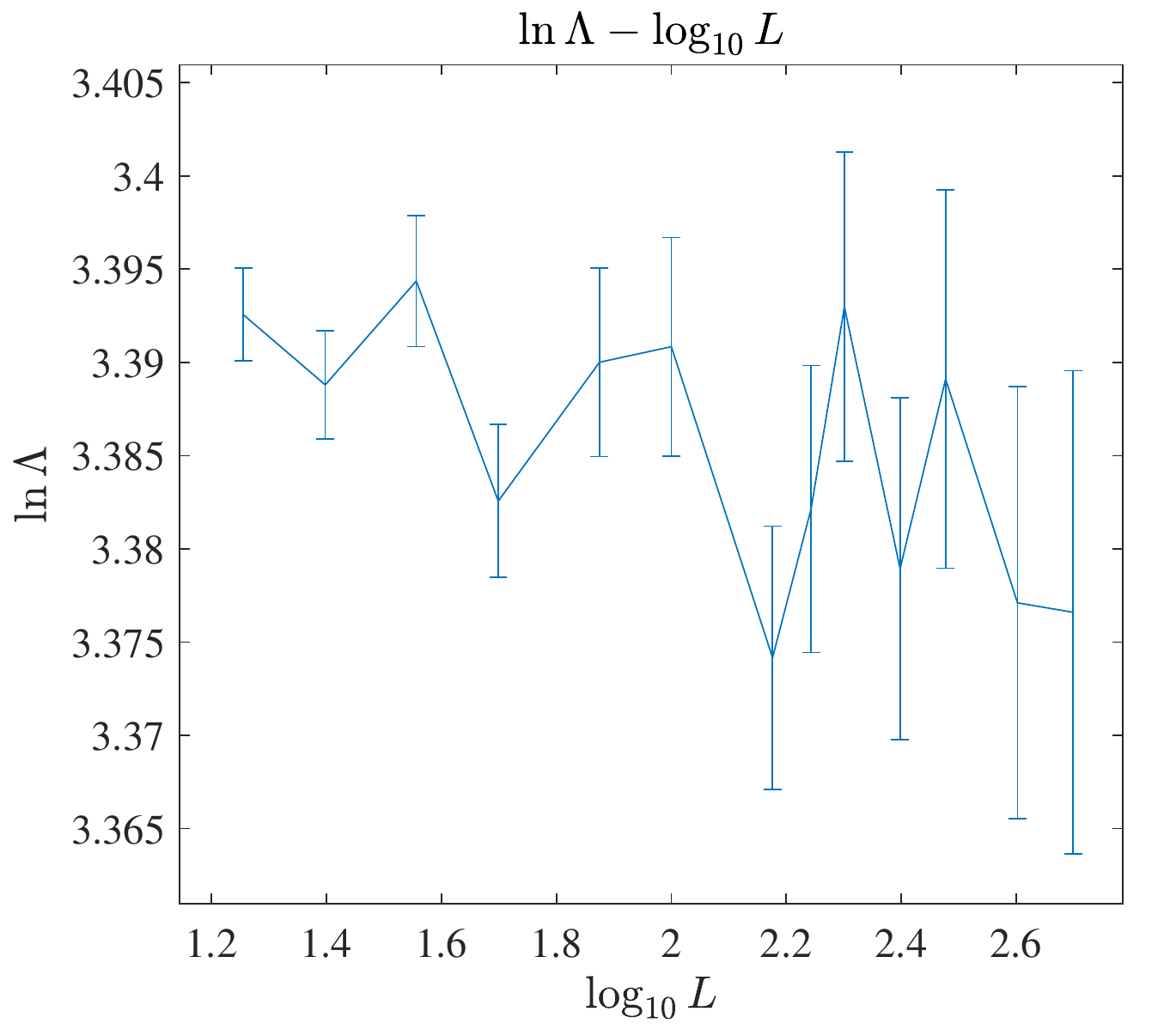}}
    \subfigure[~$A=1.2,\,\tilde{t}=1.15\pi v/2a,\,W=1.5v/a$ in $H_{8B}'$]{\label{Lambda-L-ga0.575pi-simp}
	\includegraphics[width=.48\linewidth]{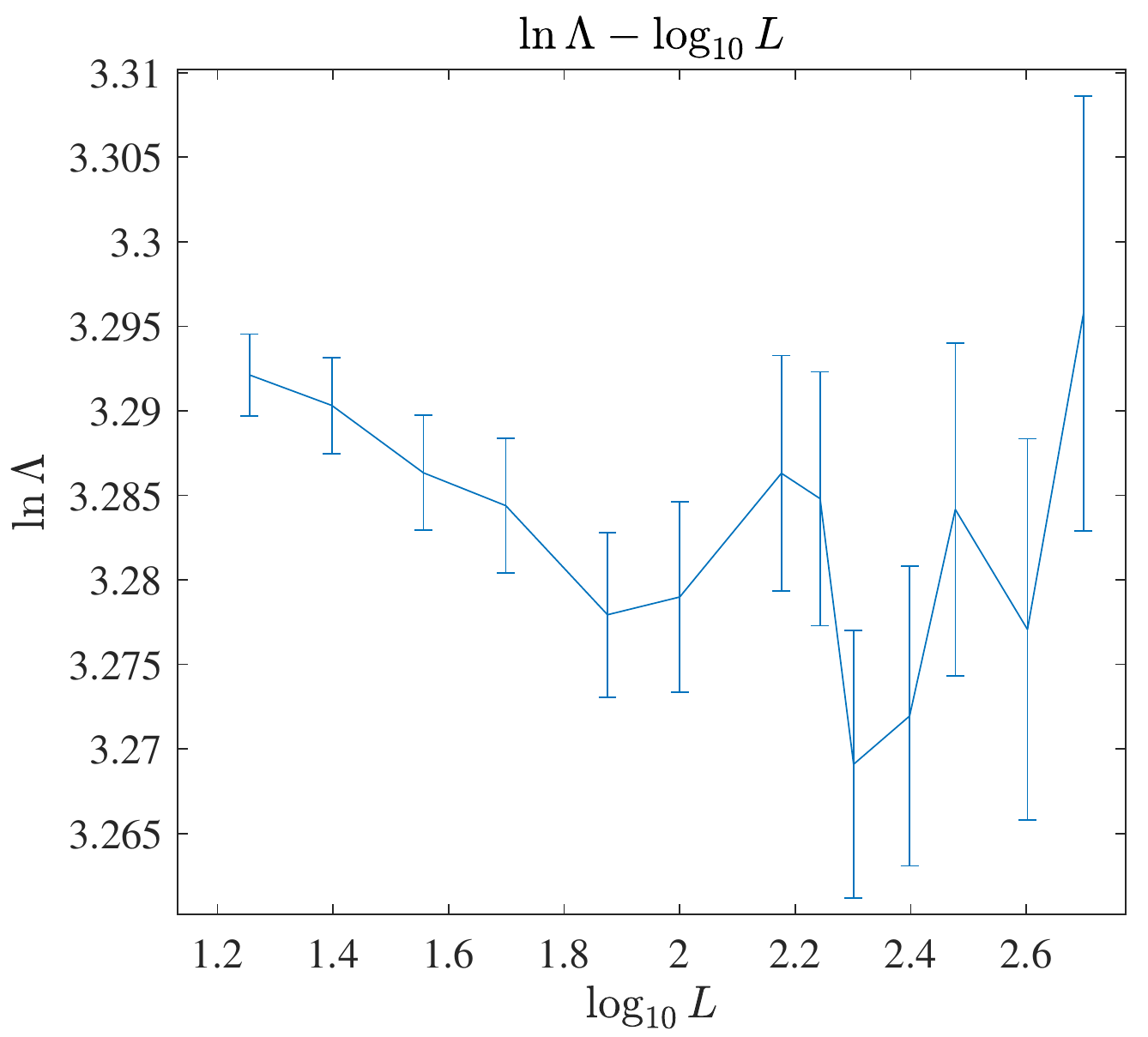}}\\
    \caption[]{The normalized localization length $\Lambda$ versus transversal size $L$ in CMPs of $H_{8B}$ ((\cref{TB-H-real})) and $H_{8B}'$ (\cref{simplified-H-reciprocal}). $\mathrm{(a),(b)}$ $\Lambda$ as the function of $L$ for $H_{8B}$ with $\tilde{t}=\pi v/2a,\,W=2v/a$ and $\tilde{t}=2v/a,\,W=2v/a$, respectively. $\mathrm{(c),(d)}$ $\Lambda$ as the function of $L$ for $H_{8B}'$ with $A=1.2,\,W=1.5v/a,\,\tilde{t}=1.1\pi v/2a$ and $A=1.2,\,W=1.5v/a,\,\tilde{t}=1.15\pi v/2a$, respectively. For all these four plots, the maximum transversal size reaches $L_{\max}=500$, the longitudinal size $M=10^8$ and the data precision ($\sigma_{\Lambda}/\Lambda$) reaches $1\%$. None of these plots shows a significant decline of $\ln \Lambda$ when $L\rightarrow\infty$, which indicates that the scale independence of $\Lambda$ in the CMPs in \cref{phase-diagram-full} and \cref{phase-diagram-simplify} is not due to finite size effects. Therefore, this figure support the existence of CMPs in $H_{8B}$ and $H_{8B}'$. }
    \label{ll-test}
\end{figure}

\begin{figure}[h]
	\centering
    \includegraphics[width=.7\linewidth]{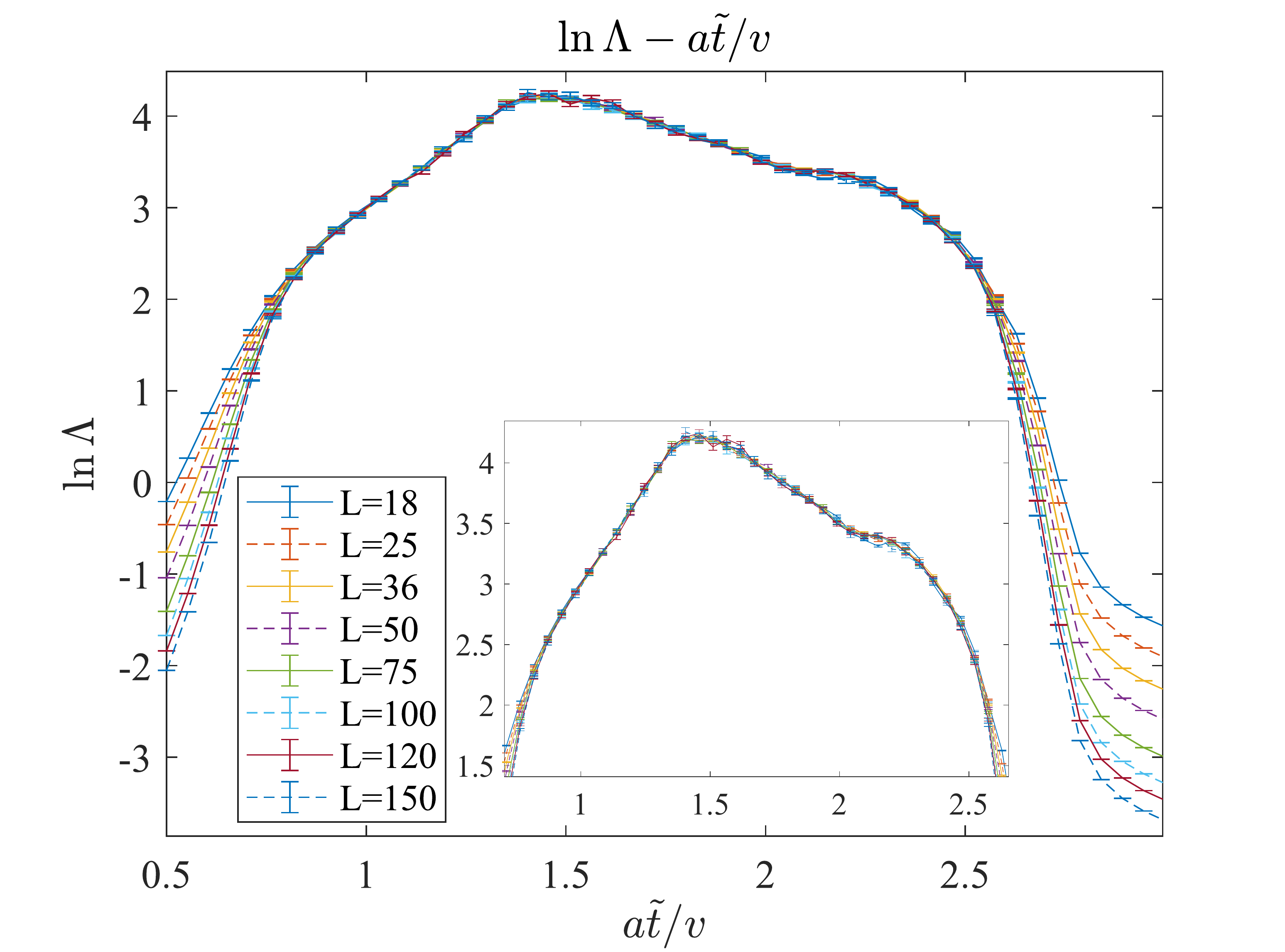}
    \caption[]{Localization length $\Lambda$ as the function of $\tilde{t}$ of $H_{8B}$ (\cref{TB-H-real}) with $W=2v/a$ and $E_F=-\pi v/10a$. The longitudinal size $M=10^7$ and the data precision ($\sigma_{\Lambda}/\Lambda$) reaches $3\%$. The middle region in this figure is a critical phase, since $\Lambda$ remains invariant as $L\rightarrow\infty$. The inset shows the zoomed critical region. This figure shows the existence of CMP in $H_{8B}$ for $E_F\neq0$. Thus, the CMP in $H_{8B}$ does not rely on the special choice $E_F=0$. }
    \label{full-EF_no_zero}
\end{figure}

\begin{figure}[h]
	\centering
    \includegraphics[width=.7\linewidth]{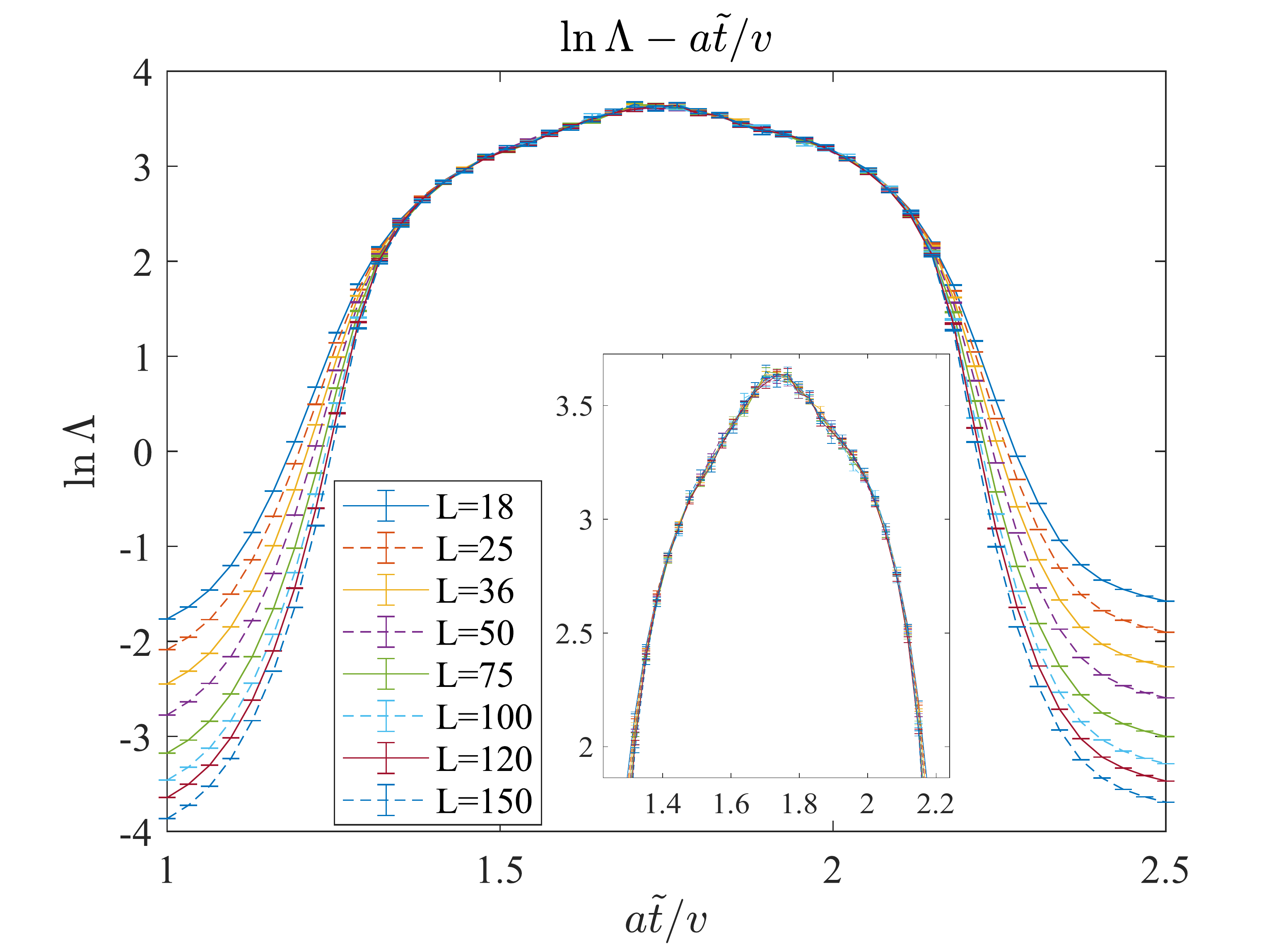}
    \caption[]{Localization length $\Lambda$ as the function of $\tilde{t}$ of $H_{8B}'$ (\cref{simplified-H-reciprocal}) with $A=1.2,\,W=1.5v/a$ and $E_F=-\pi v/10a$. The longitudinal size $M=10^7$ and the data precision ($\sigma_{\Lambda}/\Lambda$) reaches $3\%$. The middle region in this figure is a critical phase, since $\Lambda$ remains invariant as $L\rightarrow\infty$. The inset shows the zoomed critical region. This figure shows the existence of CMP in $H_{8B}'$ for $E_F\neq0$. Thus, the CMP in $H_{8B}'$ does not rely on the special choice $E_F=0$ and the \emph{averaged} chiral symmetry at $E_F=0$ (see the last but one paragraph of \cref{subsec:simplified-model}).   }
    \label{simp-no_ave_chiral}
\end{figure}

We also executed the examination of scale dependence of $\Lambda$ up to $L_{\max}=500$. The results are shown in \cref{Lambda-L-ga0.55pi-simp} \& \ref{Lambda-L-ga0.575pi-simp} and no significant scale dependence can be found. In addition, to make sure that the choice $E_F = 0$ is not special, we calculate $\Lambda$ versus $\tilde{t}$ with $E_F=-\pi v/10a$ (\cref{simp-no_ave_chiral}), which still shows the existence of CMP.

\clearpage 
\subsection{Local Chern markers of $H_{8B}$ and $H_{8B}'$}
\label{subsec:LCM}

In \cref{sec:numerical-localization-length}, we show that the network model (\cref{network-decoupled-chiral-wires} \& (\ref{network-delta-scattering-potential})), $H_{8B}$ (\cref{TB-H-real}), and $H_{8B}'$ (\cref{simplified-H-reciprocal}) all have critical metal phases. Recall that the CMP in network model represents a conductive percolation system (Fig.~\textcolor{green}{1(b)}), and the CMPs in $H_{8B}$ and $H_{8B}'$ correspond to transitions between trivial OAIs (\cref{subsec:8-bands full model} \& \ref{subsec:simplified-model}). Also, in \cref{sec:map-network-lattice}, we show a quantitative mapping from the network model to $H_{8B}$ and $H_{8B}'$. Then, it is natural to ask whether the CMPs in $H_{8B}$ and $H_{8B}'$ represent conductive percolation systems. Especially for $H_{8B}'$, the answer is far from obvious since the band structure of $H_{8B}'$ near the Fermi level $E_F=0$ (Fig.~\ref{bands-simplified-TB}) significantly differs from that of the network model with $E_F=\pi v/4a$ (\cref{bands-network} \& \ref{bands3d-lattice}). Fortunately, the local Chern marker (LCM) enables us to investigate this question directly.

Refs.~\cite{loring2011disordered,Prodan2010Entanglement} showed that the Chern number of a gapped lattice Hamiltonian can be computed from the Bott index as 
\begin{equation} \label{eq:Bott}
    C = \frac1{2\pi} \mathrm{Im} \brak{ \mathrm{Tr} \ln (\Phi_x \Phi_y \Phi_x^\dagger \Phi_y^\dagger) }
\end{equation}
where 
\begin{equation}
    \Phi_x = \hat{P} \exp(i \frac{2\pi}{L} \hat{x} ) \hat{P},\qquad 
    \Phi_y = \hat{P} \exp(i \frac{2\pi}{L} \hat{y} ) \hat{P}
\end{equation}
$\hat{x},\hat{y}$ are position operators, $P$ the projector to the occupied states, and $L$ the system size. 
\cref{eq:Bott} applies to either clean or disordered gapped systems subject to the periodic boundary condition. 
When $L\to \infty$, \cref{eq:Bott} reduces to 
\begin{equation}
\label{Def-C-real}
C = \frac{1}{2\pi i A_{\mathcal{B}}} \sum_{\vec{R}\alpha \in \mathcal{B}} \left\langle \vec{R}\alpha\right\vert \left[\hat{P} \hat{x} \hat{P},\hat{P} \hat{y} \hat{P} \right] \left\vert \vec{R}\alpha \right\rangle
= \frac{1}{\pi A_{\mathcal{B}} }\,\Im \sum_{\vec{R}\alpha \in \mathcal{B}} \left\langle \vec{R}\alpha\right\vert \hat{P} \hat{x} \hat{P} \hat{y} \hat{P} \left\vert \vec{R}\alpha\right\rangle 
\end{equation}
where $\vec{R}$ is the position vector of primitive cell and $\alpha$ indicates the orbitals in one primitive cell. 
$\mathcal{B}$ is a large but finite region where $x/L, y/L\ll 1$, $A_{\mathcal{B}}$ is the area of this region. 
In the second equation of Eq.~(\ref{Def-C-real}), we use the fact:
\begin{equation}
    \left\langle \vec{R}\alpha\right\vert \hat{P} \hat{x} \hat{P} \hat{y} \hat{P} \left\vert \vec{R}\alpha\right\rangle
=\left(\left\langle \vec{R}\alpha\right\vert \hat{P} \hat{y} \hat{P} \hat{x} \hat{P} \left\vert \vec{R}\alpha\right\rangle\right)^*
\end{equation}
where $(...)^*$ is the complex conjugate.
We can consider every term in the summation as ``local Chern number of site $(\vec{R},\alpha)$", \ie we define LCM (for a 2D system) as \cite{Bianco2011mapping}
\begin{equation}
\label{Def-LCM}
    C(\vec{R},\alpha)=\frac{4\pi n_c}{A_c} \Im \left\langle \vec{R}\alpha\right\vert \hat{P} \hat{x} \hat{P} \hat{y} \hat{P} \left\vert \vec{R}\alpha\right\rangle 
\end{equation}
where $A_c$ is the area of one primitive cell and $n_c$ is the number of orbitals per primitive cell. 
One can check that the Chern number Eq.~(\ref{Def-C-real}) can be regarded as the average of LCM Eq.~(\ref{Def-LCM}). 
Alternatively, one can use
\begin{equation}
\label{Def-LCM-cell}
    C(\vec{R})=\frac{4\pi}{A_c}\sum_{\alpha} \Im \left\langle \vec{R}\alpha\right\vert \hat{P} \hat{x} \hat{P} \hat{y} \hat{P} \left\vert \vec{R}\alpha\right\rangle 
\end{equation}
as the ``local Chern number of cell $\vec{R}\,$". From now on, LCM stands for the local Chern marker of one cell $C(\vec{R})$, unless otherwise stated.

Eq.~(\ref{Def-C-real}) can be applied to systems subject to open boundary conditions as long as the region $\mathcal{B}$ is well separated from the boundary such that $\mathcal{B}$ is well-gapped. 
Otherwise, the yielded $C$ will not reflect the bulk Chern number. 
For example, if we take $\mathcal{B}$ as the whole system including the boundaries, \cref{Def-C-real} would become zero because
\begin{equation}
    \sum_{\vec{R}\alpha}\left\langle \vec{R}\alpha\right\vert \hat{P} \hat{x} \hat{P} \hat{y} \hat{P} \left\vert \vec{R}\alpha\right\rangle=\sum_{\vec{R}\alpha}\left\langle \vec{R}\alpha\right\vert \hat{P} \hat{y} \hat{P} \hat{x} \hat{P} \left\vert \vec{R}\alpha\right\rangle
\end{equation}
due to the cyclic property of the trace operation. 
That means, for a $L\times L$ sample with bulk Chern number $C$, the LCMs near the boundary will diverge in the order of $\mathcal{O}(C \cdot L)$ to cancel the summation of the bulk LCMs.  
More generally, if the sample comprises several macroscopically (comparable to the sample size) homogeneous regions with different Chern numbers, LCMs deep inside each region still converge to the corresponding Chern number. And near the boundaries of different regions, LCMs fluctuate around the mean value of the corresponding Chern numbers. 
Therefore, \cref{Def-LCM,Def-LCM-cell} are useful local indicators of topological properties.

With respect to our lattice models, since the total Chern number is enforced to be zero by the averaged $C_{2z}T$ symmetry, no boundary singularity of LCM will present. If the percolation argument is true, the profile of Chern numbers will be mottled in CMP, and no macroscopically homogeneous region is guaranteed. 
So we expect that LCMs will strongly fluctuate in CMP. 

\begin{figure}[h]
	\centering
	\subfigure[~$\tilde{t}=\pi v/2a,\,W=2v/a$]{\label{Cmarker-fullon-site-a}
    \includegraphics[width=.45\linewidth]{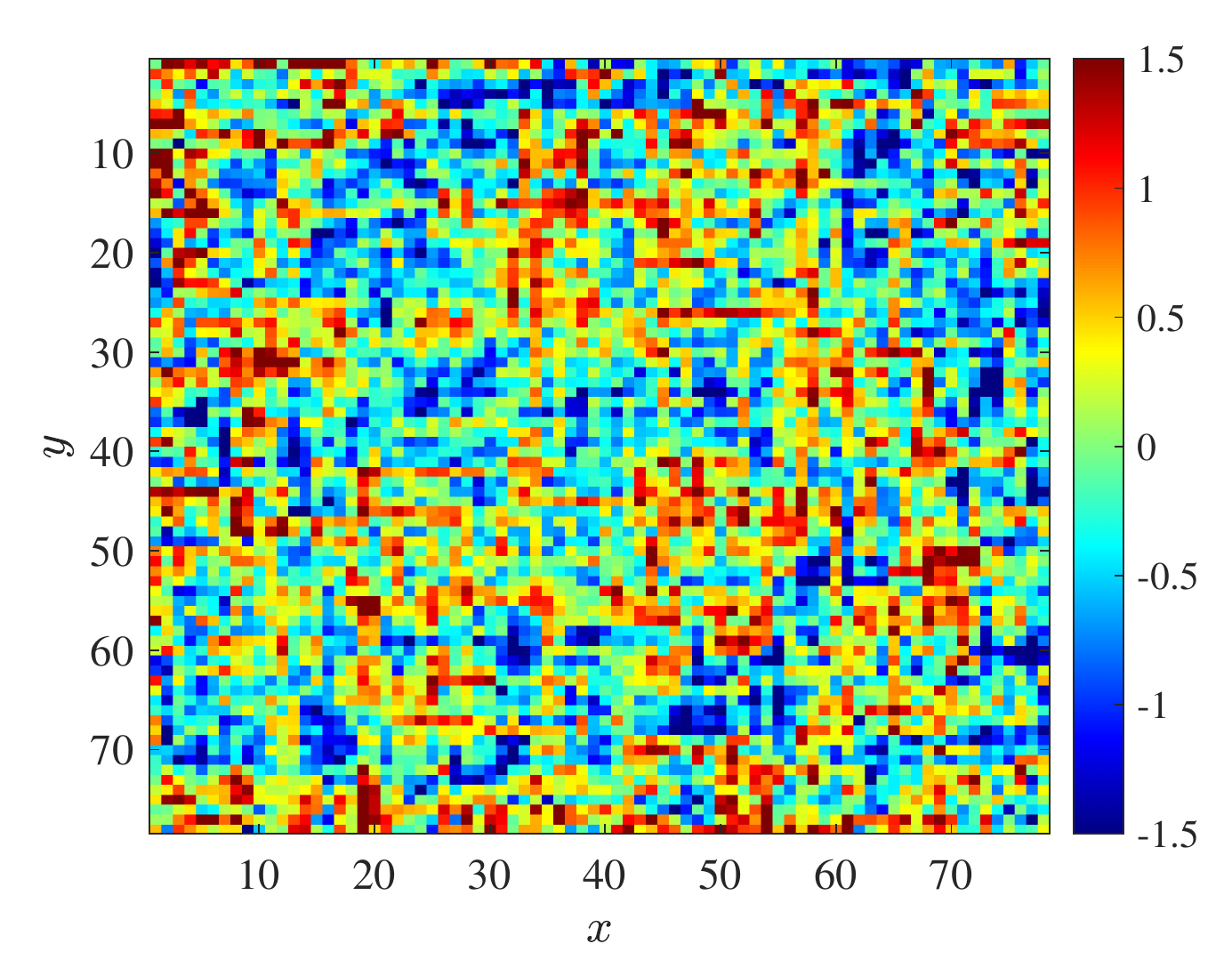}}
    \quad
	\subfigure[~$\tilde{t}=\pi v/2a,\,W=5v/a$]{\label{Cmarker-fullon-site-b}
	\includegraphics[width=.45\linewidth]{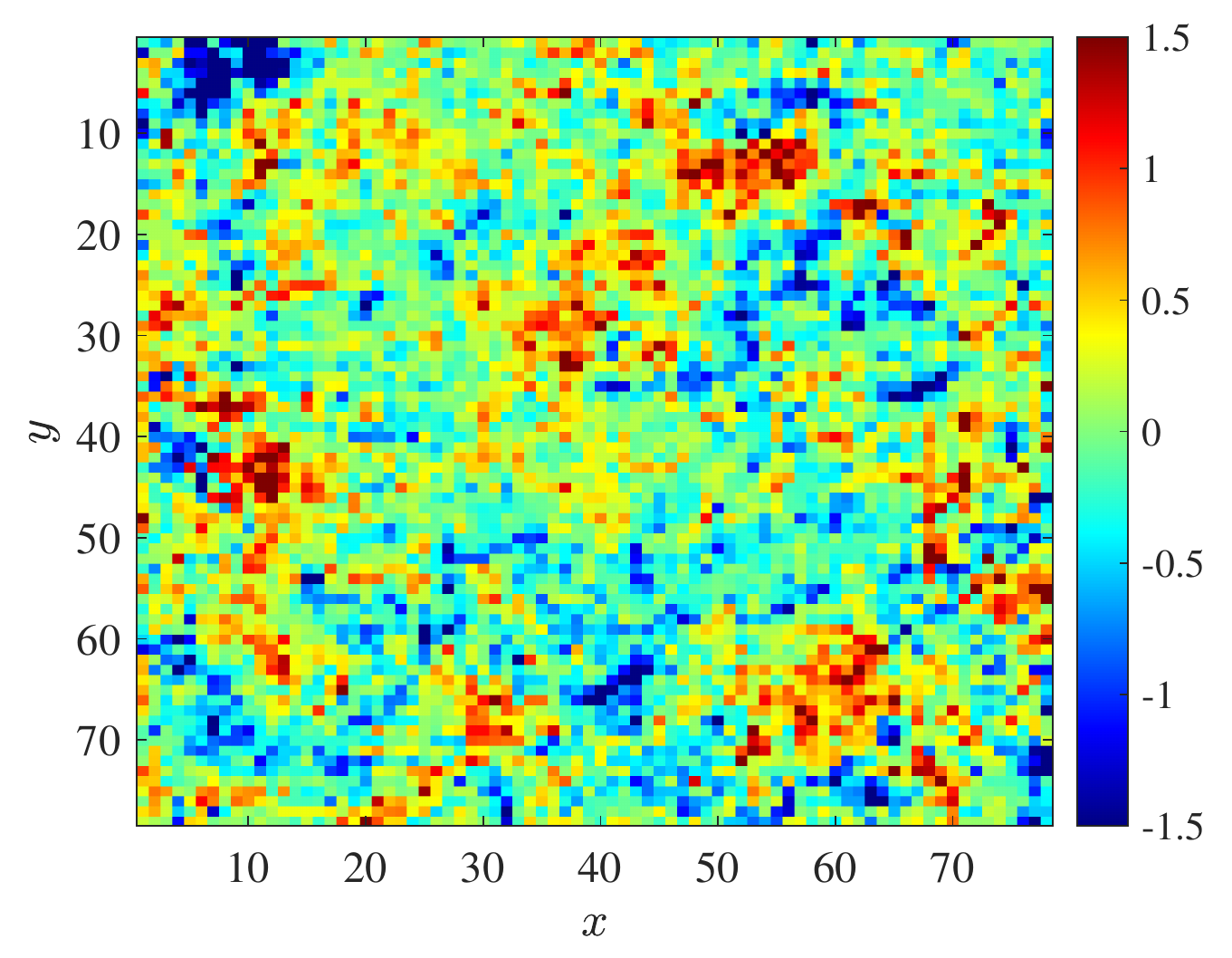}}\\
	\subfigure[~$\tilde{t}=\pi v/2a,\,W=7v/a$]{\label{Cmarker-fullon-site-c}
    \includegraphics[width=.46\linewidth]{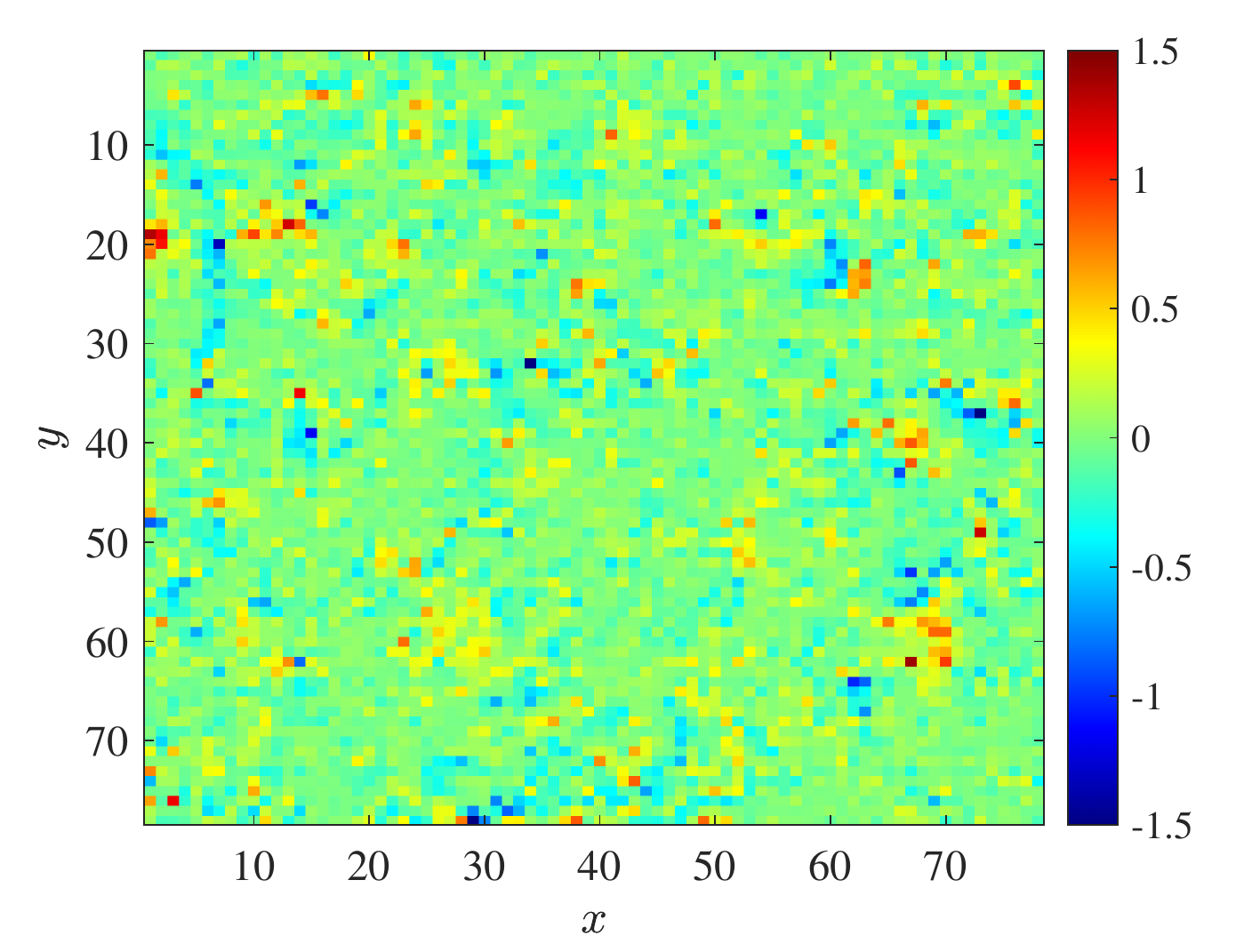}}
    \quad
	\subfigure[~$\tilde{t}=0.8v/a,\,W=2v/a$]{\label{Cmarker-fullon-site-d}
    \includegraphics[width=.45\linewidth]{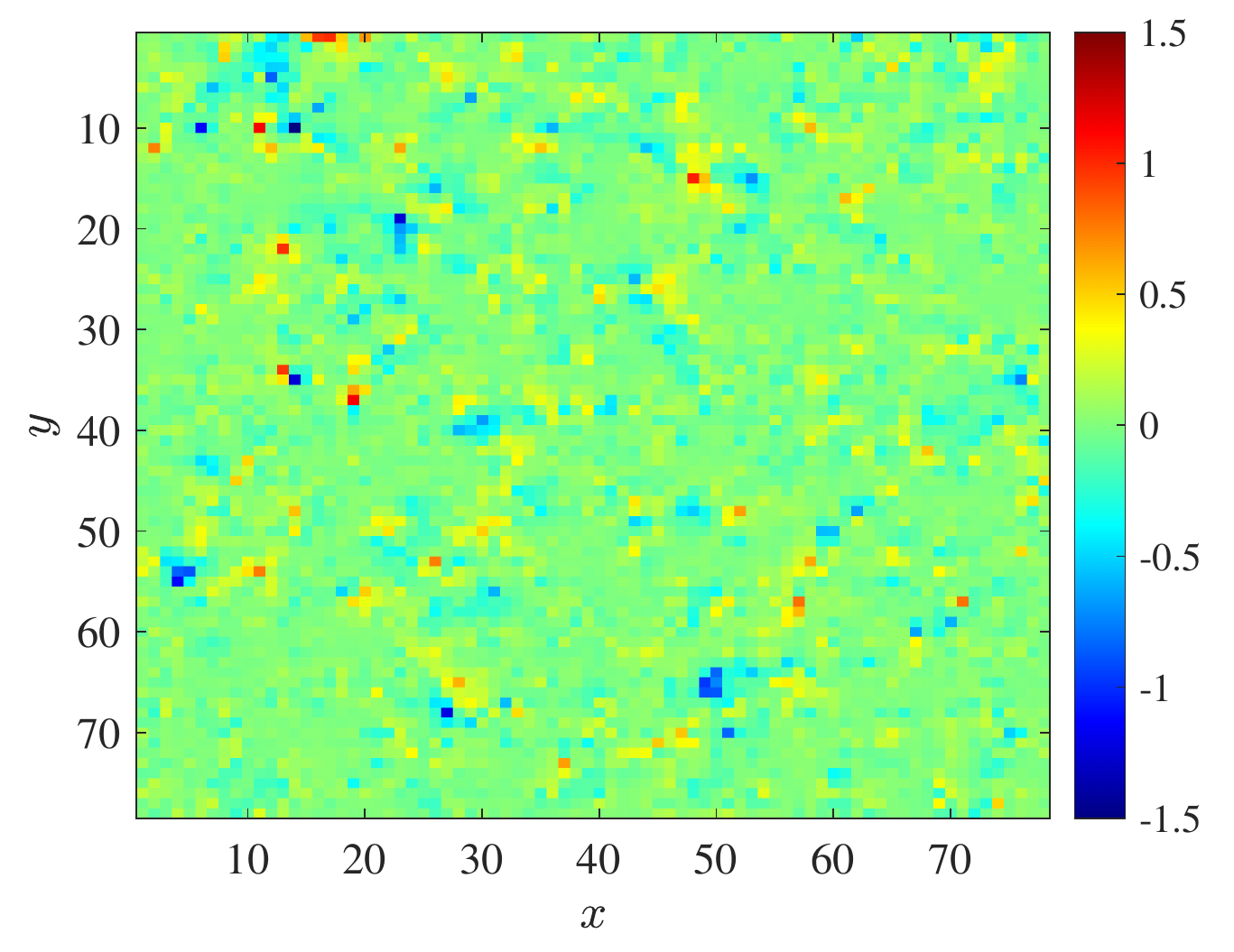}}\\
    \caption[]{LCMs of $H_{8B}$ in given disorder configurations at $E_F=0$ with $80\times80$ cells. Each pixel in the color maps corresponds to one cell. The open boundary condition is applied. LCMs with $\vert C(\vec{R})\vert>1.5$ will be mapped to $\pm1.5$ in the color maps. For $\mathrm{(a)\sim(b)}$, $\tilde{t}$ is fixed on $\pi v/2a$ and $aW/v=2,\,5,\,7$, respectively. $\mathrm{(d)}$ LCM with $\tilde{t}=0.8v/a$ and $W=2v/a$.}
	\label{Cmarker-fullon-site}
\end{figure}

This is indeed the case for $H_{8B}$. 
As shown in Fig.~\ref{Cmarker-fullon-site-a}$\sim$ \ref{Cmarker-fullon-site-c}, the fluctuation of LCM fades as $W$ increases and drives the system into LP. 
Fig.~\ref{Cmarker-fullon-site-d} shows that localizing the system by tuning $\tilde{t}$ will also fade the fluctuation. 
Therefore, LCMs indeed form a staggered pattern in CMP, which disappears once the system is localized. 
We also calculate the statistical distribution of $\left\langle C^2(\vec{R})\right\rangle^{1/2}$ which reflects the ratio of regions with nontrivial Chern number. 
Here, $\langle\cdot\cdot\cdot\rangle$ means averaging over all the primitive cells. The results are shown in Fig.~\ref{Statistic-Cmarker-full-critical}$\sim$\ref{Statistic-Cmarker-full-local}. As we can see, for all the configurations, $\left\langle C^2(\vec{R})\right\rangle^{1/2}$ exceeds $0.5$ in CMP and diminishes toward zero in LP. That means the nontrivial Chern blocks dominate ($p_{1}+p_{-1}>0.5$) in CMP and the trivial blocks dominate ($p_{1}+p_{-1}<0.5$) in LP. From these observations, we conclude that the percolation argument is indeed valid for $H_{8B}$.

\begin{figure}[h]
	\centering
	\subfigure[~$\tilde{t}=1.1\pi v/2a,\,W=1.5v/a$]{\label{Cmarker-simplifyon-site-a}
    \includegraphics[width=.45\linewidth]{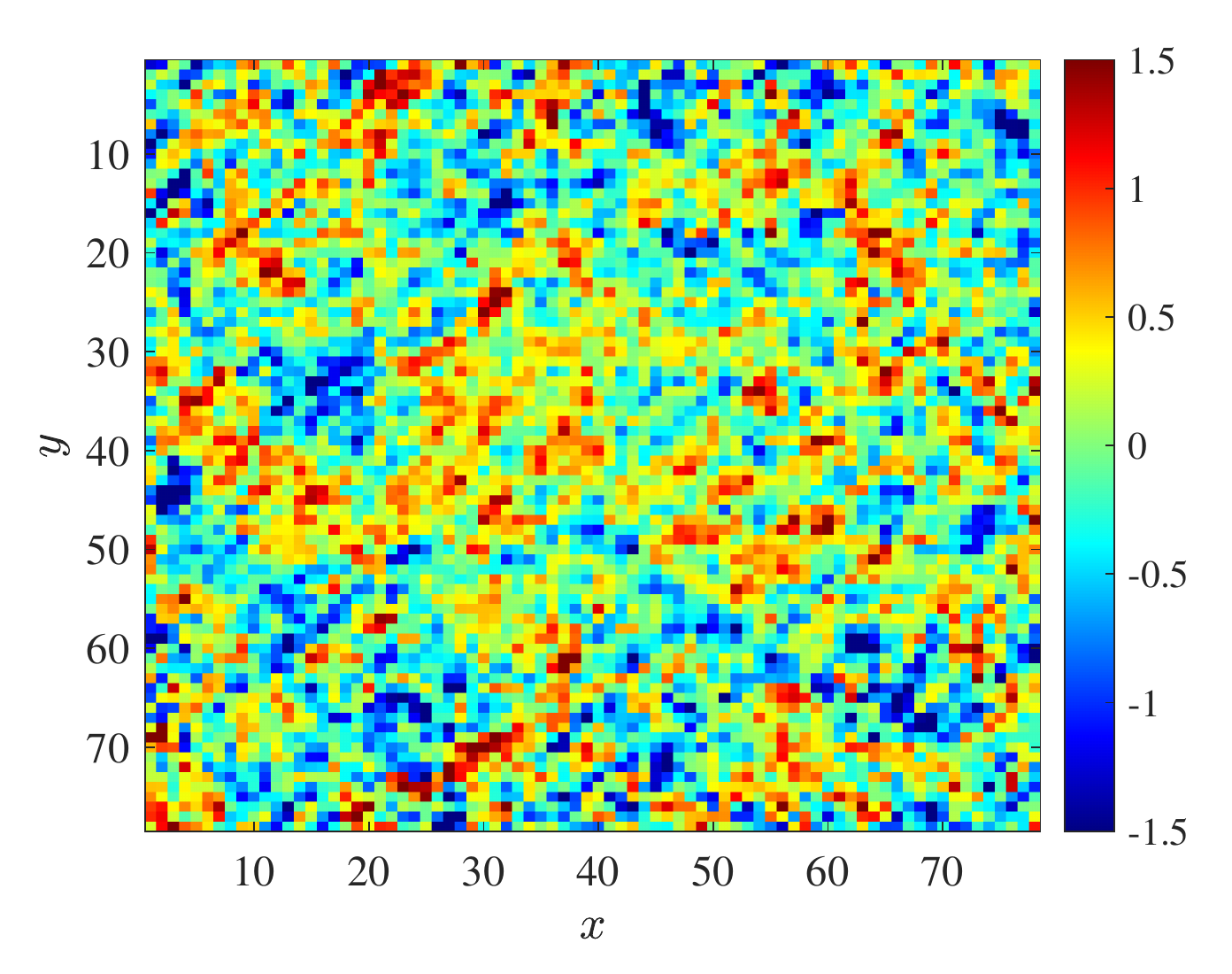}}
    \quad
	\subfigure[~$\tilde{t}=1.1\pi v/2a,\,W=3v/a$]{\label{Cmarker-simplifyon-site-b}
	\includegraphics[width=.45\linewidth]{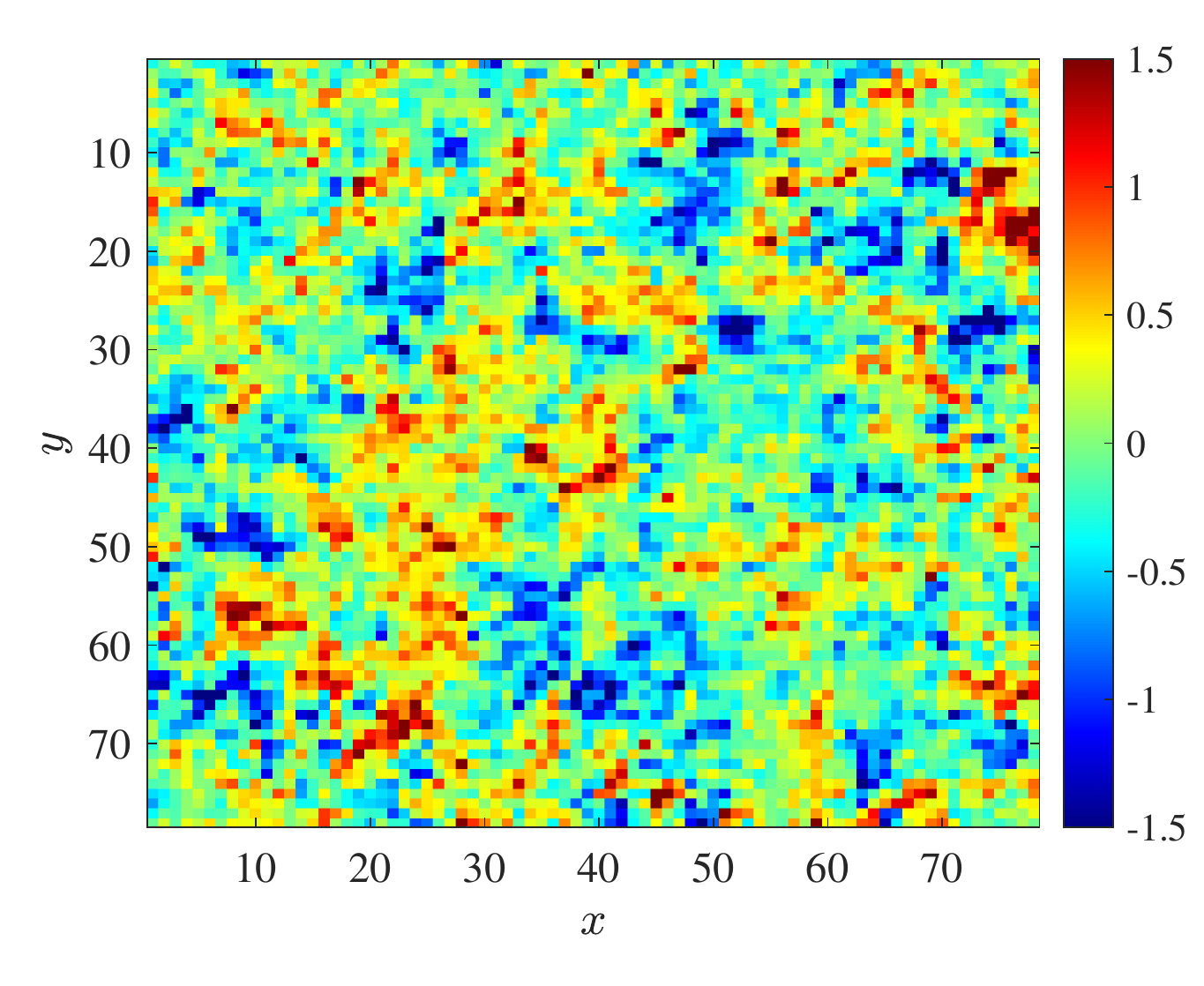}}\\
	\subfigure[~$\tilde{t}=1.1 \pi v/2a,\,W=5v/a$]{\label{Cmarker-simplifyon-site-c}
    \includegraphics[width=.45\linewidth]{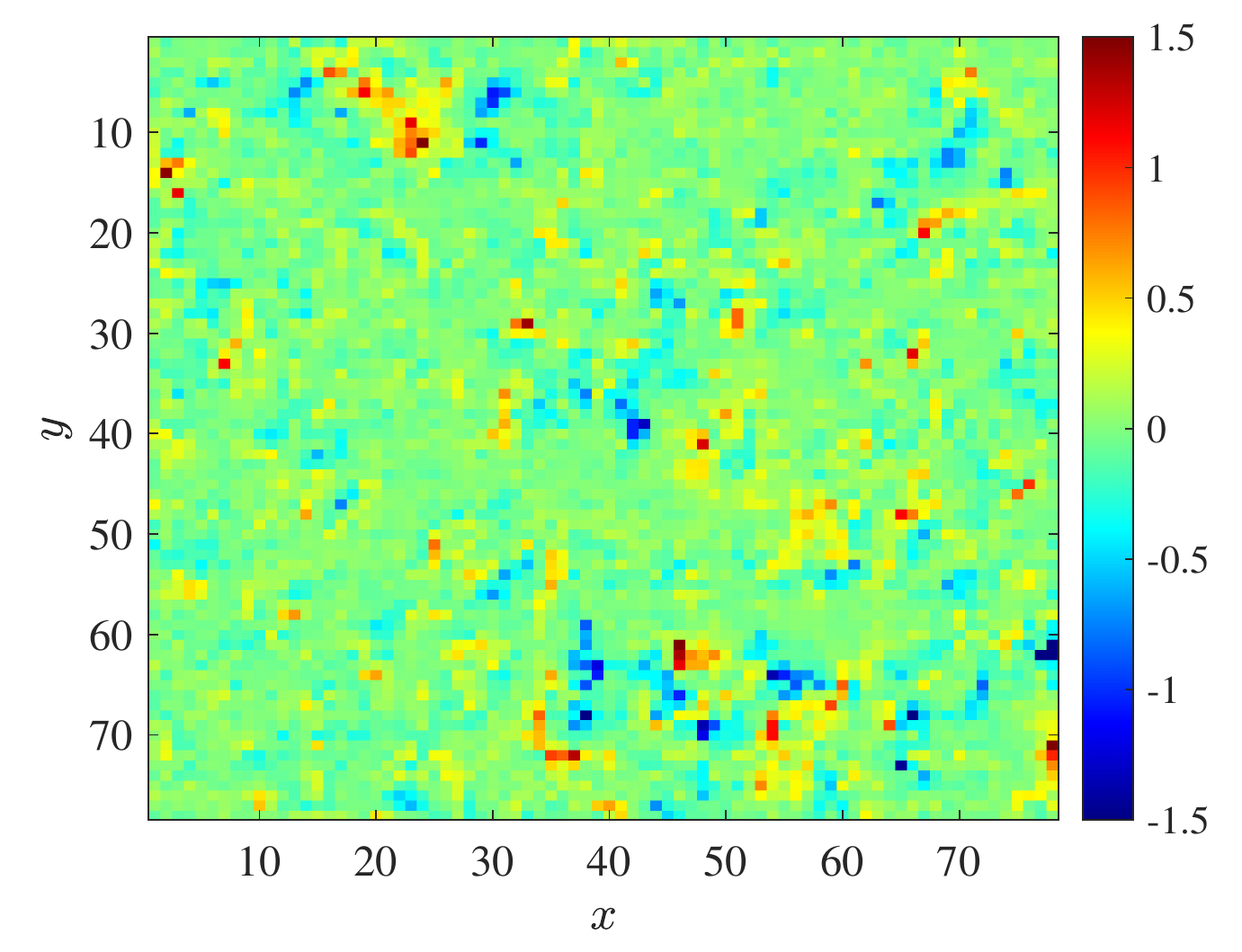}}
    \quad
	\subfigure[~$\tilde{t}=1.5 v/a,\,W=1.5v/a$]{\label{Cmarker-simplifyon-site-d}
	\includegraphics[width=.45\linewidth]{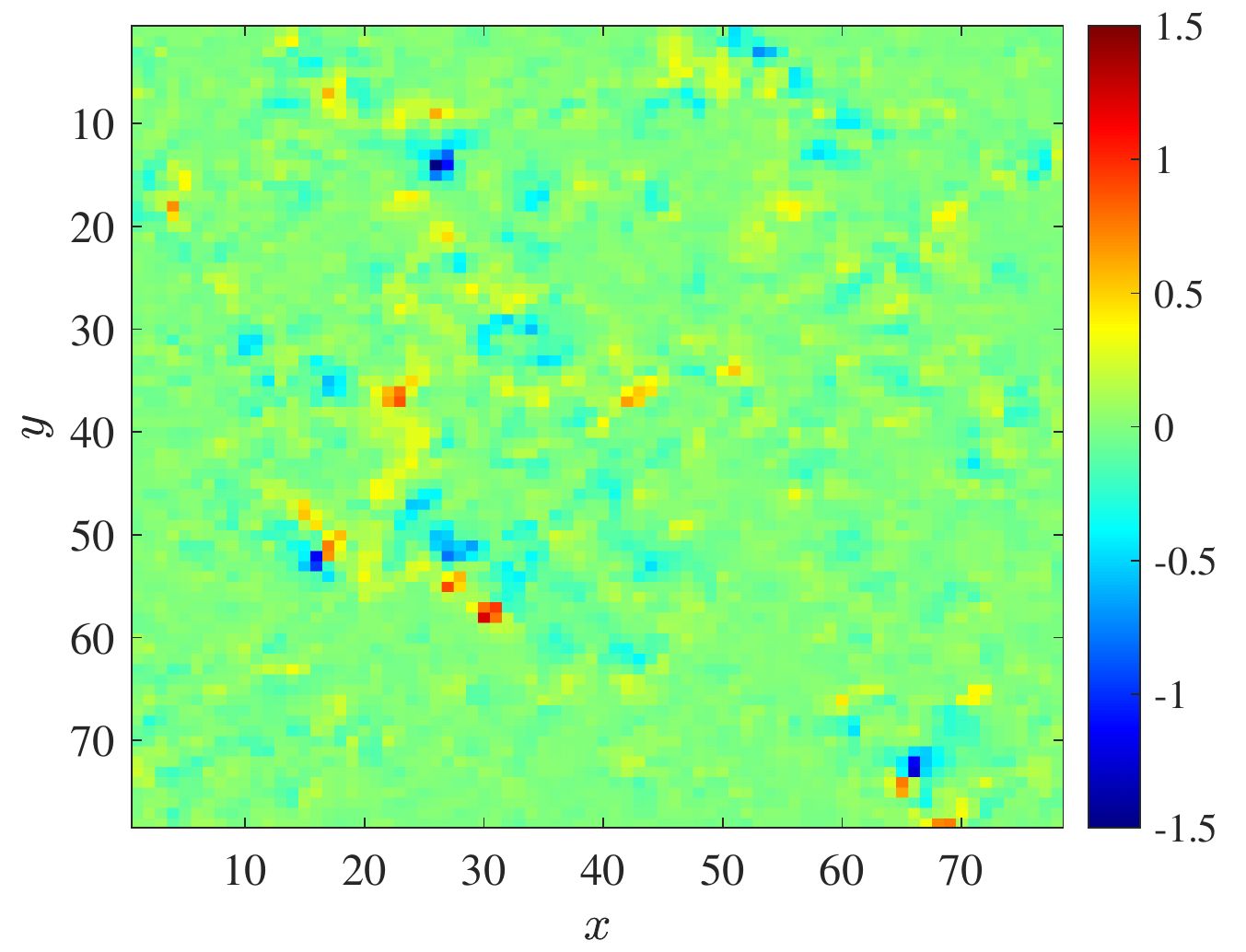}}\\
    \caption[]{LCMs of $H_{8B}'$ in given disorder configurations with $A=1.2,\,E_F=0$ and $80\times80$ cells. Each pixel in the color maps corresponds to one cell. The open boundary condition is applied. LCMs with $\vert C(\vec{R})\vert>1.5$ will be mapped to $\pm1.5$ in the color maps. For $\mathrm{(a)\sim(c)}$, $\tilde{t}$ is fixed on $1.1\pi v/2a$ and $aW/v=1.5,\,3,\,5$, respectively. $\mathrm{(d)}$ LCM with $\tilde{t}=1.5v/a$ and $W=1.5v/a$.}
	\label{Cmarker-simplifyon-site}
\end{figure}

\begin{figure}[h]
	\centering
	\subfigure[~$\tilde{t}=\pi v/2a,\,W=2v/a,\,N_{sample}=500$]{\label{Statistic-Cmarker-full-critical}
    \includegraphics[width=.44\linewidth]{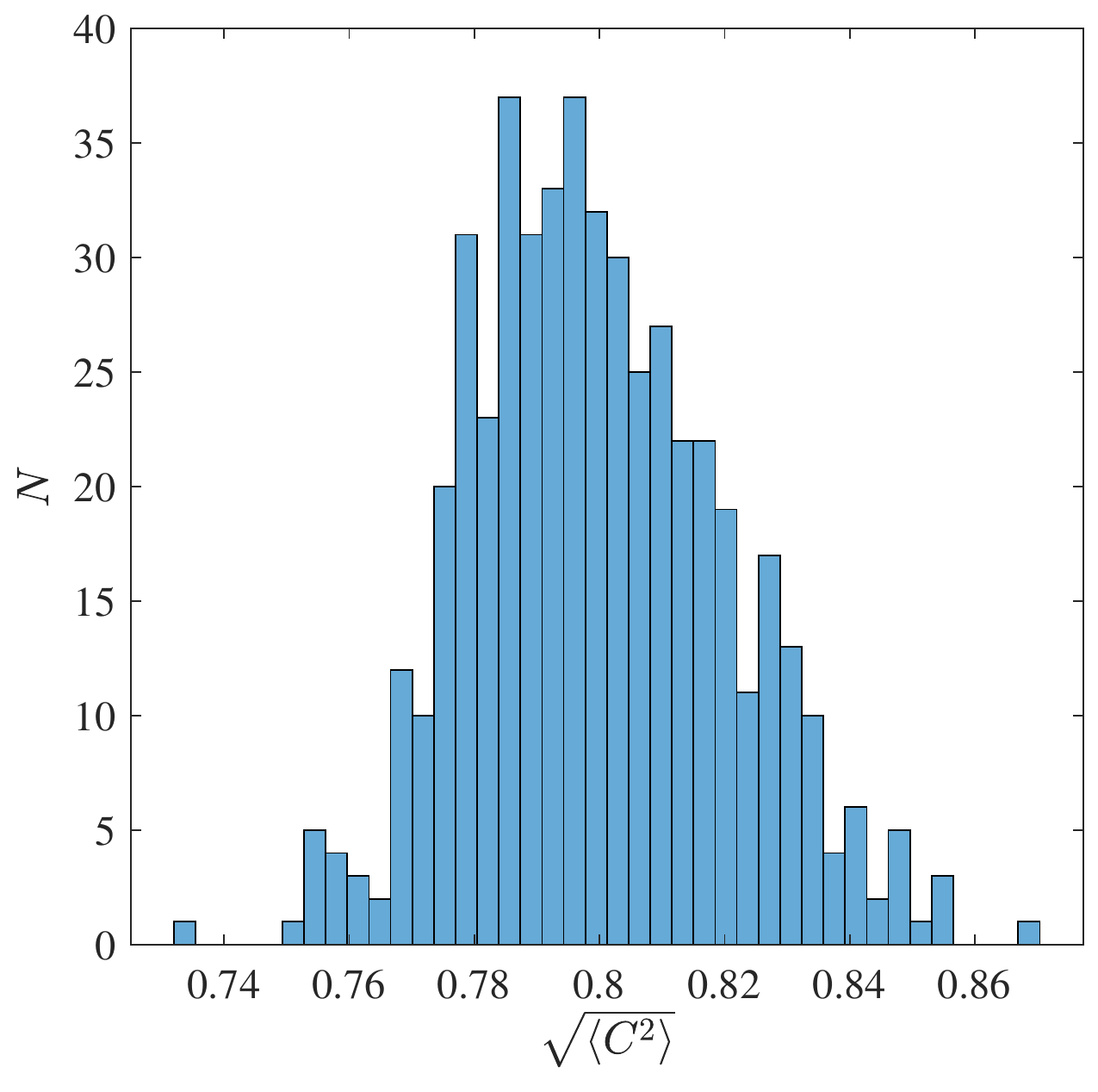}}
    \quad
	\subfigure[$~\tilde{t}=\pi v/2a,\,W=7v/a,\,N_{sample}=500$]{\label{Statistic-Cmarker-full-local}\includegraphics[width=.45\linewidth]{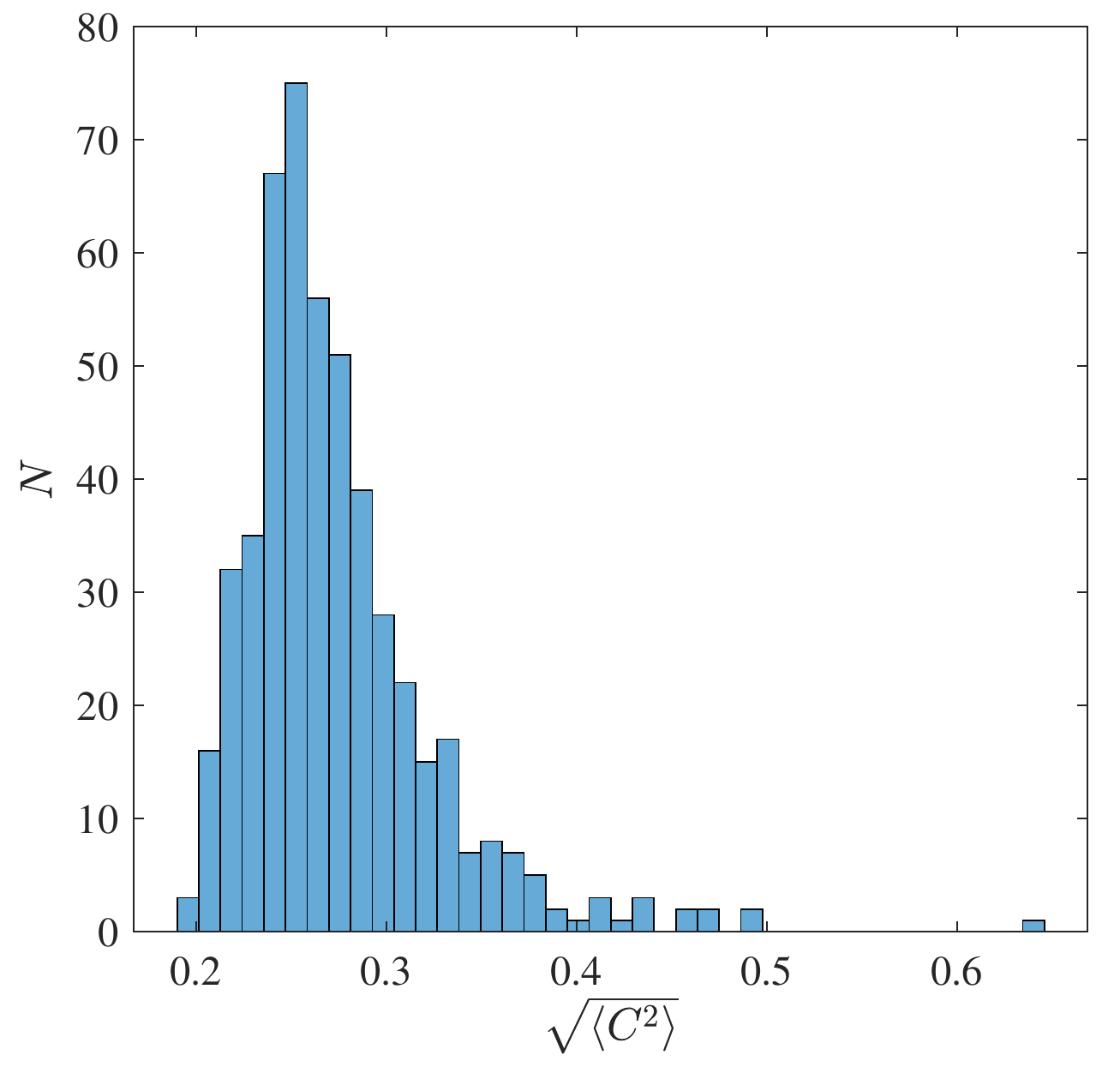}}\\
    \subfigure[$~A=1.2,\,\tilde{t}=1.1\pi v/2a,\, W=1.5v/a,\,N_{sample}=500$]{\label{Statistic-Cmarker-simp-critical}\includegraphics[width=.45\linewidth]{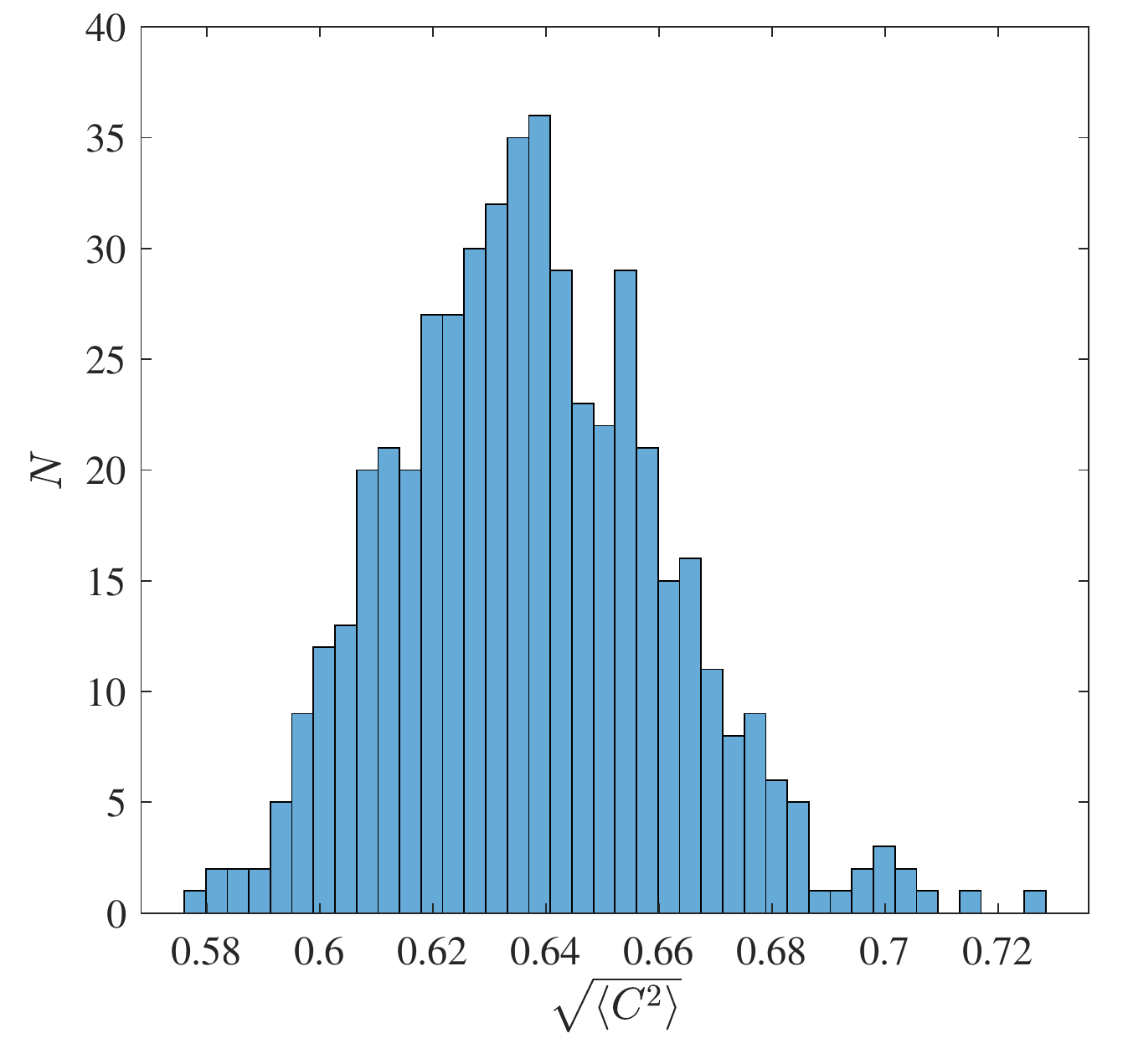}}
    \quad
	\subfigure[~$A=1.2,\,\tilde{t}=1.1\pi v/2a,\, W=5v/a,\,N_{sample}=500$]{\label{Statistic-Cmarker-simp-local}\includegraphics[width=.45\linewidth]{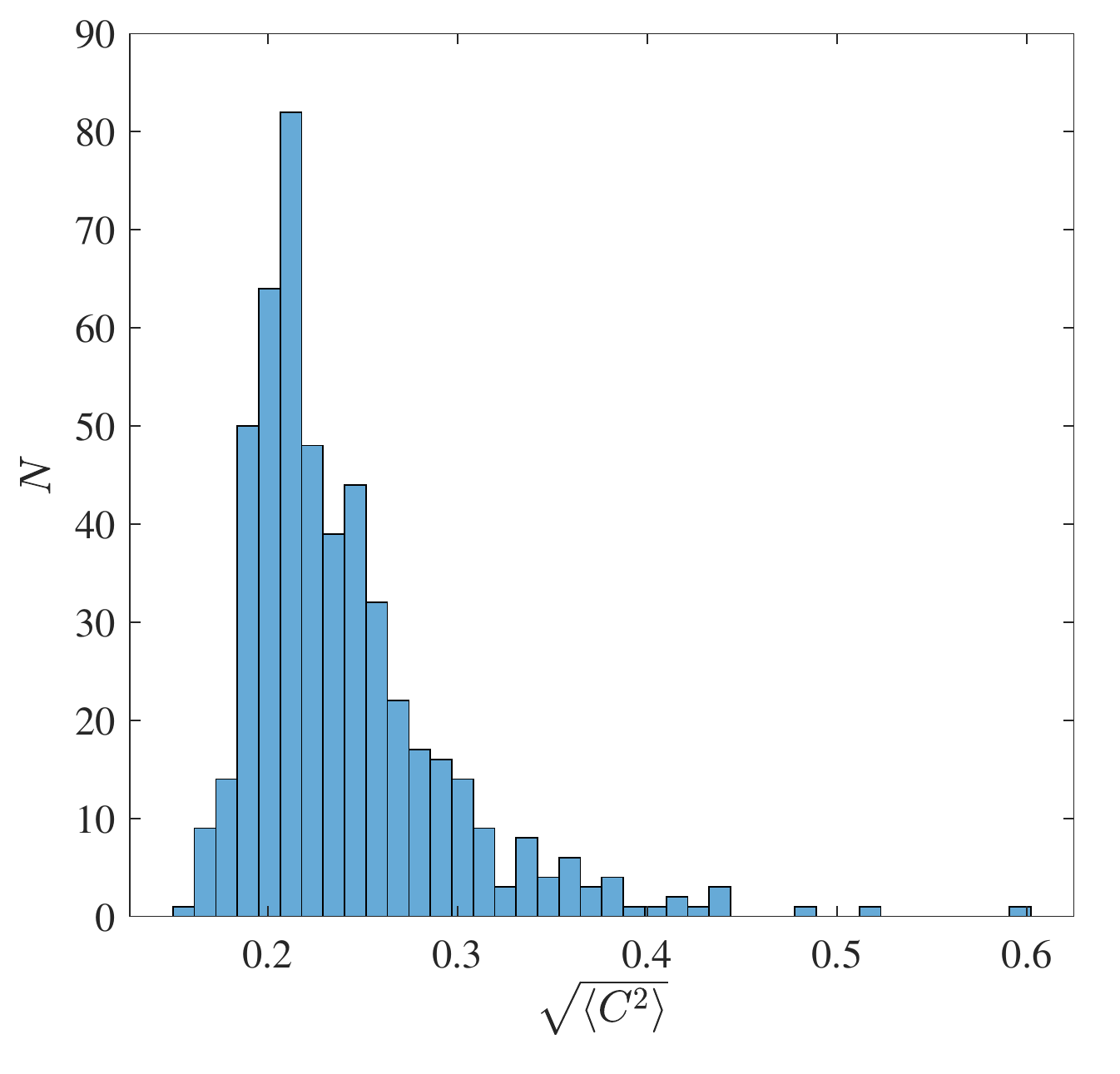}}\\
    \caption[]{Distributions of the root mean squares of LCM in lattice models with $E_F=0$, open boundary condition and different strengths of disorder. Each plot contains $500$ disorder configurations and each sample comprises $80\times80$ cells. $\mathrm{(a)}$ Distribution of $H_{8B}$ in CMP ($\tilde{t}=\pi v/2a,\, W=2v/a$). $\mathrm{(b)}$ Distribution of $H_{8B}$ in LP ($\tilde{t}=\pi v/2a,\, W=7v/a$). $\mathrm{(c)}$ Distribution of $H_{8B}'$ in CMP ($A=1.2,\,\tilde{t}=1.1\pi v/2a,\, W=1.5v/a$). $\mathrm{(d)}$ Distribution of $H_{8B}'$ in LP ($A=1.2,\,\tilde{t}=1.1\pi v/2a,\, W=5v/a$).}
	\label{Statistic-Cmarker}
\end{figure}

The results of $H_{8B}'$ are shown in Fig.~\ref{Cmarker-simplifyon-site} and Fig.~\ref{Statistic-Cmarker-simp-critical}$\sim$\ref{Statistic-Cmarker-simp-local}. As shown by Fig.~\ref{Cmarker-simplifyon-site-a}$\sim$ \ref{Cmarker-simplifyon-site-c}, the fluctuation of LCM fades as $W$ increases and drives the system into LP. Fig.~\ref{Cmarker-simplifyon-site-d} shows that localizing the system by tuning $\tilde{t}$ will also fade the fluctuation. The statistical results is also similar to that of $H_{8B}$, $\left\langle C^2(\vec{R})\right\rangle^{1/2}>1/2$ in CMP and diminishes toward zero in LP. So one can conclude that the percolation argument is also valid for $H_{8B}'$.

Before the end of this section, we can make one more comment about our LCM data. These profiles of LCM confirm our claim about CMP in an independent way from the localization length. Since electrons can propagate along the edges of Chern blocks, a sample dominated by staggered $C=\pm1$ Chern blocks (and the ratios of $C=1,-1$ regions are equal) can not be localized. On the other hand, the numerical results of the localization length (\cref{ll-full} \& \cref{ll-simplify}) show that there is no metallic phase ($\beta=\frac{\mathrm{d}ln g}{\mathrm{d}\ln L}>0$) . Hence, the only choice is to be critical.

\clearpage

\section{Critical metal phase in generic magnetic point groups}
\label{sec:CMP_beyondC2T}

The models we have extensively investigated are characterized by $C_{2z}T$ on average and transition of the Stiefel-Whitney class. Since we are motivated by the semi-classical picture of conducting percolation, which is not limited to Manhattan lattice and can be protected by average symmetries other than $C_{2z}T$, CMP should not be limited to Manhattan network, $C_{2z}T$ and the Stiefel-Whitney class (a special case of Real Space  Invariance in $C_{2z}T$). In this section, we will first show two additional models that have CMPs. The first model is a variant of $H_{8B}'$ (see \cref{simp-TB-cell}) with average symmetries $C_{4z}T$ and $m_{xy}$. The second model is a Kagome-like network model whose percolation picture is validated by average $C_{2z}T$. Then, we will discuss several simple magnetic groups that can protect the percolation picture and how they can support CMP during OAI transitions.

\subsection{Two additional models of CMP}

\begin{figure}[h]
	\centering
	\subfigure[~Unite cell of $H_{8B}''$. The orange, purple and green arrows correspond to hoppings $t_{+}'=-(1+Ai)\pi v/4a$, $t_{+}''=qt_{+}'$, and $\tilde{t}$, respectively. 2c and 4d indicate the representative positions of corresponding Wyckoff positions.]{\label{P4pmpm-a}
    \includegraphics[width=.4\linewidth]{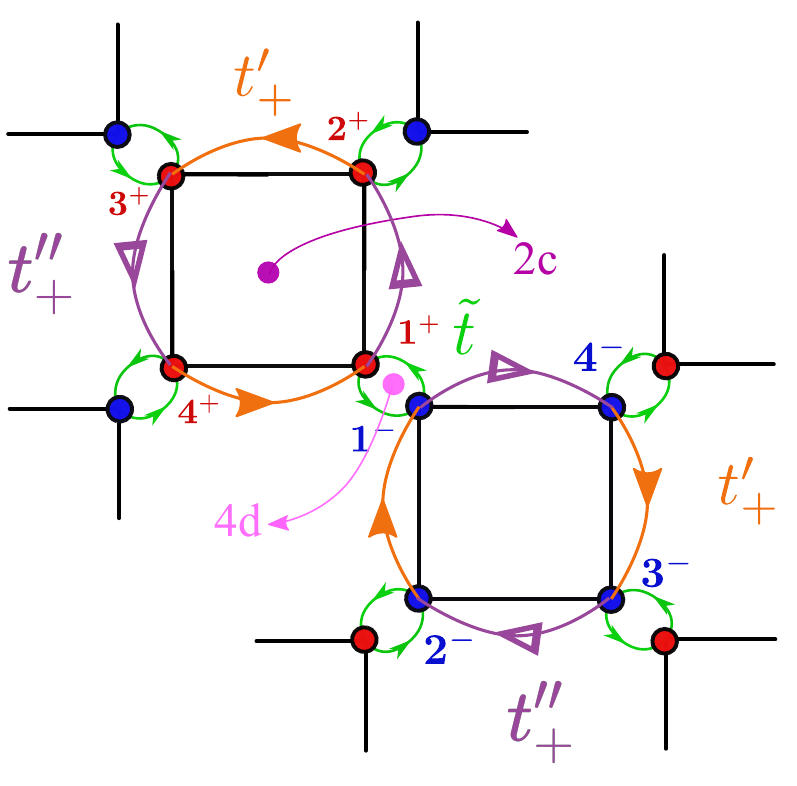}}
    \quad
	\subfigure[~Localization length $\Lambda$ as the function of $\tilde{t}$ of $H_{8B}''$ with $A=1.2$, $q=0.625\exp(-0.15\pi)$, $W=1.5v/a$ and $E_F=0$. The longitudinal size is $M=10^7$ and the data precision ($\sigma_{\Lambda}/\Lambda$) reaches $1.5\%$. The inset shows the zoomed critical region.]{\label{P4pmpm-b}
	\includegraphics[width=.5\linewidth]{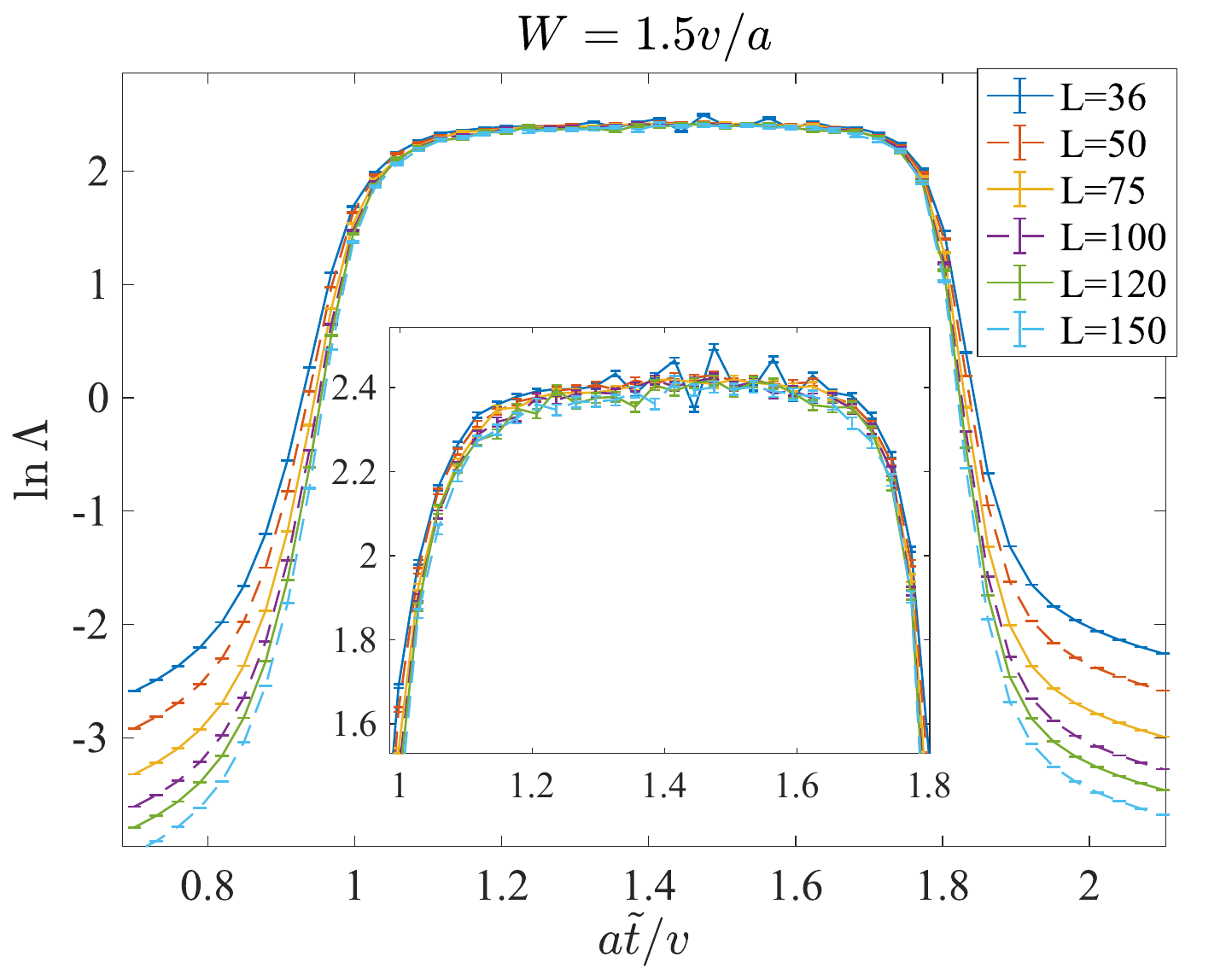}}
    \caption[]{Unit cell and localization length of the modified model $H_{8B}''$.}
	\label{P4pmpm}
\end{figure}

First, notice that more than one average symmetry in $H_{8B}'$ can protect the percolation mechanism. $C_{4z}T$, $m_{xy}$, and glides along $x$ and $y$ axes all can protect the equal ratio of $C=\pm1$. Hence, the percolation mechanism is still protected even if we break some of the symmetries. In addition, since a gapless band structure in the clean limit is necessary for a delocalized phase, we have to keep two OAI limits at $a\tilde{t}/v=0,\infty$ inequivalent so that the transition will enforce a gap closure. For example, $C_{4z}T$ and $m_{xy}$ can protect the percolation mechanism alone since both can reverse the Chern number. However, the two OAI limits are adiabatically connected if only one of them is present. Hence, to enforce a gapless region supporting CMP, we need both $C_{4z}T$ and $m_{xy}$. We design the model $H_{8B}''$ (\cref{P4pmpm-a}) by multiplying half of the edge hoppings $t'_{+}$ in $H_{8B}'$ by a complex factor $q=t_{+}''/t_{+}'$. The symmetry group of $H_{8B}''$ is $P4'm'm$ (\# 99.165 in BNS setting) containing both $C_{4z}T$ and $m_{xy}$. Similar to $H_{8B}'$, the OAI transition ($a\tilde{t}/v=0\rightarrow\infty$) of $H_{8B}''$ also induces a braiding of four Dirac points stabilized at the zero energy. The numerical data (see \cref{P4pmpm-b}) confirms that CMP survives in $H_{8B}''$.

\begin{figure}[h]
	\centering
	\subfigure[~Profile of the Kagome-like network model. The arrows indicate the chirality of chiral wires. The intersections of wires correspond to scattering nodes with effects defined in \cref{scatter matrix}. ]{\label{kagome-a}
    \includegraphics[width=.45\linewidth]{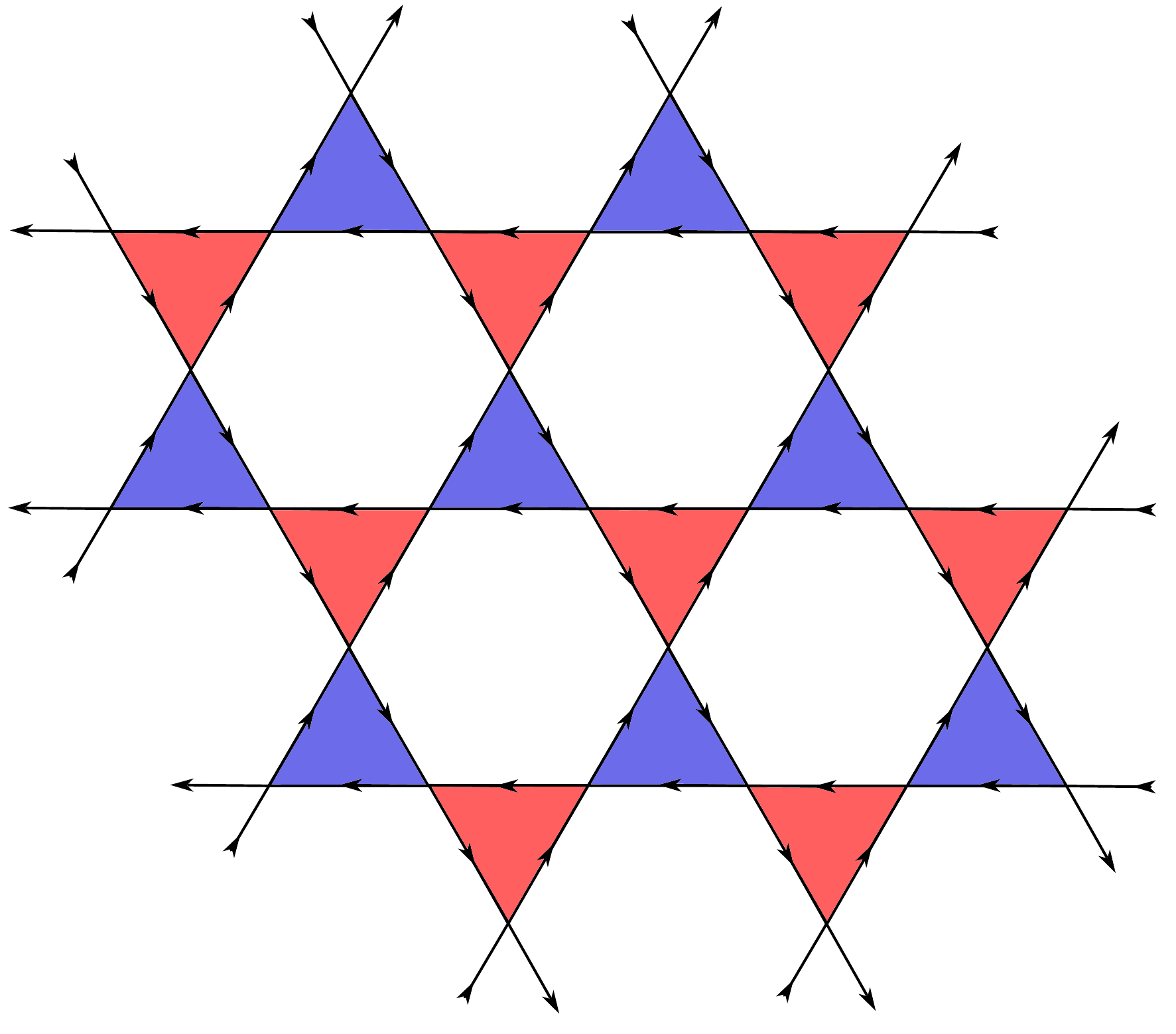}}
    \quad
	\subfigure[~Localization length $\ln \Lambda$ versus scattering angle $\theta$ with different transversal sizes $L$ in the Kagome-like network model. The longitudinal size is $M=10^7$. The data precision $\sigma_{\Lambda}/\Lambda$ reaches 1\% for $\theta<-\pi/8$, 1.5\% for $-\pi/8<\theta<-\pi/16$ and 3\% for other data points. $\sigma_{\Lambda}$ is the unbiased estimation of error mentioned in \cref{subsec:basic-localization-length}.]{\label{kagome-b}
	\includegraphics[width=.45\linewidth]{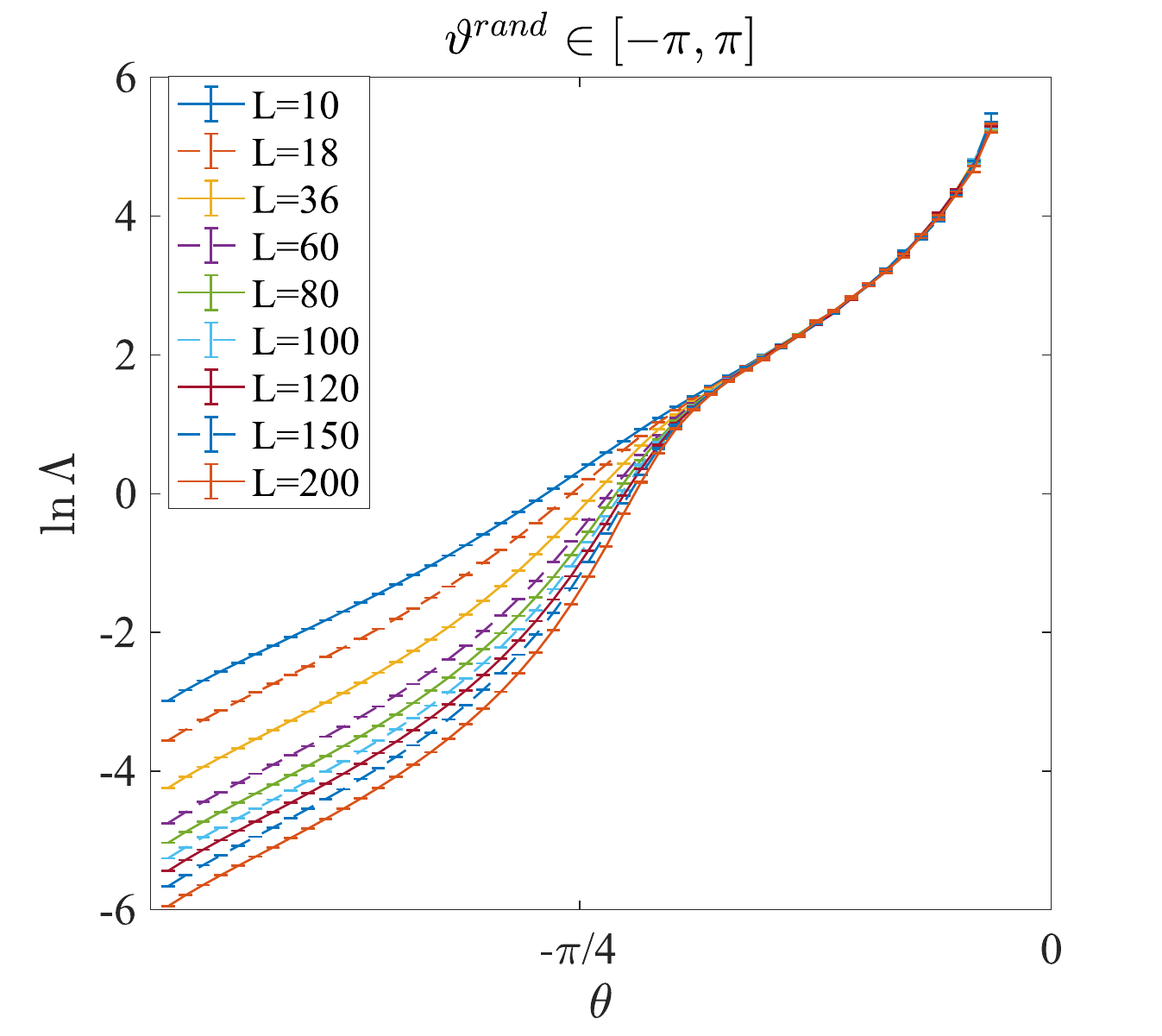}}
    \caption[]{Profile and localization length of the Kagome-like network model.}
	\label{kagome}
\end{figure}

Second, we can consider a Kagome-like network (\cref{kagome-a}) whose scattering nodes locate at the intersections of chiral wires. Each scattering node is identical to that of the Manhattan network (\cref{network-b}, \cref{scatter matrix}). The Kagome-like network respects $C_{2z}T$ and can also mimic a tricolor percolation process. As demonstrated by \cref{kagome-b}, it has a CMP similar to the Manhattan network.

\subsection{CMP in generic magnetic point groups}

According to the above results, CMP could be a general phenomenon when an average symmetry protects a conducting percolation phase during a transition between inequivalent OAIs. However, we do not claim these conditions are sufficient for CMP. As we have discussed in Sec. ${\color{red}\mathrm{D}}$ of the main text, the number of Dirac points and correlations of disorder-induced Dirac masses are also important. 
To further show the generality of CMP, we inspect the minimal OAI transitions of all the simple symmetries that can protect the percolation picture alone and how a CMP could arise. More precisely, we restrict ourselves to four simple (magnetic) space groups ($Pm$, $P2'$, $P4'$, $P6'$) with and without Spin-Orbit Coupling (SOC). The results are summarized and explained in Fig.~$\color{green}{\mathrm{6}}$ of the main text. In this section we introduce the method to obtain the results.  

The quantity we used to characterize inequivalent OAIs is the real Space Invariant (RSI) \cite{song_twisted_2020}. RSI for a given (magnetic) space group is a collection of expressions of orbital occupations at inequivalent Wyckoff positions. RSI is invariant under symmetric charge additions/subtractions on Wyckoff positions,  \ie RSI is invariant under adiabatic transitions that change orbital occupations without gap closing. Hence, for a given symmetry group, transitions between two OAIs with different RSIs must go through a gap closing, \ie they are inequivalent OAIs that cannot be connected adiabatically. With the help of RSI, we can classify OAIs for a given symmetry group. Further, by expressing RSI in terms of momentum irreps, we can deduce the band deformation and how a CMP can arise during the transition.

Let us take $P4'$ with and without SOC as pedagogical examples. The Wyckoff positions of $P4'$, except for the general positions, are 1a ($0,0$), 1b ($1/2,1/2$) and 2c ($1/2,0$)\&($0,1/2$) with site symmetries $4'$, $4'$ and $2$, respectively. Taken from \href{https://www.cryst.ehu.es/cgi-bin/cryst/programs/mbandrep.pl}{MBANDREP} program on the Bilbao Crystallographic Server \cite{elcoro_magnetic_2021}, the irreducible corepresentations of site symmetry $4'$ are $\mathrm{A}$ (even under $C_{2z}$) and $\mathrm{B\!B}$ (odd under $C_{2z}$) without SOC. And the co-irreps of site symmetry $2$ are $A$ (even under $C_{2z}$) and $B$ (odd under $C_{2z}$).

The Wannier centers of occupied states of a Hamiltonian can adiabatically move if their displacements preserve symmetry. Hence, RSI should be invariant under the symmetric addition/subtraction of electrons. For a given Wyckoff position, these electrons will form an induced representation of the site symmetry. For 1a or 1b here, the induced representation is $A\oplus A \oplus B\!B$ since four electrons related by $C_{4z}T$ form two $C_{2z}$-even and two $C_{2z}$-odd orbitals. For the same reason, the induced representation on each 2c position is $A\oplus B$. Denote $m_q(R)$ as the occupation number of  co-irrep $R$ at each Wyckoff position $q$, we can define RSI at 1a/1b as $\delta_{a/b}=m_{a/b}(A)-2m_{a/b}(B\!B)$ and RSI at 2c as $\delta_{c}=m_c(A)-m_c(B)$.
One can check that the RSI is invariant under symmetric electron addition/subtraction on 1a, 1b and 2c.

To see the band deformation during OAI transitions, we should express the above RSI in terms of momentum irreps at high-symmetry points. One can find a full derivation in the Appendix ${\color{red}\mathrm{C.3}}$ of \cite{herzog2022hofstadter}. The basic idea is that an occupied molecular orbital $R$ on the Wyckoff position $q$ will introduce an MEBR $R_{q}\!\uparrow\!G$ (a group of bands with specific irreps at high-symmetry points) in the momentum space, hence RSI originally expressed by orbital occupations can be expressed by occupations of momentum irreps.

We skip the full derivation and directly put the results of $P4'$ without SOC here

\begin{equation}
    \begin{aligned}
        \delta_{a}&=m(X_1)-m(\Gamma_2\Gamma_2)-m(M_2M_2)\\
        \delta_{b}&=n_{\mathrm{band}}-m(\Gamma_2\Gamma_2)-m(X_1)-m(M_2M_2)\\
        \delta_c &=m(M_2M_2)-m(\Gamma_2\Gamma_2)
    \end{aligned}
    \label{nsocP4'-bandind}
\end{equation}
where $m(R)$ indicates the occupation number of momentum irrep $R$, and $n_{\mathrm{band}}$ is the total number of occupied bands. One can find the definitions irreps in \cref{nsocP4'-bandind} with the help of \href{https://www.cryst.ehu.es/cgi-bin/cryst/programs/msitesym.pl}{MSITESYM} program on the Bilbao Crystallographic Server \cite{elcoro_magnetic_2021}.

In Sec.~$\color{green}{\mathrm{II.D}}$ of the main text, we have defined the concept of minimal OAI transitions which have the simplest band deformations and can be viewed as the building block of generic OAI transitions.  \cref{nsocP4'-bandind} shows that OAI transitions in $P4'$ without SOC must change the occupation numbers of $\Gamma_2\Gamma_2,\,M_2M_2,$ and $X_1$. Also, we always restrict ourselves to the transitions preserving the number of occupied bands. Hence, we have three minimal OAI transitions that changes one of these occupation individually. These three minimal transitions correspond to three rows of the ``$P4'$-NSOC'' block of Fig.~$\color{green}{\mathrm{6}}$ in the main text. The first transition changes $m(\Gamma_1)$ only. Since there are only two irreps at the $\Gamma$ point -- one dimensional $\Gamma_1$ and two dimensional $\Gamma_2\Gamma_2$, the first minimal transition will decrease $m(\Gamma_1)$ by $2$ and increase $m(\Gamma_2\Gamma_2)$ by $1$. Certainly, the inverse process is also a minimal OAI transition, but we will not discuss it since the property is the same. During this transition, two 1-dim irreps $\Gamma_1$ is replaced by one 2-dim irrep $\Gamma_2\Gamma_2$. Hence, if we control the transition by some parameter, there should be a finite parameter region where only one $\Gamma_1$ goes up across the Fermi surface and the low-energy physics is dominated by a quadratic touching from $\Gamma_2\Gamma_2$.

The $k\cdot p$ model of the quadratic touching is 
\begin{equation}
    H(\vec k)=(k_x^2-k_y^2)\sigma_x+ck_xk_y\sigma_y\quad c=\mathrm{Const},
    \label{quadra-nsocP4'}
\end{equation}
where $\vec k$ is the momentum deviation from $M$ point. We have taken $C_{4z}T=i\sigma_y\mathcal{K}$ since $\Gamma_2\Gamma_2$ comes from the corepresentation $B\!B$ where $C_{2z}=(C_{4z}T)^2=-1$. 

Now, suppose we introduce slow-varying disorders that respect $P4'$ on average, \cref{quadra-nsocP4'} can locally open a mass gap, \ie 
\begin{equation}
    H(\vec k, \vec r)=(k_x^2-k_y^2)\sigma_x+ck_xk_y\sigma_y+m(\vec r) \sigma_z,
\end{equation}
which results in a Chern number $\propto \mathrm{sign}(cm(\vec r))$. Since $C_{4z}T$ reverses the sign of Chern number and protects the percolation picture, we can expect a CMP in this OAI transition. 

As to the minimal change of orbital occupation, \cref{nsocP4'-bandind} tells us that the first minimal OAI transition will change RSI by $\Delta\delta_a=\Delta\delta_b=\Delta\delta_c=-1$. According to RSI in terms of orbital occupations, one can check that $\Delta m_{a}(A)=\Delta m_{b}(A)=-1,\,\Delta m_{c}(B)=1$ is the minimal orbital transition realizing the RSI change and preserving the particle number. Both the band deformation and real space orbital transition are illustrated in the first row of block ``$P4'$-NSOC'' of Fig.~$\color{green}{\mathrm{6}}$ in the main text.  

The second minimal OAI transition will replace two 1-dim irreps $M_1$ by one 2-dim irrep $M_2M_2$, and hence is similar to the first minimal transition. Both the band deformation and real space orbital transition are illustrated in the second row of block ``$P4'$-NSOC'' of Fig.~$\color{green}{\mathrm{6}}$ in the main text.  

The third transition replaces a 1-dim irrep $X_2$ by a 1-dim irrep $X_1$. Due to symmetry $C_{4z}T$, there will also be an exchange at the $Y$ point. Hence, the band gap will closed at a single parameter point where two Dirac cones appear at $X$ and $Y$ points. According to the argument in Sec.~$\color{green}\mathrm{II.D}$ of the main text, it can be delocalized if the two disorder-induced Dirac masses are positively correlated. Both the band deformation and real space orbital transition are illustrated in the third row of block ``$P4'$-NSOC'' of Fig.~$\color{green}{\mathrm{6}}$ in the main text.

Now we consider $P4'$ with SOC. In this case, the only site symmetry irrep of $4'$ is 2-dim $^1\!E^2\!E$. Hence the induced representation of symmetric electron addition is $^1\!E^2\!E\oplus\!^1\!E^2\!E$ and we should define $\delta_{a/b}=\!\mod(m_{a/b}(^1\!E^2\!E),2)$. The site symmetry irreps of $2$ are 1-dim $^1\!E$ and $^2\!E$ where the $C_{2z}$ action are $i$ and $-i$, respectively. Hence, the induced representation of symmetric electron addition on each c position is $^1E\oplus\!^2E$ and the RSI of 2c should be $\delta_c=m_c(^1\!E)-m_c(^2\!E)$

However, there is a significant difference to the non-SOC case, not all the components of an RSI can be expressed in terms of momentum irreps. The reason is that the MEBRs of different orbitals can behave the same in terms of momentum irreps. More precisely, we can view MEBRs as vectors whose components are occupation numbers of momentum irreps. Then we can consider a matrix $R_{MEBR}$ comprises column vectors of all the involved MEBRs, \eg $^1\!E^2\!E_a\!\uparrow\!G$, $^1\!E^2\!E_b\!\uparrow\!G$, $^1\!E_c\!\uparrow\!G$, and $^2\!E_c\!\uparrow\!G$ for $P4'$ with SOC. In general, the involved MEBRs are linear dependent, \ie $\mathrm{rank}(R_{MEBR})<n_{\mathrm{MEBR}}$ where $n_{\mathrm{MEBR}}$ is the number of MEBRs. Denote the number of components of RSI as $n_{RSI}$, only $\min\{\mathrm{rank}(R_{MEBR}),n_{RSI}\}$ (linear combinations of) RSIs can be expressed by momentum irreps (see Appendix ${\color{red}\mathrm{C.3}}$ \cite{herzog2022hofstadter} for details).   
\begin{equation}
\begin{aligned}         
    \delta_{a+b}&=\mathrm{mod}\left(m_a(^1E^2E)+m_b(^1E^2E),2\right)=\mathrm{mod}(m(X_4),2)\\
    \delta_c&=2n_{\mathrm{band}}-m(X_4)
\end{aligned}
    \label{socP4'-bandind}
\end{equation}

Again, one can find the definitions of irreps $X_4$ by \href{https://www.cryst.ehu.es/cgi-bin/cryst/programs/msitesym.pl}{MSITESYM} program on the Bilbao Crystallographic Server \cite{elcoro_magnetic_2021}. Now we have two kinds of minimal OAI transitions preserving the number of occupied bands. See the ``$P4'$-SOC'' block of Fig.~${\color{green}\mathrm{6}}$ for an illustration, the first one increases $m(X_4)$ by 1 while the second one is invisible to quantities in \cref{nsocP4'-bandind}. 

The first minimal OAI transition is similar to the third minimal transition of $P4'$ without SOC. Since there are only two 1-dim irreps $X_3$, $X_4$ at the $X$ points, the first minimal OAI transition will induce a band inversion at the $X$ point replacing irrep $X_3$ by $X_4$. Due to $C_{4z}T$, there is also a band inversion at the $Y$ point. Hence, the first transition will go through a parameter point where two Dirac cones appear at the $X$ and $Y$ points. According to the discussion in Sec. ${\color{green}\mathrm{II.D}}$, it can be a critical point when the disorder-induced two Dirac masses are positively correlated. Both the band deformation and real space orbital transition are illustrated in the first row of block ``$P4'$-SOC'' of Fig.~$\color{green}{\mathrm{6}}$ in the main text. 

The second transition is different from previous transitions. It increases $\delta_a$ by $1$ and decreases $\delta_b$ by 1, and hence is invisible to the quantities in \cref{socP4'-bandind}. This transition will not induce band inversions at high-symmetry points, rather, the gap will close at general k-points. Due to symmetry $C_{4z}T$, there will be at least four touching points. Also note that, if the transition is controlled by some parameter, the touching points in general can only exist at a parameter point. The low-energy physics of these touching points can be described by a Hamiltonian of four Dirac cones related by $C_{4z}T$. Such a Hamiltonian can open a gap perturbatively without breaking $C_{4z}T$. Hence, rather than going through a braiding of Dirac points, the band structure will go through a gap closure followed by an immediate reopening during the transition. Again, according to the discussion in Sec.~$\color{green}{\mathrm{II.D}}$ in the main text, the parameter point with four Dirac cones can be delocalized. Both the band deformation and real space orbital transition are illustrated in the second row of block ``$P4'$-SOC'' of Fig.~$\color{green}{\mathrm{6}}$ in the main text.

Apply the above procedure to other groups under consideration ($Pm,\,P2',\,P6'$), we can identify the features of the minimal OAI transitions and how could they support CMPs. Since both the reciprocal and real space information of these transitions are summarized in Fig.~$\color{green}\mathrm{6}$ in the main text, we will only list RSIs of these groups here.

For $Pm$, we take the reflection as $m_{y}$ ($(x,y)\rightarrow(x,-y)$) and the Wyckoff positions are 1a ($x,0$) and 1b ($x,1/2$).  The site symmetry for both 1a and 1b is $m_{y}$ whose irreps are $A'$ (even under reflection) and $A''$ (odd under reflection) without SOC. The symmetric electron addition/subtraction will induce representation $A'\oplus A''$ on each Wyckoff position. Hence, we have RSI:
\begin{equation}
    \begin{aligned}
        \delta_{a}&=m_a(A')-m_a(A'')=m(X_1)-m(M_2)\\
        \qquad \delta_{b}&=m_b(A')-m_b(A'')=m(X_1)+m(M_2)-n_{\mathrm{band}}.
    \end{aligned}
    \label{nsocPm-rsi}
\end{equation}

Here we use a different notation to that of Bilbao Crystallographic Server which denotes $X_a,\,M_a$ as $Y_a,\,C_a$, respectively. Although the minimal OAI transitions of $Pm$ are realized by band inversions, the transition will go through a gapless parameter region rather than a point. The first minimal transition (first row of ``$Pm$-NSOC'' block of Fig.~$\color{green}{6}$ in main text) replaces one $X_2$ by $X_1$. According to \href{https://www.cryst.ehu.es/cgi-bin/cryst/programs/msitesym.pl}{MSITESYM} program on the Bilbao Crystallographic Server \cite{elcoro_magnetic_2021}, $X_1$ only comes from MEBR $A'_{a,b}\!\uparrow\!G$ while $X_2$ only comes from $A''_{a,b}\!\uparrow\!G$. Hence, mirror reflection action is $1$ ($-1$) on $X_1$ ($X_2$), and the first minimal transition will change the parity of $X$ point under reflection. Also notice that all points along $k_x$ axis have well-defined and identical reflection parities. Hence, to change the parity of the $X$ point, the band inversion should create at least a pair of Dirac points that moving along the $k_x$ axis and change its parity. Hence, this transition can induce a finite parameter region where two Dirac appears at the $k_x$ axis and a CMP is supported when the disorder-induced Dirac mass terms are positively correlated. 

The situation is similar for transitions that changes $m(M_2)$ only. Although $M_2$ comes from both reflection parities ($A_a''\!\uparrow\!G,\,A_b'\!\uparrow\!G$), RSI in terms of orbitals \cref{nsocPm-rsi} suggests that the transition is realized by $A_{a/b}'\leftrightarrow A_{b/a}''$. Hence the transition will create at least a pair of Dirac points moving along the $M\!-\!Y\!-\!M$ line and change its parity. Also, it turn out that $Pm$ with SOC gives out the same results as no SOC.

For $P2'$, the Wyckoff positions are 1a ($0,0$), 1b ($1/2,0$), 1c ($0,1/2$) and 1d ($1/2,1/2$). They have the same site symmetry $2'$ and irrep $A$ without SOC. Note that SOC is irrelevant in $P2'$ since we have $(C_{2z}T)^2=\,^{d}\!1\,T^2=1$ for spinful electrons, and hence SOC will not affect the representation of $P2'$. Now the induced representation of symmetric electron addition/subtraction is $A\oplus A$ and the RSI should be
\begin{equation}
    \delta_{a/b/c/d}=\mathrm{mod}(m(A_{a/b/c/d}),2)
\end{equation}

Only their summation  $\delta_{a+b+c+d}=\mathrm{mod}_2 (m(A_a)+m(A_b)+m(A_c)+m(A_d),2)$ can be expressed in the momentum space
\begin{equation}
    \delta_{a+b+c+d}=\mathrm{mod}(n_{\mathrm{band}},2) 
\end{equation}
Hence, the minimal OAI transition respecting the number of occupied bands will close the band gap at a general k-point. The Dirac Hamiltonian respecting $C_{2z}T$ cannot open a gap perturbatively. Hence, the transition will go through a finite parameter region with at least two Dirac points in contrast to the case of $C_{4z}T$.

For $P6'$, the Wyckoff positions are 1a ($0,0$), 2b ($1/3,2/3$) \& ($2/3,1/3$) and 3c ($1/2,0$) \& ($0,1/2$) \& ($1/2,1/2$). The site symmetry of 1a is $6'$ and the irreps are $A_1$ and $^1\!E^2\!E$ without SOC. $A_1$ is one-dimensional and the $C_{3z}=(C_{6z}T)^2$ action is trivial. $^1\!E^2\!E$ is two-dimensional and the $C_{3z}$ action is $\mathrm{diag}(e^{2\pi i/3}, e^{-2\pi i/3})$. Hence, the symmetric electron addition/subtraction will change both $m_a(A_1)$ and $m_a(^1\!E^2\!E)$ by $2$. As to 2b, the site symmetry is $3$ and irreps are $A_1$ ($C_{3z}=1$), $^1\!E$ ($C_{3z}=e^{-2\pi i/3}$), and $^2\!E$ ($C_{3z}=e^{2\pi i/3}$). All of these irreps are one-dimensional. Hence, the symmetric electron addition/subtraction will change all of $m_b(A_1),\,m_b(^1\!E),\,m_b(^2\!E)$ by $1$   
As to 3c, the site symmetry is $2'$ and the only 1-dim irrep is A without SOC. Hence the symmetric electron addition/subtraction will change $m_c(A)$ by 2. From the above, we can conclude the RSI: 

\begin{equation}
    \begin{aligned}
    \delta_{a1}&=m_a(A_{1})-m_a(^1\!E^2\!E)\\
    \delta_{a2}&=\mathrm{mod}(m_a(A_{1}),2)\\
    \delta_{b1}&=m_b(A_{1})-m_b(^1\!E)\\
    \delta_{b2}&=m_b(A_{1})-m_b(^2\!E)\\
    \delta_c&=\mathrm{mod}(m_c(A),2)
    \end{aligned}
    \label{nsocP6'-orb}
\end{equation}

Only four (conbinations) of them can be translated to band indicators:
\begin{equation}
\begin{aligned}
\delta_{a1}&=n_{\mathrm{band}}-m(\Gamma_2\Gamma_3)-m(K_2K_3)-m(K_2'K_3')\\
\delta_{b1}&=m(K_2'K_3') -m(\Gamma_2\Gamma_3)\\
\delta_{b2}&=m(K_2K_3)-m(\Gamma_2\Gamma_3)\\
\delta_{a2+c}&=\mathrm{mod}(m_a(A_1)+m_c(A),2)=\mathrm{mod}(n_{\mathrm{band}},2)
\end{aligned}
\label{nsocP6'-bandind}
\end{equation}

There are four minimal OAI transitions as illustrated by the ``$P6'$-NSOC'' block of Fig.~$\color{green}\mathrm{6}$ in the main text. The first three correspond to band inversions at $\Gamma$, $K$ and $K'$ points respectively. As the first minimal transition of $P4'$-NSOC, all of them will go through finite parameter regions with quadratic touchings from 2-dim irreps near the Fermi level. The last minimal transition decreases (increases) $m_a(A_1)$, $m_a(^1E^2E)$ by $1$ and increase (decreases) $m_c(A)$ by $1$. Hence, the last transition is invisible to quantities in \cref{nsocP6'-bandind} and will close band gap at (at least) 6 general k-points enforced by $C_{6z}T$. Since $(C_{6z}T)^3=C_{2z}T$ should be respected, the effective Dirac Hamiltonian cannot open a gap perturbatively. According to the discussion in Sec.~$\color{green}\mathrm{II.D}$ of the main text, a CMP can exist in the parameter region where the touching points braid each other. 

The situation of $P6'$ with SOC is the same to $P6'$ without SOC. When spin is under consideration, $(C_{6z}T)^{6}=1,\,(C_{6z}T)^2=-C_{3z}$, and hence the representations of double group is the same to that of the original group. 

We would like to end this section here. One can directly generalize the above argument to more complicated magnetic groups. In a summary, by analyzing band deformations of OAI transitions in several magnetic space groups, we have seen that CMPs are commonly supported by average symmetries protecting the percolation mechanism. Therefore, CMPs in 2D class A systems have a general relevance and whose nature deserves further investigation.

\clearpage

\section{Network model}\label{sec:Network-model}

\subsection{Network model on the Manhattan lattice}\label{subsec:Manhattan}

\begin{figure}[h]
	\centering
	\subfigure[Profile of the network model. A vertical (horizontal) wire labeled by $n,\alpha$ ($m,\alpha$) lies in $x=2a(n+\frac{1+\alpha}4)$ ($y=2a(m+\frac{1+\alpha}4)$). Here $a$ is the side length of the colored squares, and the lattice constant is $2a$. ]{\label{network-a}
    \includegraphics[width=.5\linewidth]{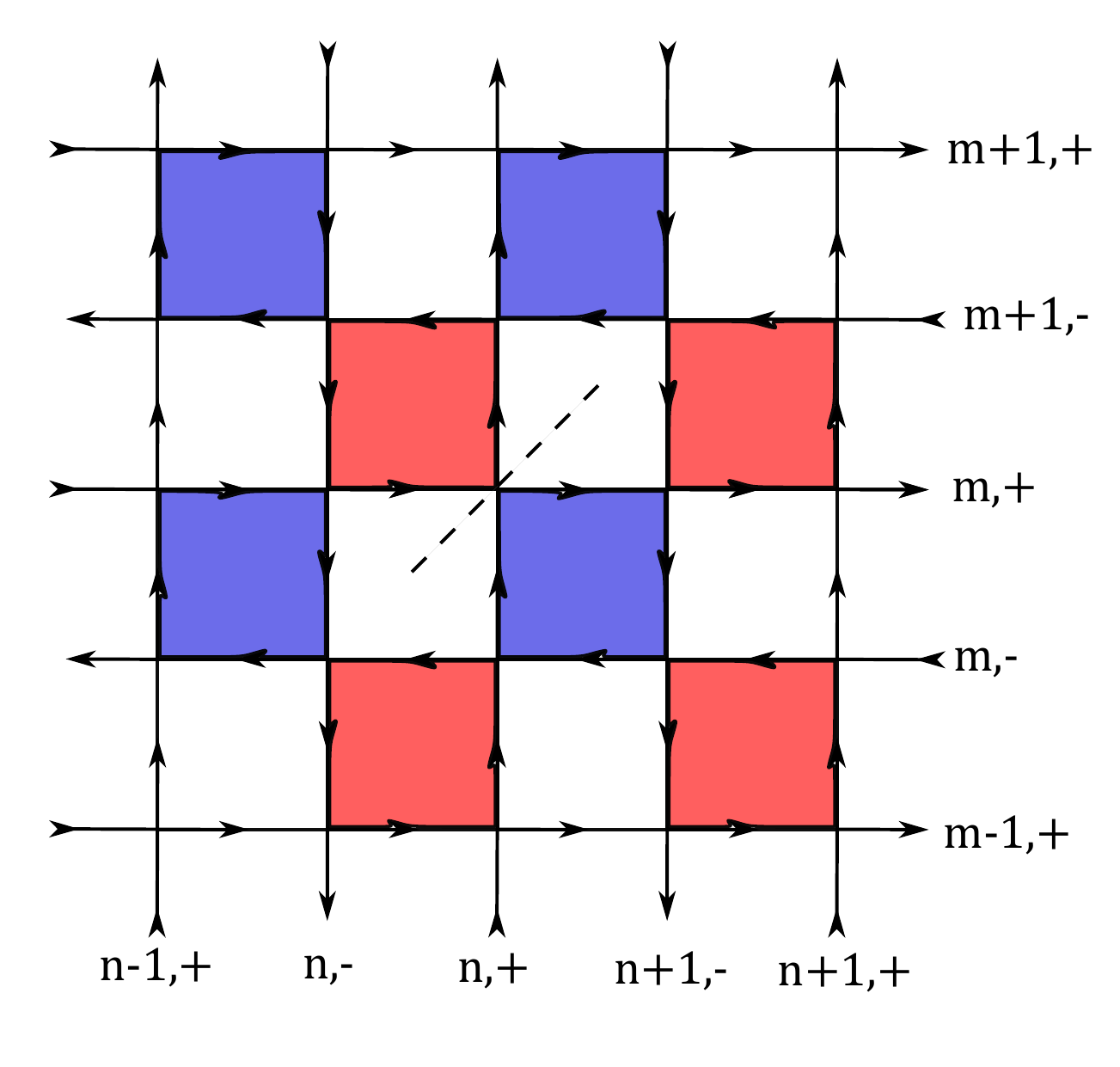}}
    \quad
	\subfigure[Channels of a scattering node]{\label{network-b}
	\includegraphics[width=.25\linewidth]{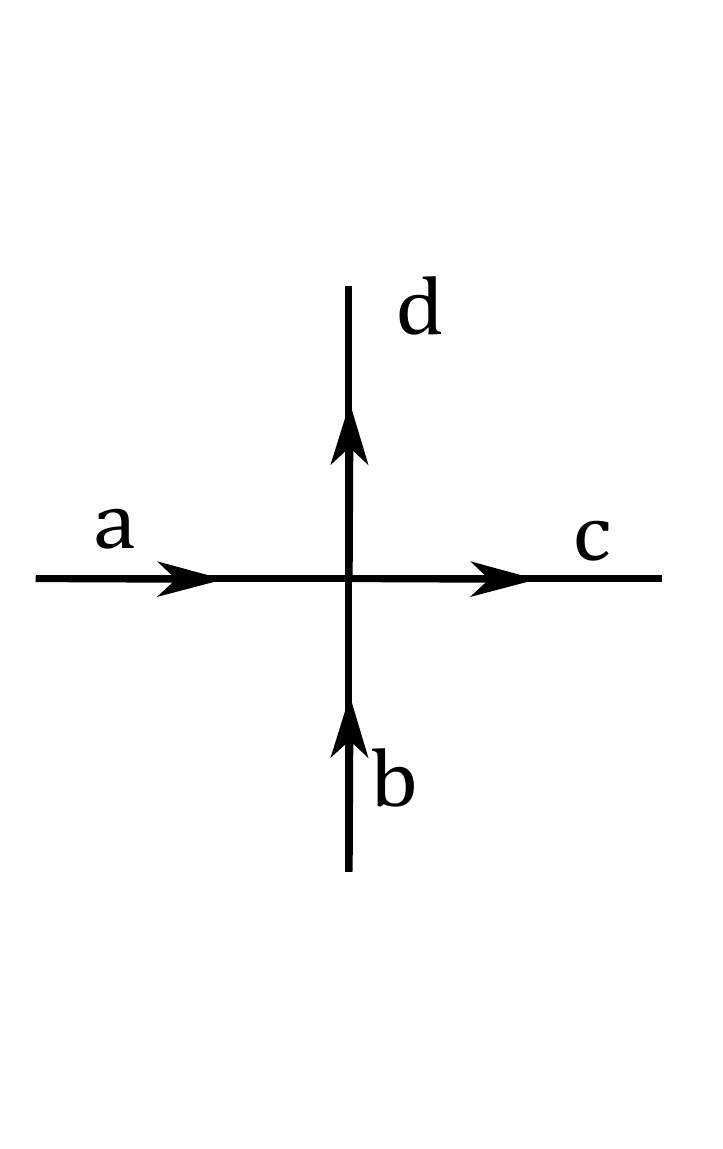}}
    \caption[]{Network and the scattering node}
	\label{network}
\end{figure}

The wires in Fig.~\ref{network-a} represent the 1D chiral modes with directions indicated by the arrows.
There is a $\delta$ scattering potential at every intersection of the wires.
We can view the red (blue) squares as Chern blocks with Chern number $C=1(-1)$ since the edges of colored squares can be viewed as the edge state of local Chern insulators.
The scattering potentials give rise to tunnelings between Chern blocks. 

The network model has $C_{2z}T = \{2'_{001}|0,0,0\}$, $C_{4z} =\{4_{001}^+ | 0, \frac12, 0 \}$ and $M_{xy} = \{ m_{1\bar10} | 0, 0, 0\}$ symmetries, where the fractional translations are in units of the lattice constant $2a$ (\cref{network-a}), \ie $\{R| \alpha,\beta,\gamma\}$ represents a rotation or/and reflection operation $R$ followed by a translation operation $\vec{t}=(2\alpha a, 2\beta a, 2\gamma a)$. As to the rotation/reflection part, $2'_{001}$ represents the combination of a $\pi$-rotation along the $z$-direction and a time-reversal operation; $4^+_{001}$ represents an anticlockwise $\pi/2$-rotation viewed from the positive z-direction; $m_{1\bar{1}0}$ represents a reflection whose normal is along the $(1,-1,0)$-direction. 
The $C_{2z}T$ and $C_{4z}$ centers locate at the intersections and Chern block centers, respectively. 
The mirror plane of $M_{xy}$ is indicated by the dashed line in \cref{network-a}. 
$C_{4z}$ rotations transform the red (blue) squares into red (blue) squares while $C_{2z}T$ transform red (blue) into blue (red) ones. Notice that the $C_{2z}T $ centers and $C_{4z}$ centers are not coincident, hence the joint action $C_{4z}^2C_{2z}T\neq T$. 
Actually, these symmetries generate a type-IV magnetic space group $P_{C}4bm$ (\#100.177 in BNS setting)  with the magnetic translation given by $\widetilde{T}=\{ 1'|\frac12, \frac12, 0 \} = C_{2z}T \cdot C_{4z}^2 \cdot \{1 | 0,1,0 \}$, where $1'$ represents the time reversal operation. 
(Here, we adopt a different convention of the origin point as the \href{https://www.cryst.ehu.es/cgi-bin/cryst/programs/magget_gen.pl}{MGENPOS} program on the Bilbao Crystallographic Server \cite{gallego_magnetic_2012}. 
To obtain our symmetry operations, one should specify the ``origin shift'' as $(\frac14,-\frac14,0)$ when using the MGENPOS program.)

We label the four channels involved in a scattering node as a, b, c, and d. (see Fig.~\ref{network-b}). One can regard $a$ \& $b$ as the incident channels and $c$ \& $d$ as the outcoming channels. The scattering potential is defined by its action on the channels:
\begin{equation}
     \begin{aligned}
        \begin{pmatrix}
           \psi_c(0) \\
           \psi_{d}(0)
        \end{pmatrix}
       =
        \begin{pmatrix}
        \cos\theta & -i\sin\theta \\
        -i\sin\theta & \cos\theta
        \end{pmatrix}
       & \begin{pmatrix}
        \psi_a(0)\\
        \psi_b(0)
        \end{pmatrix}
        \quad
        \theta \in [-\pi/2,\pi/2],
    \end{aligned}
    \label{scatter matrix}
\end{equation}
where $\psi_j(0)$ is the amplitude of channel $i$ near the scattering node. Notice that the scattering effect of $\theta+\pi$ is merely that of $\theta$ multiplied by $-1$ which can be removed by redefining $\psi_c$ and $\psi_d$. It is compatible to take such a redefinition on every intersection, and hence the effect of replacement $\theta\rightarrow\theta+\pi$ is a pure gauge transformation. Therefore, we take $\theta\in[-\pi/2,\pi/2]$ rather than $\theta\in[-\pi,\pi]$. Also, as we will see in \cref{subsec:numerical-network}, $\theta\in[-\pi/2,\pi/2]$ is enough to realize a localized-delocalized-localized percolation transition.

When $\theta=-\pi/2$, the network decouples to disconnected red ($C=1$) and blue ($C=-1$) squares. The electrons then form local loop currents surrounding them. According to Eq.~(\ref{scatter matrix}), an electron in a local current will obtain a $\pi/2$ phase at each corner when going along the arrows and accumulate a $2\pi$ phase after going around. In other words, an electron will obtain a phase $2\pi\,(-2\pi)$ after going around the red (blue) square anticlockwise, showing that the red (blue) squares indeed has Chern number $1\,(-1)$. Hence one can view the local orbitals indicated by red squares as having ``angular momentum" $L=1$, and the blue ones have $L=-1$.  
When $\theta=\pi/2$, we also have a group of decoupled squares, but the phase jump will be reversed, leading to ``angular momentum'' $L=-1$ for red squares and $1$ for blue ones. 
So in both limits ($\theta=\pm \pi/2$), the system comprises a group of well-separated local currents and hence is localized, but the two limits have opposite ``angular momentum" at each square. 

The disorder in this paper is the random size of Chern blocks which can be realized by the random phases on square edges in the network. More discussion about the disorder can be found in \cref{subsec:disorder}. In this section, we focus on the clean limit, in which the system has to go through a metallic phase when we change $\theta$ from $-\pi/2$ to $\pi/2$.
When $\theta=0$, the network is merely two bundles of decoupled chiral wires states and is hence gapless.

\subsection{Effective Hamiltonian $H_{N}$ and band structure of the network model}\label{subsec:H-network}

We use the wires in the limit $\theta=0$ as the basis of the network model and denote the fermion annihilation operator on horizontal and vertical wires as $\psi_{h}(x;m,\alpha)$ and $\psi_{v}(y;n,\alpha)$, respectively. Note that we use different notations from Eq.~$\color{green}{(2)}$ except for the direction subscript $d=h,v$. 
Here, $x\,(y)$ is the coordinate inside a horizontal (vertical) wire, and $m\,(n)$ \& $ \alpha=\pm1$ together label different horizontal (vertical) wires (see \cref{network-a}, where $(m,n)$ indicates the unit cell and $\alpha$ indicates the chirality). The indices $\xi$ and $l$ in Eq.~$\color{green}{(2)}$ correspond to $x$ or $y$ and $(2m+\alpha)a$ or $(2n+\alpha)a$, respectively. 
We choose a global coordinate system in which $\psi_{h}(x;m,\alpha)$, $\psi_{v}(y;n,\alpha)$ act on $(x,y=(2m+(1+\alpha)/2)a),\; (x=(2n+(1+\alpha)/2)a,y)$, respectively. 
Here $a$ is the side length of the Chern squares. The unit cell $(m,n)$ corresponds to the intersection region of two horizontal wires $(m,-)\,\&\,(m,+)$ and two vertical wires $(n,-)\,\&\,(n,+)$.
These annihilation and creation operators are subject to the anti-commutation relations
\begin{equation}
    \{ \psi_{h}(x;m,\alpha), \psi_{h}^\dagger(x';m',\alpha') \} = \delta_{m,m'} \delta_{\alpha,\alpha'} \delta(x-x'),\qquad 
    \{ \psi_{v}(y;m,\alpha), \psi_{v}^\dagger(y';m',\alpha') \} = \delta_{m,m'} \delta_{\alpha,\alpha'} \delta(y-y')\ . 
\end{equation}

\begin{figure}[!h]
	\centering
	\subfigure{
    \includegraphics[width=.7\linewidth]{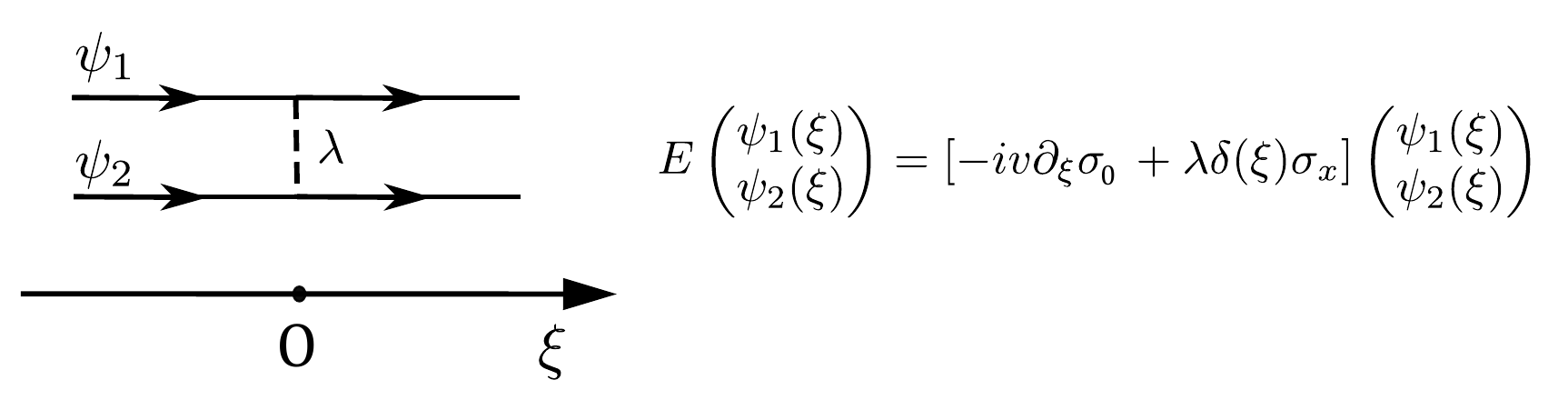}
    }
    \caption[]{Scattering problem of a single node}
	\label{scatter2}
\end{figure}

The Hamiltonian of the network model $H_{N}$ has two parts. The first part is the decoupled wires:
\begin{equation}
\label{network-decoupled-chiral-wires}
    H_{N,0}=\sum_{m,\alpha} \alpha v \int \mathrm{d}x \thickspace \psi_{h}^{\dagger}(x;m,\alpha) (-i \partial_x ) \psi_{h}(x;m,\alpha)+\sum_{n,\alpha} \alpha v \int \mathrm{d}y \thickspace \psi_{v}^{\dagger}(y;n,\alpha) (-i \partial_y ) \psi_{v}(y;n,\alpha)\ .
\end{equation}
The scattering effect in Eq.~(\ref{scatter matrix}) can be realized by the second part: \emph{real} $\delta$-scattering potentials:
\begin{equation}
\label{network-delta-scattering-potential}
    H_{N,1}=\sum_{m,n,\alpha,\beta} \lambda \thinspace  \psi_{h}^{\dagger}(2na+\frac{1+\beta}{2}a;m,\alpha) \thinspace \psi_{v}(2ma+\frac{1+\alpha}{2}a;n,\beta) \thinspace + \thinspace h.c.
\end{equation}
where $\lambda$ has the dimension of velocity. 
For a scattering node shown in Fig.~\ref{network-b}, we can rotate the vertical wire to $x$-direction and view the scattering as a 1D  problem shown in Fig.~\ref{scatter2}. 
By solving this problem, we will obtain the relation between $\lambda$ and $\theta$ of \cref{scatter matrix}.

We can define $\begin{pmatrix} \psi_1(\xi) \\ \psi_2(\xi) \end{pmatrix}=\psi_+(\xi) \mathbf{e_+}+\psi_-(\xi)\mathbf{e_-}$ on the basis basis $\mathbf{e_+}=\begin{pmatrix} 1\\1 \end{pmatrix},\,\mathbf{e_-}=\begin{pmatrix} 1\\-1 \end{pmatrix}$ and decouple the Schrodinger equation to
\begin{equation}
\label{SE1-scattering-node}
E\psi_{\pm}(\xi)=[-iv\thinspace\frac{\mathrm{d}}{d{\xi}}\pm\lambda\delta(\xi)]\psi_{\pm}(\xi)\ .
\end{equation}
Divide both sides by $\psi_{\pm}(\xi)$, we obtain
\begin{equation}
\label{SE2-scattering-node}
    -iv \frac{\mathrm{d} \ln \psi_{\pm}(\xi)}{\mathrm{d} \xi}= E \mp \lambda \delta(\xi).
\end{equation}
For $\xi\neq0$, there must be
\begin{equation}
\label{scattering-away-from-zero}
    \psi_{\pm}(\xi)=
    \left\{ \begin{aligned}
    \psi_{\pm}(0^-)\exp\left(i\frac{E}{v} \xi\right) \quad (\xi<0)\\
    \psi_{\pm}(0^+)\exp \left(i\frac{E}{v} \xi\right)\quad (\xi>0) \\
    \end{aligned}
    \right. \ . 
\end{equation}
Integrate Eq.~(\ref{SE2-scattering-node}) near zero, \ie $\int_{0^-}^{0^+}$, we have
\begin{equation}
\label{new-scattering-result}
    \psi_{\pm}(0^+)=e^{\mp i\lambda/v} \psi_{\pm}(0^-) 
    \Longrightarrow 
    \begin{pmatrix}
    \psi_1(0^+)\\
    \psi_2(0^+)
    \end{pmatrix}
    =
    \begin{pmatrix}
    \cos\frac{\lambda}{v}& -i \sin \frac{\lambda}{v}\\
    -i \sin\frac{\lambda}{v} & \cos\frac{\lambda}{v}
    \end{pmatrix}
    \begin{pmatrix}
    \psi_1(0^-) \\
    \psi_2(0^-)
    \end{pmatrix}.
\end{equation}
Now we can conclude that $\theta=\lambda/v$. Recall that only $\theta\in[-\pi/2,\pi/2]$ is considered because $\theta\rightarrow\theta+\pi$ is merely a gauge transformation that will not affect the spectrum. We should take $\lambda/v \in[-\pi /2,\pi /2]$. Certainly, $\lambda/v \in [(2n-1/2)\pi,(2n+1/2)\pi]$ is also feasible, but numerical calculation with smaller $\vert\lambda/v\vert$ will be more accurate (see \cref{network-H-reciprocal} and the discussion below \cref{eq:sym-reciprocal}, a larger $|\lambda/v|$ will introduce a larger cutoff error).

The actions of $C_{2z}T$, $C_{4z}$ and $M_{xy}$ are:
\begin{equation}
\label{eq:sym-real}
    \begin{aligned}
    (C_{2z}T) \psi_{h}(x;m,\alpha) (C_{2z}T)^{-1}&= \psi_{h}(-x;(-m-\frac{1+\alpha}{2}),\alpha) \\
    (C_{2z}T) \psi_{v}(y;n,\alpha) (C_{2z}T)^{-1}&= \psi_{v}(-y;(-n-\frac{1+\alpha}{2}),\alpha)\\
    C_{4z} \psi_{h}(x;m,\alpha) C_{4z}^{-1}&= \psi_{v}(x+a;(-m-\frac{1+\alpha}{2}),\alpha)\\ 
    C_{4z} \psi_{v}(y;n,\alpha) C_{4z}^{-1}&= \psi_{h}(-y;(n+\frac{1+\alpha}{2}),-\alpha)\\ 
     M_{xy} \psi_{h}(x;m,\alpha) M_{xy}^{-1}&= \psi_{v}(x;m,\alpha)\\ 
    M_{xy} \psi_{v}(y;n,\alpha) M_{xy}^{-1}&= \psi_{h}(y;n,\alpha)\\ 
    \end{aligned}
\end{equation}
It is direct to check that 
\begin{equation}
(C_{2z}T) (H_{N,0}+H_{N,1}) (C_{2z}T)^{-1}=C_{4z}(H_{N,0}+H_{N,1})C_{4z}^{-1}=m_{xy}(H_{N,0}+H_{N,1})m_{xy}^{-1}=H_{N,0}+H_{N,1}
\end{equation}

In addition, we should note that this model has an accidental particle-hole symmetry
\begin{equation}
    P \psi_h(x;m,\alpha) P^{-1}=\psi_h^{\dagger}(x;m,\alpha) \quad P \psi_v(y;n,\alpha) P^{-1}= -\psi_v^{\dagger}(y;n,\alpha)
\end{equation}
However, this will be broken after introducing the disorder (in the form of vector potential, see \cref{subsec:disorder}).

To obtain the band structure of the network model, we need to rewrite the Hamiltonian in reciprocal space. 
Considering the discrete translation symmetry of $H_{N,1}$, we write down the Fourier transform of $\psi_h(x), \thinspace \psi_v (y)$ in this form:
\begin{equation}
\label{Fourier-network}
    \begin{aligned}
    & \psi_h(x;m,\alpha)=\frac{1}{\sqrt{NL}}\sum_{k_x,k_y}\sum_{G_x} \exp\left[i (k_x+G_x)x+ ik_y \left(2m+\frac{1+\alpha}{2}\right)a\right] \phi_{h}(k_x,k_y;\alpha,G_x)\\
    & \psi_v(y;n,\alpha)=\frac{1}{\sqrt{NL}}\sum_{k_x,k_y}\sum_{G_y} \exp\left[i (k_y+G_y)y+ ik_x \left(2n+\frac{1+\alpha}{2}\right)a\right] \phi_{v}(k_x,k_y;\alpha,G_y),
    \end{aligned}
\end{equation}
where $L=2Na$ is the system size in each direction, and the system contains $N\times N$ unit cells.
Since $x,\thinspace y$ are continuous variables, their Fourier co-variables should be unbounded. 
We use $k_j=\frac{2\pi n_j}{2Na}$ ($n_j=0\cdots N-1$) to fit the lattice periodicity (the ``first" BZ) and the boundlessness of $k_j$ is compensated by the reciprocal vector $G_j=\frac{2\pi N_j}{2a}$ ($N_j=0,\pm1,\pm2 \cdots$). 
The prefactor $\frac1{\sqrt{NL}}$ is used to normalize the operator such that 
\begin{equation}
    \{ \phi_{h/v}(k_x,k_y;\alpha,G_x), \phi_{h/v}^\dagger(k_x',k_y';\alpha',G_x') \}
    = \delta_{k_x,k_x'} \delta_{k_y,k_y'} \delta_{\alpha,\alpha'} \delta_{G_x,G_x'}\ .
\end{equation}
The inverse transformation can be written as 
\begin{equation}
\label{inverse-Fourier-network}
\begin{aligned}
\phi_h(k_x,k_y; \alpha, G_x) =& \frac1{\sqrt{N L}} \sum_{m} \int dx \ \exp\brak{ - i (k_x+G_x)x -ik_y \pare{2m + \frac{1+\alpha}2}a } \psi_h(x;m,\alpha) \\
\phi_v(k_x,k_y; \alpha, G_y) =& \frac1{\sqrt{N L}} \sum_{n} \int dy \ \exp\brak{ - i (k_y+G_y)y -ik_x \pare{2n + \frac{1+\alpha}2}a } \psi_v(y;n,\alpha) 
\end{aligned} \ . 
\end{equation}

The Hamiltonian in reciprocal space can be written as
\begin{equation}
\label{network-H-reciprocal}
    \begin{aligned}
    H_{N,0}=&\sum_{k_{x},k_{y}}\sum_{G_{x},\alpha} \alpha  v  (k_{x}+G_{x}) \phi_{h}^{\dagger}(k_{x},k_{y};\alpha,G_{x})  \phi_{h}(k_{x},k_{y};\alpha,G_{x})\\
    &+\sum_{k_{x},k_{y}}\sum_{G_{y},\alpha} \alpha  v  (k_{y}+G_{y}) \phi_{v}^{\dagger}(k_{x},k_{y};\alpha,G_{y})  \phi_{v}(k_{x},k_{y};\alpha,G_{y})\\
    H_{N,1}=&\frac{\lambda}{2a}\sum_{k_{x},k_{y}}\sum_{G_{x},G_{y}}\sum_{\alpha,\beta} \exp\left[i(\frac{1+\alpha}{2}) G_{y}a - i(\frac{1+\beta}{2})G_{x}a\right] \phi_{h}^{\dagger}(k_{x},k_{y};\alpha,G_{x}) \phi_{v}(k_{x},k_{y};\beta,G_{y}) \thinspace + \thinspace h.c.
    \end{aligned}
\end{equation}
The symmetry operations on the reciprocal basis can also be derived as
\begin{equation}
\begin{aligned}
C_{2z}T \phi_{h}(k_x,k_y; \alpha,G_x) (C_{2z}T)^{-1}
=& \phi_{h}(k_x,k_y; \alpha,G_x) \\
C_{2z}T \phi_{v}(k_x,k_y; \alpha,G_y) (C_{2z}T)^{-1}
=& \phi_{v}(k_x,k_y; \alpha,G_y) \\
C_{4z} \phi_{h}(k_x,k_y; \alpha,G_x) C_{4z}^{-1}
=& e^{i (k_x+G_x)a} \phi_{v}(-k_y,k_x; \alpha,G_x) \\
C_{4z} \phi_{v}(k_x,k_y; \alpha,G_y) C_{4z}^{-1}
=& e^{ik_x a} \phi_{h}(-k_y,k_x; -\alpha,-G_y) \\
M_{xy} \phi_{h}(k_x,k_y; \alpha,G_x) M_{xy}^{-1}
=& \phi_{v}(k_y,k_x; \alpha,G_x) \\
M_{xy} \phi_{v}(k_x,k_y; \alpha,G_y) M_{xy}^{-1}
=& \phi_{h}(k_y,k_x; \alpha,G_y)
\end{aligned}\ .
\label{eq:sym-reciprocal}
\end{equation}
Here the phase factors $e^{ia(k_x+G_x)}$ and $e^{ia k_x}$ result from the fractional translation in the $C_{4z}$ operator. Take $\phi_{h}(k_x,k_y;\alpha,G_x)$ for an example:
{\small
\begin{equation}
\begin{aligned}
    C_4\phi_{h}(k_x,k_y;\alpha,G_x)C_4^{-1} &\overset{\mathrm{\cref{inverse-Fourier-network}}}{=}\frac{1}{\sqrt{NL}} \sum_m \int dx \exp\left[-i(k_x+G_x)x-ik_y(2m+\frac{1+\alpha}{2})a\right]C_4\psi_h(x;m,\alpha)C_4^{-1}\\
    & \overset{\mathrm{\cref{eq:sym-real}}}{=}\frac{1}{\sqrt{NL}}\sum_m \int dx \exp\left[-i(k_x+G_x)x-ik_y(2m+\frac{1+\alpha}{2})a\right]\psi_v(x+a;(-m-\frac{1+\alpha}{2}),\alpha)\\
    & =\frac{e^{i(k_x+G_x)a}}{\sqrt{NL}}\sum_m \int dx' \exp\left[-i(k_x+G_x)x'+ik_y\left(-2(m+\frac{1+\alpha}{2})+\frac{1+\alpha}{2}\right)a\right]\psi_v(x';(-m-\frac{1+\alpha}{2}),\alpha)\\
    &=e^{i(k_x+G_x)a} \phi_v(-k_y,k_x;\alpha,G_x)
\end{aligned}
\end{equation}}%

When we numerically calculate the band structure of the network model, we have to adopt a cutoff $\Lambda_1$ for the $G_x, G_y$ indices.
The hybridization introduced by $H_{N,1}$ between any two $G$ indices is on the order of $\lambda/a$.
Therefore, as long as the kinetic energy at the cutoff, \ie $ v \Lambda_1 $, is much larger than $\lambda/a$, the off-diagonal terms between high/low lying bands and the middle bands are negligible. Hence, a high enough cutoff $\Lambda_1$ should not affect the structure of the middle bands. 
However, simply discarding terms in Eq.~(\ref{network-H-reciprocal}) that involve $\vert{G}\vert>\Lambda_1$ is problematic (see \cref{subsec:cutoff} for details).
Another cutoff parameter $\Lambda_2$ on $H_{N,1}$ should be considered. The scattering between two modes with a momentum difference larger than $\Lambda_2$ will be strongly suppressed, which corresponds to a broadened $\delta$-potential with a characteristic length proportional to $1/\Lambda_2$.

In practice, we will (i) use a basis set up to the cutoff $\Lambda_1$ and (ii) omit the $\lambda$ coupling between modes with momentum difference larger than $\Lambda_2$.
Therefore, the actual Hamiltonian with finite $\Lambda_{1,2}$ can be written as 
\begin{equation}
\label{network-H-reciprocal-cutoff}
    \begin{aligned}
    H_{N,0}=&\sum_{k_{x},k_{y}}\sum_{\alpha}\sum_{|G_{x}|<\Lambda_1} \alpha  v  (k_{x}+G_{x}) \phi_{h}^{\dagger}(k_{x},k_{y};\alpha,G_{x})  \phi_{h}(k_{x},k_{y};\alpha,G_{x})\\
    &+\sum_{k_{x},k_{y}}\sum_{\alpha}\sum_{|G_{y}|<\Lambda_1} \alpha  v  (k_{y}+G_{y}) \phi_{v}^{\dagger}(k_{x},k_{y};\alpha,G_{y})  \phi_{v}(k_{x},k_{y};\alpha,G_{y})\\
    H_{N,1}=&\frac{\lambda}{2a}\sum_{k_{x},k_{y}}\sum_{\alpha,\beta} \sum_{|G_{x}|<\Lambda_1}\sum_{|G_{y}|<\Lambda_1}\Xi_{\Lambda_2}(\alpha (k_x+G_x)-\beta (k_y+G_y))\exp\left[i(\frac{1+\alpha}{2}) G_{y}a - i(\frac{1+\beta}{2})G_{x}a\right] \\ & \times \phi_{h}^{\dagger}(k_{x},k_{y};\alpha,G_{x}) \phi_{v}(k_{x},k_{y};\beta,G_{y}) \thinspace + \thinspace h.c.
    \end{aligned}
\end{equation}
where $\Xi_{\Lambda_2}$ is a hard-cutoff factor
\begin{equation} \label{eq:truncation}
    \Xi_{\Lambda_2}(\tau)=\left\{
    \begin{aligned}
        0 \quad (\vert\tau\vert >\Lambda_2)\\
        1 \quad  (\vert\tau\vert <\Lambda_2)\\
        \frac{1}{2} \quad  (\vert\tau\vert =\Lambda_2)
    \end{aligned}
    \right.\ . 
\end{equation}
In all the calculations, we choose $\Lambda_1\gg \Lambda_2$ such that the degrees of freedom on a chiral wire inside the $\delta$ potential region ($\sim 1/\Lambda_2$) can be treated as continuous. 
We leave further discussions on $\Lambda_{1,2}$ to \cref{subsec:cutoff}.

\begin{table}[t]
\centering{
\begin{tabular}{|lccccc|lccccc|lc|}
\hline
 & $\Gamma_1$ & $\Gamma_2$ & $\Gamma_3$ & $\Gamma_4$ & $\Gamma_5$ &  & $M_1$ & $M_2$ & $M_3$ & $M_4$ & $M_5$ &  & $X_1$ \\ \hline
\multicolumn{1}{|l|}{$\{1\vert 0,0,0\}$} & 1 & 1 & 1 & 1 & 2 & \multicolumn{1}{l|}{$\{1\vert 0,0,0\}$} & 1 & 1 & 1 & 1 & 2 & \multicolumn{1}{l|}{$\{1\vert 0,0,0\}$} & 2 \\
\multicolumn{1}{|l|}{$C_{2z} = \{2_{001}\vert - \frac12, \frac12, 0\}$} & 1 & 1 & 1 & 1 & -2 & \multicolumn{1}{l|}{$C_{2z} = \{2_{001}\vert - \frac12, \frac12, 0\}$} & $-1$ & $-1$ & $-1$ & $-1$ & 2 & \multicolumn{1}{l|}{$C_{2z} = \{2_{001}\vert - \frac12, \frac12, 0\}$} & 0 \\
\multicolumn{1}{|l|}{$C_{4z}=\{4_{001}^+ \vert 0,\frac12,0\}$} & 1 & -1 & -1 & 1 & 0 & \multicolumn{1}{l|}{$C_{4z}=\{4_{001}^+ \vert 0,\frac12,0\}$} & $i$ & $-i$ & $-i$ & $i$ & 0 & \multicolumn{1}{l|}{$\{m_{100}\vert 0,\frac{1}{2},0\}$} & 0 \\
\multicolumn{1}{|l|}{$M_{xy}=\{m_{1\bar{1}0}\vert 0,0,0\}$} & 1 & -1 & 1 & -1 & 0 & \multicolumn{1}{l|}{$M_{xy}=\{m_{1\bar{1}0}\vert 0,0,0\}$} & -1 & 1 & -1 & 1 & 0 & \multicolumn{1}{l|}{$\{m_{010}\vert\frac{1}{2},0,0\}$} & 0 \\ \hline
\end{tabular}
}
\caption[]{Character table of irreps at high-symmetry momenta in magnetic space group $P_C4bm$ ($\#100.177$ in BNS setting), taken from the  \href{https://www.cryst.ehu.es/cgi-bin/cryst/programs/corepresentations.pl}{COREPRESENTATIONS} program on the Bilbao Crystallographic Server \cite{elcoro_magnetic_2021}. 
Characters of the listed symmetry operations can uniquely determine the irreps
One should notice that we use a different convention of the origin point as the Bilbao Crystallographic Server, as explained in the second paragraph in \cref{subsec:Manhattan}. 
Our $C_{2z} = \{2_{001}\vert - \frac12, \frac12, 0\}$, $C_{4z}=\{4_{001}^+ \vert 0,\frac12,0\}$, $M_{xy}=\{m_{1\bar{1}0}\vert 0,0,0\}$, $\{m_{100}\vert 0,\frac{1}{2},0\}$, and  $\{m_{010}\vert\frac{1}{2},0,0\}$ correspond to $\{2_{001} | 0,0,0\}$, $\{4_{001}^+ | 0,0,0\}$, $\{m_{1\bar{1}0} | \frac12, - \frac12,0\}$, $\{m_{100}\vert \frac12,\frac{1}{2},0\}$, and $\{m_{010}\vert \frac12, - \frac{1}{2},0\}$ in the standard convention of the Bilbao Crystallographic Server, respectively. 
}
\label{Irreps}
\end{table}

\begin{table}[t]
\centering{
\begin{tabular}{|l|c|c|c|c|c|}
\hline
Wyckoff pos. & \multicolumn{4}{c|}{$\mathrm{2b}\, (\frac34 \frac14 0),\,(\frac{1}{4} \frac{3}{4} 0)$} & \begin{tabular}[c]{@{}c@{}}$\mathrm{4c} \,(0 0 0),\,(0 \frac12 0),$\\  $(\frac12 0 0), (\frac12 \frac12 0)$ \end{tabular} 
\\ \hline
Site sym.   &\multicolumn{4}{c|}{$4m^{\prime}m^{\prime},\,4$}         & $2^{\prime}m^{\prime}m,\,m$\\
\hline 
MEBR &\multicolumn{1}{c|}{$\mathrm{A_b\!\uparrow\!G}$} & \multicolumn{1}{c|}{$\mathrm{B_b\!\uparrow\!G}$} & \multicolumn{1}{c|}{$\mathrm{^{1}E_b\!\uparrow\!G} $} & $\mathrm{^{2}E_b\!\uparrow\!G} $ & $\mathrm{A^{\prime \prime}_c\!\uparrow\!G}$ \\ \hline
Orbital & \multicolumn{1}{c|}{$1$}&\multicolumn{1}{c|}{$ d_{x^2-y^2} + i d_{xy} $}& \multicolumn{1}{c|}{$p_x+i p_y$}  & $p_x-ip_y$ & $ p_y$ \\ \hline
Irreps at $\Gamma$ & $\Gamma_1\oplus\Gamma_4$ & $\Gamma_2\oplus\Gamma_3$ & $\Gamma_5$ & $\Gamma_5$ & $\Gamma_2\oplus\Gamma_4\oplus \Gamma_5$ \\
\hline
Irreps at $M$ & $M_5$ & $M_5$ & $M_2\oplus M_3$ & $M_1\oplus M_4$ & $M_2\oplus M_4\oplus M_5$ \\
\hline
Irreps at $X$ & $X_1$ & $X_1$ & $X_1$ & $X_1$ & $2 X_1$ \\
\hline
\end{tabular}
}
\caption{MEBRs of $P_C4bm $ ($\#100.177$ in BNS setting) involved in this work, taken from the \href{https://www.cryst.ehu.es/cgi-bin/cryst/programs/mbandrep.pl}{MBANDREP} program on the Bilbao Crystallographic Server \cite{elcoro_magnetic_2021}. 
The real space orbital character of each MEBR is shown in the ``Orbital'' row.
For example, the MEBR $\mathrm{ B_b \uparrow G}$ can be generated by an $d_{x^2-y^2}+i d_{xy}$ type orbital at the first 2b position $(\frac34,\frac14,0)$. Note that $\mathrm{A^{\prime \prime}_c\!\uparrow\!G}$ is the only decomposable MEBR, although bands with this MEBR are always connected in this work.
One should notice that we use a different convention of the origin point as the Bilbao Crystallographic Server, as explained in the second paragraph in \cref{subsec:Manhattan}. 
}
\label{EBRs}
\end{table}

\begin{figure}[h!]
	\centering
	\subfigure[$\theta=-\pi/2$]{\label{band-network-a}
    \includegraphics[width=.45\linewidth]{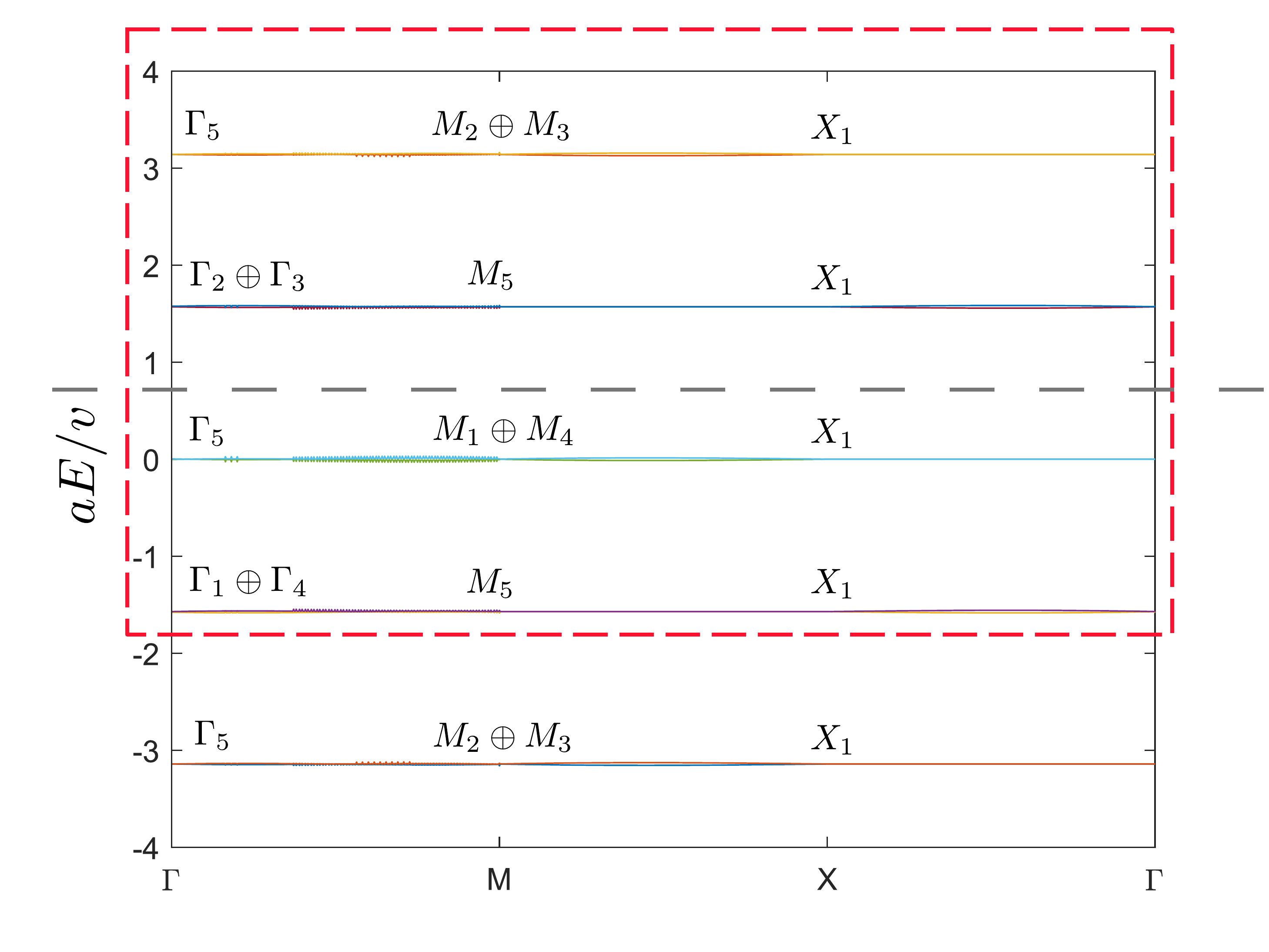}}
	\subfigure[$\theta=-\pi/4$]{\label{band-network-b}
	\includegraphics[width=.45\linewidth]{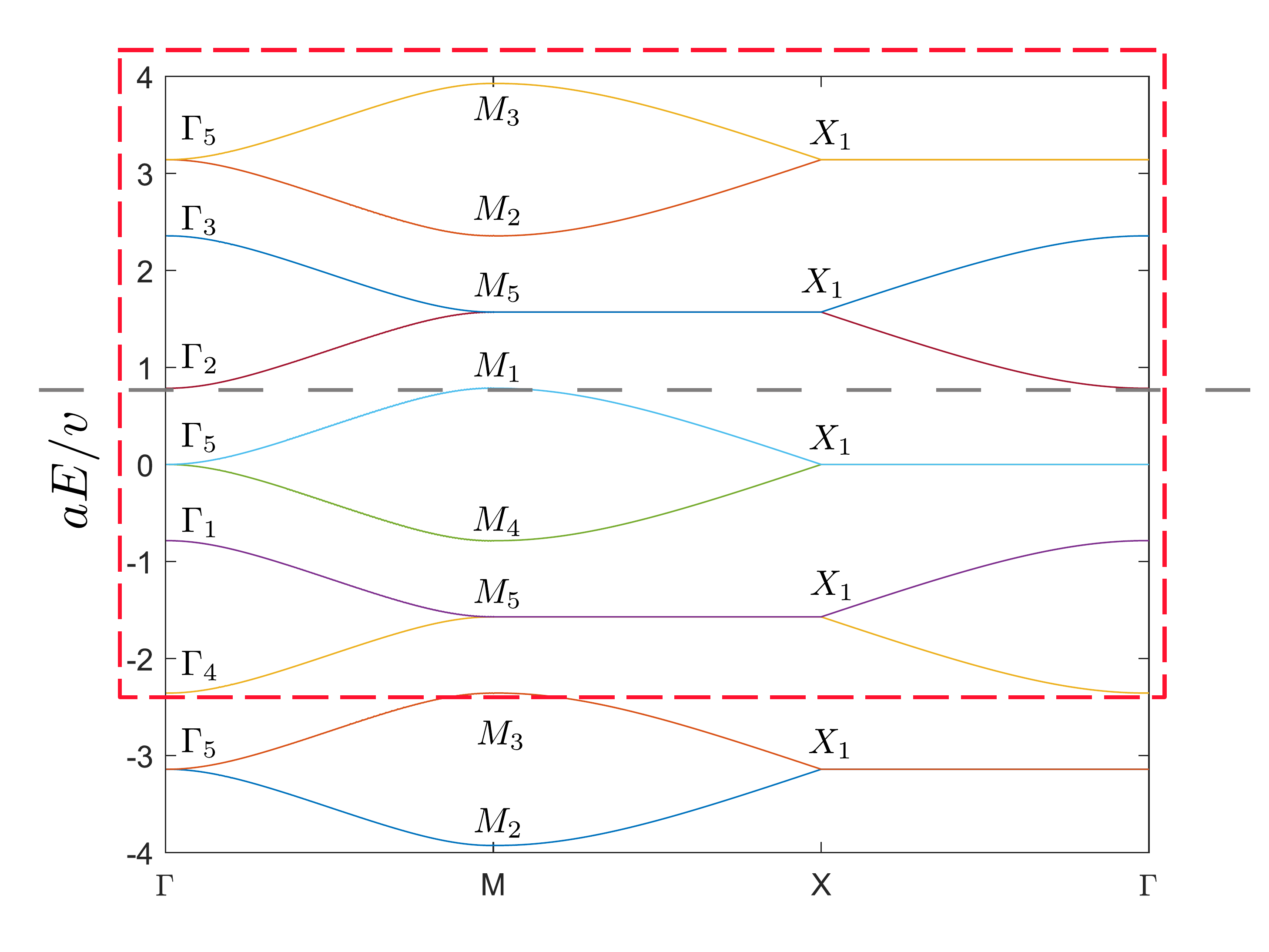}}\\
	\subfigure[$\theta=0$]{\label{band-network-c}
	\includegraphics[width=.4\linewidth]{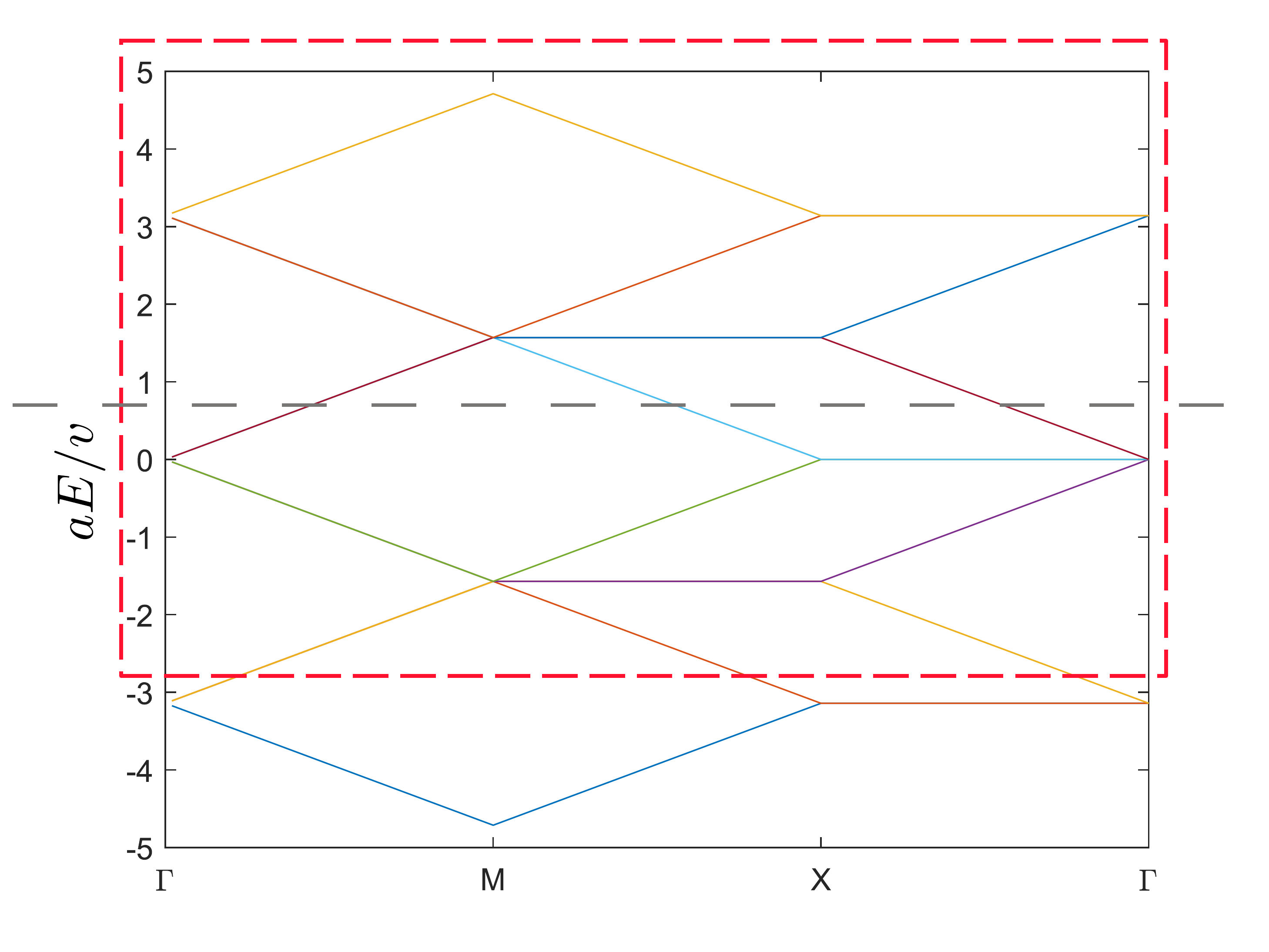}}\\
    \subfigure[$\theta=\pi/4$]{\label{band-network-d}
	\includegraphics[width=.45\linewidth]{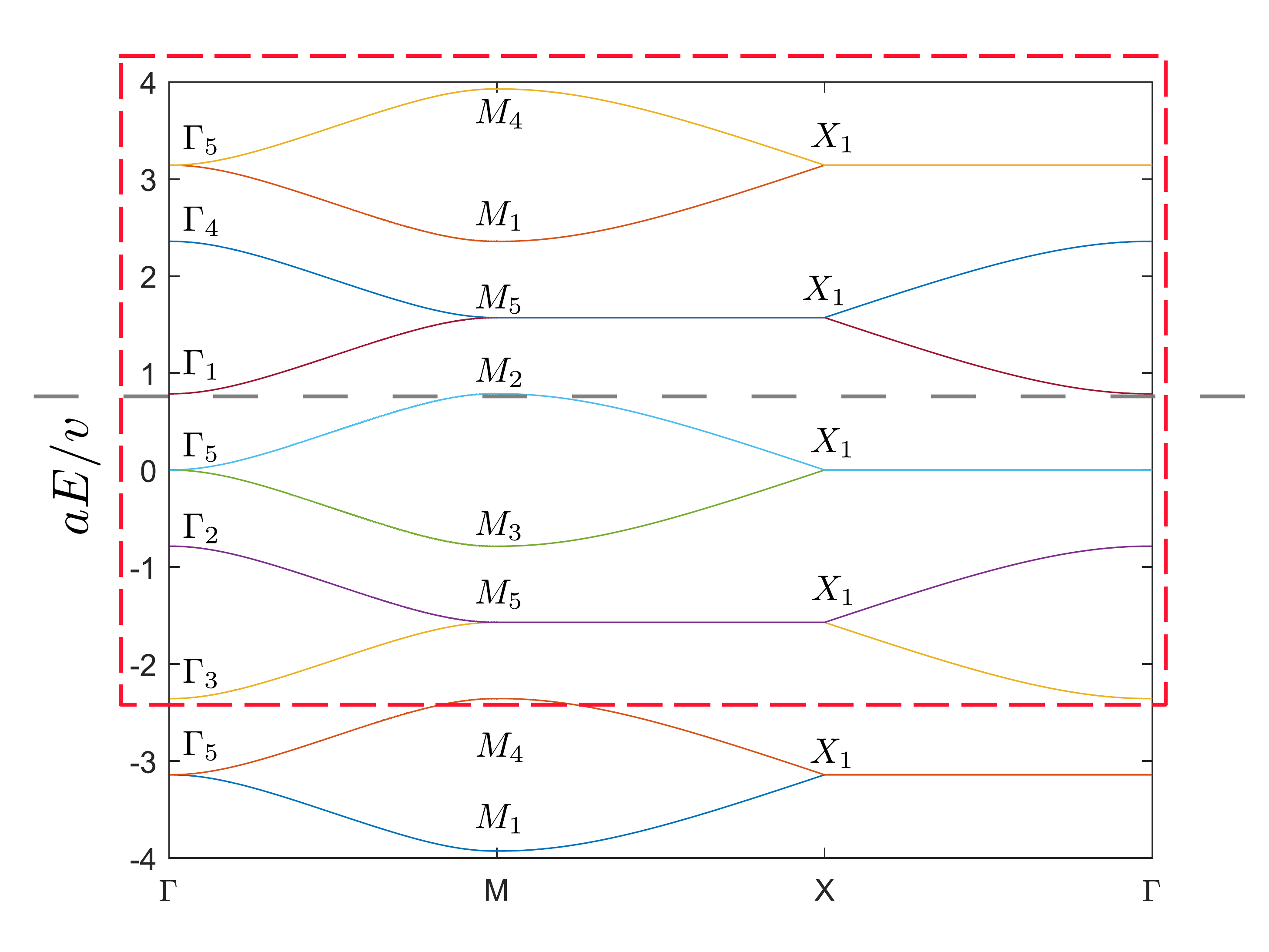}}
	\subfigure[$\theta=\pi/2$]{\label{band-network-e}
	\includegraphics[width=.45\linewidth]{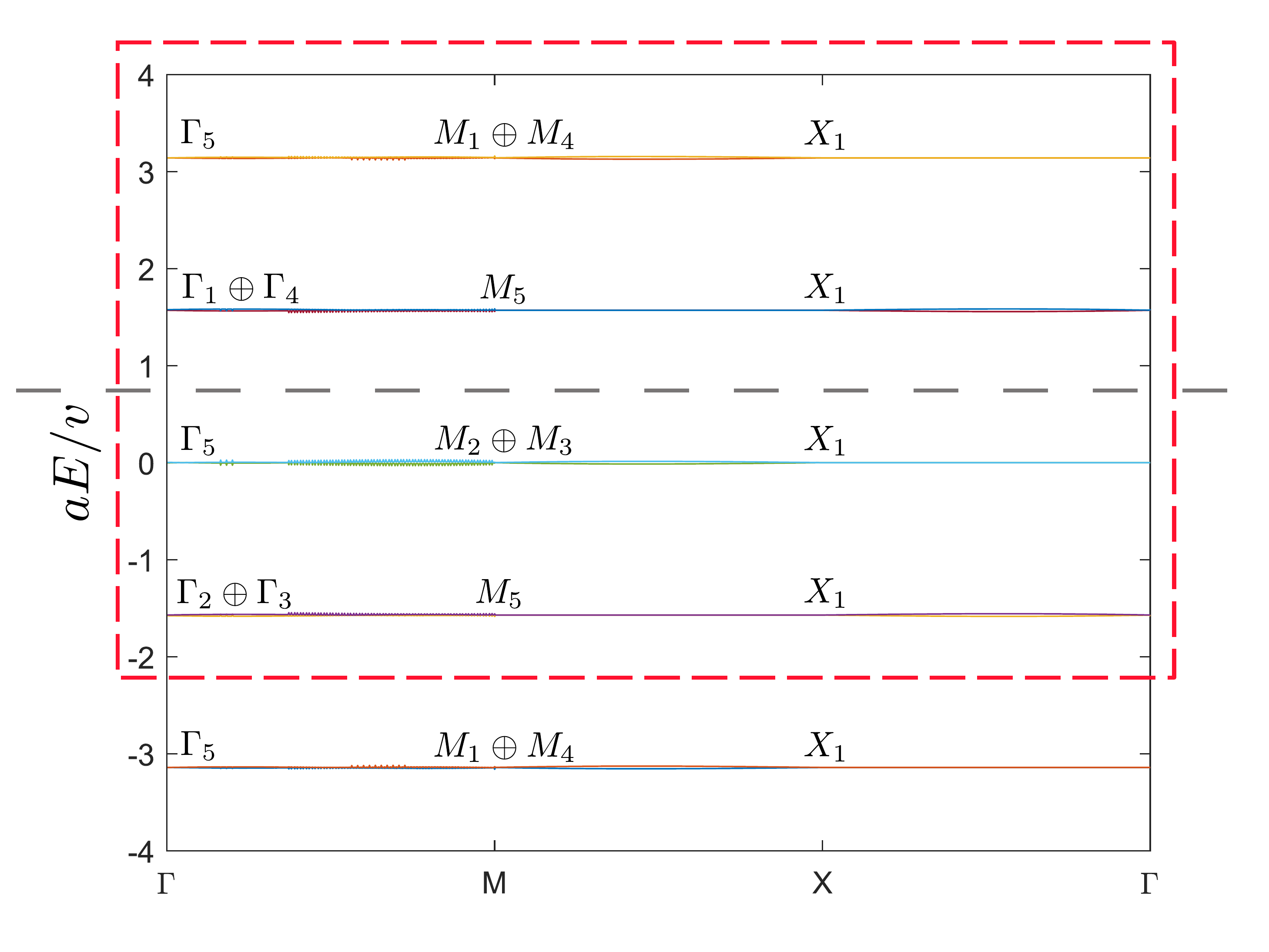}}
    \caption[]{Band representations and structures of the network model with different scattering strengths ($\Lambda_1=50\pi/a,\,\Lambda_2=10\pi/a$). The red dashed panes indicate the repeating units of the band structures. The gray dashed lines indicate the Fermi level used in later sections. Note that the corresponding main text plots, i.e., Fig.~$\color{green}{1 \text{(g), (h), (i)}}$, take $\pi v/4a$ here as the zero energy.}
	\label{bands-network}
\end{figure}

The band structures and the corresponding irreps (irreducible representations) of the middle ten bands with $v\!=\!a\!=\!1,\Lambda_1=50\pi,\,\Lambda_2=10\pi $, and various $\theta=\lambda/v$'s are shown in Fig.~\ref{bands-network} 
Irreps are defined in \cref{Irreps}.
The degeneracy at the $\mathrm{X}$ point comes from the magnetic translation $\widetilde{T}=\{1'\vert 1/2,1/2,0\}$ (here the length unit is lattice constant 2a), which squares to $\{1 | 1,1,0\} = -1$ at $X$ and hence protects Kramers' pairs. 
One may notice the periodicity in energy - the only difference between the $\rm{n}$th and the $\rm{n+8}$th band is a constant energy shift (the red panes in \cref{bands-network} indicate the repeating unit).
We will prove this periodicity at the end of this subsection. Nevertheless, we should note that this periodicity is not essential for the main conclusion of our work (the appearance of critical metal phase). In fact, we build 8-band lattice models in later sections, and they also have critical metal phases as the network model.

Now let us analyze the representations of the band structure.
We take the upper eight of the ten bands as the repeating unit in energy and focus on it. 
Then we notice that the band structure comprises disconnected branches, each of them containing two bands. 
One branch forms one of the following four MEBRs \cite{elcoro_magnetic_2021} (magnetic element band representations) defined in \cref{EBRs}: $\mathrm{A_{b}\! \uparrow \! G,\, B_{b}\! \uparrow \! G,\,^{1}E_{b}\!\uparrow \! G,\, ^{2}E_{b}\! \uparrow \! G}$.
One MEBR is the minimal trivial group of bands formed by symmetric local orbitals in real space, and the left part of a MEBR notation indicates the site symmetry representation of orbitals.
These four MEBRs ($\mathrm{A_{b}\! \uparrow \! G,\, B_{b}\! \uparrow \! G,\,^{1}E_{b}\!\uparrow \! G,\, ^{2}E_{b}\! \uparrow \! G}$) are formed by effective $s$, $d_{x^2-y^2}+i d_{xy}$ (or equivalently $d_{x^2-y^2}- i d_{xy}$), $p_x-ip_y$, $p_x+ip_y$ orbitals, respectively. All these orbitals center at the Wyckoff position $\mathrm{2b}$ ($(\frac34\frac14 0)$ and $(\frac14\frac34 0)$  ). Since $2\mathrm{b}$ has multiplicity $2$ (two $C_{4z}$ centers per unit cell), each MEBR contains two bands. 
The transition between two localized limits at $\theta=\pm\pi/2$ manifests in the transitions of MEBRs.
When $\lambda < 0$ ( $\theta < 0 $), the bands are gapped and the representation of the lower four bands in one repeating window is a direct sum of two MEBRs: $\mathrm{^{2}E_{b}\! \uparrow \! G \oplus A_{b}\! \uparrow \! G}$. 
When $\lambda=0$ ($\theta=0$), the gaps close and the band structure is indeed that of ballistic 1D metals with linear dispersion relations. 
As $\lambda$ increases across $0$, if one focus on the energy level indicated by the gray dashed lines in \cref{bands-network} (the middle of the chosen window), one will find that irreps $ \Gamma_1,\,M_1$ ($\Gamma_2,\,M_2$) go up (down) across the energy level. Similar irrep exchanges also happen beyond and below the window.  
After the transition ($\lambda>0$), the gaps reopen and the MEBRs of the lower four bands in one repeating window change to $\mathrm{^{1}E_{b}\! \uparrow \! G \oplus B_{b}\! \uparrow \! G}$.
The transition is reversed for the upper four bands. Hence, $\theta=\pm\pi/2$ corresponds to two different trivial phases.

\newpage
The band structure periodicity in energy can be understood heuristically by the locality of $H_{N,1}$. For a given wave vector $\vec{k}$, we take the eigenstates of $H_{N,0}(k_x,k_y)$ as a basis $\left\{\Ket{\phi_{d}(k_x,k_y;\alpha,G)}\vert d=h,v,\,\alpha=\pm1,\,G=\mathbb{Z}\pi/a\right\}$. Since $H_{N,1}$ corresponds to $\delta$-scattering potentials at square corners, it can only perceive the phases (of states) at square corners. According to \cref{inverse-Fourier-network}, $\Ket{\phi_d(k_x,k_y;\alpha,G)}$ has the same phases as $\Ket{\phi_d(k_x,k_y;\alpha,G+2\mathbb{Z}\pi/a)}$ at square corners. Therefore, the scattering amplitude (caused by $H_{N,1}$) between $\Ket{\phi_d(k_x,k_y;\alpha,G)}$ \& $\Ket{\phi_{d'}(k_x,k_y;\alpha',G')}$ is the same as that of $\Ket{\phi_d(k_x,k_y;\alpha,G+2\pi/a)}$ \& $\Ket{\phi_{d'}(k_x,k_y;\alpha',G'+2\pi/a)}$. The only difference between these two pairs is an overall energy shift $2\pi v/a$ caused by $H_{N,0}$. Hence, the band structure should have a period of $2\pi v/a$ in energy. Notice that the spectrum of $H_{N,0}$ is continuous and there are eight bands between $E=vG$ and $vG+2\pi v/a$ correspond to $\{\phi_h(k_x,k_y;+,G),\,\phi_h(k_x,k_y;-,-G),\,\phi_h(k_x,k_y;+,G+\pi v/a),\,\phi_h(k_x,k_y;-,-G-\pi v/a); \, h\leftrightarrow v\}$. (We ignore the dispersion energy $\alpha vk_{x/y}$ for clarity since the energy order of bands only depends on $\alpha G$ when $\vec{k}$ is restricted to the first BZ.) The scattering induced by $H_{N,1}$ will open gaps among the connected bands of $H_{N,0}$, so there are still eight bands between $E=vG$ and $vG+2\pi v/a$. Therefore, the band structure of $H_{N,0}+H_{N,1}$ has an energy period of $2\pi v/a$, and each period contains eight bands.

To prove the periodicity more quantitatively, we introduce the unitary transformation (we ignore $\Lambda_1,\,\Lambda_2$ or take $\Lambda_1,\,\Lambda_2\rightarrow \infty$ in this proof)
\begin{equation}
    U \phi_{h}(k_x,k_y;\alpha, G_x) U^\dagger = \phi_{h}(k_x,k_y;\alpha, G_x - \alpha 2\pi/a ),\qquad 
    U \phi_{v}(k_x,k_y;\alpha, G_x) U^\dagger = \phi_{v}(k_x,k_y;\alpha, G_y - \alpha 2\pi/a )\ . 
\end{equation}
Or in the representation of the invariant subspace spanned by $\left\{\Ket{\phi_{h/v}(k_x,k_y;\alpha,G)}\vert G\in \mathbb{Z} \pi/a,\,\alpha=\pm1 \right\}$
\begin{equation}
    \left(U(k_x,k_y;h/v)\right)_{\alpha G,\alpha' G'}=\delta_{\alpha,\alpha'}\delta_{G,G'-\alpha 2\pi/a}
\end{equation}
It is direct to verify that, when the cutoff $\Lambda_1$ is infinity, $U$ transforms $H_{N,0}$ to itself plus a density term
\begin{align}
    U H_{N,0} U^\dagger =& H_{N,0} + 2\pi \frac{v}a 
    \sum_{k_x,k_y,\alpha} \sum_{G} \pare{ \phi_v^\dagger(k_x,k_y;\alpha,G) \phi_v(k_x,k_y;\alpha,G) +  \phi_h^\dagger (k_x,k_y;\alpha,G) \phi_h(k_x,k_y;\alpha,G) } \nonumber\\
    =& H_{N,0} + 2\pi \frac{v}a \hat{N}
\end{align}
with $\hat{N}$ being the particle number operator. 
One can also verify that 
\begin{equation}
    U H_{N,1} U^\dagger =  H_{N,1}\ .
\end{equation}
Notice that the momentum shifted by $U$ is chosen as twice the minimal reciprocal lattice length, \ie $2\times 2\pi/(2a)$, such that the phase factors in $H_{N,1}$ are invariant under the transformation $G_x \to G_x - \alpha \frac{2\pi}{a}$, $G_y \to G_y - \beta \frac{2\pi}{a}$, \ie
\begin{align}
& \Xi_{\Lambda_2}( \alpha (k_x+G_x) - \beta (k_y + G_y) )
\exp\left[i(\frac{1+\alpha}{2}) G_{y}a - i(\frac{1+\beta}{2})G_{x}a\right] \nonumber\\
\to & \Xi_{\Lambda_2}( \alpha (k_x+G_x - \alpha 2\pi/a) - \beta (k_y + G_y - \beta 2\pi/a) )
\exp\left[i(\frac{1+\alpha}{2}) (G_{y} - \beta 2\pi/a) a - i(\frac{1+\beta}{2}) (G_{x} - \alpha 2\pi/a) a\right] \nonumber\\
=& \Xi_{\Lambda_2}( \alpha (k_x+G_x) - \beta (k_y + G_y) )
\exp\left[i(\frac{1+\alpha}{2}) G_{y}a - i(\frac{1+\beta}{2})G_{x}a\right]\ .
\end{align}
In summary there is 
\begin{equation}
    U (H_{N,0} + H_{N,1}) U^\dagger = H_{N,0} + H_{N,1} + \frac{2\pi v}{a} \hat{N}\ .
\end{equation}
Suppose $\ket{\psi_{\kk,n}}$ is a single-particle eigenstate of $H_{N,0} + H_{N,1}$ with the energy $E_n(\kk)$, then $U^\dagger \ket{\psi_{\kk,n}}$ is a single-particle state with the energy $E_n(\kk) + \frac{2\pi v}{a}$. 
Therefore, we have proven the energy periodicity. 

Now, we argue that every energy window $[E, E+\frac{2\pi v}{a})$ contains eight bands by counting the Hilbert space dimension. 
Suppose there are $N_G$ number of $G$'s satisfying $|G|<\Lambda_1$. 
Then the Hilbert space dimension at each $\kk$ is $4 N_G$ since for each $G$ there are $\alpha=\pm1$ horizontal and vertical modes, \ie $\phi_{h}(k_x,k_y;\alpha,G)$ and $\phi_{v}(k_x,k_y;\alpha,G)$ ($\alpha=\pm1$). 
As $U$ shifts $G$ by twice the minimal reciprocal lattice, $U$ can act in the Hilbert space at most $\frac{N_G}2$ times.
To be particular, $U^{\frac{N_G}2}$ sends the largest negative $G$ to the largest postive $G$. 
Therefore, $U$ divides the total Hilbert space into $N_G/2$ pieces.  
If we act $U$ on the energy bands, it will yield $ N_G/2 $ groups, with every group $4N_G/(N_G/2)=8$ bands. 
Nearby groups have an energy difference of $2\pi v/a$ according to the last paragraph. 
Thus, eight nearby bands form the smallest repeating unit in energy.

In practice, finite cutoff factors $\Lambda_1,\,\Lambda_2$ will be introduced. The above argument, in fact, fails when applied to high-energy bands close to the cutoff. 
Nevertheless, when the cutoff is sufficiently large, it applies to low-energy states far away from the cutoff.

\newpage
\subsection{Further discussions on the cutoff} \label{subsec:cutoff}

The effect of $\Lambda_1$ is straightforward: it discretizes the in-line coordinates of chiral wires, \ie the $x$ in $\psi_{h}(x;m,\alpha)$ and $y$ in $\psi_{v}(y;n,\alpha)$. We denote the granularity as $b\sim\frac{1}{\Lambda_1}$. 

To see the role of $\Lambda_2$, we begin with the scattering potential with a (hard) cutoff in reciprocal space and ignore $\Lambda_1$ temporarily
\begin{equation}
\label{H1-cutoff}
    \begin{aligned}
    H_{N,1}=\!\sum_{\vec{k}}\sum_{G_xG_y}\sum_{\alpha\beta} \frac{\lambda}{2 a} \,\Xi_{\Lambda_2}&\left[\alpha(k_x+G_x)-\beta(k_y+G_y)\right] \\\times&\exp\left[i(\frac{1+\alpha}{2}) G_{y}a - i(\frac{1+\beta}{2})G_{x}a\right] \phi_{h}^{\dagger}(k_{x},k_{y};\alpha,G_{x}) \phi_{v}(k_{x},k_{y};\beta,G_{y})+h.c.\\
    \end{aligned}
\end{equation}
The cutoff factor $\Xi_{\Lambda_2}$ is defined in \cref{eq:truncation}. Then we carry out the Fourier transform \cref{inverse-Fourier-network}:
\begin{equation}
    \begin{aligned}
    H_{N,1}=\!\sum_{nm\alpha\beta}
    \iint \!\mathrm{d}x \mathrm{d}y \, \delta [\alpha(x-2na-&\frac{1+\beta}{2}a)-\beta(y-2ma-\frac{1+\alpha}{2}a)] \\
    \times & \frac{\lambda\sin\left[\Lambda_2(x-2na-\frac{1+\beta}{2}a)\right]}{\pi(x-2na-\frac{1+\beta}{2}a)}\psi_h^{\dagger}(x;m\alpha)\psi_v(y;n\beta)\space +h.c.
    \end{aligned}
\end{equation}
We obtain a broadened $\delta$-scattering potential in the form of $sinc$ function.
Other (hard or soft) momentum cutoff factors will result in different broadening profiles. Regardless of the concrete formula of cutoff factor, $\Lambda_2$ always characterizes the width $\zeta\sim\frac{1}{\Lambda_2}$ of broadened $\delta$-potential.   

Now let us reinspect \cref{new-scattering-result}.
The derivation of Eq.~(\ref{new-scattering-result}) includes an integration of Eq.~(\ref{SE2-scattering-node}) near zero ($\int_{0^-}^{0^+}$). 
Since the generalized function $\delta(\xi)$ is usually understood as the limit of some ordinary function sequence, integration of Eq.~(\ref{SE2-scattering-node}) should be understood in this way:
\begin{equation}
\label{actuall-meaning-intergrate-delta}
    \lim_{\xi_0\rightarrow0^+} \int_{-\xi_0}^{\xi_0} \mathrm{d} \xi \left[-iv \frac{\mathrm{d} \ln \psi_{\pm}(\xi)}{\mathrm{d} \xi}\right]=\lim_{\xi_0\rightarrow0^+}\left\{\lim_{\zeta\rightarrow0^+} \int_{-\xi_0}^{\xi_0} \mathrm{d} \xi \left[E\mp \lambda \delta_{\zeta}(\xi)\right]\right\}
\end{equation}
where $\delta_{\zeta}$ is a broadened $\delta$ function with characteristic length $\zeta \sim \frac1{\Lambda_2}$.
The order of limits on the right side of Eq.~(\ref{actuall-meaning-intergrate-delta}) tells us that $\zeta\ll\xi_0$, since it takes $\zeta \rightarrow 0 $ first and then $\xi_0 \rightarrow 0 $. We should further notice that, when taking these two limits, $\xi$ is viewed as a continuous integral variable, \ie the granularity of $\xi$ is much smaller than $\zeta$. Hence, we have the hierarchy $b\ll \zeta\ll\xi_0$ ($\Lambda_2\ll \Lambda_1$). 
Therefore, if we take $\Lambda_2\geq \Lambda_1$, Eq.~(\ref{new-scattering-result}) is invalid.

 We can now understand why introducing $\Lambda_2$ is necessary. If we only take a cutoff $\Lambda_1$ on $G_{x,y}$, then effectively $\Lambda_2\to \infty$. In this case, Eq.~(\ref{new-scattering-result}) is invalid. Physically, it is equivalent to discretizing the chiral wires without broadening the $\delta$-potentials.

\begin{figure}[h]
	\centering
	\subfigure[~Continuous wire with shrinking scattering potential]{\label{twolimit-a}
    \includegraphics[width=.45\linewidth]{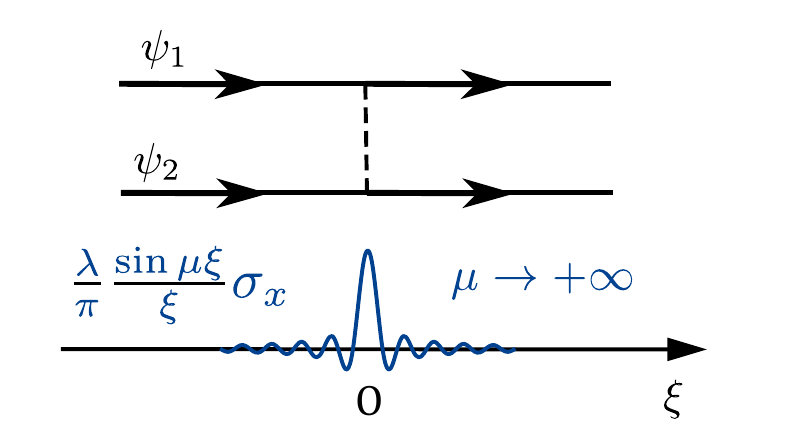}}
    \quad
	\subfigure[~One-point scattering hopping between discrete wires that are more and more fine-grained]{\label{twolimit-b}
	\includegraphics[width=.45\linewidth]{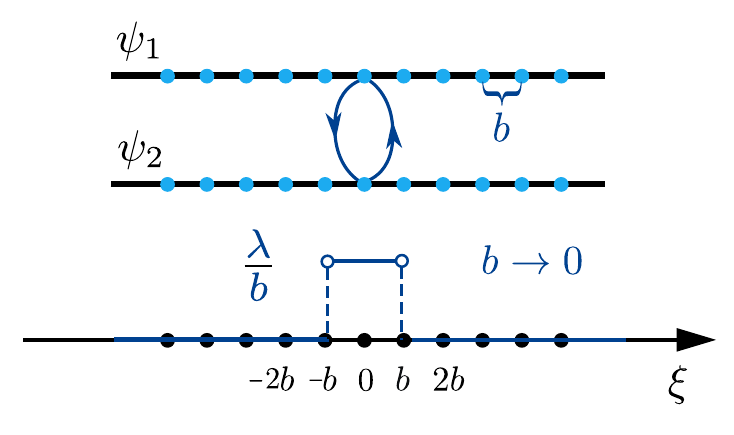}}
    \caption[]{Two methods to approach a $\delta$ scattering problem}
	\label{twolimit}
\end{figure}

The above analysis informs us that although Eq.~(\ref{new-scattering-result}) is an inevitable result from the perspective of the differential equation, the situation is subtle in the numerical calculation that involves $\Lambda_{1,2}$. More concretely, we should distinguish two cases of $b$ versus $\zeta$, \ie $b\ll\zeta$ ($\Lambda_1\gg\Lambda_2$) and $b\geq\zeta$ ($\Lambda_1\leq\Lambda_2$). To validate \cref{new-scattering-result}, we should take the former case.
Nevertheless, it is beneficial to show how such two cases are different. In fact, these two cases of $b$ vs. $\zeta$ correspond to two system sequences (Fig.~\ref{twolimit}) that both take a $\delta$-potential on a continuous chiral wire (Fig.~\ref{scatter2}) as their limitations. However, such two system sequences lead to different scattering results. 

When $b\ll\zeta$, \ie $b$ is much smaller than any scale, $\xi$ can be viewed as a continuous variable, and the scattering process can be effectively depicted by the system sequence shown in \cref{twolimit-a}. In \cref{twolimit-a}, we set continuous wires \& ordinary-function-type scattering potential first and then shrinks the potential. (We should mention that the ordinary function sequence is not limited to $\{\frac{1}{\pi}\sin\frac{\mu\xi}{\xi}\vert \mu \rightarrow \infty\}$ shown in \cref{twolimit-a}, any sequence of ordinary functions that converges to $\delta(\xi)$ is allowed.) This system sequence validates Eq.~(\ref{new-scattering-result}). And if we integrate Eq.~(\ref{SE1-scattering-node}) near zero, we will obtain
\begin{equation}
\label{integrate-deltapsi-across-zero}
\begin{aligned}
    \int_{0^-}^{0^+}d\xi E\psi_{\pm}(\xi)&=\int_{0^-}^{0^+}d\xi[-iv\frac{d}{d\xi}\pm\lambda\delta(\xi)]\psi_{\pm}(\xi)\\
    0&=-iv[\psi_{\pm}(0^+)-\psi_{\pm}(0^-)]\pm\lambda\int_{0^-}^{0^+}\delta(\xi) \psi_{\pm}(\xi)
\end{aligned}
\end{equation}

Since now \cref{SE2-scattering-node} is valid, we can use its corollary $\psi_{\pm}(0^+)=e^{\mp i\lambda/v} \psi_{\pm}(0^-)$, then we have  
\begin{equation}
\label{new-int-delta-psi}
\begin{aligned}
   \int_{0^-}^{0^+} \mathrm{d} \xi \,\delta(\xi) \psi_{\pm}(\xi)&=\pm i\frac{v}{\lambda} (e^{\mp i\lambda/2v}-e^{\pm i\lambda/2v})\sqrt{\psi_{\pm}(0^-)\psi_{\pm}(0^+)}\\
   &= \frac{\sin(\lambda/2v)}{(\lambda/2v)}\sqrt{\psi_{\pm}(0^-)\psi_{\pm}(0^+)}.
   \end{aligned}
\end{equation}

For the second sequence (Fig.~\ref{twolimit-b}), we first discretize the chiral wire (which results in a discrete wire with infinite-long range hopping), then add one-point exchange hopping between two discretized chiral wires, and finally take the granularity of the discretized wires to zero. Hence, the second sequence corresponds to $b\geq \zeta$ and has a different integral result from \cref{new-int-delta-psi}: 
\begin{equation}
\label{old-int-delta-psi}
    \int_{0^-}^{0^+} \mathrm{d} \xi \, \delta(\xi) \psi_{\pm}(\xi)=\frac{\psi_{\pm}(0^+)+\psi_{\pm}(0^-)}{2}.
\end{equation}
which can be numerically confirmed by calculating a single node system defined in Fig.~\ref{twolimit-b}. 
Combining \cref{integrate-deltapsi-across-zero}and \cref{old-int-delta-psi}, the scattering result of the second sequence is
\begin{equation}
\label{old-scattering-result}
    \begin{pmatrix}
    \psi_1(0^+)\\
    \psi_2(0^+)
    \end{pmatrix}
    =
    \begin{pmatrix}
    \frac{1-\widetilde{\lambda}^2}{1+\widetilde{\lambda}^2}& \frac{-2i\widetilde{\lambda}}{1+\widetilde{\lambda}^2}\\
    \frac{-2i\widetilde{\lambda}}{1+\widetilde{\lambda}^2}& \frac{1-\widetilde{\lambda}^2}{1+\widetilde{\lambda}^2}\\
    \end{pmatrix}
    \begin{pmatrix}
    \psi_1(0^-) \\
    \psi_2(0^-)
    \end{pmatrix}
    \qquad \widetilde{\lambda}=\frac{\lambda}{2v},
\end{equation}
which can reproduce \cref{scatter matrix} when $\lambda\in[-2v,2v]$. However, \cref{old-scattering-result} lacks the periodicity on $\lambda$ compared to \cref{new-scattering-result}. More importantly, the second sequence discretizes chiral wires, which results in infinite-long-range hopping on each discretized wire. In Sec.~\ref{subsec:standing-wave}, we need to redistribute degrees of freedom in the crossed two wires into two corners (see \cref{lattice-b}). For the second sequence, such redistribution will result in infinite-long-range hopping between corners (rather than inside each corner). This feature of the second sequence make the situation complicated. On the other hand, the first sequence does not discretize chiral wires and has no such a problem. Therefore, we take the first sequence ($b\ll\zeta$) in this section and use it for constructing the lattice models in \cref{sec:map-network-lattice}.

\begin{figure}[h]
    \centering
    \subfigure[$\lambda=-\pi v/2$]{\label{two-delta-a}
	\includegraphics[width=.3\linewidth]{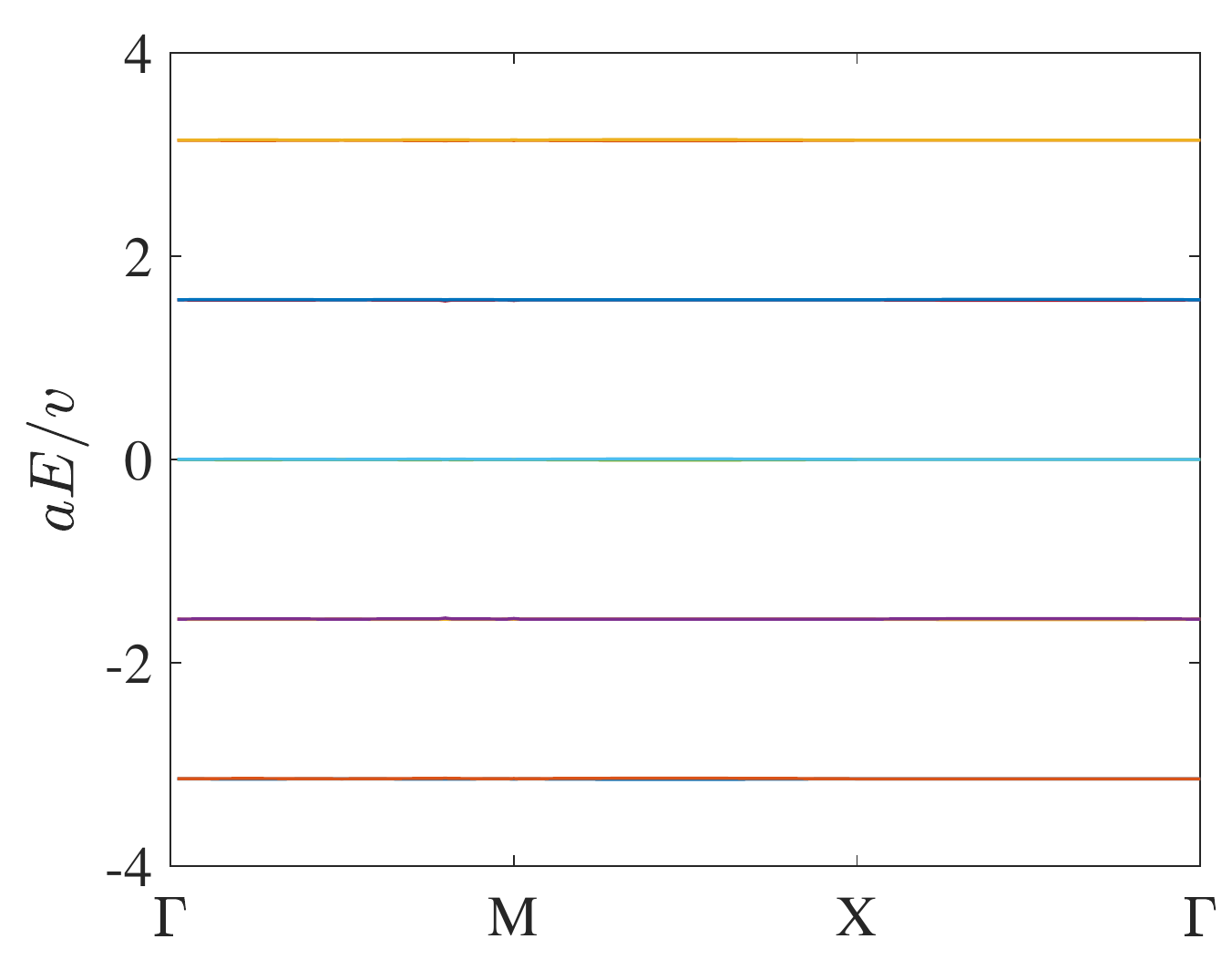}}
    \subfigure[$\lambda=-2v$]{\label{two-delta-b}
	\includegraphics[width=.3\linewidth]{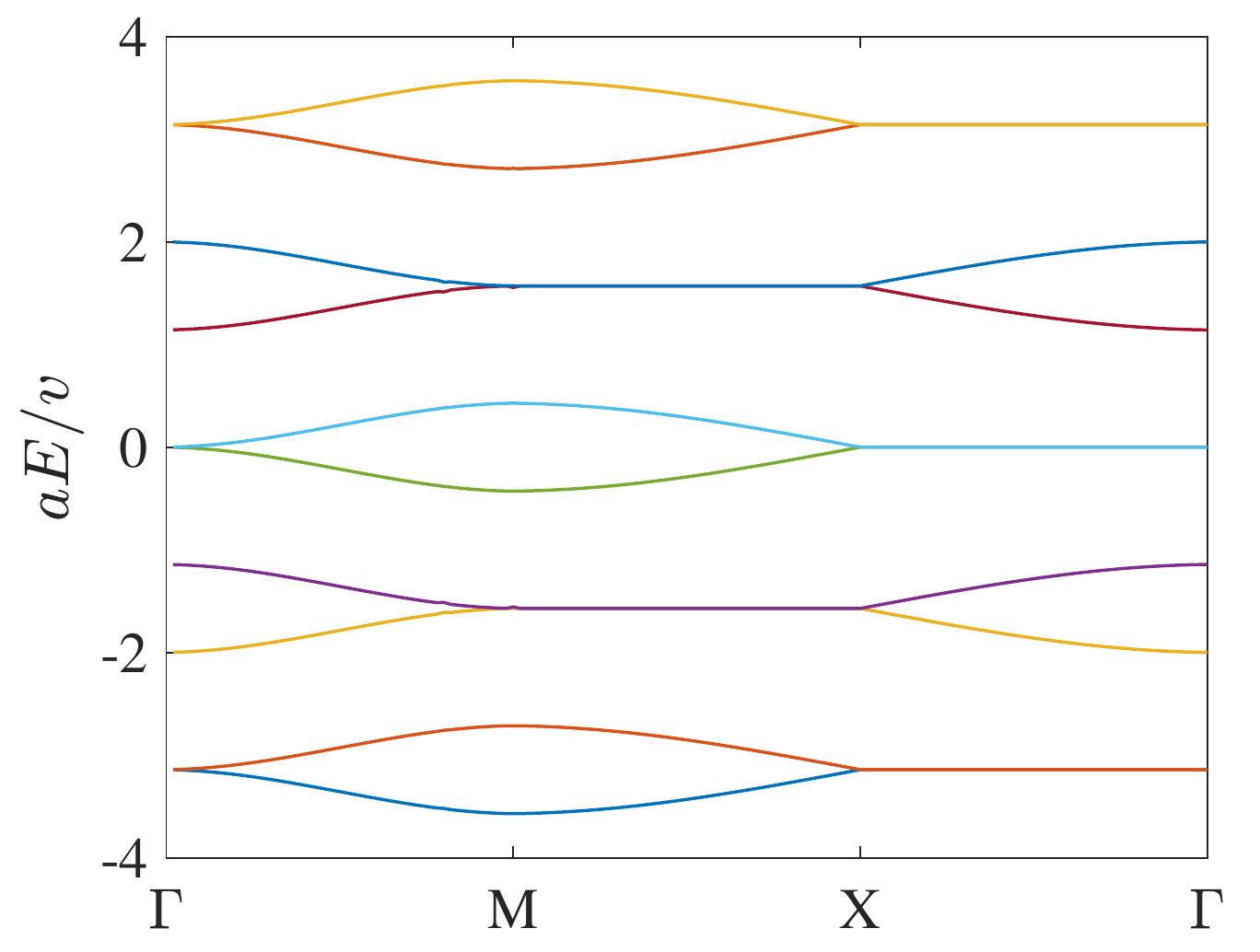}}
     \subfigure[$\lambda=-\pi v$]{\label{two-delta-c}
	\includegraphics[width=.3\linewidth]{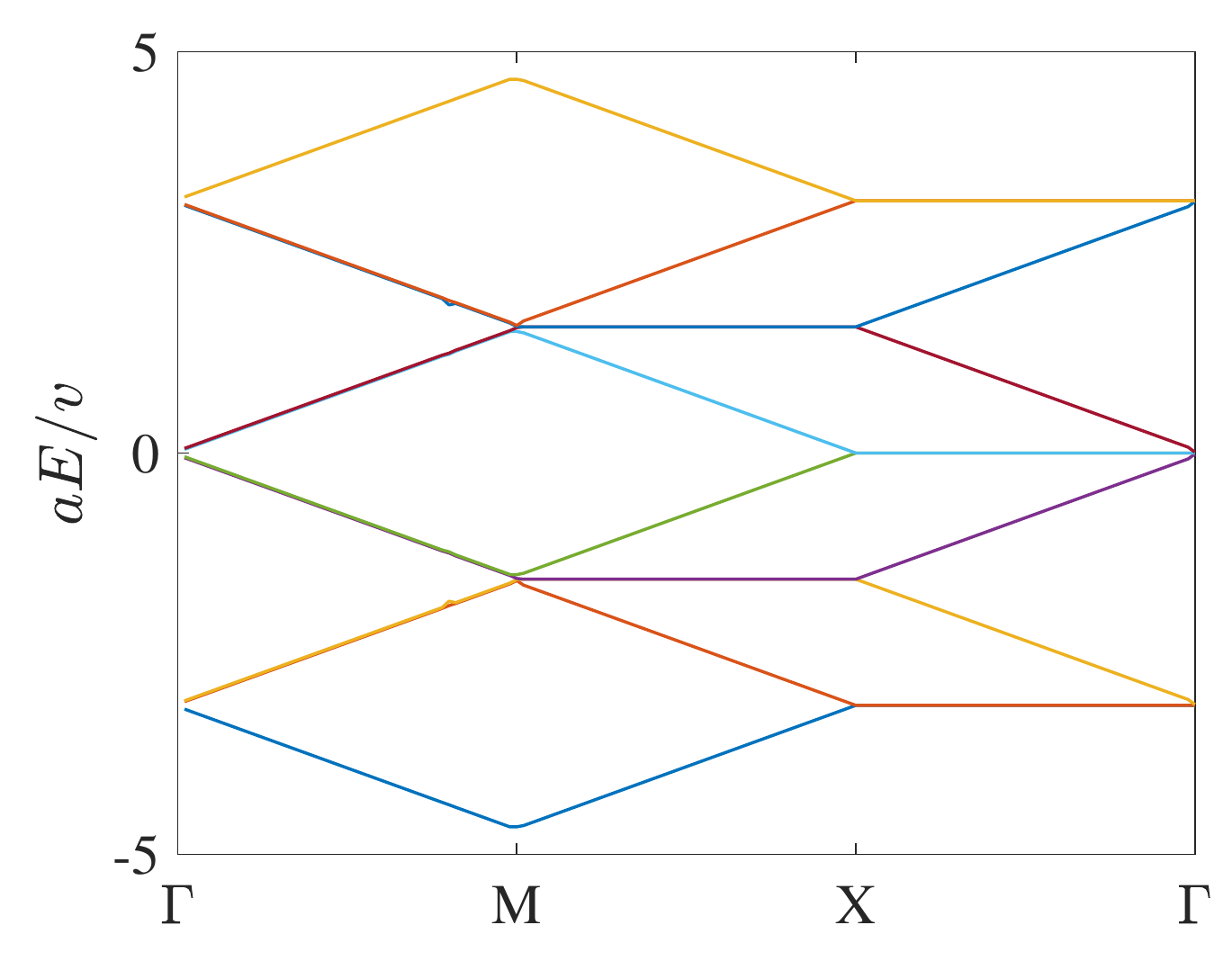}}\\
    \subfigure[$\lambda=-\pi v/2$]{\label{two-delta-numeric-d}
     \includegraphics[width=.3\linewidth]{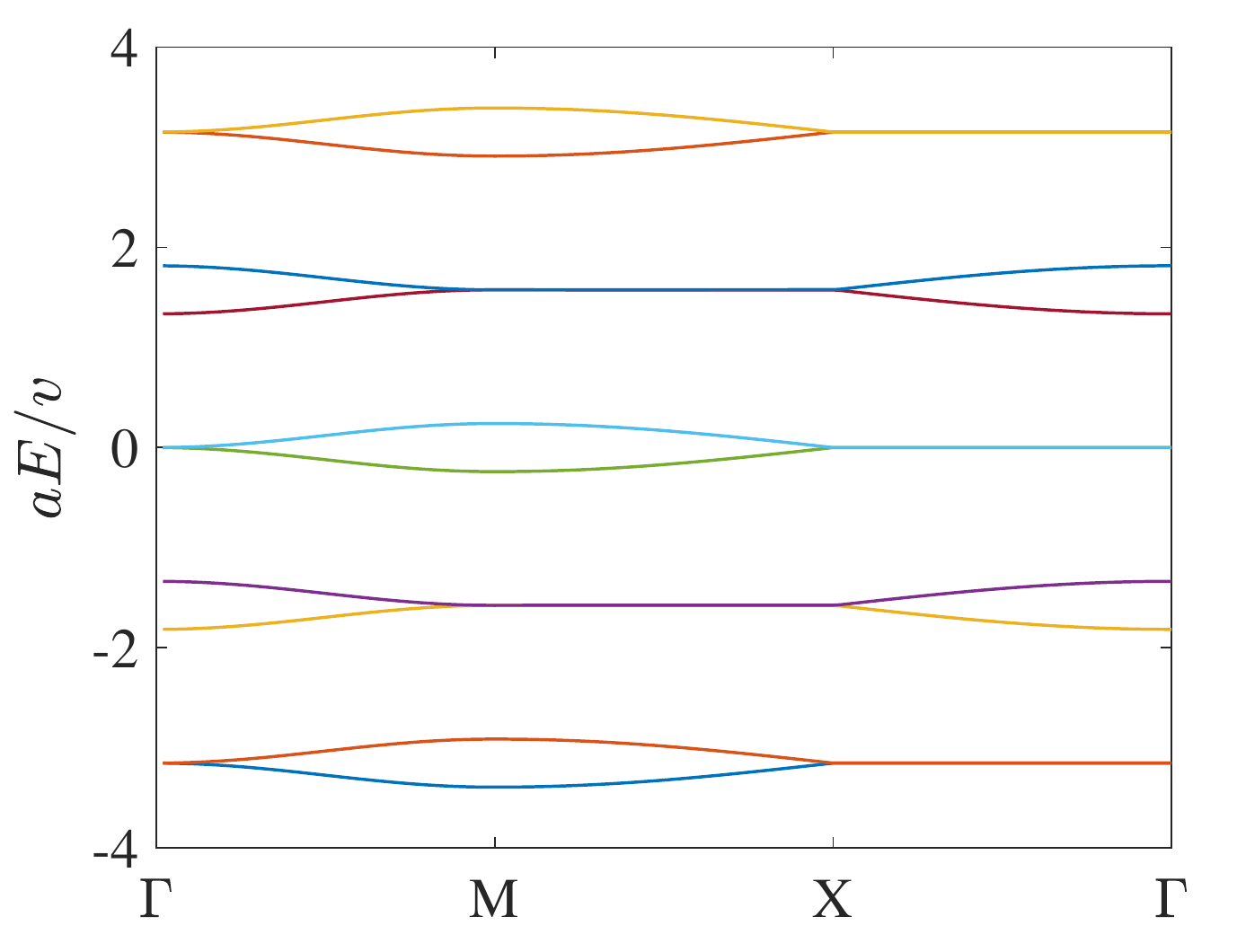}}
	\subfigure[$\lambda=-2 v$]{\label{two-delta-e}
	\includegraphics[width=.3\linewidth]{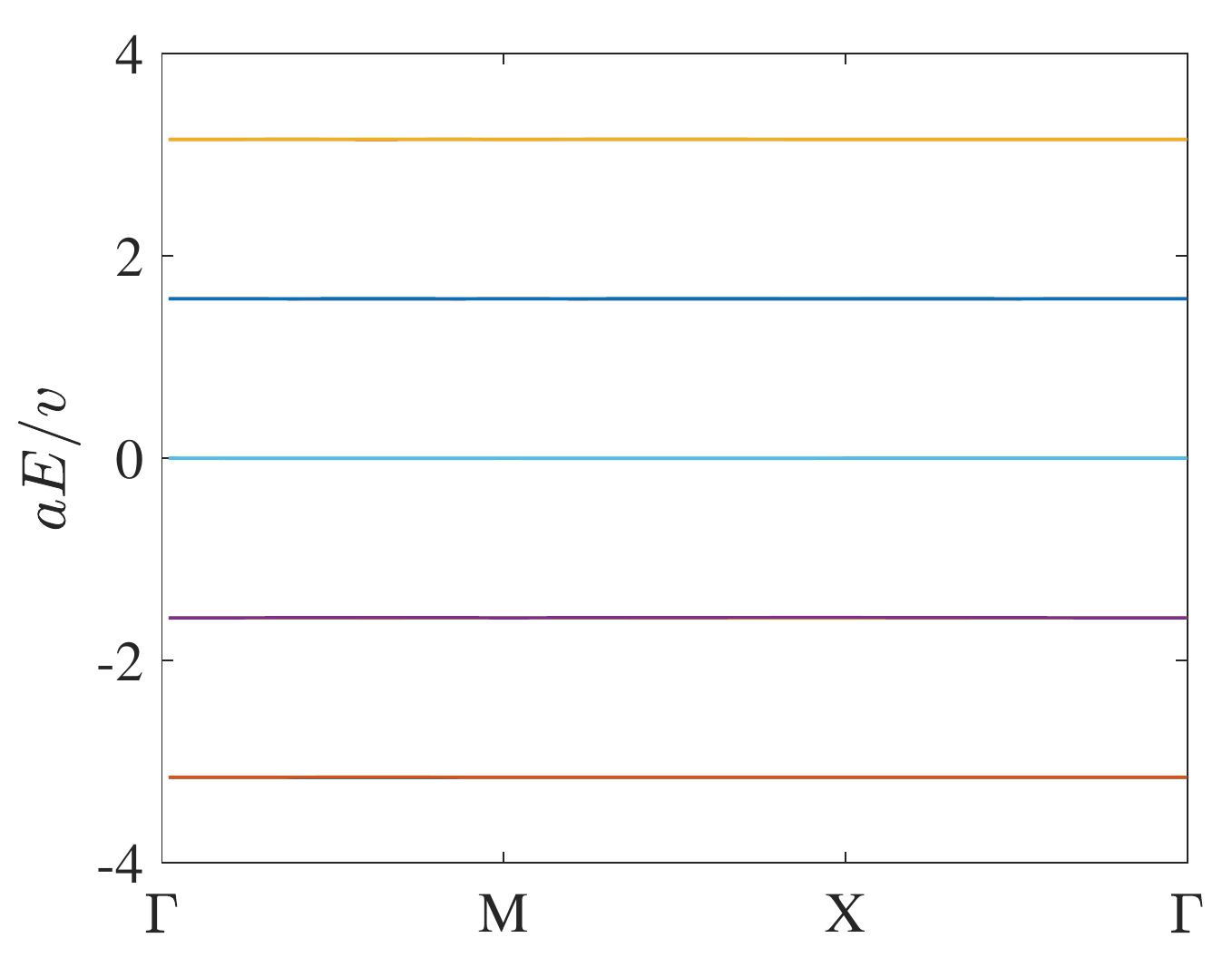}}
	\subfigure[$\lambda=-20v$]{\label{two-delta-f}
	\includegraphics[width=.3\linewidth]{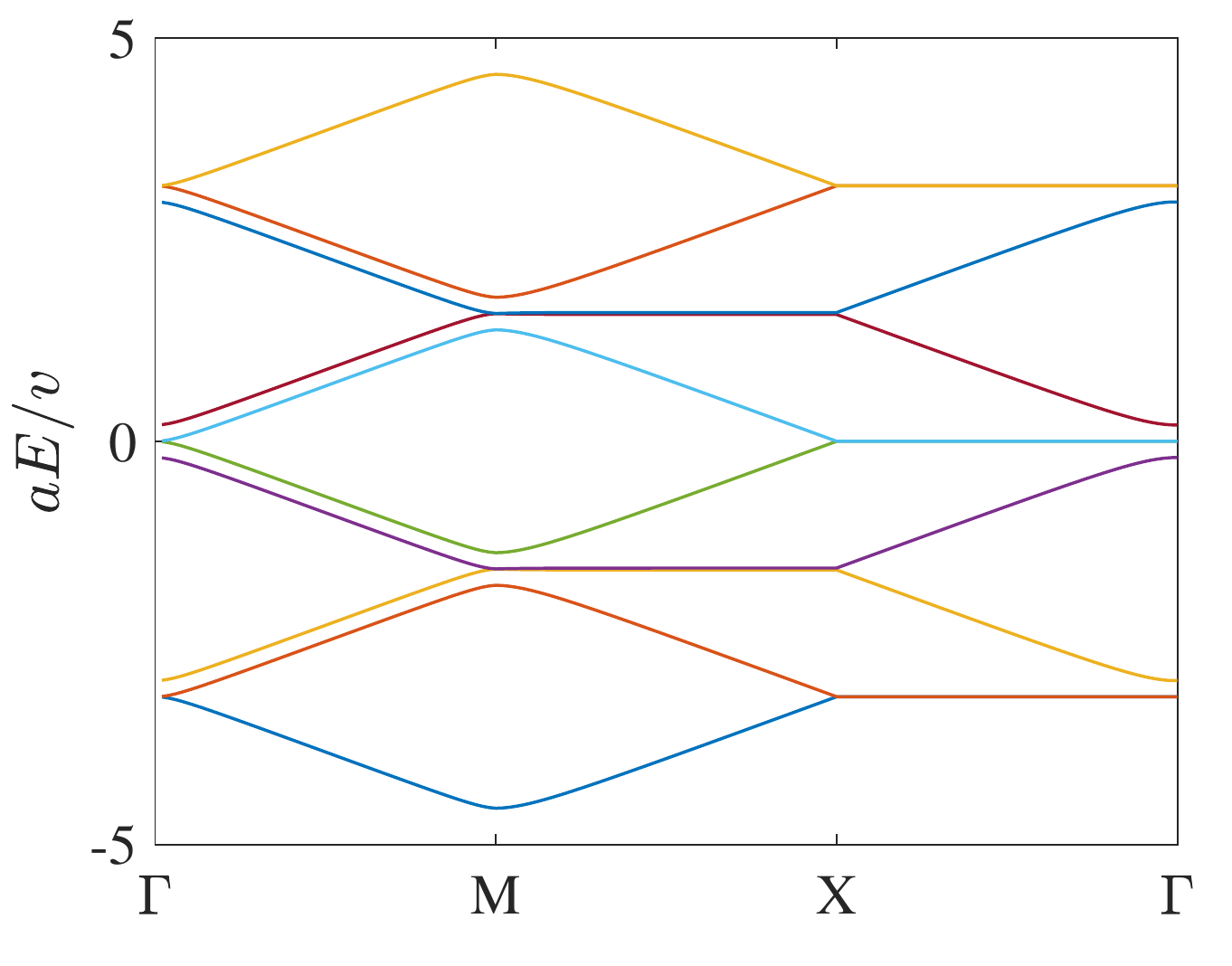}}\\
    \caption[]{Band structures of \cref{network-H-reciprocal-cutoff} with $\Lambda_1=50\pi/a,\,\Lambda_2=25\pi/a$ for (a) $\sim$ (c) and $\Lambda_1=50\pi/a,\,\Lambda_2=100\pi /a$ for (d) $\sim$ (f). (a) $\sim$ (c) correspond to \cref{twolimit-a}, which subjects to \cref{new-scattering-result}. According to \cref{new-scattering-result}, the eigenstates are decoupled local currents at $\lambda=(\mathbb{Z}+\frac12)\pi v$ and are decoupled horizontal \& vertical wires at $\lambda=\mathbb{Z}\pi v$. These predictions are confirmed by (a) and (c), respectively.
    (d) $\sim$ (f) correspond to \cref{twolimit-b}, which subjects to \cref{old-scattering-result}. According to \cref{old-scattering-result}, the eigenstates are decoupled local currents at $\lambda=\pm2 v$ and are decoupled horizontal \& vertical wires at $\lambda=0,\pm\infty$. These are consistent with (d) $\sim$ (f). Though not illustrated in this figure, it is obvious that the dispersion relations for these two systems at $\lambda=0$ are quasi-1D linear as (c), since it is just $H_{N,0}$. }
	\label{two-delta-numeric}
\end{figure}

In terms of numerical program, where we first have discrete wires, the first sequence corresponds to $\Lambda_1,\Lambda_2 \rightarrow \infty, \,\Lambda_2/\Lambda_1\rightarrow0$ ($\zeta,b\rightarrow0,\zeta/b\rightarrow\infty$), the second case corresponds to $\Lambda_1,\Lambda_2 \rightarrow \infty, \,\Lambda_2\geq\Lambda_1$ ($\zeta,b\rightarrow0,\zeta\leq b$). 
Numerical works on the band structure of $H_{N,0}+H_{N,1}$ proved the above analysis. When $\Lambda_1,\Lambda_2 > 10 \,\&\, \Lambda_2/\Lambda_1 \leq 1/2$, the bands are flat at $\lambda=(\mathbb{Z}+1/2)\pi v$ and are quasi-1D linearly dispersed at $\lambda=\mathbb{Z}\pi v$ (see \cref{two-delta-a} \& \ref{two-delta-c}). This corresponds to \cref{twolimit-a} which subjects to Eq.~(\ref{new-scattering-result}). Since Eq.~(\ref{new-scattering-result}) implies that the eigenstates are decoupled local currents when $\lambda=(\mathbb{Z}+1/2)\pi v$ and are decoupled horizontal \& vertical wires when $\lambda=\mathbb{Z}\pi v$. When $\Lambda_1,\Lambda_2 > 10 \,\&\, \Lambda_2/\Lambda_1 \geq 2$, the bands are flat at $\lambda/v\!=\!\pm2$ (see \cref{two-delta-e}) and are quasi-1D linearly dispersed at $\lambda=0,\,\pm\infty$ (see \cref{two-delta-f}). This corresponds to \cref{twolimit-b} which subjects to \cref{old-scattering-result}, since Eq.~(\ref{old-scattering-result}) implies that the eigenstates are decoupled local currents when $\lambda=\pm2v$ and are decoupled horizontal \& vertical wires when $\lambda=0,\,\pm\infty$. 

\vspace{-2ex}
\vspace{-2ex}

\clearpage
\section{Mapping the network model to lattice models}\label{sec:map-network-lattice}

\subsection{Un-truncated lattice model: $H_{N}$ on standing wave basis}

\label{subsec:standing-wave}

The network model was introduced through the wire construction; we
can equivalently rewrite it in a local orbital basis.
Without loss of generality, we can choose the basis as the local current loop states in the limit $\theta=-\pi/2$ discussed in \cref{subsec:Manhattan}, \ie standing waves going around the red/blue squares. 
We emphasize that the existence of the local current loop basis means our model has no symmetry anomaly associated to the crystalline symmetries or the accidental particle-hole symmetry discussed in \cref{subsec:H-network}.
(Notice that the accidental particle-hole symmetry will be broken when the disorder is considered.)

\begin{figure}[h]
	\centering
	\subfigure[Unit cell and its coordinate convention. The green dot indicates the origin of unit cell used in the network model. One should distinguish it from the origin of coordinates $x_{\pm}$.]{\label{lattice-a}
    \includegraphics[width=.3\linewidth]{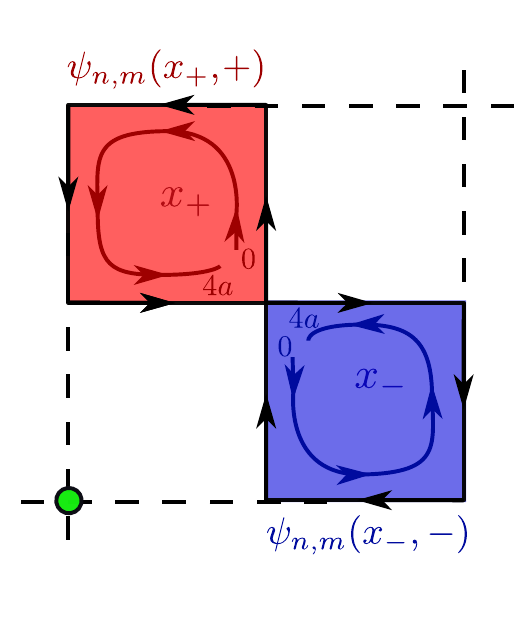}}
    \quad\quad\quad\quad
	\subfigure[Mapping of a scattering node (left: network, right: lattice). $\lambda$ and $\vec\gamma$ are constants represent the scattering strengths. The basis of the $\sigma_x$ on the left is $\psi_1,\psi_2$ and that of the $\vec \sigma$ on the right is $\psi'_1,\psi'_2$]{\label{lattice-b}
    \includegraphics[width=.55\linewidth]{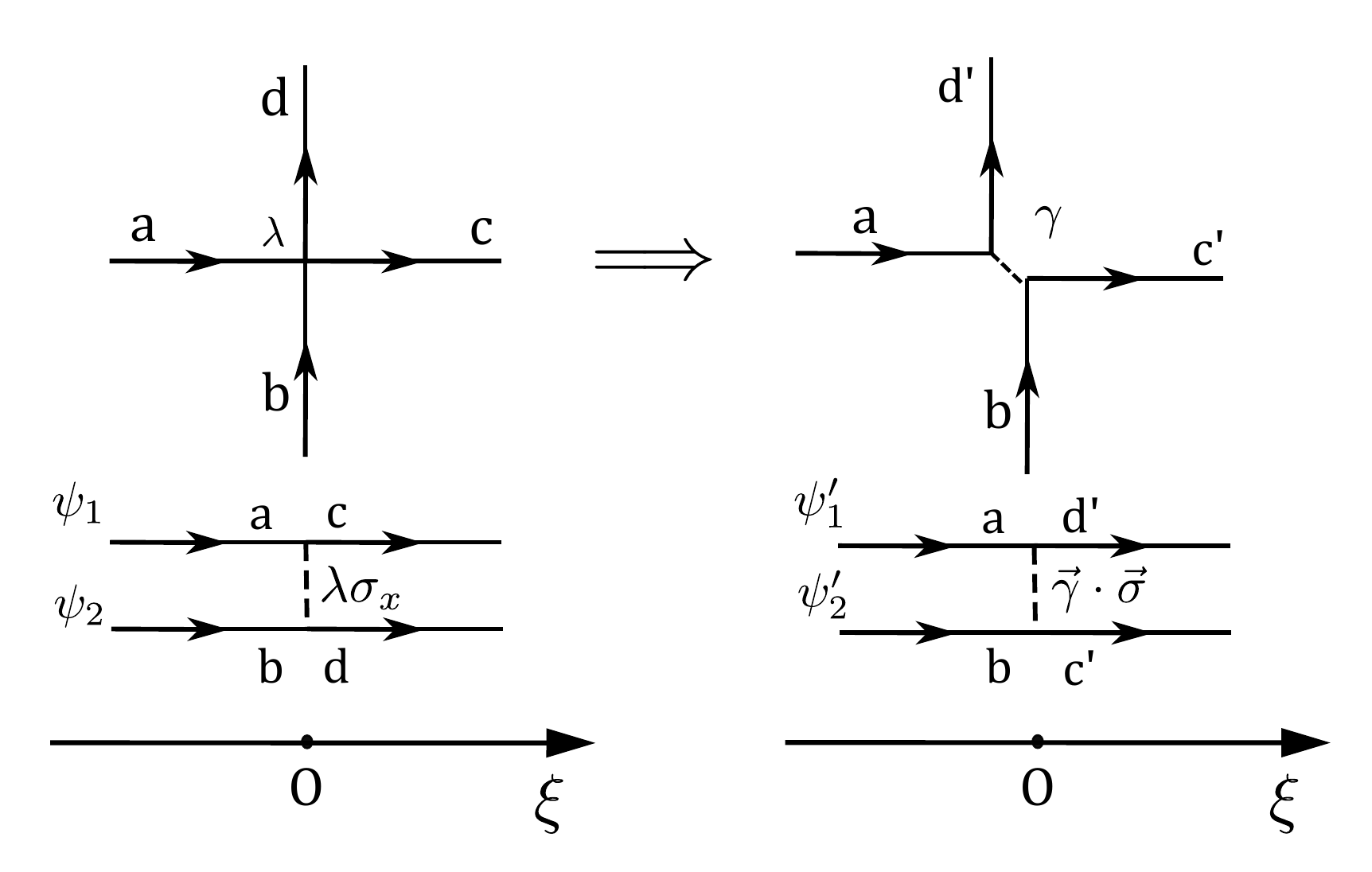}}
    \caption[]{Unit cell and scattering potential of un-truncated lattice model.}
	\label{lattice}
\end{figure}

In order to obtain the standing wave representation, we first introduce the circular chiral states $\psi_{n,m}(x_\alpha,\alpha)$ going around Chern blocks centered at $(2n+1 - \frac{\alpha}{2}, 2m+1 + \frac{\alpha}{2})a$, as shown in Fig.~\ref{lattice-a}. Here
$\alpha$ represents the Chern number of the associated Chern block, and
$x_\alpha\in [0,4a)$ is a local coordinate of the circular chiral states, which goes in the anti-clockwise directions for both $\alpha=+$ and $-$ blocks. One should not confuse $x_{\alpha}$ here with the global coordinate $(x,y)$ defined in \cref{subsec:H-network}. 
By definition, there is $\psi_{n,m}(x_\alpha,\alpha) = \psi_{n,m}(x_\alpha+4a,\alpha) $. We also remind the reader to distinguish $\psi_{n,m}$ from the fields in the wire construction $\psi_{h/v}$.
The fields $\psi_{n,m}(x_\alpha,\alpha)$ are basically re-combinations of the vertical and horizontal wires: on each edge of a Chern block, $\psi_{n,m}(x_\alpha,\alpha)$ equals to some $\psi_h$ or $\psi_v$. For instance, $\{\psi_{n,m}(x_+,+)\vert x_{+}\in [0,a)\}$ equals to $\{\psi_v(y;n,+)\vert y\in[(2m+1)a,(2m+2)a) \}$.
However, one must be careful about how different segments of $\psi_h$ and $\psi_v$ are connected in $\psi_{n,m}(x_\alpha,\alpha)$. 

To see the connection of different segments, we consider a scattering potential $v \theta \sigma_x$ between two chiral modes $\psi_1$ and $\psi_2$ (which is also the basis of $\sigma_x$), as shown on the left side of \cref{lattice-b}. 
Here $v$ is the velocity of the chiral modes. 
As established in \cref{subsec:H-network}, $\theta$ is the scattering angle, \ie $\psi_c = \cos\theta \psi_a -i \sin\theta \psi_d$, $\psi_d = \cos\theta \psi_b - i \sin\theta \psi_a$. 
($\psi_a = \psi_1(0^-)$, $\psi_b = \psi_2(0^-)$, $\psi_c = \psi_1(0^+)$, $\psi_d = \psi_2(0^+)$.) 
In order to investigate the scattering potential in circular chiral state basis $\psi_{n,m}$, we need to recombine $\psi_{1,2}$ to $\psi_{1,2}'$ (see the right side of \cref{lattice-b}) such that 
\begin{equation}
\psi_{1}'( \xi ) = \psi_{1}( \xi ),\qquad 
\psi_{2}'( \xi ) = \psi_{2}( \xi ),\qquad (\xi < -\xi_0) 
\end{equation}
\begin{equation}
\psi_{1}'( \xi ) \propto \psi_{2}( \xi ),\qquad 
\psi_{2}'( \xi ) \propto \psi_{1}( \xi ),\qquad (\xi > \xi_0) 
\end{equation}
where $\xi_0\ll a$ is a small positive quantity. 
In other words, $\psi_{1}$ and $\psi_2$ are interchanged (up to some phase factors) on the ``right side" of the $\delta$-potential. 
To realize such an interchange, we use a smooth transformation 
\begin{equation}
\label{mapping-network-lattice}
\begin{aligned}
    &\begin{pmatrix}
    \psi_1'(\xi)\\
    \psi_2'(\xi)
    \end{pmatrix}
    =\begin{pmatrix}
    \cos\Phi(\xi) & -i\sin\Phi(\xi)\\
    -i\sin\Phi(\xi) & \cos\Phi(\xi)
    \end{pmatrix}
    \begin{pmatrix}
    \psi_1(\xi)\\
    \psi_2(\xi)
    \end{pmatrix}
    \\
    &\Phi(\xi)=\left\{
    \begin{aligned}
    &0\quad (\xi\leq -\xi_0)\\
    &\begin{aligned}&\mathrm{quickly\,and\,monotonicly}\\
    &\mathrm{becomes}\, (2n+1/2)\pi
    \end{aligned}\quad (-\xi_0<\xi\leq\xi_0)\\
    &(2n+1/2)\pi \quad (\xi>\xi_0)
    \end{aligned}
    \right.
\end{aligned}
\end{equation}
where $n$ is an arbitrary integer. (Now we keep $n$ for generality. Nevertheless, we will later show that $n$ has no effect and can be set as $0$.) From the perspective of probability conservation, both $-i\sin\Phi$ and $i\sin\Phi$ are permitted. But since we choose the eigenstates of the network model at $\theta=-\pi/2$ as the basis of desired lattice model, and these eigenstates have phase jumps $i$ at corners ($\psi_{1,2}(0^+)=i\psi_{2,1}(0^-)$), we should choose $-i\sin\Phi$ to remove the phase jump in the representation of $\psi_{1,2}'$. More explicitly, taking $-i\sin\Phi$ means $\psi_{1,2}'(0^+)=-i\psi_{2,1}(0^+),\,\psi_{1,2}'(0^-)=\psi_{1,2}(0^-)$. The eigenstates of $\theta=-\pi/2$ satisfy $\psi_{1,2}(0^+)=i\psi_{2,1}(0^-)$. Combine these together, we have $\psi_{1,2}'(0^+)=-i\psi_{2,1}(0^+)=\psi_{1,2}(0^-)=\psi_{1,2}'(0^-)$, \ie the eigenstats at $\theta=-\pi/2$ are smooth functions of $\xi$ in the representation of $\psi'_{1,2}$.

As shown in the next equation, the transformed Hamiltonian has the same form as the original one but has a different scattering angle.
The scattering Hamiltonian in terms of $\psi'_{1,2}$ reads
\begin{equation}
 \begin{aligned}
         &\begin{pmatrix}
         {\psi}^{\dagger}_1(\xi) & {\psi}^{\dagger}_2(\xi)
         \end{pmatrix}
        \left(-iv \partial_{\xi}\sigma_0+v\theta\delta(\xi) \sigma_x\right)
        \begin{pmatrix}
        {\psi}_1(\xi)\\
        {\psi}_2(\xi)
        \end{pmatrix}\\
        = &\begin{pmatrix}
         {\psi'}^{\dagger}_1(\xi) & {\psi'}^{\dagger}_2(\xi)
         \end{pmatrix}
        \begin{pmatrix}
        \cos\Phi(\xi)& -i\sin\Phi(\xi)\\
        -i \sin\Phi(\xi) & \cos\Phi(\xi)
        \end{pmatrix}
        \left(-iv \partial_{\xi}\sigma_0+v\theta\delta(\xi) \sigma_x\right)
        \begin{pmatrix}
          \cos\Phi(\xi)& i\sin\Phi(\xi)\\
        i \sin\Phi(\xi) & \cos\Phi(\xi)
        \end{pmatrix}
        \begin{pmatrix}
        {\psi}_1'(\xi)\\
        {\psi}_2'(\xi)
        \end{pmatrix}\\
        =&\begin{pmatrix}
         {\psi'}^{\dagger}_1(\xi) & {\psi'}^{\dagger}_2(\xi)
         \end{pmatrix}
        \left[-iv\partial_{\xi}\sigma_0 + v\partial_{\xi} \Phi(\xi) \sigma_x+v\theta\delta(\xi)\sigma_x\right]        
        \begin{pmatrix}
        {\psi}_1'(\xi)\\
        {\psi}_2'(\xi)
        \end{pmatrix}\\
        =&\begin{pmatrix}
         {\psi'}^{\dagger}_1(\xi) & {\psi'}^{\dagger}_2(\xi)
         \end{pmatrix}
         \left[ -iv\partial_{\xi}\sigma_0 +v(\theta+ 2n\pi + \pi/2) \delta(\xi)\sigma_x \right]
         \begin{pmatrix}
        {\psi}_1'(\xi)\\
        {\psi}_2'(\xi)
        \end{pmatrix}
    \end{aligned}
\end{equation}
We made use of the steep growth of $\Phi(\xi)$ in the last equation, \ie $\partial_\xi \Phi (\xi) = (2n\pi + \pi/2) \delta(\xi)$. 
The result of the transformation is a shift of the scattering angle: $\theta \to \theta + \pi/2 + 2\pi n$. One should notice that the scattering problem in terms of $\psi'_{1,2}$ is similar to that of $\psi_{1,2}$. Hence, the scattering effect is also periodic on the potential strength, and the ambiguity of integer $n$ does not make any difference. We take $n=0$ hereafter.

The transformed scattering potential can be understood in some limits. The local orbital basis $\mathcal{B}$ we use in this lattice model comprises the eigenstates of the network model at $\theta=-\pi/2$. Therefore, there must be no scattering between the local orbitals when $\theta=-\pi/2$, \ie the scattering potential in the representation of $\mathcal{B}$ should be $0$. When $\theta=\pi/2$, the eigenstates are also local currents, but the phase jumps are reversed (see \ref{scatter matrix}, when $\theta=\pm\pi/2$, the phase jump is $\mp i$ from the incoming to the outgoing channel). Effectively, in the representation of $\mathcal{B}$, these eigenstates are local currents with phase jumps $-1$ at corners. Hence the effective scattering angle in terms of $\mathcal{B}$ should be $\pi$. When $\theta=0$, the eigenstates are decoupled horizontal and vertical wires. Since a complete transmission between two wires (electron goes straight along the vertical or horizontal wire at the intersection without phase jump) can be viewed as a complete reflection between two local orbitals (electron completely jumps to the adjacent square at the intersection with a phase jump $-i$), the effective scattering angle for $\mathcal{B}$ should be $\pi/2$. One can directly check that the shifting $\theta\rightarrow \theta+\pi/2$ satisfies all these conditions.

After applying the above transformation to every corner of Chern blocks, the scattering potential $v \theta \delta_x$ between horizontal and vertical wires will be mapped to $v (\theta+\pi/2) \delta_x$ between the circular chiral modes at nearby squares (choosing $n=0$). 
Therefore, the network model Hamiltonian can be equivalently written as 
\begin{equation}
    \begin{aligned}
    H_{N,0} =\sum_{nm}\sum_{\alpha}& \int \mathrm{d}x \thinspace \psi_{n,m}^{\dagger}(x,\alpha) (-i\alpha v \partial_x) \psi_{n,m}(x,\alpha)\\
    H_{N,1} = \sum_{nm}\sum_{\alpha}\thickspace &a\tilde{t}\, [ \psi_{n,m}^{\dagger}(0,\alpha) \psi_{n,m}(0,\bar{\alpha})+\psi_{n,m}^{\dagger}(a,\alpha) \psi_{n(m+\alpha)}(a,\bar{\alpha})\\ &\;+\psi_{n,m}^{\dagger}(2a,\alpha) \psi_{(n-\alpha)(m+\alpha)}(2a,\bar{\alpha})
    +\psi_{n,m}^{\dagger}(3a,\alpha) \psi_{(n-\alpha)m}(3a,\bar{\alpha}) ] \quad \quad  
    \end{aligned}
    \label{lattice-H-real}
\end{equation}
where $\bar{\alpha}=-\alpha$, and $a\tilde{t} = v(\theta + \pi/2)$ is the transformed scattering potential. 
For $\alpha=1$ ($\alpha=-1$) the four terms in $H_{N,1}$ couples $\psi_{n,m}^{\dagger}$ to its neighboring circular chiral states in the right lower (left upper), right upper (left lower), left upper (right lower), left lower (right upper) directions, respectively. 
Again, the reader should not confuse the `$x$' in Eq.~(\ref{lattice-H-real}) with the global coordinate $(x,y)$ defined in Sec.~\ref{subsec:H-network}. `$x$' in Eq.~(\ref{lattice-H-real}) is the coordinate inside one square, \ie the $x_{\alpha}$ defined in Fig.~\ref{lattice-a}, and we have ignored the subscript $\alpha$ for simplicity . 

When $\theta=-\frac{\pi}2$, we can see that $\tilde{t} = \frac{v}{a}(\theta+\pi/2)=0$ and hence the circular chiral modes are decoupled from each other. 
Then the kinetic energy Hamiltonian $H_{N,0}$ can be diagonalized in representation of the circular chiral modes,
\begin{equation}
      \phi_{n,m}^{\dagger}(K,\alpha) =\frac{1}{\sqrt{4a}}\int \mathrm{d} x_{\alpha} \mathrm{exp} \left[ i\alpha \frac{2\pi K}{4a} x_{\alpha}\right]  \psi_{n,m}^{\dagger}(x_{\alpha},\alpha) \quad\quad\quad\quad \alpha=\pm 1,\thickspace K\in \mathbb{Z}\ .
     \label{standing wave}
\end{equation}
$\phi_{n,m}(K,\alpha)$ can be thought of as a standing wave going around the $C=\alpha$ Chern block in the unit cell $(n,m)$.
$K$ being an integer is required by the periodicity $\psi_{n,m}(x_{\alpha},\alpha) = \psi_{n,m}(x_{\alpha}+4a,\alpha)$. 
Note that we use $\phi_{n,m}(K,\alpha)$ to denote the standing wave with phase factor $\propto\alpha K x_{\alpha}$. Such a special choice has the convenience that kinetic energy of $\phi_{n,m}(K,\alpha)$ depends on $K$ only, and this is due to the special direction choice of $x_{\alpha}$. For the red square, $x_+$ increases along the direction of chiral wires, so that a standing wave with phase factor $\propto Kx_{+}$ has kinetic energy $vK\pi/2a$. For the blue square, $x_-$ increases against the chiral direction, so that a standing wave with phase factor $\propto -Kx_{-}$ also has kinetic energy $vK\pi/2a$.

The inverse transformation is 
\begin{equation}
\psi_{n,m}^{\dagger}(x_{\alpha},\alpha)
= \frac1{\sqrt{4a}} \sum_{K}  \exp\brak{ -i\alpha \frac{2\pi K}{4a} x_\alpha } \phi_{n,m}^{\dagger}(K,\alpha)\ .
\label{co-standing wave}
\end{equation}
After the Fourier transform, $H_{N,0}$ can be written as 
\begin{equation} \label{eq:H0-standing-wave}
    H_{N,0} = \sum_{n,m,\alpha,K} \frac{\pi v K}{2a} \phi_{n,m}^{\dagger}(K,\alpha) \phi_{n,m}(K,\alpha)\ .
\end{equation}
And $H_{N,1}$ introduces hoppings between the standing waves: 
\begin{align}\label{eq:H1-standing-wave}
H_{N,1} =& \frac{\tilde{t}}4 \sum_{n m \alpha} \sum_{KK'}  \Xi_{K_{\max}'} \left[\alpha K + \bar{\alpha} K' \right] \bigg[     
    \phi_{n,m}^\dagger(K,\alpha)\phi_{n,m}(K',\bar\alpha)
   + e^{-i\alpha \frac{\pi}2 K + i\bar \alpha \frac{\pi}2 K'} \phi_{n,m}^\dagger(K,\alpha)\phi_{n,m+\alpha}(K',\bar\alpha) \nonumber \\
  & +  e^{-i\alpha \pi K + i\bar \alpha \pi K'} \phi_{n,m}^\dagger(K,\alpha)\phi_{n-\alpha,m+\alpha}(K',\bar\alpha)
  +  e^{-i\alpha \frac{3\pi}a K + i\bar \alpha \frac{3\pi}a K'} \phi_{n,m}^\dagger(K,\alpha)\phi_{n-\alpha,m}(K',\bar\alpha) \bigg]\ .
\end{align}
Here we introduce a cutoff factor $\Xi_{K_{\max}'} \left[\alpha K + \bar{\alpha} K' \right]$ (defined in \cref{eq:truncation}) to describe the broadening of the $\delta$ scattering potential.
Note that $\frac{\pi}{2a}(\alpha K + \bar{\alpha} K')$ is the transferred momentum in the scattering.
$K_{\max}'$ should have the order of $ 2a \Lambda_2 /\pi $, with $\Lambda_2$ being the broadening truncation in the wire construction introduced in \cref{subsec:H-network}. 
Therefore, we have successfully rewritten the network model as a lattice model with an infinite number of orbitals per cell (we have not introduced the cutoff $\Lambda_1$ on $K$), where $H_{N,0}$ sets the on-site energy and $H_{N,1}$ are the hopping terms. 

\cref{eq:H0-standing-wave,eq:H1-standing-wave} should have the same symmetries as the original network model. 
The actions of $C_{2z}T$, $C_{4z}$, and $M_{xy}$ (defined in \cref{subsec:Manhattan}) act on the circular chiral basis as  %
\begin{equation}
    \begin{aligned}
    (C_{2z}T) \psi_{n,m}(x,\alpha) (C_{2z}T)^{-1}&= \psi_{-n-1,-m-1}(x,\bar{\alpha}) \\
    C_{4z} \psi_{n,m}(x,\alpha) C_{4z}^{-1}&= \psi_{-m-1,n+1 - (1+\alpha)/2}(x+a,\alpha)\\ 
    M_{xy} \psi_{n,m}(x,\alpha) M_{xy}^{-1}&= \psi_{m,n}(4a-x,\bar{\alpha})\\ 
    \end{aligned}\ .
    \label{symmetry-action-lattice0}
\end{equation}
Applying these actions to the standing wave basis, one should obtain 
\begin{equation}
    \begin{aligned}
    (C_{2z}T) \phi_{n,m}(K,\alpha) (C_{2z}T)^{-1}&= \phi_{-n-1,-m-1}(K,\bar{\alpha}) \\
    C_{4z} \phi_{n,m}(K,\alpha) C_{4z}^{-1}&= e^{i \frac{\pi}2 \alpha K} \phi_{-m-1,n+1 - (1+\alpha)/2}(K,\alpha) \\ 
    M_{xy} \phi_{n,m}(K,\alpha) M_{xy}^{-1}&= \phi_{m,n}(K,\bar{\alpha})\\ 
    \end{aligned}\ .
    \label{symmetry-action-lattice}
\end{equation}
It is direct to check that they commute with \cref{eq:H0-standing-wave,eq:H1-standing-wave}. 

In order to obtain the band structure of \cref{eq:H0-standing-wave,eq:H1-standing-wave}, we introduce the Bloch basis of the standing wave basis as 
\begin{equation}
    \begin{aligned}
    \phi_{n,m}(K,\alpha)=\frac{1}{N}\sum_{\vec{q}}\thinspace \exp\left(i\vec{q}\cdot(\vec{R}_{nm}+\vec{t}_{\alpha})\right) \thinspace \phi_{\vec{q}}\thinspace(K,\alpha)
    \end{aligned}
\end{equation}
where $N$ is the number of unit cells in each direction, and $\vec{R}_{nm}=(2na,2ma)$ and $\vec{t}_{\alpha}=\left((2-\alpha)a/2 , (2+\alpha)a/2\right)$ are the cell position and the relative position of square center under the \emph{global} coordinate system (see the first paragraph in \cref{subsec:H-network}), respectively. 
$\vec{q}$ takes values in the first Brillouin zone $[-\pi/(2a), \pi/(2a) ) \times [-\pi/(2a), \pi/(2a) )$.
(Note that the lattice constant is $2a$.)
We then obtain the Hamiltonian in reciprocal space: 
\begin{equation}
    \begin{aligned}
    H_{N,0} &=\sum_{\vec{q}}\sum_{\alpha K}\thickspace \frac{\pi v K}{2a}\thinspace \phi_{\vec{q}}^{\dagger}\thinspace(K,\alpha) \phi_{\vec{q}}(K,\alpha)\\
    H_{N,1} &=\frac{\tilde{t}}{4} \sum_{\vec{q}}\sum_{\alpha}\sum_{K_1,K_2}\thickspace  
    \Xi_{K_{\max}'} \left[\alpha K_1 + \bar{\alpha} K_2 \right] \left[ \, 
    e^{i\alpha(q_x a-q_y a)}
    +(-\alpha i)^{(K_1+K_2)}\thinspace e^{i\alpha(q_x a+q_y a)}
    \right.\\ & \qquad \qquad \qquad\qquad\quad\left.
    +(-1)^{(K_1+K_2)}\thinspace e^{-i\alpha (q_x a -q_y a)} +(\alpha i)^{(K_1+K_2)}\thinspace e^{-i\alpha(q_x a+q_y a)} \, \right]\thickspace \phi^{\dagger}_{\vec{q}}\thinspace(K_1,\alpha) \phi_{\vec{q}}\thinspace(K_2,\bar{\alpha})
    \end{aligned}\ . 
    \label{lattice-H-reciprocal}
\end{equation}
As in \cref{eq:H1-standing-wave}, for $\alpha=1$ ($-1$), the four terms in above $H_{N,1}$ correspond to couplings between neighbor circular chiral states in the right lower (left upper), right upper (left lower), left upper (right lower), left lower (right upper) directions, respectively. 

In actual calculations, we should take a cutoff on $K$ ($K\in[K_{\min}, K_{\max}]$), and there will be $2(K_{\max}-K_{\min}+1)$ bands in total. Note that the network model and un-truncated lattice model represent the same system on different bases. \cref{lattice-H-reciprocal} will follow the band periodicity of the network model when $(K_{\max}-K_{\min})\rightarrow\infty$. Hence, only $K_{\max}-K_{\min}$ matters, and we can take $K_{\min}=0$ for notation convenience. We notice that $K_{\max}$ plays the same role as $\Lambda_1$ in the network model.
So in principle, the cutoff $K_{\max}'$ of $\delta$-potential in $H_{N,1}$, should be (i) much smaller than $K_{\max}$ such that the degrees of freedom in chiral edges can be treated as continuous and (ii) much larger than one such that the characteristic length of the $\delta$-potential is much smaller than the unit cell size. 
(See the discussions in \cref{subsec:cutoff} for details).
Numerical calculation confirms that the middle eight bands of Eq.~(\ref{lattice-H-reciprocal}) can approximately reproduce the band structure of the network model (Fig.~\ref{bands-network}) in one repeating unit when $K_{\max} = 7$ and $K_{\max}'=2$. To reproduce more repeating units, a higher cutoff is needed, \eg $K_{\max}=12,\,K_{\max}'=2$ for two repeating units. 
\subsection{Eight-band lattice model \texorpdfstring{$H_{8B}$}{H8B}}
\label{subsec:8-bands full model}
The above analysis is a basis transformation of \cref{network-H-reciprocal-cutoff}. \cref{lattice-H-reciprocal} will reproduce the band structure of the network model when $K_{\max},\,K_{\max}'\rightarrow\infty$ and $K_{\max}'/K_{\max} \rightarrow 0$. This verifies that the un-truncated lattice model is an equivalent description of the network model with local basis. However, the un-truncated lattice model still contains infinite bands and numerically requires a high cutoff to reproduce the band structure of the network model. In the end of \cref{subsec:standing-wave}, we mentioned that 15 bands are needed to reproduce the network model's band structure in one repeating window. In this subsection, we attempt to construct a truncated lattice model that contains fewer bands but still capture the low energy physics of the network model. 

In this paragraph, we sketch the procedure that leaded us to the desired model and show its properties in later paragraphs. We started from a large $K_{\max}$ such as $200$ and fixed $K'_{\max}$ to some finite integer such as $10$. Then we decreased $K_{\max}$ until the low energy physics is essentially changed. ``Low energy physics" here means the physics near the Fermi surface which is chosen as the middle of the bands to minimize the effects of cutoff on $K$. When $K_{\max}\gg K'_{\max}$, we had basically the same band structure as the network model except for several highest and lowest bands that are significantly influenced by the cutoff on $K$. As $K_{\max}$ decreased, the band structure gradually changed away from that of the network model. Such a deviation is not uniform for all the bands, the middle bands deviate less while the high-lying and low-lying bands deviate more. When $K_{\max}\gtrsim K'_{\max}$, the 8-band periodicity of the network model cannot be reproduced, but some low energy properties of the network model (will shown later) still remain. Finally, we got the minimal model with $K_{\max}=3$ (eight bands) that can reproduce the low energy physics of the network model and we will justify this result later. We refer to the resulting Hamiltonian as $H_{8B}$

One may notice that the condition $\Lambda_1\gg\Lambda_2$ in the network model is now violated. Cutoff $K_{\max}$ discretizes the chiral edges and $K'_{\max}$ broaden the $\delta$ scattering potential. In the network model and its mapping to the un-truncated lattice model, we require the granularity of chiral edges to be much smaller than the broaden of the $\delta$ potential. But this requirement is violated now, since we take $K_{\max}<K'_{\max}$ (and $K'_{\max}=\infty$ effectively) on the way to $H_{8B}$, although it finally turns out that $H_{8B}$ can reproduce the network model's low energy physics. Nevertheless, we can offer two reasons about why $H_{8B}$ with $K_{\max}=3,\,K'_{\max}=\infty$ is a valid approximation. 
First, the repeating unit of the the network model's band structure comprises eight bands, so it is reasonable to presume that the essential physics of the network model can be realized by only eight bands, \ie $K_{\max}=3$. 
Second, when $K_{\max}=3$, the condition $K_{\max}\propto \Lambda_1 \rightarrow \infty$ is not satisfied, \ie the granularity of chiral edges is too coarse to discriminate \cref{twolimit-a} and \cref{twolimit-b}. And numerical calculation confirmed that, given $K_{\max}=3$, different $K'_{\max}$ choices result in similar band structures. Therefore, we can take $K_{\max}'=\infty$, \ie ignore the truncation factor $\Xi_{K_{\max}'}$ for simplicity. 
Now, we will show the numerical results and illustrate that $H_{8B}$ indeed captures the low energy features of the network model's band structure.

\begin{figure}[h]
	\centering
	\subfigure[$\tilde{t}=0$]{\label{band-lattice-a}
    \includegraphics[width=.3\linewidth]{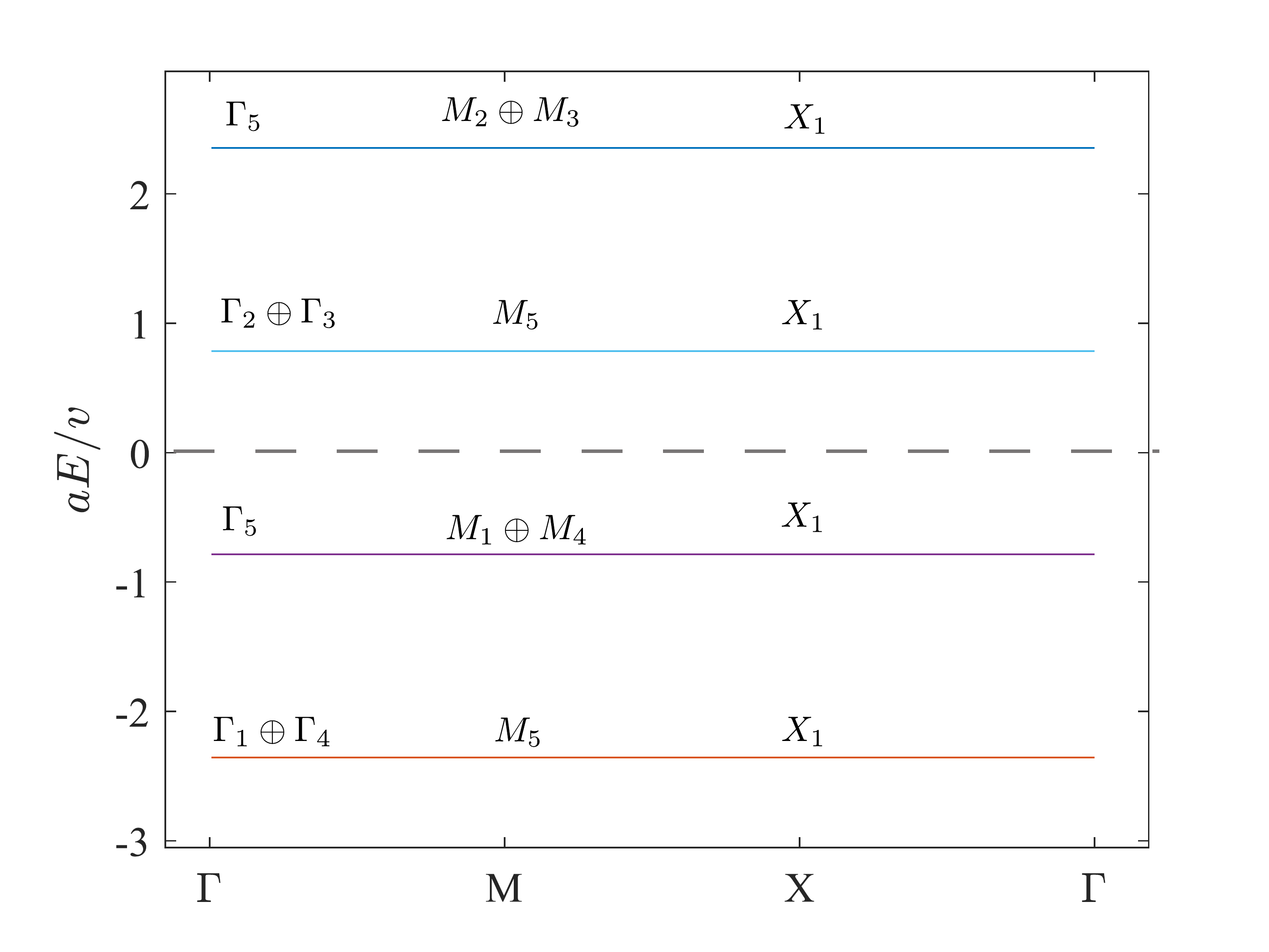}}
	\subfigure[$\tilde{t}=0.3v/a$]{\label{band-lattice-b}
	\includegraphics[width=.3\linewidth]{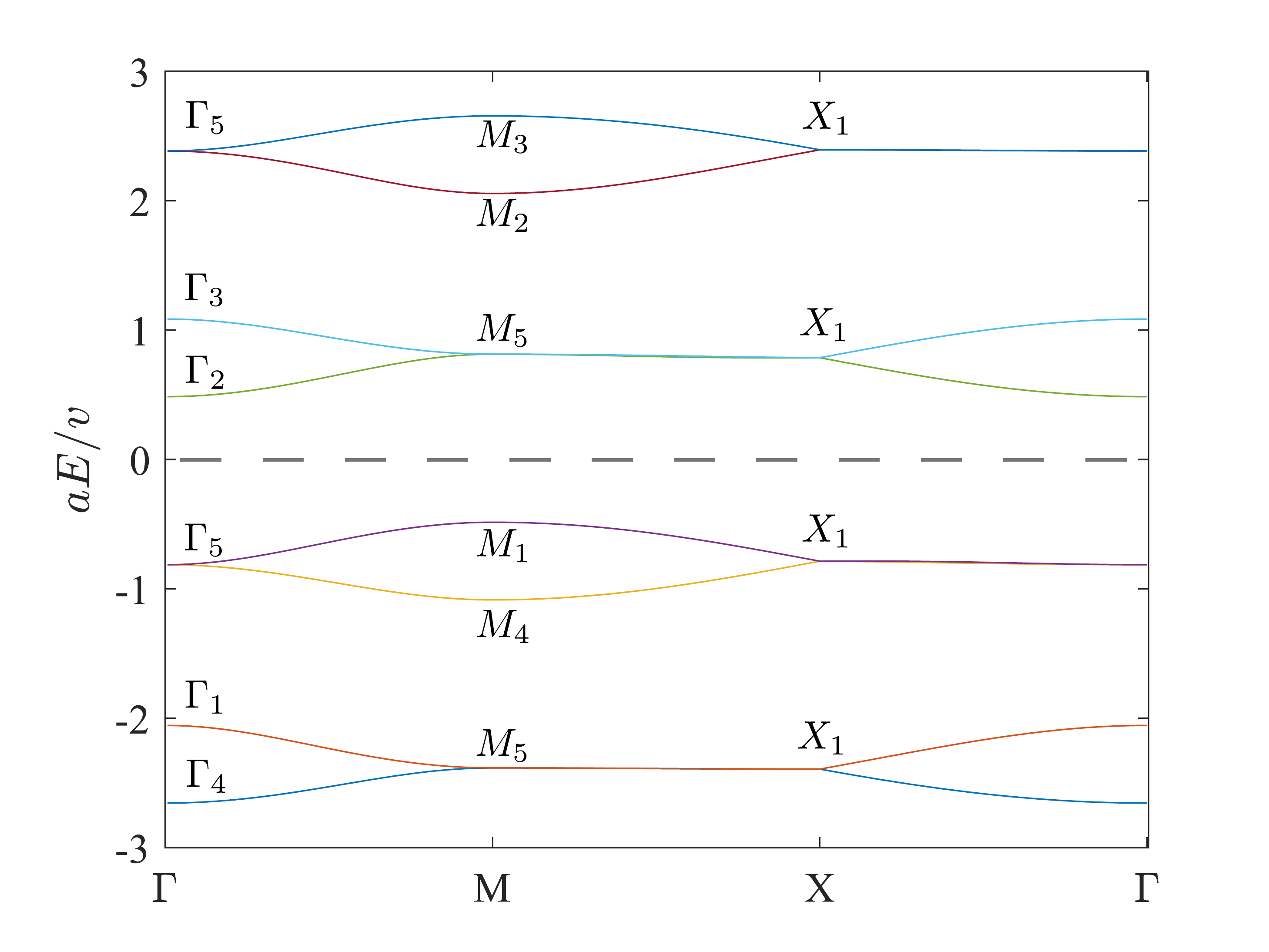}}
	\subfigure[$\tilde{t}=\pi v/4a$]{\label{band-lattice-c}
	\includegraphics[width=.3\linewidth]{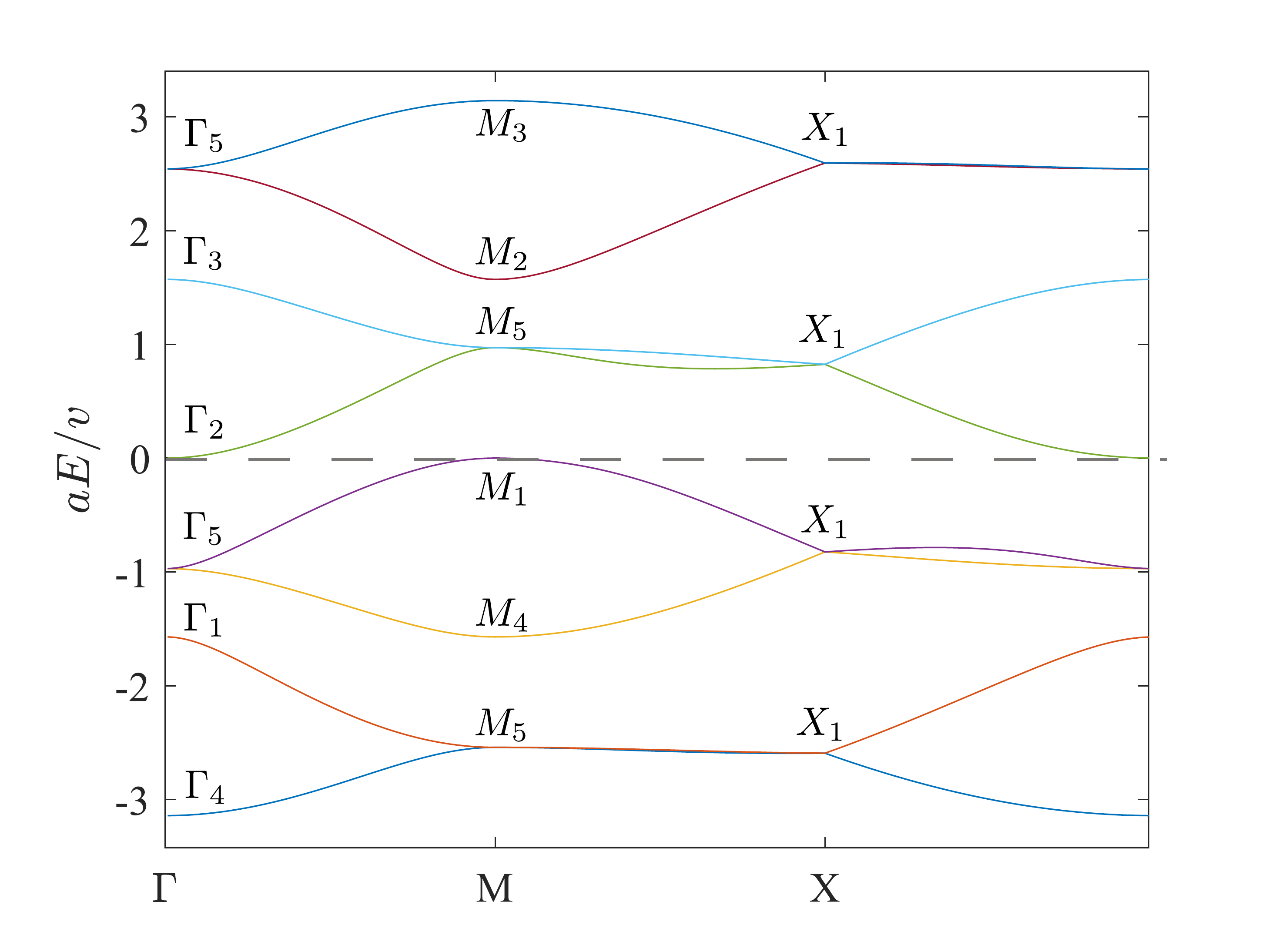}} \\
    \subfigure[$\tilde{t}=3\pi v/8a$]{\label{band-lattice-d}
	\includegraphics[width=.3\linewidth]{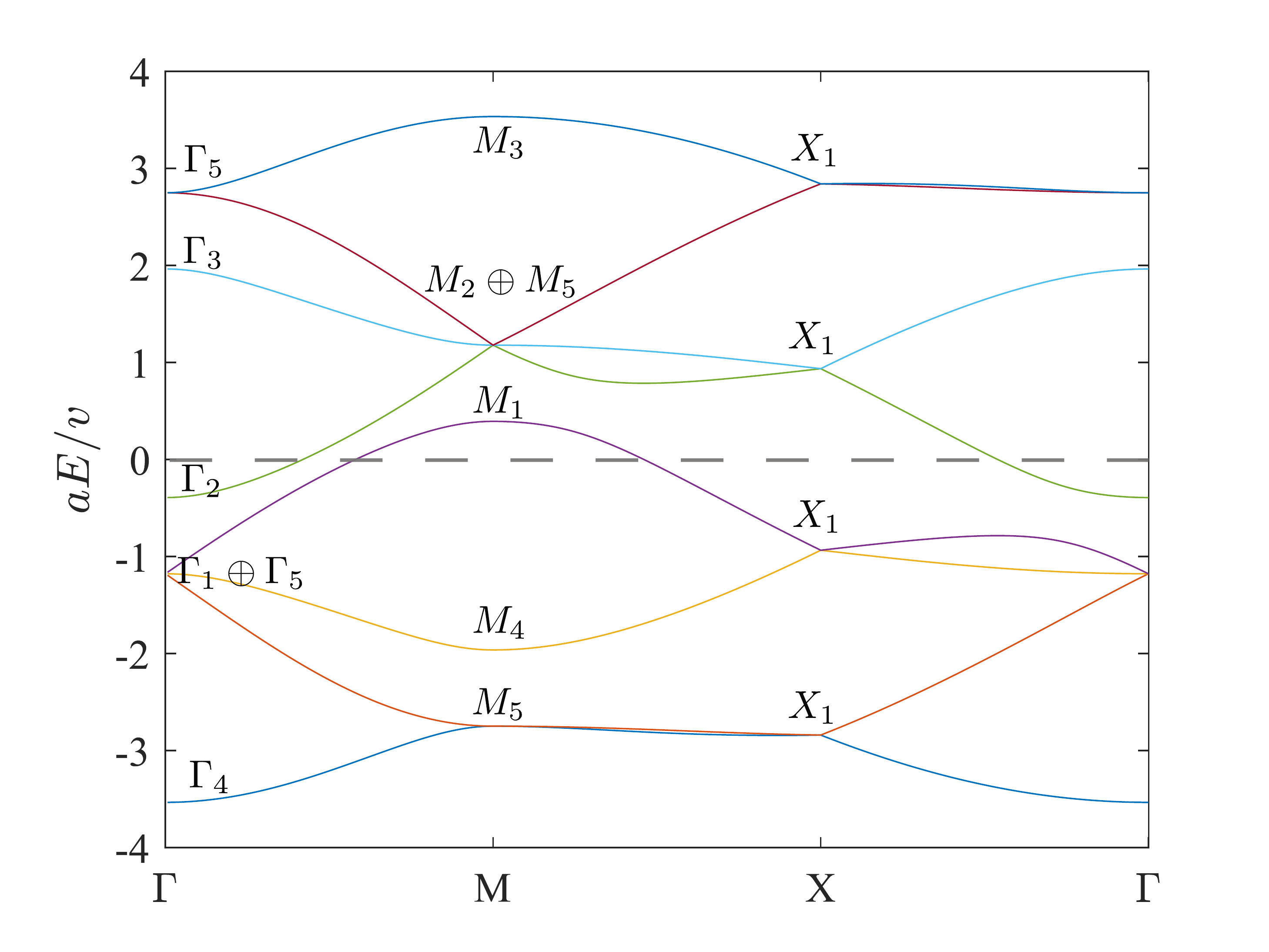}}
	\subfigure[$\tilde{t}=1.4 v/a$]{\label{band-lattice-e}
	\includegraphics[width=.3\linewidth]{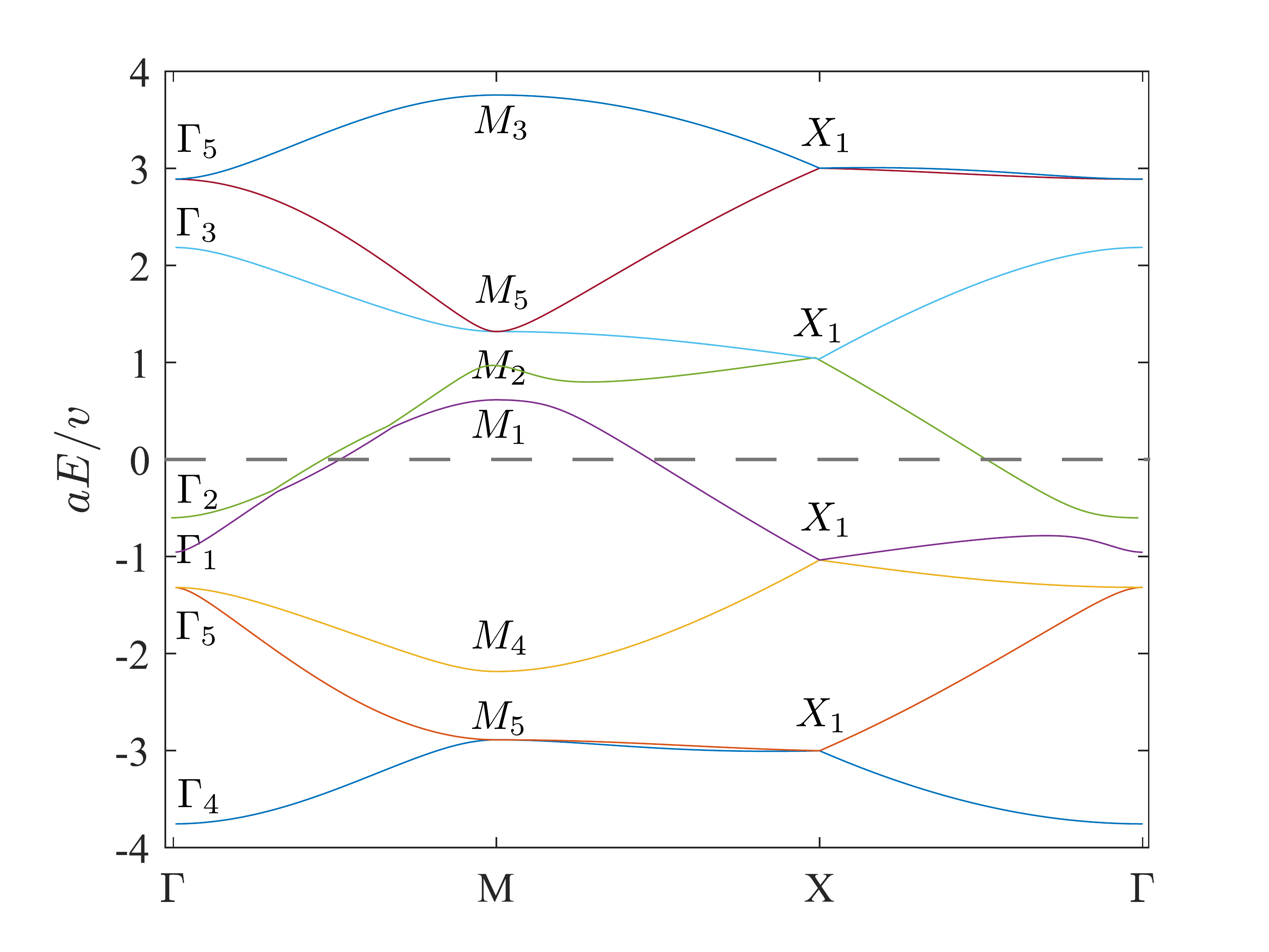}}
	\subfigure[$\tilde{t}=\pi v/2a$]{\label{band-lattice-f}
	\includegraphics[width=.3\linewidth]{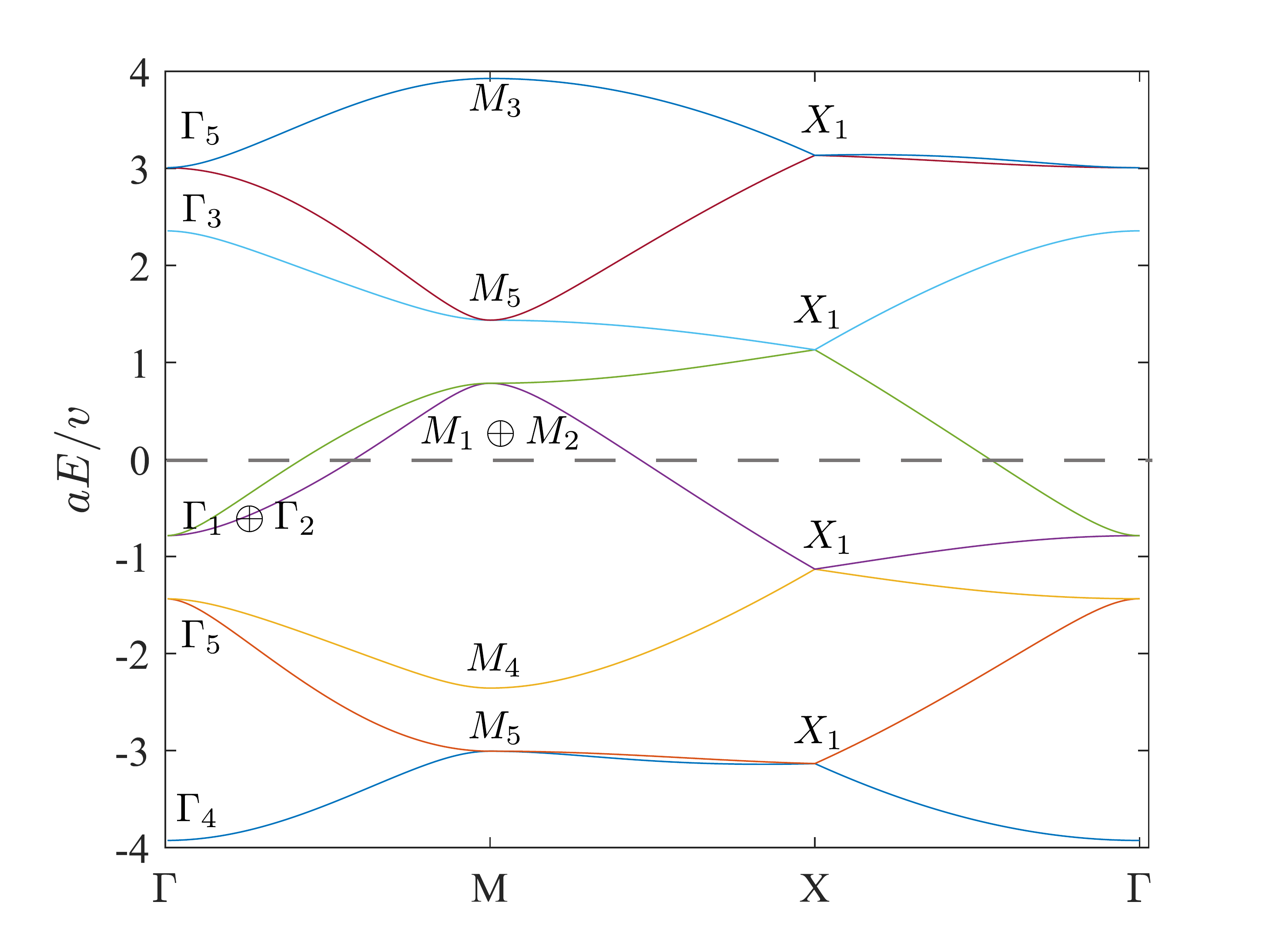}}
	\\	
	\subfigure[$\tilde{t}=3\pi v/4a$]{\label{band-lattice-g}
	\includegraphics[width=.3\linewidth]{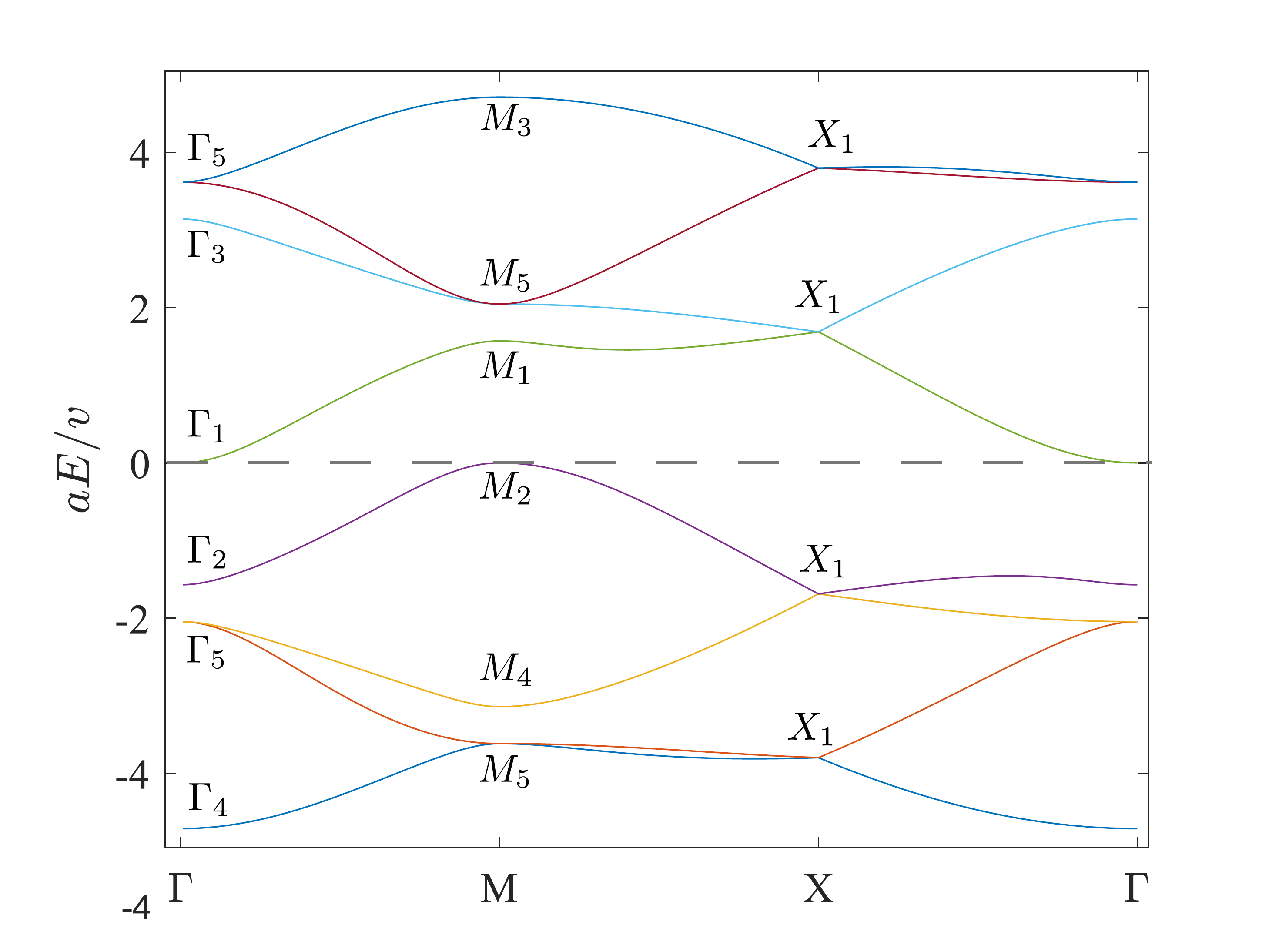}}
 	\subfigure[$\tilde{t}=\pi v/a$]{\label{band-lattice-h}
	\includegraphics[width=.3\linewidth]{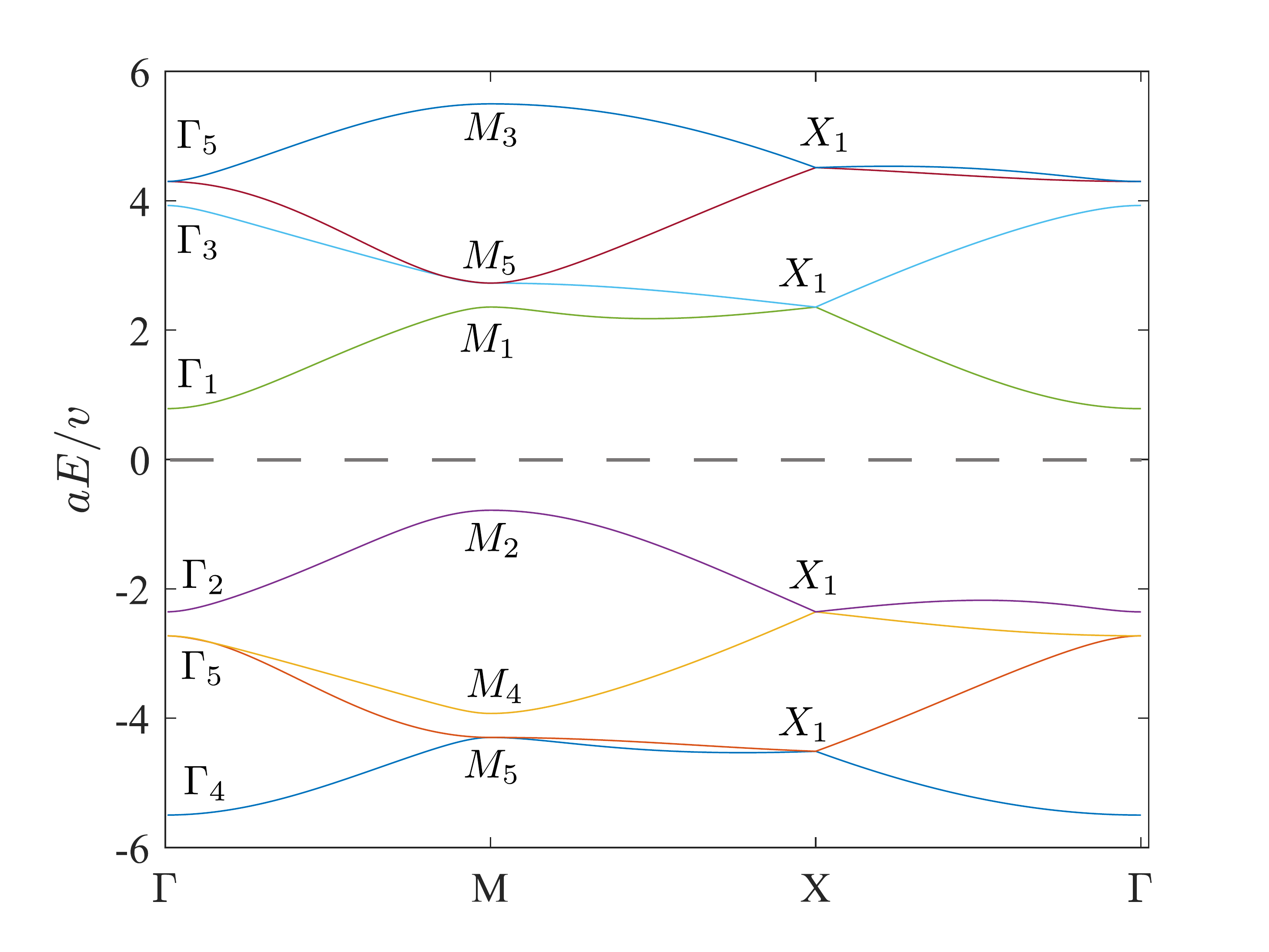}}
    \caption[]{Band structures of lattice model $H_{8B}$ with different scattering strengths $\tilde{t}$. The gray dashed lines indicate the Fermi level $E_F=0$. Note that we have shifted the zero energy to the middle of the eight bands. }
	\label{bands-lattice}
\end{figure}
\begin{figure}[h]
	\centering
     \subfigure[]{
    \includegraphics[width=.45\linewidth]{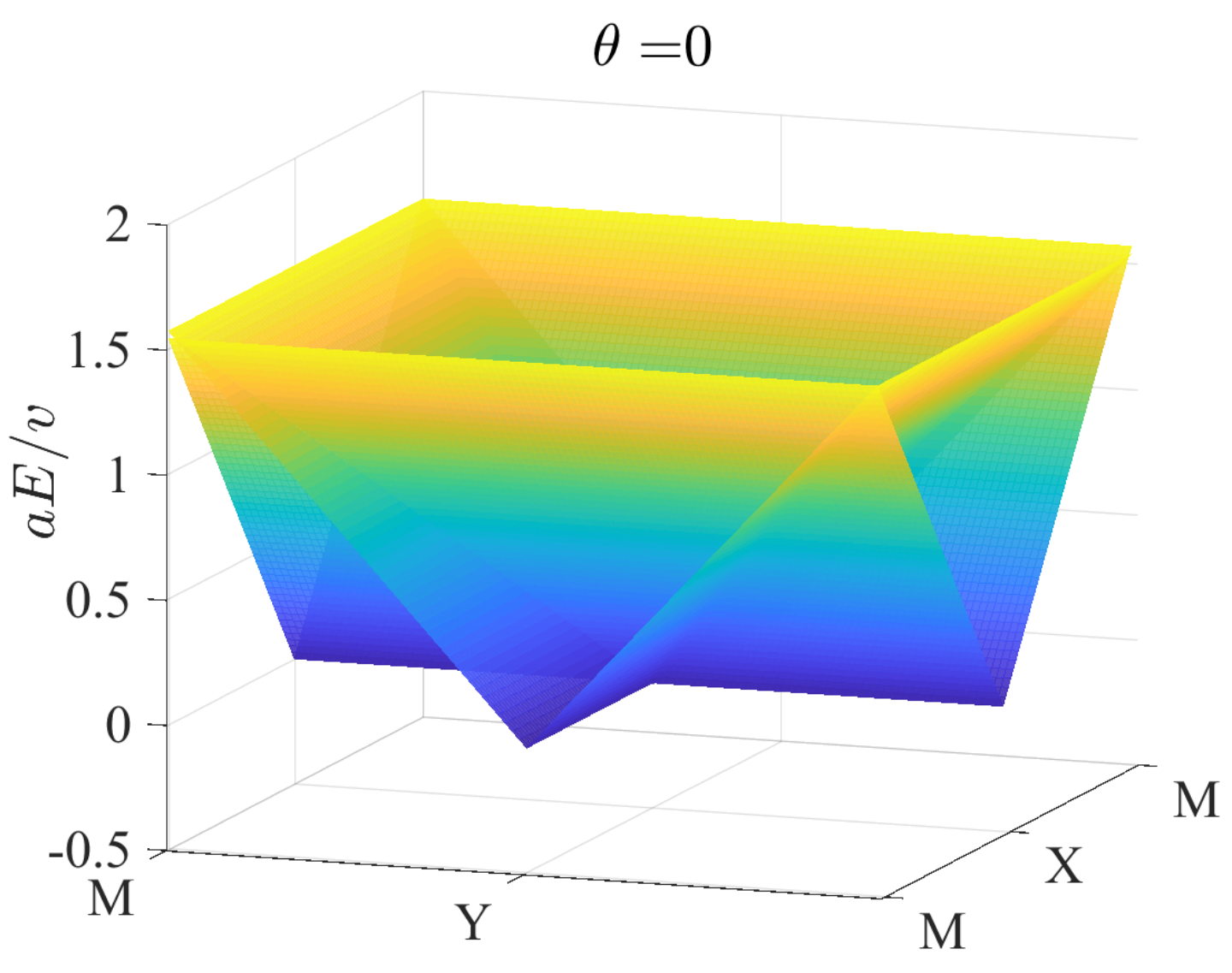}}
	\subfigure[]{
    \includegraphics[width=.45\linewidth]{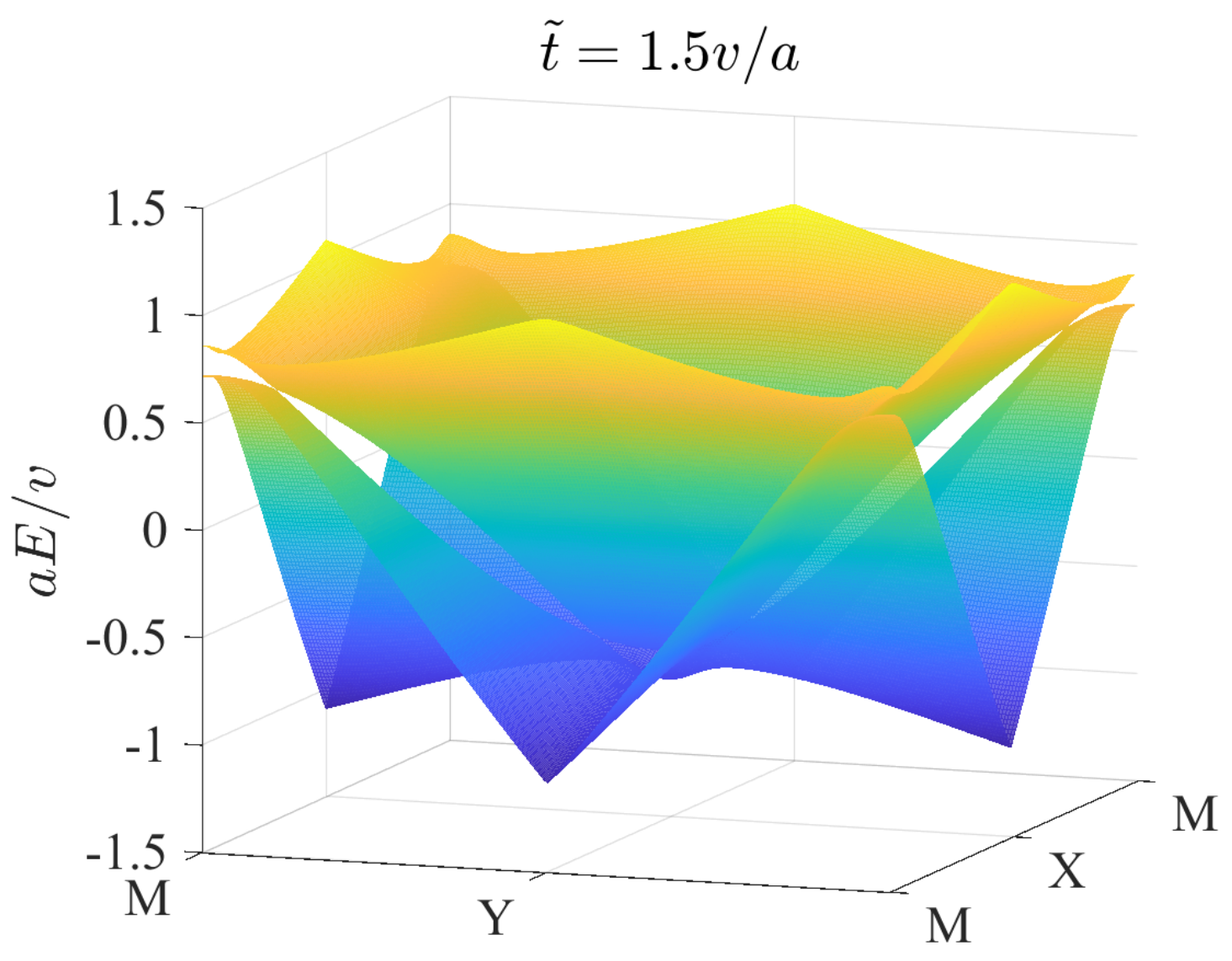}}
    \caption[]{(a) The quasi-1D band structure of the network model in the metal limit ($\theta=0$) and near the Fermi level ($E_{F-N}=\pi v/4a$). It is exactly a superposition of two 1D bands with linear dispersion in $x$ \& $y$ directions (chiral wires in $x$ \& $y$ directions), respectively. The corresponding plot in the main text (the inset of Fig.~\textcolor{green}{1(h)}) chooses $E_{F-N}$ here as the zero energy.  (b) The quasi-1D band structure of $H_{8B}$ in the metal limit ($\tilde{t}\approx \pi v/2a$) and near the Fermi level ($E_{F-8B}=0$). It is similar to (a) but with a minor coupling between two 1D subsystems. Note that we have shifted the zero energy of $H_{8B}$ to the middle of the eight bands. And the corresponding plot in the main text (inset of Fig.~\textcolor{green}{2(d)}) chooses the same zero energy.
    }
	\label{bands3d-lattice}
\end{figure}

We calculate the band structure and representations at high-symmetry points of $H_{8B}$ and show the results in Fig.~\ref{bands-lattice}. (We also shift the zero energy to the middle of the eight bands.)
First, if we take the Fermi energy in the middle of the eight bands (the gray dashed lines in \cref{bands-lattice}), \ie $E_{F-8B}=0$, the system is insulating when $\tilde{t}\rightarrow 0$ and $\tilde{t} \rightarrow \pi v/a$. (Even though bands are not flat for $\tilde{t}=\pi v/a$ as they were in the network model.) 
This is indeed the case for the network model with a Fermi level $E_{F-N}=\pi v/4a$ laid in the middle of a repeating unit (the gray dashed lines in \cref{bands-network}). 
Second, for $\tilde{t}\approx\pi v/2a$, the band structure of the two bands closest to the Fermi level $E_{F-8B}$ is similar to a superposition of two 1D bands with linear dispersion in $x$ \& $y$ direction, respectively (see Fig.~\ref{bands3d-lattice}). 
Such a superposition is approximately the quasi-1D band structure of the network model with $\theta=0$ ($\tilde{t}=\pi v/2a$), \ie which consists of decoupled chiral wires put along $x$ \& $y$ directions, respectively (see the inset of Fig.~\textcolor{green}{1(h)}). 

Another supporting evidence is the change of representations near the Fermi level in the phase transition. 
As $\tilde{t}$ increases from $0$ to $\pi v/a$, $\Gamma_1,\,M_1$ rise across the Fermi level while $\Gamma_2,\,M_2$ fall below the Fermi level. In the network model, if we take the red dashed pane (contains eight bands) in \cref{bands-network} as a repeating unit and set the Fermi level at the gray dashed line in \cref{bands-network} (the middle of the unit), during the transition ($\theta=-\pi/2\rightarrow \pi/2$), the same pairs of representations will switch across the Fermi level in the same way. One may ask why we choose the upper eight bands in \cref{bands-network} rather than the lower eight as the `repeating unit. The reason is that the standing waves with $K\in[0,3]$ at $\tilde{t}=0$ (\cref{band-lattice-a}) correspond to the upper eight bands in \cref{band-network-a} at $\theta=-\pi/2$ (\cref{band-network-a}).
Based on these observations and the numerical results that will be presented in Sec.~\ref{sec:numerical-localization-length}, we claim that $H_{8B}$ correctly captures the low energy physics of the network model in Sec.~\ref{sec:Network-model}.  

Despite similarities in low energy physics, we must point out that the BR transition of $H_{8B}$ differs from that of the network model. In the network model with a sufficient number of bands, the four bands below the Fermi level and inside a repeating unit (inside the red pane and below the gray dashed line in \cref{bands-network}) not only exchange representations with the bands above them but also with bands (in another repeating unit) below them. During the transition ($\theta=-\pi/2\rightarrow\pi/2$), these four bands change from $\mathrm{^{2}E_b\! \uparrow \!G \oplus A_b\! \uparrow\!G}$ to $\mathrm{^{1}E_b\! \uparrow \!G \oplus B_b\! \uparrow\!G}$. As we explained, such four bands corresponds to the lower four bands of $H_{8B}$. However, in $H_{8B}$, there is no band below these four bands.
As a result, different from the network model, the lower four bands of $H_{8B}$  change from $\mathrm{^{2}E_b\! \uparrow \!G \oplus A_b\! \uparrow\!G}$ to $\mathrm{A''_c\!\uparrow\! G}$ during the transition ($\tilde{t}=0\rightarrow\pi v/a$).
The Wyckoff position of $\mathrm{A''_c\!\uparrow\! G}$ is $\mathrm{4c}$, which is the $C_{2z}T$ center rather than the $C_{4z}$ center (\cref{EBRs}). 
This difference of molecular orbital transition is unavoidable since the transition in the network model involves exchanges of representations between different repeating units, while $H_{8B}$ only has one unit. Nevertheless, since the origin of this difference is well below the Fermi level, it is not a difference in the low energy physics. And we have seen that $H_{8B}$ and the network model have the same representation exchange near the Fermi surface; therefore, $H_{8B}$ still reproduces the low energy physics of the network model.

It is worth mentioning that the transition in $H_{8B}$ changes the position of the MEBRs from $C_{4z}$-centers (2b) to $C_{2z}T$-centers (4c) (see \cref{TB-cell}). No $C_{2z}T$ center is occupied before the transition, and the Real Space Invariant $\delta_{w}=0$. Given that there are four $C_{2z}T$ centers per cell and four occupied bands, every $C_{2z}T$ center is occupied by one electron after the transition, and the system has Real Space Invariant $\delta_{w}=1$. Therefore, the second Stiefel-Whitney class $w_2$ \cite{ahn_failure_2019,fang_topological_2015} must change from 0 to 1. 
The transition process must involve braiding of the Dirac points. 
We will discuss this in detail in \cref{subsec:dirac-evolution}. 
In addition, although the lower four bands form $\mathrm{A''_c\!\uparrow\! G}$ are always connected in our models, in general cases, $\mathrm{A''_c\!\uparrow\! G}$ can be decomposed into two fragile topological bands ($\Gamma_5,\,M_5,\,X_1$) and two trivial bands (forming MEBR $A_2\!\uparrow\!G$ with Wyckoff position $2a\,(\frac14\,\frac14,0),\,(\frac34\,\frac34,0)$, \ie the centers of white squares in \cref{network-a}), which is expected from $w_2=1$.

\subsection{\texorpdfstring{$H_{8B}$}{H8B} on corner state basis}
\label{subsec:corner-state}

\begin{figure}[h]
	\centering
	\subfigure{
    \includegraphics[width=.35\linewidth]{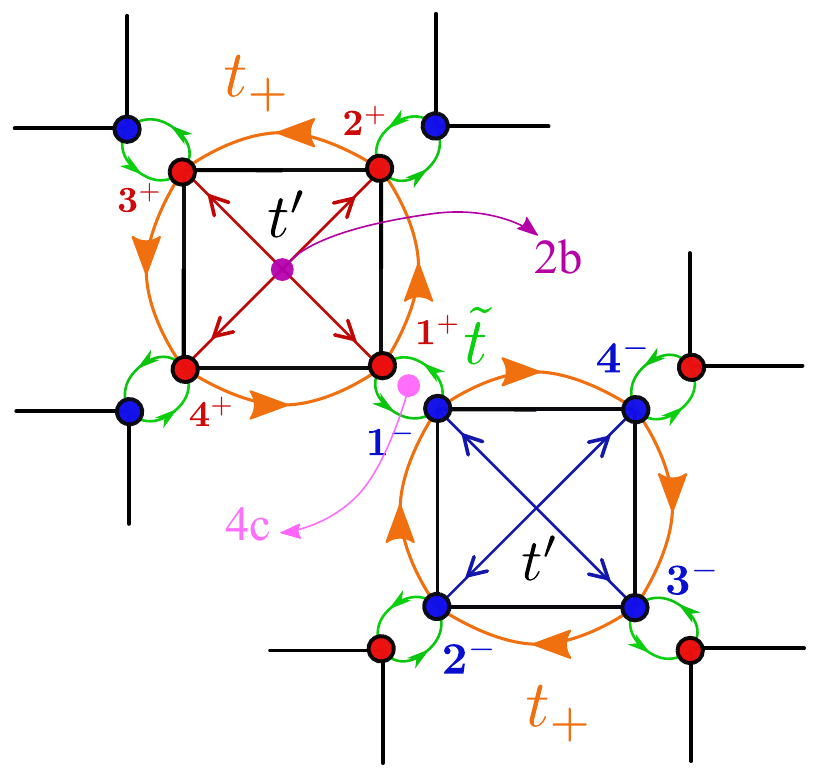}}
    \caption[]{Unit cell of $H_{8B}$ and the corner state basis. The orange and green curves with arrows illustrate the hoppings of \cref{TB-H-real}. Inside one square, $t_+=-(1+i)\pi v/4a$ denote the square edge hopping and $t'=-\pi v/4a$ denotes the square diagonal hopping. The direction of $t_+$ is indicated by the orange arrows surrounding the square. $\tilde{t}$ denotes the nearest neighbor (NN) hopping between squares. The red \& blue circles and decorated numbers $w^{\pm}$ ($w=1,2,3,4$) indicate the positions of corner states $c_{n,m}(w,\pm)$. 4c and 2b indicate the (representative of) corresponding Wyckoff positions.}
	\label{TB-cell}
\end{figure}

In previous sections, we have seen that the low energy physics of the network model can be reconstructed using four standing waves per square. In order to obtain a more local tight-binding model, we can convert the standing waves into four ``corner states" $ c_{n,m}(w,\alpha)$ $ (w=1,2,3,4)$, which are wave-packets centered at four corners of a square (see Fig.~\ref{TB-cell}).
The relation between the ``corner states" and the standing waves is the Fourier transform on each square:
\begin{equation}
    \begin{pmatrix}
    c^{\dagger}_{n,m}(1,\alpha)\\
    c^{\dagger}_{n,m}(2,\alpha)\\
    c^{\dagger}_{n,m}(3,\alpha)\\
    c^{\dagger}_{n,m}(4,\alpha)
    \end{pmatrix}    
    =\frac{1}{2}
    \begin{pmatrix}
    1 & 1 & 1 & 1 \\
    1 & -\alpha i & -1 & \alpha i \\
    1 & -1 & 1 & -1 \\
    1 & \alpha i & -1 & -\alpha i \\
    \end{pmatrix}
    \begin{pmatrix}
    \phi^{\dagger}_{n,m}(0,\alpha) \\
    \phi^{\dagger}_{n,m}(1,\alpha) \\
    \phi^{\dagger}_{n,m}(2,\alpha) \\
    \phi^{\dagger}_{n,m}(3,\alpha) \\
    \end{pmatrix}
    =U_{\alpha}
    \begin{pmatrix}
    \phi^{\dagger}_{n,m}(0,\alpha) \\
    \phi^{\dagger}_{n,m}(1,\alpha) \\
    \phi^{\dagger}_{n,m}(2,\alpha) \\
    \phi^{\dagger}_{n,m}(3,\alpha) \\
    \end{pmatrix}\ ,
    \label{standing-wave-to-corner-dot}
\end{equation}
where the transform coefficient between $ c_{n,m}^\dagger(w,\alpha)$ and $\phi^{\dagger}_{n,m}(K,\alpha)$ is $(U_{\alpha})_{w,K+1}=e^{-i \frac{\pi}2 \alpha K (w-1) }$. 
More explicitly, in the real space representation
\begin{equation}
\label{standing-wave-to-coner-dot-realspace}
    \langle0\vert \psi_{n,m}(x_{\alpha}) c^{\dagger}_{n,m}(w,\alpha)\vert 0\rangle= \frac{1}{2}\sum_{K=0}^{3}e^{i\frac{\pi}{2}\alpha K[x_{\alpha}-(w-1)a]}.
\end{equation}

The right side does not contain $n,m$ because $x_{\alpha}$ particularly refers to the coordinate around the $\alpha$ square in the $(n,m)$ cell as defined in \cref{lattice-a}.
From \cref{standing-wave-to-coner-dot-realspace}, we can see that $c_{n,m}(w,\alpha)$ represents a superposition of standing wave states with different wave vectors $K=[0,3]$ and the same phase origin $x_{\alpha,w}=(w-1)a$ in the $\alpha$ square of $(n,m)$ cell. (Further explanation of $x_{\alpha,w}$: it represents a specific point in the $\alpha$ square with coordinate value $(w-1)a$ in terms of coordinate $x_{\alpha}$.) Hence, $c_{n,m}(\alpha,w)$ should be a wave packet centered at $x_{\alpha,w} = (w-1)a$, \ie at a corner of the $\alpha$ square of $(n,m)$ cell. The four $c_{n,m}(\alpha,w)$ defined in \cref{standing-wave-to-corner-dot} respectively center at four $\alpha$ square corners according to \cref{TB-cell}, and we call them ``corner states". Since each corner state is a summation over only four waves, it has a considerable broadening that results in various hoppings shown in \cref{TB-H-real}.

Substituting Eq.~(\ref{standing-wave-to-corner-dot}) into \cref{eq:H0-standing-wave,eq:H1-standing-wave} we obtain a tight-binding Hamiltonian on the corner state basis. Note that the notation in \cref{TB-H-real} is different from that of Eq.~\textcolor{green}{(4)}. The site indices $p,q$ in Eq.~\textcolor{green}{(4)} become $(n,m,\alpha,w)$ here. Square edge and diagonal hopping, \ie $\langle\!\langle\cdot\rangle\!\rangle$ and $\langle\!\langle\!\langle\cdot\rangle\!\rangle\!\rangle$, are described by $B_{\alpha}$ in the following $H_{\mathrm{8B},0}$. And the nearest neighbor hopping $\langle\cdot\rangle$ corresponds to the following $H_{\mathrm{8B},1}$.
\begin{equation}
    \begin{aligned}
    H_{\mathrm{8B},0} &=\sum_{n m} \sum_{\alpha} \sum_{w_1 w_2} \thickspace (B_{\alpha})_{w_1,w_2}\thinspace c_{n,m}^{\dagger}\thinspace(w_1,\alpha) c_{n,m}(w_2,\alpha)\\
    H_{\mathrm{8B},1} &= \sum_{n m}\sum_{\alpha}\thickspace \tilde{t} \left[ c^{\dagger}_{n,m}(1,\alpha)c_{n,m}(1,\bar{\alpha})+c^{\dagger}_{n,m}(2,\alpha)c_{n,(m+\alpha)}(2,\bar{\alpha})+\right. \\ & \left. \qquad \qquad \qquad c^{\dagger}_{n,m}(3,\alpha)c_{(n-\alpha),(m+\alpha)}(3,\bar{\alpha})+c^{\dagger}_{n,m}(4,\alpha)c_{(n-\alpha),m}(4,\bar{\alpha})\right]
    \end{aligned}
    \label{TB-H-real}
\end{equation}
where 
\begin{equation}
    B_{\alpha} =U_{\alpha}^{\dagger}\mathbb{K} U_{\alpha}=
    \frac{\pi v}{4a}\begin{pmatrix}
    0 & \alpha i-1 & -1 & -\alpha i-1 \\
    -\alpha i-1 & 0 & \alpha i -1 & -1 \\
    -1 & -\alpha i-1 & 0 & \alpha i-1 \\
    \alpha i-1 & -1 & -\alpha i-1 & 0
    \end{pmatrix}\qquad \quad
    \mathbb{K}=
    \begin{pmatrix}
    0 & 0 & 0 & 0\\
    0 & \frac{\pi v }{2a}  & 0 &0 \\
    0 & 0  & \frac{\pi v}{a} &0 \\
    0 & 0 & 0 & \frac{3\pi v}{2a} \\ 
    \end{pmatrix} - \frac{3\pi v}{4a}\mathbb{I}_4\ .
\end{equation}
Here $\mathbb{K}$ represents the on-site energies of the standing wave basis ($K=0,1,2,3$) in \cref{eq:H0-standing-wave}. 
We have shifted these on-site energies by a constant $-\frac{3\pi v}{4a}$ such that the energy bands are centered at the zero energy. 
Eq.~(\ref{TB-H-real}) can be visualized by Fig.~\ref{TB-cell}. Inside one square, there are both the square diagonal and square edge hoppings: $t'=-\frac{\pi v} {4a} $ and $t_{+}=-(1+i)\frac{\pi v}{4a}$ (along the directions indicated by the orange arrows around the square). 
For different squares, there are real hoppings $\tilde{t}$ between adjacent corners (green arrows in Fig.~\ref{TB-cell}). 

In the corner state basis, the nature of the phase transition becomes more conspicuous. 
When $\tilde{t}$ approaches $ 0$, electrons mainly go around the squares, \ie around $C_{4z}$ centers. 
As $\tilde{t} \rightarrow +\infty$, electrons are trapped in corners \ie $C_{2z}T$ centers.

The actions of $C_{2z}T$, $C_{4z}$,  $M_{xy} $ on the corner state basis are
\begin{equation}
    \begin{aligned}
    (C_{2z}T) c_{n,m}(w,\alpha) (C_{2z}T)^{-1}&= c_{-n-1,-m-1}(w,\bar{\alpha}) \\
    C_{4z} c_{n,m}(w,\alpha) C_{4z}^{-1}&= c_{-m-1,n+1 - (1+\alpha)/2}(w+1\ \mathrm{mod}\ 4,\alpha)\\ 
    M_{xy} c_{n,m}(w,\alpha) M_{xy}^{-1}&= c_{m,n}(4-w\ \mathrm{mod}\ 4,\bar{\alpha})\\ 
    \end{aligned}\ .
\end{equation}
Basically, they are the same as the actions on the circular chiral basis (\cref{symmetry-action-lattice0}), but now the continuous coordinate $x_\alpha$ is discretized to $w$. 
One can check that the Hamiltonian in \cref{TB-H-real} respects the crystalline symmetries.

\subsection{Evolution of the Dirac points} \label{subsec:dirac-evolution}

\begin{figure}[h]
    \centering
    \includegraphics[width=0.7\linewidth]{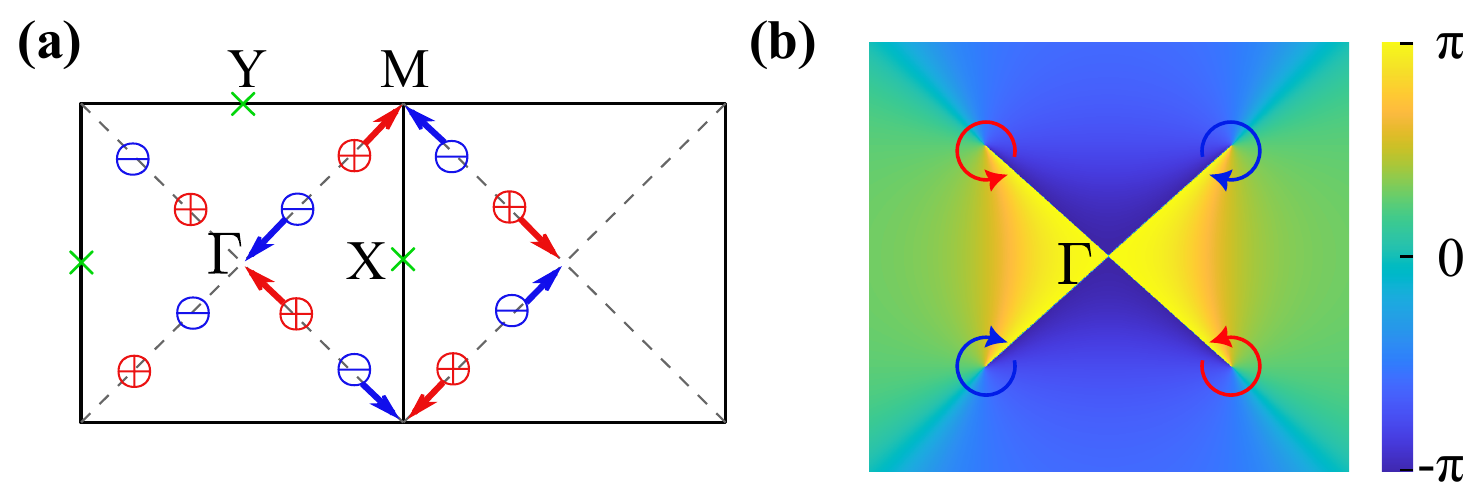}
    \caption{Evolution of Dirac points. (a) The trajectory of the Dirac points, where red and blue circles represent Dirac points with positive and negative chiralities, respectively. (b) An illustration of the phase $\varphi_\kk$ in the k$\cdot$p expansion around $\Gamma$ point. Anti-clockwise (clockwise) winding of $\varphi_\kk$ around a Dirac point indicates a positive (negative) chirality.}
    \label{fig:DiracEvolution}
\end{figure}

As shown in \cref{bands-lattice}, as we tune $\tilde{t}$ from $0$ to $\frac{\pi v}{a}$, a pair of Dirac points is created on the $\Gamma M$ pave. 
Due to the $C_{4z}$ symmetry, there should be four pairs of Dirac points created along equivalent paths. 
As $\lambda$ continues to increase, four Dirac points move to $\Gamma$ and then annihilate each other, and the other four move to $M$ and then annihilate each other (shown in \cref{fig:DiracEvolution}\textcolor{red}{(a)}). 

In a generic multi-band system with $C_{2z}T$ symmetry, the only topological charge of a Dirac point is its $\pi$ Berry's phase, which is $Z_2$-valued.  
However, if the considered two bands are disconnected from other bands, the Dirac points instead carry $Z$-valued topological charges, \ie chiralities, characterized by the non-Abelian Berry's connection  \cite{Wu2019Science,ahn_failure_2019,bouhon2020non}. 
Even if the considered two bands are connected to other bands, locally two Dirac points can only annihilate each other if they have opposite chiralities. 
Thus, given that we are only interested in a small region of the Brillouin zone where the two bands are well separated from other bands, we can still talk about this $Z$-valued topological charge \cite{Wu2019Science,ahn_failure_2019,bouhon2020non}.
That means if in the evolution (i) the Dirac points do not move outside this region and (ii) the two bands in this region are always well separated from other bands, then two Dirac points will annihilate each other only if they carry opposite chiralities.

We first look at the four Dirac points annihilated at $\Gamma$. 
When they are close to $\Gamma$, they can be described by a k$\cdot$p model around $\Gamma$.
We will derive the chiralities of the Dirac points by studying such a k$\cdot$p model.  
The two involved energy levels at $\Gamma$ are $\Gamma_1$ and $\Gamma_2$.
According to the character table in \cref{Irreps}, we can write the symmetry operators as 
\begin{equation}
    C_{4z} = \sigma_z,\qquad M_{xy} = \sigma_z,\qquad C_{2z}T = K,\ 
\end{equation}
where $K$ is the complex conjugation. 
$C_{2z}T$ restricts the Hamiltonian to be a real matrix. 
Thus, we parameterize the Hamiltonian as 
\begin{equation}
    H(\kk) = \epsilon_0(\kk) \sigma_0 + Z(\kk) \sigma_z + X(\kk) \sigma_x
\end{equation}
with $\epsilon_0(\kk), Z(\kk), X(\kk)$ being real valued parameters to be determined. 
The $C_{4z}$ symmetry implies
\begin{equation}
\epsilon_0(-k_y,k_x) = \epsilon_0(k_x,k_y),\qquad 
Z(-k_y,k_x) = Z(k_x,k_y),\qquad 
X(-k_y,k_x) = -X(k_x,k_y)\ . 
\end{equation}
The mirror symmetry $M_{xy}$ implies 
\begin{equation}
\epsilon_0(k_y,k_x) = \epsilon_0(k_x,k_y),\qquad 
Z(k_y,k_x) = Z(k_x,k_y),\qquad 
X(k_y,k_x) = -X(k_x,k_y)\ . 
\end{equation}
To second order of $\kk$, $\epsilon_0, Z, X$ must hence have the forms
\begin{equation}
\epsilon_0 (\kk) = E_0 + C \kk^2,\qquad 
Z(\kk) = - \Delta + A\kk^2,\qquad 
X(\kk) = B (k_x^2 - k_y^2)\ ,
\end{equation}
respectively. 
Here $A, B, C, E_0, \Delta$ are all real parameters. $Z(\kk)$ is the term that creates band inversion at the $\Gamma$ point
The band energies are given by 
\begin{equation}
    E_\pm(\kk) = E_0+C\kk^2 \pm \sqrt{ (\Delta-A\kk^2)^2 + B^2 (k_x^2-k_y^2)^2  }\ .
\end{equation}
If $A\cdot\Delta >0$, then there are four Dirac points locating at $k_y=\pm k_x$, $k_x = \pm \sqrt{\Delta/A}$. 
When $\Delta \to 0$, the four Dirac points move to $\Gamma$ to annihilate each other. 
Some quantitative constraints on $A,B,C,\Delta$ can be inferred from \cref{bands-lattice}.
First, the Dirac points exist when $\Gamma_2$ has a higher energy than $\Gamma_1$, \ie $\Delta>0$. 
Then there must be $A>0$ due to the existence condition $A\cdot \Delta>0$ of Dirac points.  
Second, since the sign of the off-diagonal coefficient $B$ can be changed by a gauge transformation $e^{i\frac{\pi}2\sigma_z}$, the sign of $B$ is a gauge choice and we choose $B>0$.
Third, when $\Delta=0$ (\cref{band-lattice-e}), both the two bands increase in energy with increasing $\kk$ in the $\Gamma M$ direction, implying $C>A>0$.
Fourth, when $\Delta=0$, one band increases in energy while the other decreases in energy along the $\Gamma X$ direction, implying $C<\sqrt{A^2+B^2}$. 

The eigenstates corresponding to $E_{\pm}(\kk)$ are 
\begin{equation}
    \ket{u_{+}} = e^{i\varphi(\kk)/2} \begin{pmatrix}
    \cos \frac12 \varphi(\kk) \\ \sin \frac12 \varphi(\kk) 
    \end{pmatrix},\qquad 
    \ket{u_{-}} = e^{i\varphi(\kk)/2} \begin{pmatrix}
    \sin \frac12 \varphi(\kk) \\ -\cos \frac12 \varphi(\kk) 
    \end{pmatrix}
\end{equation}
with $\varphi_\kk$ given by 
\begin{equation}
    \varphi_\kk = \arccos \frac{ -\Delta + A\kk^2 }{ \sqrt{  (\Delta-A\kk^2)^2 + B^2 (k_x^2-k_y^2)^2 }  } \ .
\end{equation}
We illustrate $\varphi_\kk$ in \cref{fig:DiracEvolution}\textcolor{red}{(b)}.
One can see that at the Dirac points, $\varphi_\kk$ is not well-defined, and there is a nontrivial winding of $\varphi_\kk$ around a Dirac point. 
The prefactors $e^{i\varphi(\kk)/2}$ in $\ket{u_{\pm}(\kk)}$ are needed for $\ket{u_{\pm}(\kk)}$ (away from Dirac points) to be single-valued because the entries of the two-by-one wavefunction vectors transform as $\sin \frac{\varphi}2 \to -\sin \frac{\varphi}2$, $\cos \frac{\varphi}2 \to -\cos \frac{\varphi}2$ upon $\varphi\rightarrow\varphi+2\pi$. 
The sewing matrix of $C_{2z}T$ is given by  
\begin{equation}
    C_{2z}T \ket{ u_n(\kk) } = \ket{ u_n(\kk) } e^{i\theta_n(\kk)}, \qquad \theta_{n}(\kk) = - \varphi(\kk),\qquad (n=\pm )\ .
\end{equation}
The non-Abelian Berry's connection can be calculated as
\begin{equation} \label{eq:nonAbelA}
    \vec{\mathcal{A}}_{m,n}(\kk) = i \bra{ u_m(\kk) } \partial_\kk \ket{ u_n(\kk) }
    = \begin{pmatrix}
    -\frac12 \partial_\kk \varphi(\kk)  & i \frac12 \partial_\kk \varphi(\kk) \\
    -i \frac12 \partial_\kk \varphi(\kk) &  -\frac12 \partial_\kk \varphi(\kk)  
    \end{pmatrix}_{m,n}\ .
\end{equation}
We compare it to the generic form of non-Abelian Berry's connection with $C_{2z}T$ symmetry \cite{ahn_failure_2019}
\begin{equation}\label{eq:nonAbelA-generic}
    \vec{\mathcal{A}}(\kk) 
    = \begin{pmatrix}
    \frac12 \partial_\kk \theta_+(\kk)  & i \vec{a}(\kk) e^{ i\frac{\theta_+(\kk) - \theta_-(\kk)}2 }  \\
    -i \vec{a}(\kk) e^{ - i\frac{\theta_+(\kk) - \theta_-(\kk)}2 } &  \frac12 \partial_\kk \theta_-(\kk)  
    \end{pmatrix}\ ,
\end{equation}
where $\vec{a}$ is the real-valued off-diagonal Berry's connection and $\theta_{\pm}(\kk)$ are the phase factors appearing in the $C_{2z}T$ sewing matrix. 
$\vec{a}$ is gauge invariant up to a global ambiguity of $\pm$ sign. 
After the global sign is fixed (as will be done in \cref{eq:a-gauge-fixing}), the chirality of the $i$th Dirac point can be defined through $\vec{a}(\kk)$ as 
\begin{equation}
    \chi_{i} = \frac1{\pi} \oint_{\partial D_i} d\kk \cdot \vec{a}(\kk) \ ,
\end{equation}
where $D_i$ is an infinitely small disk containing the $i$th Dirac point and $\partial D_i$ is its boundary. 
Comparing \cref{eq:nonAbelA} to \cref{eq:nonAbelA-generic}, we choose $\theta_+(\kk) = \theta_-(\kk) = \varphi(\kk)$ and  
\begin{equation} \label{eq:a-gauge-fixing}
    \vec{a}(\kk) = \frac12 \partial_\kk \varphi_\kk \ . 
\end{equation}
For this gauge choice, the chirality of a Dirac point is simply the winding number of $\varphi_\kk$
\begin{equation}
   \chi_{i} = \frac1{2\pi} \oint_{\partial D_i} d\kk \cdot \partial_\kk \varphi_\kk\ .
\end{equation}
(Another gauge choice can be obtained by replacing $\ket{u_{-}(\kk)} \to -\ket{u_{-}(\kk)} $, then there would be $\vec{a}(\kk) = - \frac12 \varphi(\kk)$ and $\chi_i$ would be reversed.) 
As shown in \cref{fig:DiracEvolution}, the Dirac points at $(k_0,k_0)$, $(-k_0,k_0)$, $(-k_0,-k_0)$, $(k_0,-k_0)$ ($k_0=\sqrt{\Delta/A})$ have the chiralities $-1$, $+1$, $-1$, $+1$, respectively. 
If the considered two bands were disconnected from other bands, then their Euler class was given by $\frac12 \sum_i \chi_i$ \cite{ahn_failure_2019}.
Even if the two bands are connected to other bands, $\chi_i$ is still a {\it locally} well-defined $Z$-valued quantity in the sense that two Dirac points can {\it locally} annihilate each other only if they carry opposite $\chi$'s. 

One can see that two Dirac points related by the $C_{4z}$ operation have opposite chiralities.
This is true even in the absence of the mirror symmetry $M_{xy}$, which allows $X(\kk)$ to have the form $B (k_x^2-k_y^2) + B' 2k_xk_y$ to the second order of $\kk$.  
By a proper re-definition of the coordinate ($\kk'$), this term can be rewritten as $B'' (k_x^{\prime 2}-k_y^{\prime 2} )$ for some $B''$.
Then the analysis in the above paragraph applies. 
Therefore, we conclude that as long as the two bases of the k$\cdot$p expansion have $C_{4z}$ eigenvalues $1$ and $-1$, a pair of $C_{4z}$-related Dirac points should have opposite chiralities.

The k$\cdot$p theory at the $M$ point is equivalent to the one at $\Gamma$. 
The two involved levels at $M$ are $M_1$ and $M_2$.
According to \cref{Irreps}, we can write the symmetry operators as 
\begin{equation}
    C_{4z} = i\sigma_z,\qquad  M_{xy} = -\sigma_z, \qquad C_{2z}T = K\ .
\end{equation}
It is worth mentioning that, at the $M$ point, $C_{2z}T$ anti-commutes with $C_{4z}$ due to the translation part of $C_{2z}T$.
Nevertheless, $C_{4z}$, $M_{xy}$, and $C_{2z}T$ impose the same constraints on the k$\cdot$p Hamiltonian as those at $\Gamma$.
Therefore, the analyses in the two paragraphs above also apply to $M$, and the $C_{4z}$-related Dirac points must have opposite chiralities. 
Suppose the four Dirac points locate at $(\pi-p_0, \pi - p_0)$, $(\pi+p_0, \pi - p_0)$, $(\pi + p_0, \pi + p_0)$, $(\pi-p_0, \pi + p_0)$, respectively, for some small positive $p_0$.
The first Dirac point ($(\pi-p_0, \pi - p_0)$) must have $\chi=1$ because it is emerged together with the Dirac point at $(k_0,k_0)$, which has $\chi=-1$ according to the last two paragraphs.
Since $C_{4z}$-related Dirac points have opposite chiralities, the four Dirac points should have the chiralities $1$, $-1$, $1$, $-1$, respectively. 

One can see that, upon the phase transition form the initial gapped state to the final gapped state, the trajectories of Dirac points form a closed path $\mathcal{C}$ separating $X$ from $Y$ (\cref{fig:DiracEvolution}).
As shown in \cref{bands-lattice}, $X$ has a Dirac point between the third and fourth bands; $Y$ must also have a Dirac point according to $C_{4z}$. 
Drawing a path connecting $X$ and $Y$, it must cross $\mathcal{C}$ odd times.
According to Refs.~\cite{Wu2019Science,ahn_failure_2019,bouhon2020non}, the relative chirality between the Dirac points at $X$ and $Y$ changes after such a phase transition, as expected because the transition changes $w_2$.

\subsection{Simplified eight-band lattice model \texorpdfstring{$H_{8B}'$}{H8B'}}
\label{subsec:simplified-model}

The motivation of a further simplification on $H_{8B}$ is to obtain a more convincing critical metal phase. We leave the detailed discussion of localization for \cref{sec:analytical-localization-length} \& \ref{sec:numerical-localization-length} and the disorder potential for \cref{subsec:disorder}. Here we merely quote the result: both the network model and $H_{8B}$ have a critical metal phase. However, in the clean limit, $H_{8B}$ has a finite density of states at the Fermi level ($DOS_F$) $E_F=0$ in the critical region ($\tilde{t}\approx \pi v/2a$). Finite $DOS_F$ in the clean limit may lead to a large localization length that exceeds the numerical accessible transversal size. As explained in \cref{subsec:basic-localization-length}, such a large localization length will weaken the validity of our proof of the critical metal phase in $H_{8B}$. Although we offer some evidences (see \cref{subsec:basic-localization-length}) that the critical phase in $H_{8B}$ is reliable, a critical phase with $DOS_F=0$ in the clean limit will be more convincing. 

We first simplify $H_{8B}$ by removing the square diagonal hopping. 
The motivation is that the square diagonal hopping is not relevant for the phase transition: (i) without $t'$, $\tilde{t}=0$ and $\tilde{t}=\infty$ still represent the molecular orbital limit with charge centers at $C_{4z}$-centers and the bonding state limit with charge centers at $C_{2z}T$-centers, respectively, (ii) omitting $t'$ does not change the symmetry class because time-reversal-symmetry (complex conjugation) is already broken by the complex square edge hopping $t_{+}$. 
Therefore, in the absence of $t'$, changing $\tilde{t}=0$ to $\tilde{t}=\infty$ still realizes a phase transition of $w_2$ in the symmetry class A, where time-reversal symmetry is broken.

\begin{figure}[h]
	\centering
	\subfigure[$\tilde{t}=0.4\pi v/a$]{\label{band-TBA1-a}
    \includegraphics[width=.28\linewidth]{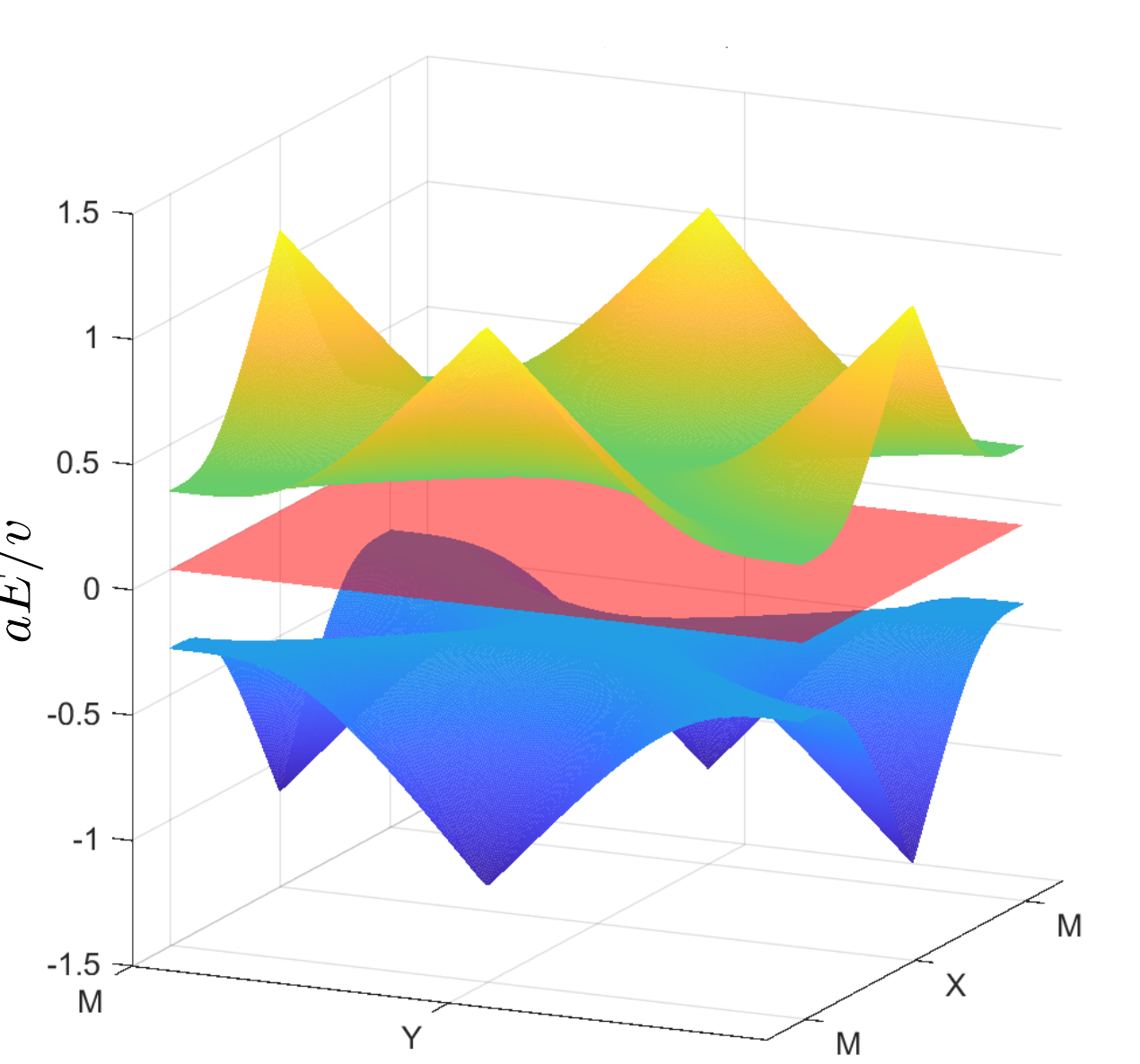}}
	\quad\subfigure[$\tilde{t}=0.5\pi v/a$]{\label{band-TBA1-b}
	\includegraphics[width=.28\linewidth]{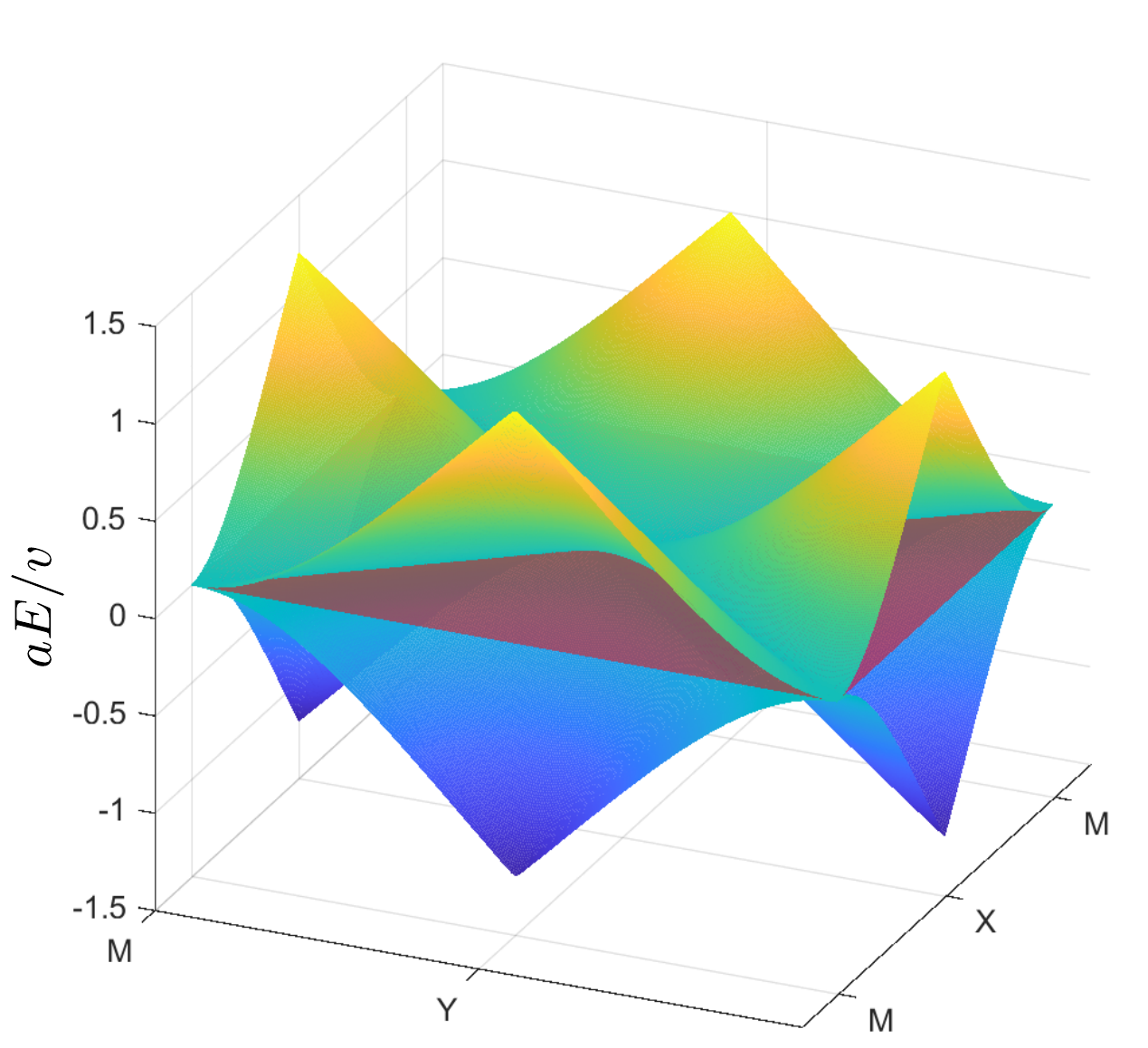}}
	\quad\subfigure[$\tilde{t}=0.6\pi v/a$]{\label{band-TBA1-c}
	\includegraphics[width=.24\linewidth]{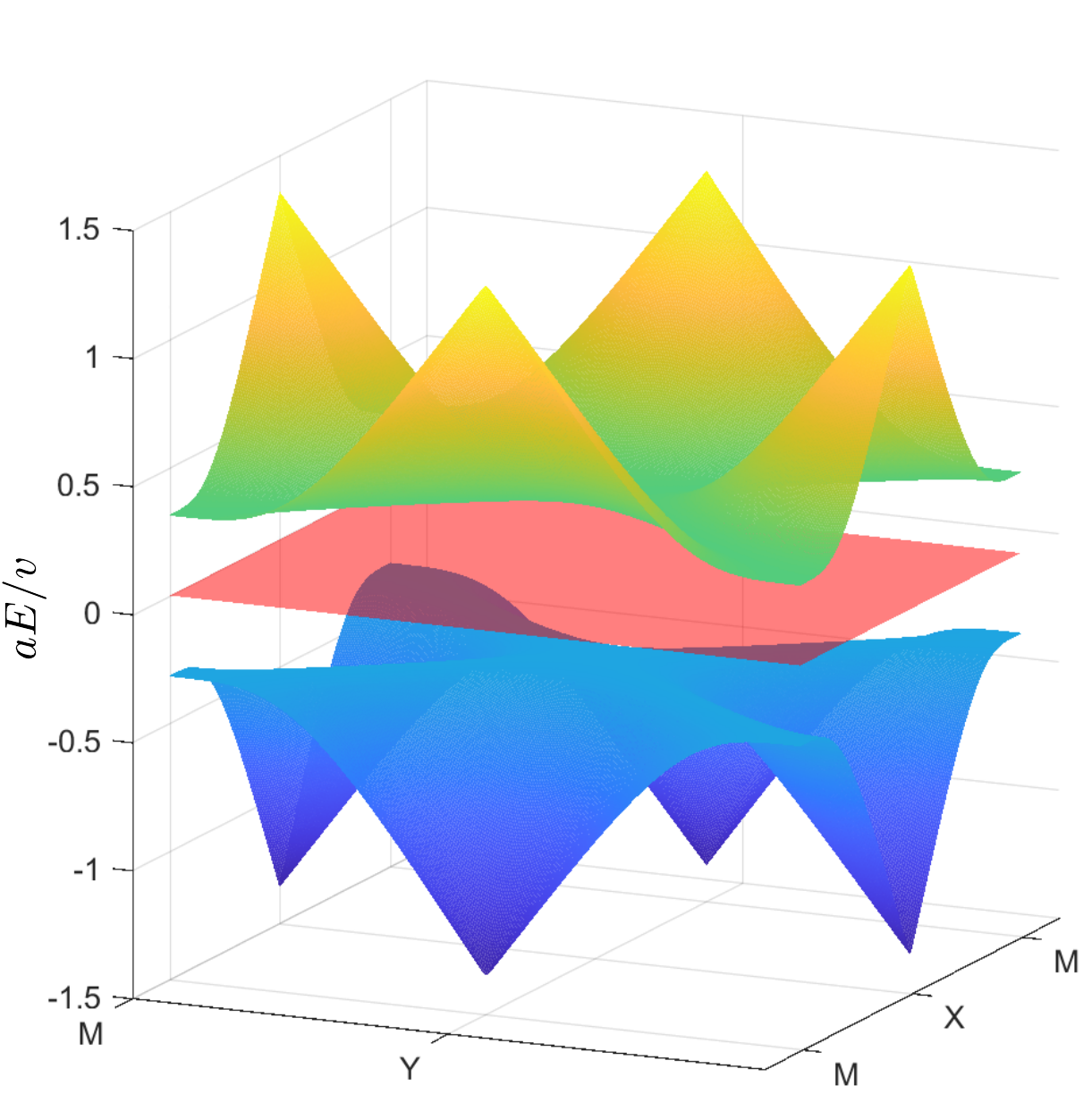}}
    \caption[]{3D plot of the band structure of $H_{8B}'$ with $A=1$ (merely remove square diagonal hopping from $H_{8B}$) near the Fermi level $E_F=0$ (the red plane). The two bands closest to the Fermi level touch at $\tilde{t}=\pi v/2a$ and are separated for other $\tilde{t}$. Such an evolution of band structure ($A=1$) can be viewed as a special case of the evolution in \cref{bands-simplified-TB} where $A\neq1$. }
	\label{bands-simplified-TB-A1}
\end{figure}

\begin{figure}[h]
	\centering
	\subfigure{
    \includegraphics[width=.35\linewidth]{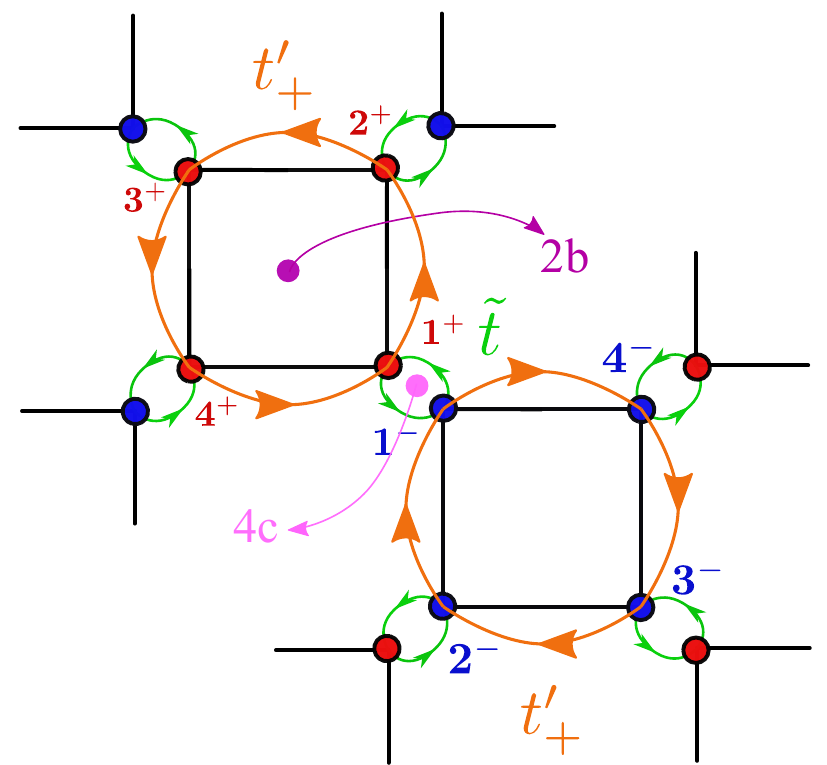}}
    \caption[]{Unit cell of $H_{8B}'$. The orange and green curves with arrows illustrate the hoppings of \cref{simplified-H-reciprocal}. Inside one square, $t_+'=-(1+A i)\pi v/4a$ denotes the modified square edge hopping. And the direction of $t_+'$ is indicated by the orange arrows. $\tilde{t}$ denotes the hopping between squares. The red \& blue circles and decorated numbers $w^{\pm}$ ($w=1,2,3,4$) indicate the positions of corner states $c_{n,m}(w,\pm)$. 4c and 2b indicate the (representative of) corresponding Wyckoff positions.}
	\label{simp-TB-cell}
\end{figure}

 Removing square diagonal hopping changes the band structure near the Fermi level but $DOS_F$ is still finite (see \cref{bands-simplified-TB-A1}). Thus, in addition to omitting square diagonal hopping, we will replace the square edge hopping $t_{+} = (1+ i)$ by  $t = (1+ A i)$ for some real parameter $A$.
(One should not confuse it with the vector potential $\mathcal{A}$.)
 We denote the eight-band lattice model with these modifications as $H_{8B}'$. We leave the numerical localization calculation of $H_{8B}'$ for \cref{sec:numerical-localization-length}. Here we merely mention that $H_{8B}'$ also has a critical phase. In the following paragraphs,  we will show that $H_{8B}'$ indeed has vanishing $DOS_F$ when $A\neq1$, and $A$ will determine the range of the critical phase. Hence, we obtain a convincing critical metal phase in $H_{8B}'$. 

\begin{figure}[p]
	\centering
	\subfigure[$\tilde{t}=0$]{\label{band-TB-a}
    \includegraphics[width=.35\linewidth]{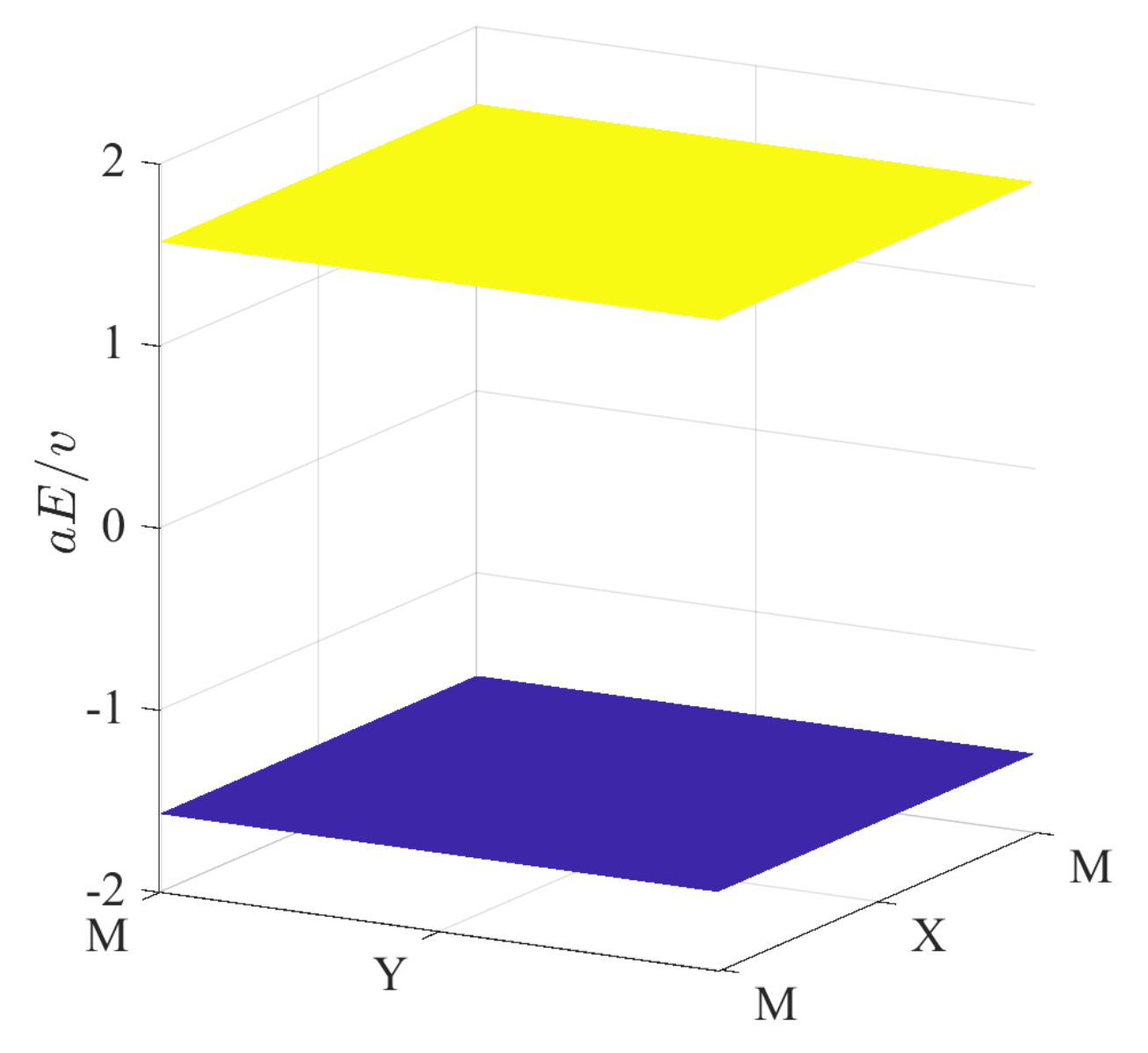}}
	\subfigure[$\tilde{t}=\pi v/4a$]{\label{band-TB-b}
	\includegraphics[width=.35\linewidth]{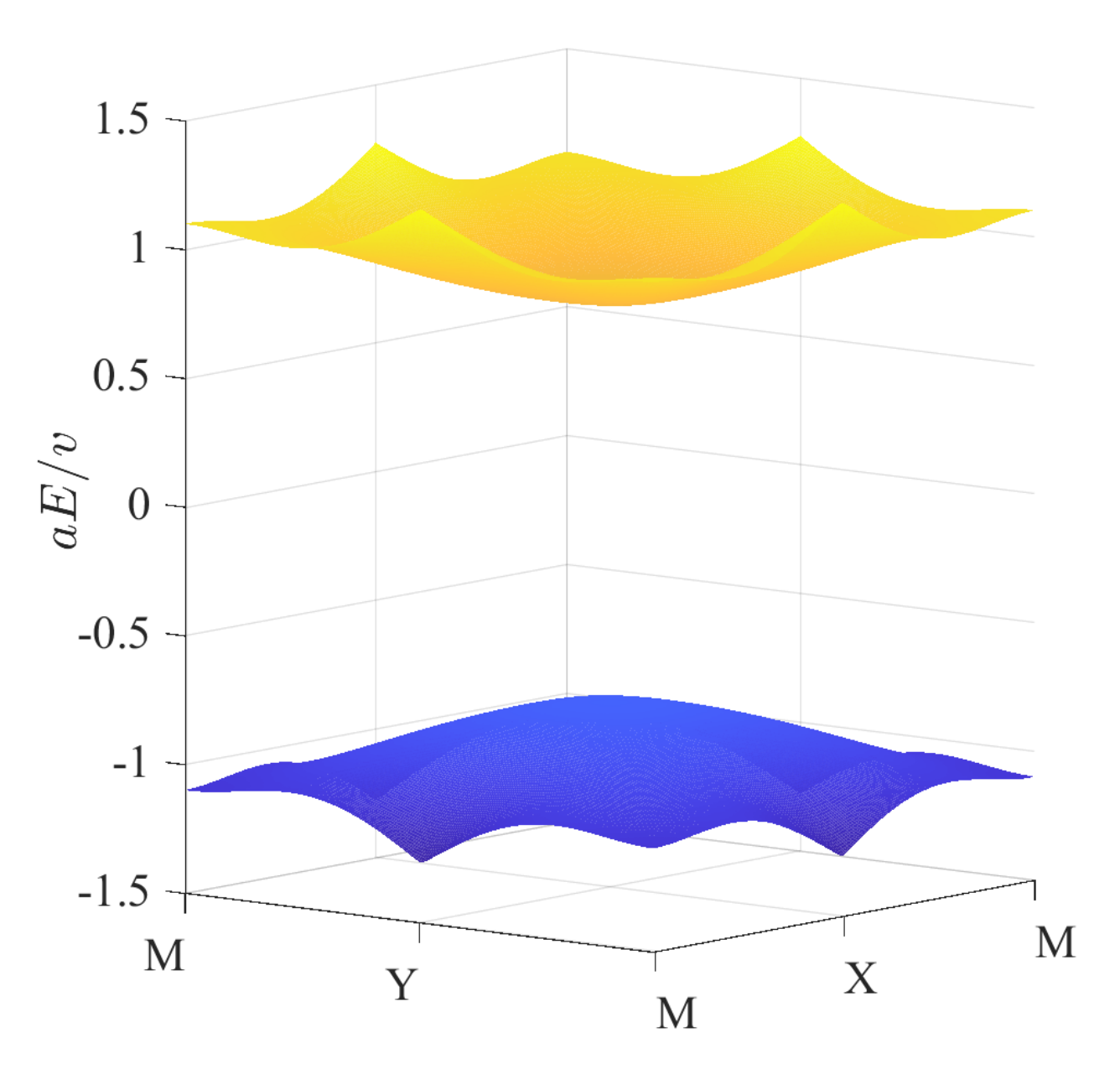}}\\
	\subfigure[$\tilde{t}=\pi v/2a$]{\label{band-TB-c}
	\includegraphics[width=.35\linewidth]{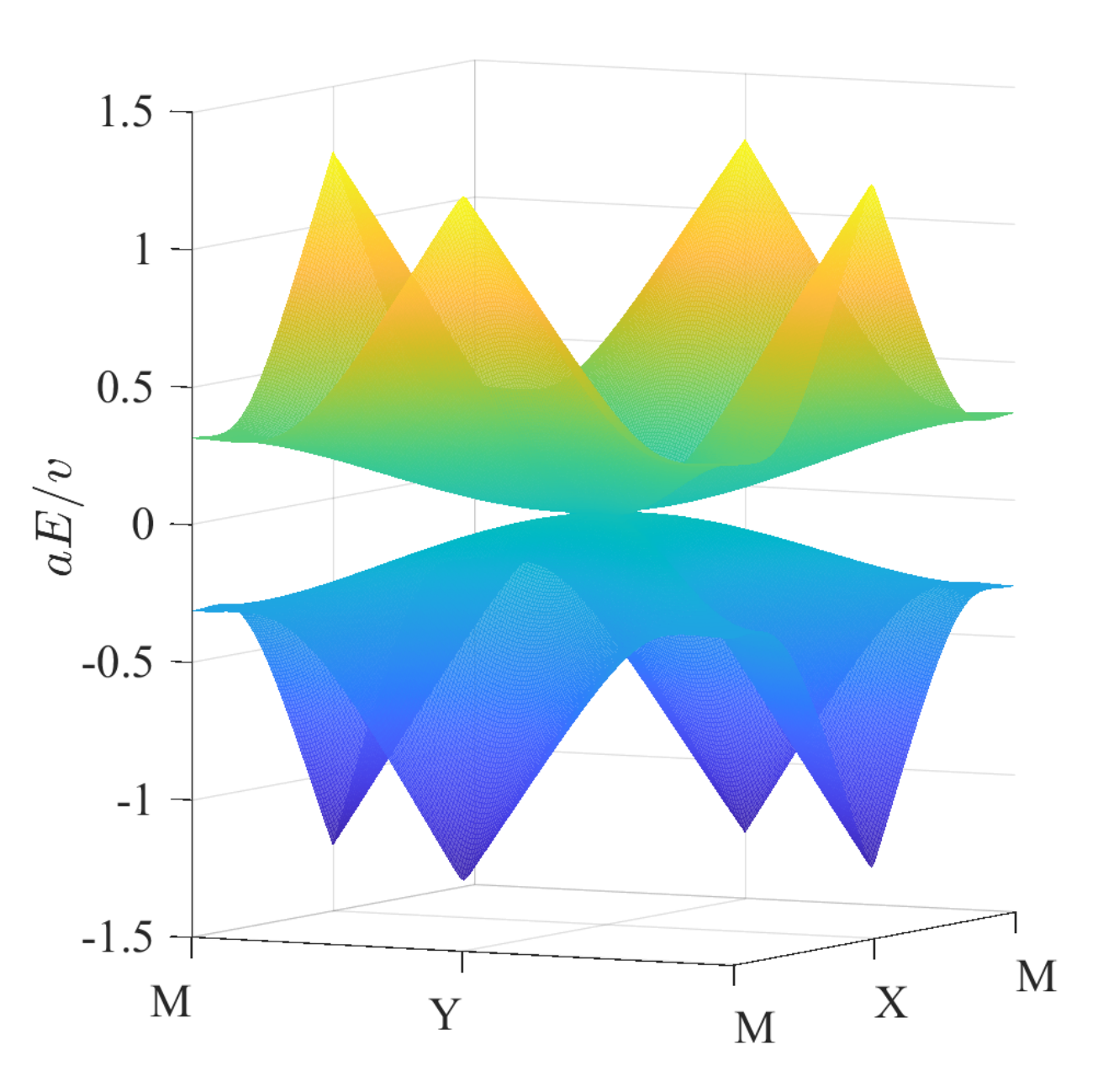}}
    \subfigure[$\tilde{t}=1.1\pi v/2a$]{\label{band-TB-d}
	\includegraphics[width=.35\linewidth]{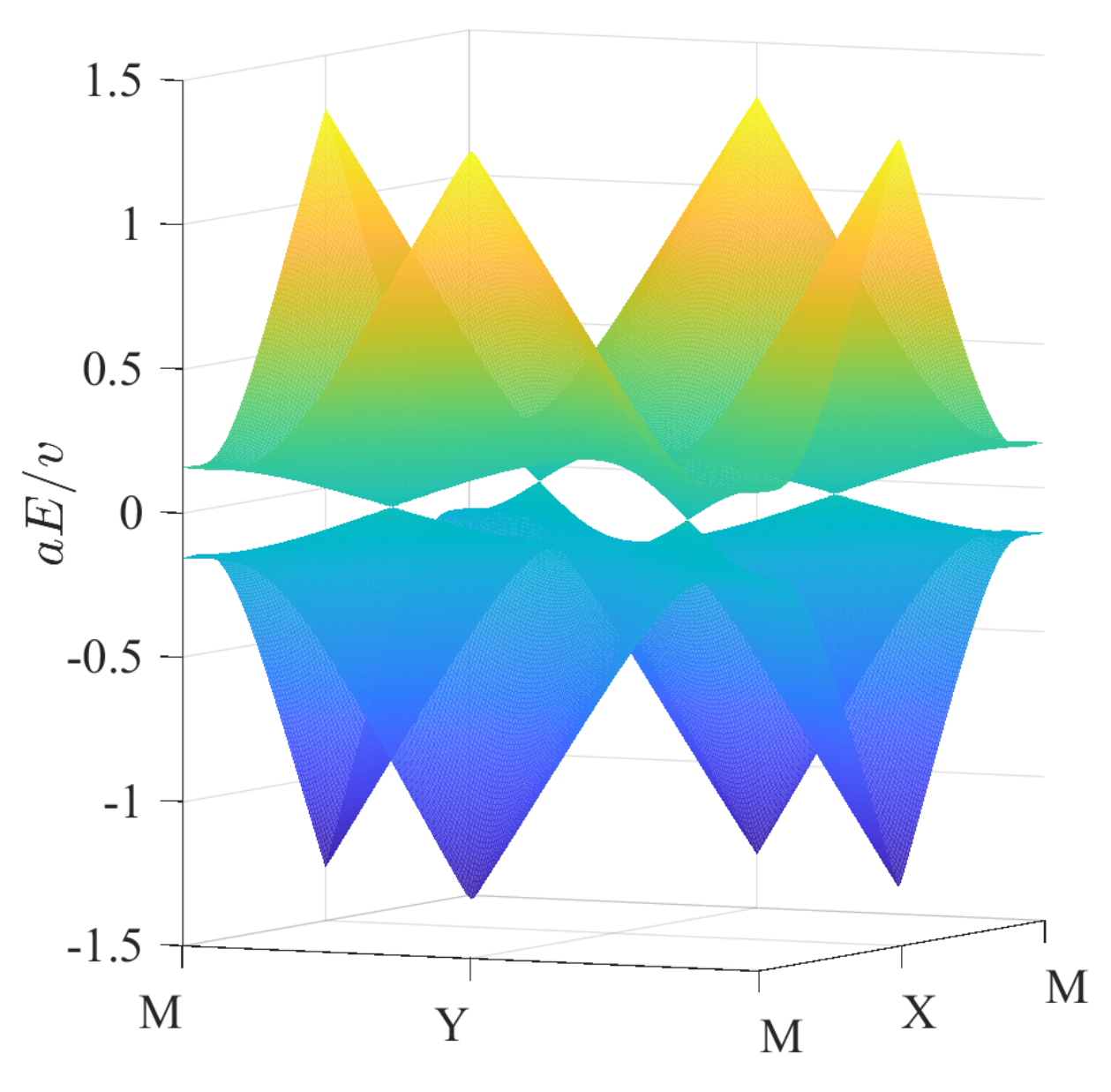}}\\
	\subfigure[$\tilde{t}=1.2\pi v/2a$]{\label{band-TB-e}
	\includegraphics[width=.35\linewidth]{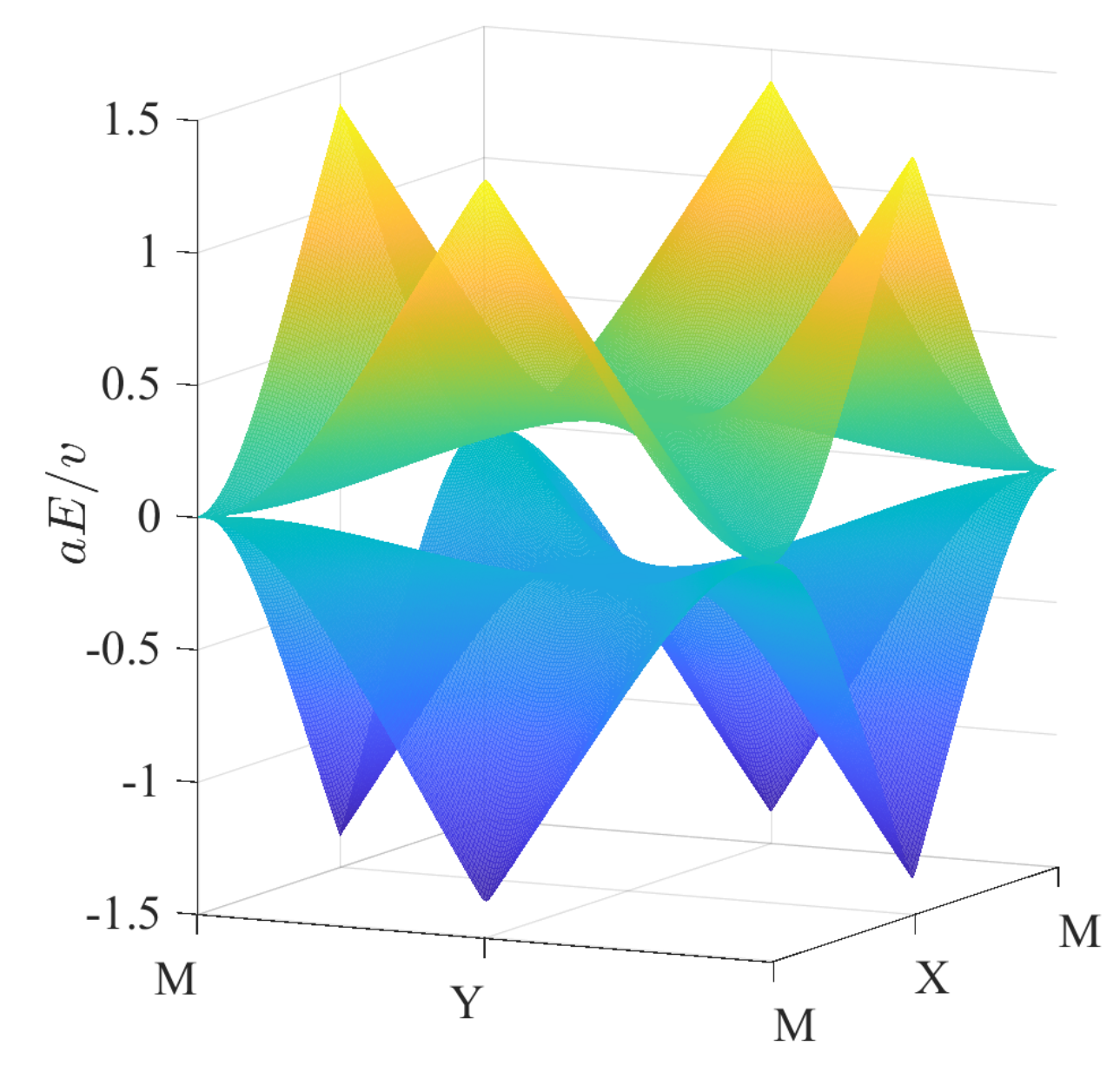}}
	\subfigure[$\tilde{t}=\pi v/a$]{\label{band-TB-f}
	\includegraphics[width=.35\linewidth]{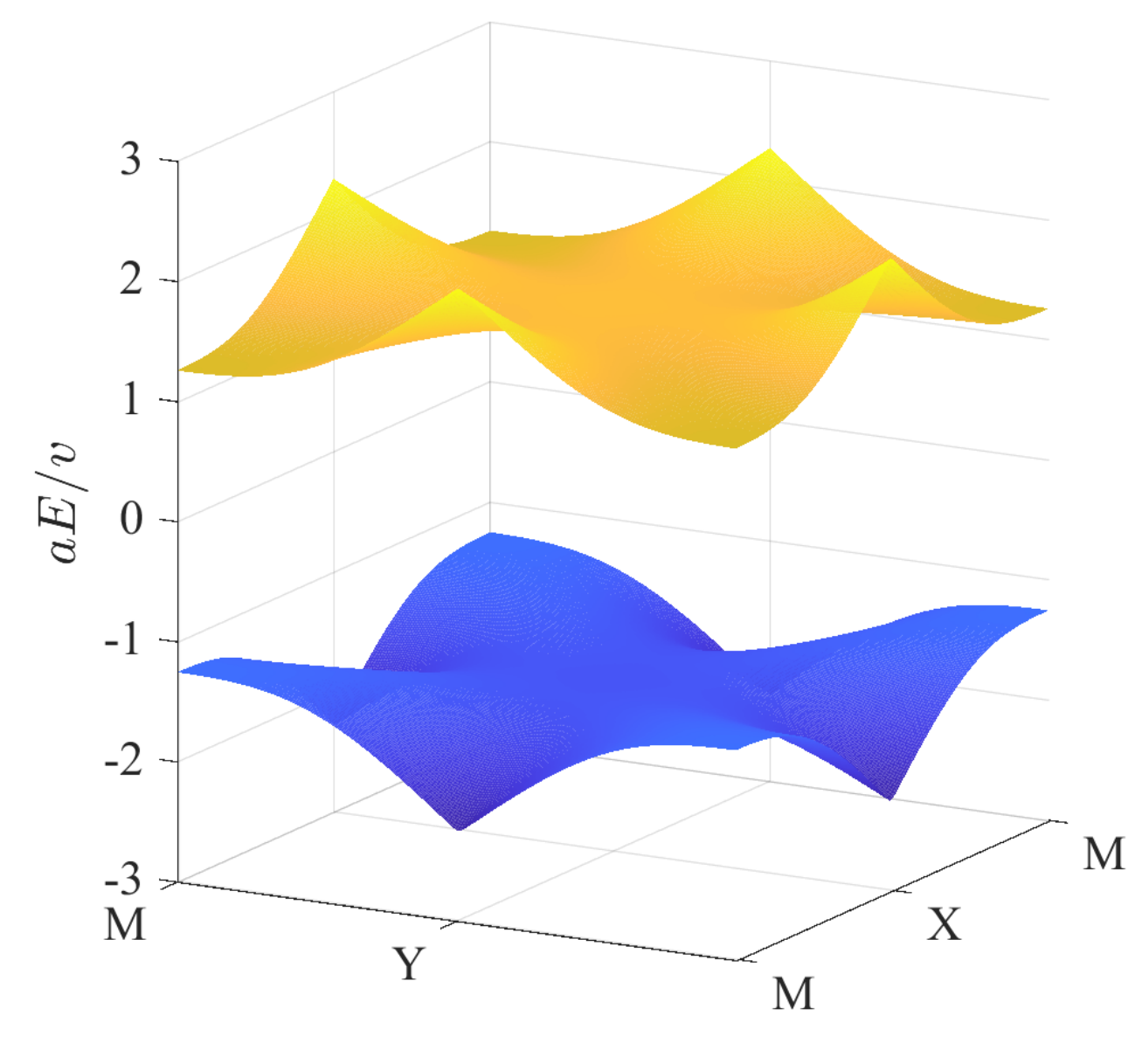}}
    \caption[]{3D Band structure of simplified lattice model near the Fermi level ($A=1.2$).
    }
	\label{bands-simplified-TB}
\end{figure}

\begin{figure}[h!]
	\centering
	\subfigure{
    \includegraphics[width=.6\linewidth]{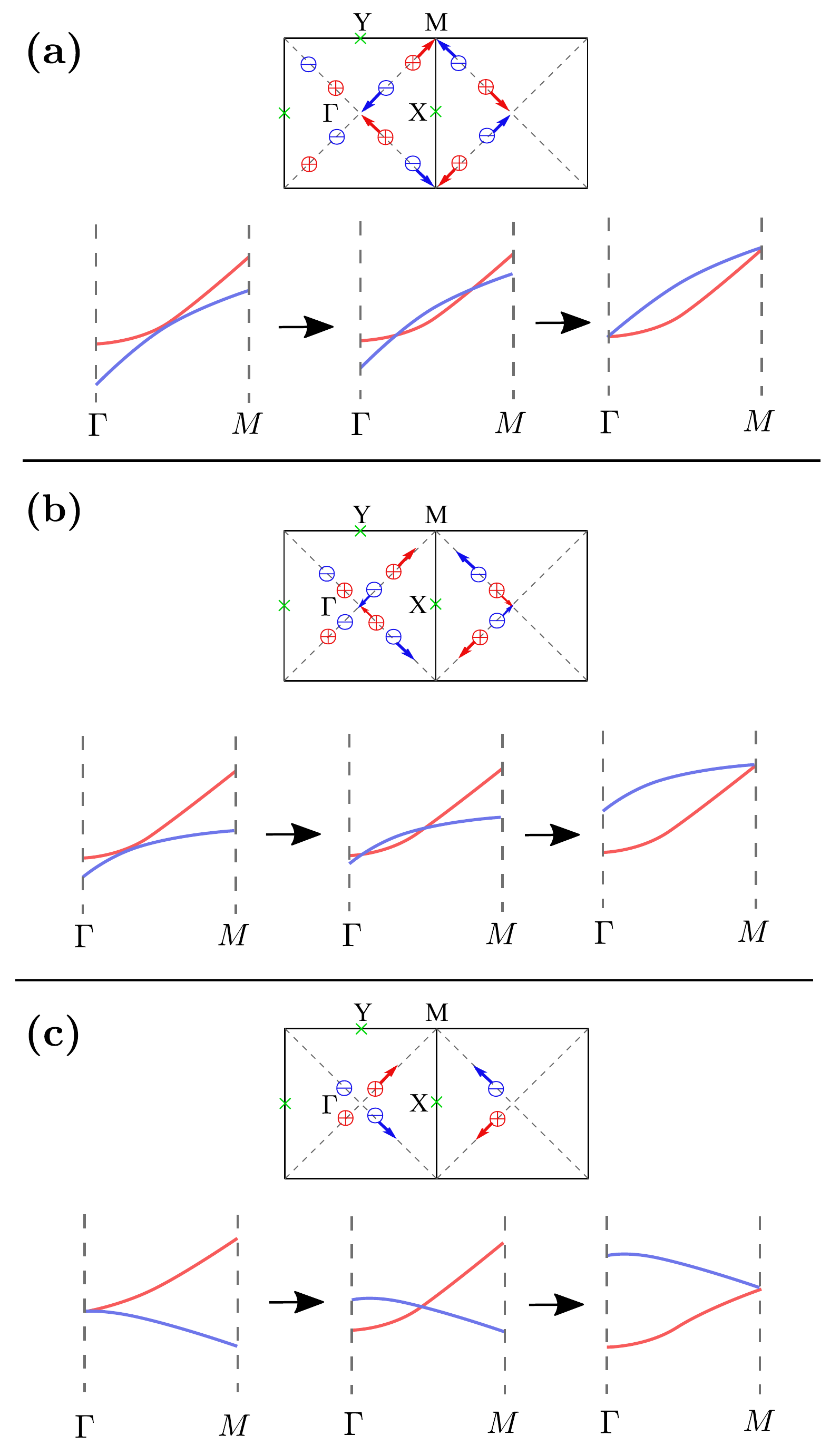}}
    \caption[]{Continuous deformation from the transition process of $H_{8B}$ to that of $H'_{8B}$. (a) The evolution of Dirac points of $H_{8B}$ and the corresponding band structure evolution according to \cref{band-lattice-d}$\sim$\ref{band-lattice-e}. Dirac points appear approximately at the middles of the $\Gamma$-$M$ lines. (b) An intermediate state between $H_{8B}$ and $H_{8B}'$. Dirac points appear at the positions that are closer to the $\Gamma$ point. (c) The evolution of Dirac points of $H'_{8B}$ and the corresponding band structure evolution according to \cref{bands-simplified-TB}. All these transition processes are equivalent to one positive Dirac point going around the $X$ point clockwise once.}
	\label{8-to-4-dirac}
\end{figure}

In order to obtain the band structure of $H_{8B}'$, we introduce the Fourier transform:
\begin{equation}
\begin{aligned}
    &c_{\vec{q}}\thinspace(w,\alpha)=\frac{1}{N} \sum_{nm} e^{-i\vec{q}\cdot(\vec{R}_{nm}+\alpha\thinspace \vec{t}_w+\vec{\Delta})} c_{n,m}(w,\alpha)\\
    &\vec{t}_1=(0,0) \qquad \vec{t}_2=(0,a)\qquad \vec{t}_3=(-a,a)\qquad \vec{t}_4=(-a,0)\qquad \vec{\Delta}=(a,a)
    \end{aligned}
\end{equation}
where $\alpha\vec{t}_{w}+\vec{\Delta}$ are the relative positions of the corner states in each unit cell.
We obtain the simplified tight-binding Hamiltonian $H_{8B}'$ in reciprocal space:
\begin{equation}
    \label{simplified-H-reciprocal}
    \begin{aligned}
    H_{8B,0}' &=\sum_{\vec{q}} \sum_{\alpha} \sum_{w_1 w_2} \thickspace (\widetilde{B}_{\alpha})_{w_1,w_2}\thinspace c_{\vec{q}}^{\dagger}\thinspace(w_1,\alpha) c_{\vec{q}}\thinspace(w_2,\alpha),\qquad 
    H_{8B,1}' &= \sum_{\vec{q}}\sum_{\alpha}\sum_{w}\thickspace \tilde{t} \thickspace c^{\dagger}_{\vec{q}}\thinspace(w,\alpha) c_{\vec{q}}\thinspace(w,\bar{\alpha})
    \end{aligned}
\end{equation}
\begin{equation}
 \label{simplified-H-reciprocal-coefficient}
    \widetilde{B}_{\alpha} =
    \frac{\pi v}{4a}
    \begin{pmatrix}
    0 & -(1-\alpha A i)e^{i\alpha q_y} & 0 & -(1+\alpha A i)e^{-i\alpha q_x} \\
    -(1+\alpha A i)e^{-i\alpha q_y} & 0 & -(1-\alpha A i)e^{-i\alpha q_x} & 0 \\
    0 & -(1+\alpha A i)e^{i\alpha q_x} & 0 & -(1-\alpha A i)e^{-i\alpha q_y} \\
    -(1-\alpha A i)e^{i\alpha q_x} & 0 & -(1+\alpha A i) e^{i\alpha q_y} & 0
    \end{pmatrix}
\end{equation}
Fig.~\ref{bands-simplified-TB} shows the two bands closest to the Fermi level with $A=1.2$ and various $\tilde{t}$. 
Comparing \ref{bands-simplified-TB} with \cref{bands3d-lattice}, we can see that removing the square diagonal hopping and introducing $A$ indeed vanish the density of state at the Fermi level while the picture of phase transition is similar. 
For $A>1$, bands touch only at four Dirac points when $\pi v/2a \tilde{t}< A \pi v/2a$ and are separated for other $\tilde{t}$. These Dirac points appear at $\Gamma$ point when $\tilde{t}=\pi v/2a$, move along the $\Gamma-M$ line as $\tilde{t}$ increases, and merge at $M$ point when $\tilde{t}=A \pi v/2a$. Thus the system is metallic for $\pi v/2a \tilde{t}< A \pi v/2a$ and insulated for $\tilde{t}< \pi v/2a$ \& $\tilde{t}> A\pi v/2a$.  The number of Dirac points here is four, which seems different from the transition process of $H_{8B}$ that has eight. However, as shown in \cref{8-to-4-dirac}, these two transition processes can continuously deform to each other by changing the dispersion of bands. During the deformation, four of the eight Dirac points become more and more close to the $\Gamma$ point; finally, they merge at the $\Gamma$ point, leaving the other four Dirac points moving toward the $M$ point. Hence, we can view $H_{8B}'$ as having eight Dirac points, four of which have zero life spans. Additionally, the Dirac point evolution of $H_{8B},\,H_{8B}'$ and their intermediate states are equivalent to one positive Dirac point going clockwise around the $X$ point one circle. Hence, we regard the transition processes of $H_{8B}$ and $H'_{8B}$ to be equivalent.

If $A<1$, the situation is reversed: Dirac points exist when $A\pi v/2a < \tilde{t} <\pi v/2a$ and appear (merge) at $M$ ($\Gamma$) point when $\tilde{t}=A\pi v/2a$ ($\tilde{t}=\pi v/2a$). It is straightforward to see that this process can also connect to $H_{8B}$ by deforming the band dispersion in an opposite way, thus also equivalent to $H_{8B}$.
If $A=1$, bands touch at two lines that form an``X-shape" when $\tilde{t}=\pi v/2a$ and are separated for other $\tilde{t}$ (see Fig.~\ref{bands-simplified-TB-A1}). We call the region where the bands touch at the Fermi level as the semi-metal region. Only in the semi-metal region, a critical metal can arise.
$\det \left[H_{8B}'(\vec{q})\right]=0$ shows that the reciprocal coordinates of Dirac points are given by 
\begin{equation}
    \vert q_{0x}\vert=\vert q_{0y}\vert=  \frac{1}{2a}\arccos \left[\frac{A^2-\tilde{\gamma}^4}{\tilde{\gamma}^2(A^2-1)}\right] \qquad\qquad \tilde{\gamma} =\frac{a\tilde{t}}{\pi v/2}
\end{equation}

We should also mention that the BR transition of $H_{8B}'$ is the same as that of $H_{8B}$. The four bands forming $\mathrm{A^{\prime \prime}_c\!\uparrow\!G}$ are also connected, even though $\mathrm{A^{\prime \prime}_c\!\uparrow\!G}$ is decomposable for general cases. Thus, the k$\cdot$p analysis of the Dirac points presented in \cref{subsec:dirac-evolution} also applies to the four Dirac points here because the gap closes at $\Gamma$ and $M$ involve the same irreps as in the $H_{8B}$ model. 
Hence, two $C_{4z}$-related Dirac points here also have opposite chiralities.

An additional chiral symmetry in the clean limit arises from the simplification: The Hamiltonian in momentum space anti-commutes with the diagonal matrix $C=\tau_z\otimes \xi_0 \otimes \sigma_z$, where $\tau_z$ is a Pauli matrix representing the two Chern blocks, $\xi_0 $ and $ \sigma_z$ are Pauli matrices representing the four corner state basis within each Chern block. More concretely, $\xi_0\otimes\sigma_z$ is a rank-$4$ diagonal matrix that multiplies $1$ to corners $w=1,3$ and $-1$ to $w=2,4$. 
Nevertheless, the disorder we considered (\cref{subsec:disorder}) will break it. We also perform numerical works in \cref{subsec:numerical-lattice-simplified} with $E_F\neq0$ where the chiral symmetry \emph{on average} is broken (see \cref{simp-no_ave_chiral}) and the critical phase still exist. 

To end this subsection, we summarize all the models described above. We start from the Manhattan network model (\cref{network-H-reciprocal-cutoff}), which can describe the percolation process on chiral edge states (Fig.~\textcolor{green}{1(a), (b)}). We notice that the eigenstates of the network model are circular chiral states when $\theta=\pm\pi/2$. Hence, we use the circular states of $\theta=-\pi/2$ as a basis and obtain a un-truncated lattice model (\cref{lattice-H-reciprocal}). The network and the un-truncated lattice model describe the same system from different bases, and we no longer distinguish them hereafter unless otherwise stated. Then we take a low truncation on the lattice model and obtain the eight-band lattice model $H_{8B}$. We claim that $H_{8B}$ with $E_F=0$ reproduces the low energy physics of the network model with $E_F=\pi v/4a$ and show evidence for this. $H_{8B}$ can be described by both (truncated) standing wave basis (\cref{standing wave}) and corner state basis (\cref{standing-wave-to-corner-dot}). According to \cref{sec:numerical-localization-length}, $H_{8B}$ with $E_F\approx0$ has a similar critical metal phase to the network model. However, when $H_{8B}$ is critical, the corresponding band structure in the clean limit has a finite $DOS_F$ (\cref{bands3d-lattice}) which makes the critical phase harder to distinguish from a very large localization length Anderson insulator  (see \cref{subsec:basic-localization-length} for an explanation). In order to obtain a more convincing critical phase, we modify $H_{8B}$ to $H_{8B}'$ which also has a critical metal phase when $E_F=0$(\cref{subsec:numerical-lattice-simplified}) while the corresponding $DOS_F=0$ (\cref{bands-simplified-TB}). As explained in \cref{subsec:disorder}, the criticality on the network model can be directly related to those in random flux models \cite{xie_kosterlitz-thouless-type_1998,cerovski_critical_2001,xiong_metallic_2001}. On the other hand, the criticality of $H_{8B}$ and $H_{8B}'$ is due to the $w_2$ transition between two OAIs. The analysis in the above subsections (\cref{subsec:standing-wave}, \cref{subsec:8-bands full model}, \ref{subsec:corner-state}, \ref{subsec:simplified-model}) offers a quantitative mapping between these criticalities. 
In this paper, we focus on specific Fermi levels. We choose $E_{F-8B}=E_{F-8B}'=0$ for $H_{8B}$ and $H_{8B}'$ because they sit in the middle of the eight bands. We choose $E_{F-N}=\pi v/4a$ for the network model because its band structure (in the metal limit) near $E_{F-N}$ is similar to that of $H_{8B}$ near $E_{F-8B}$ (\cref{bands3d-lattice}).  However, additional numerical works show that these choices of Fermi levels are not essential for the critical metal: first, as explained in \cref{subsec:transfer-network}, $E_{F-N}$ is not essential because the localization behavior of a network model is blind to the Fermi level; second, see \cref{full-EF_no_zero} and \cref{simp-no_ave_chiral}, $E_{F-8B}$ and $E_{F-8B}'$ are not essential because $H_{8B}$ and $H_{8B}'$ still have critical phases at $E_F\neq 0$ as long as the OAI limits are intact.%

\clearpage
\subsection{Disorder potential}
\label{subsec:disorder}

Previous sections do not involve the disorder explicitly. In order to study the localization, we should add disorder to our models (especially the transfer matrices in \cref{sec:analytical-localization-length}). As mentioned in \cref{subsec:Manhattan}, in order to depict the percolation process in Fig.~\textcolor{green}{1(a), (b)}, the disorder should be the random size of the Chern blocks. Since an electron will accumulate phase when propagating along a chiral wire, the random size of Chern blocks will manifest in the random phases along the edges of Chern blocks.  For the network model, these random phases can be viewed as random fluxes through the colored squares in \cref{network-a}. And the flux disorder can be directly added to the transfer matrix (see \cref{subsec:transfer-network}). In the un-truncated lattice model (\cref{lattice-H-real}), the Chern blocks are also represented by the red and blue squares (see \cref{lattice-a}), and the random fluxes can be realized by random vector potentials around these squares. Therefore, under the circular chiral basis, \ie $\psi_{n,m}(x_{\alpha},\alpha)$ in Eq.~\ref{co-standing wave}, $H_{N,0}$ (in \cref{lattice-H-real}) and the disorder potential can be written as 
\begin{equation}
    H_{0}+H_{rand}=\sum_{nm}\sum_{\alpha}\sum_{s=1}^4\int_{(s-1)a}^{sa} \mathrm{d}x\  \psi^{\dagger}_{n,m}(x,\alpha)(-i\alpha v\partial_x-\alpha v \mathcal{A}_{nm\alpha s} )\psi_{n,m}(x,\alpha),
    \label{random vector potential}
\end{equation}
where $s=1,2,3,4$ is the index of the four edges of one square. 
$(n,m,\alpha,s)$ indicates the $s$th side, \ie $x_{\alpha}\in[(s-1)a,sa)$, of the $\alpha$ square in the $(n,m)$ cell. 
$\mathcal{A}_{n m \alpha s}$ is the random vector potential on the associated edge, and it is assumed to be uniform within each edge. In this paper, the random vector potentials on different edges are independent.

Then, we project $H_{rand}$ in \cref{random vector potential} into the space of truncated lattice model $H_{8B}$ and $H_{8B}'$, \ie the space spanned by standing waves defined in Eq.~(\ref{standing wave}) with $K=0,\,1,\,2,\,3$. And we have

\begin{equation}
\begin{aligned}
    H_{8B,rand}&=\sum_{nm}\sum_{\alpha s}\sum_{K_1K_2}\frac{1}{4a}\int_{(s-1)a}^{sa}\mathrm{d}x \exp\left[i\alpha\frac{\pi(K_2-K_1)}{2a}x\right] [-\alpha v\mathcal{A}_{nm\alpha s}]\phi^{\dagger}_{n,m}(K_1,\alpha)\phi_{n,m}(K_2,\alpha)\\
    &=\sum_{nm}\sum_{\alpha s}[-\alpha v\mathcal{A}_{nm\alpha s}]\sum_{K_1K_2}\left[\frac{\delta_{K_1,K_2}}{4}+(1-\delta_{K_1,K_2})\frac{(\alpha i)^{s(K_2-K_1)}(1-(\alpha i)^{(K_1-K_2)})}{2\pi\alpha i(K_2-K_1)}\right]\phi^{\dagger}_{n,m}(K_1,\alpha)\phi_{n,m}(K_2,\alpha),
\end{aligned}
\end{equation}
where $K_1,K_2$ are limited to $0,1,2,3$.
We define a $4$-by-$4$ matrix
\begin{equation}
    \begin{aligned}
        (C_{\alpha s})_{I_1,I_2}=\left\{
        \begin{aligned}
             &\frac{(\alpha i)^{s(I_2-I_1)}(1-(\alpha i)^{(I_1-I_2)})}{2\pi\alpha i(I_2-I_1)}  & (I_1 \neq I_2) \\
             &\qquad\qquad\qquad \frac{1}{4} & (I_1=I_2)
        \end{aligned}\right. \quad (I_1,I_2=1,2,3,4).
    \end{aligned}
\end{equation}
Since the corner state basis describes the system in a more local way and the transfer matrix used for further numerical calculations (see \cref{subsec:transfer-lattice}) is simpler on this basis.  We now represent $H_{8B,rand}$ on the corner state basis. Making use of Eq.~(\ref{standing-wave-to-corner-dot}), we have
\begin{equation}
\label{H-rand-corner-state}
    \begin{aligned}
    H_{8B,rand}&=\sum_{nm}\sum_{\alpha s}[-\alpha v\mathcal{A}_{nm\alpha s}] (\vec{c}_{nm\alpha})^{\dagger}U_{\bar{\alpha}}C_{\alpha s}U_{\alpha}\vec{c}_{nm\alpha}\\
        &=\sum_{nm}\sum_{\alpha w_1w_2}\left[\sum_{s}-\alpha v\mathcal{A}_{nm\alpha s}D_{\alpha s}(w_1,w_2)\right]c^{\dagger}_{n,m}(w_1,\alpha)c_{n,m}(w_2,\alpha).
    \end{aligned}
\end{equation}
Here, $\vec{c}_{nm\alpha}$ is a $4$-by-$1$ column vector of $c_{n,m}(w,\alpha),\,(w=1,2,3,4)$. $U_{\alpha}$ is the $4$-by-$4$ transform matrix defined in Eq.~(\ref{standing-wave-to-corner-dot}) and $\bar{\alpha}=-\alpha$. One should not confuse the transform matrx $U_{\alpha}$ with the random vector potential $\mathcal{A}_{nm\alpha s}$. $D_{\alpha s}$ is a $4$-by-$4$ matrix with indices $w_1,w_2$. More concretely,
\begin{equation}
\begin{aligned}
        D_{-1}=\frac{1}{12\pi} \begin{pmatrix}
            8+3\pi & 3(1+i) & 4i & 7(i-1)\\
            3(1-i) & -8+3\pi & 1+i & 4i\\
            -4i & 1-i & -8+3\pi & 3(1+i)\\
            -7(1+i) & -4i & 3(1-i) & 8+3\pi
        \end{pmatrix} \ ,\\
        D_{-2}=\frac{1}{12\pi} \begin{pmatrix}
            -8+3\pi & 1+i & 4i & 3(1-i)\\
            1-i & -8+3\pi & 3(1+i) & -4i\\
            -4i & 3(1-i) & 8+3\pi & -7(1+i)\\
            3(1+i) & 4i & 7(i-1) & 8+3\pi
        \end{pmatrix}\ ,\\
        D_{-3}=\frac{1}{12\pi} \begin{pmatrix}
            -8+3\pi & 3(1+i) & -4i & 1-i\\
            3(1-i) & 8+3\pi & -7(1+i) & -4i\\
            4i & 7(i-1) & 8+3\pi & 3(1+i)\\
            1+i & 4i & 3(1-i) & -8+3\pi
        \end{pmatrix}\ ,
        \\
        D_{-4}=\frac{1}{12\pi} \begin{pmatrix}
            8+3\pi & -7(1+i) & -4i & 3(1-i)\\
            7(i-1) & 8+3\pi & 3(1+i) & 4i\\
            4i & 3(1-i) & -8+3\pi & 1+i\\
            3(1+i) & -4i & 1-i & -8+3\pi
        \end{pmatrix} \ ,
\end{aligned}
\label{full_disorder}
\end{equation}
and $D_{+s}=D_{-s}^*\,(s=1,2,3,4)$. 

\begin{figure}[h]
	\centering
	\subfigure{
    \includegraphics[width=.6\linewidth]{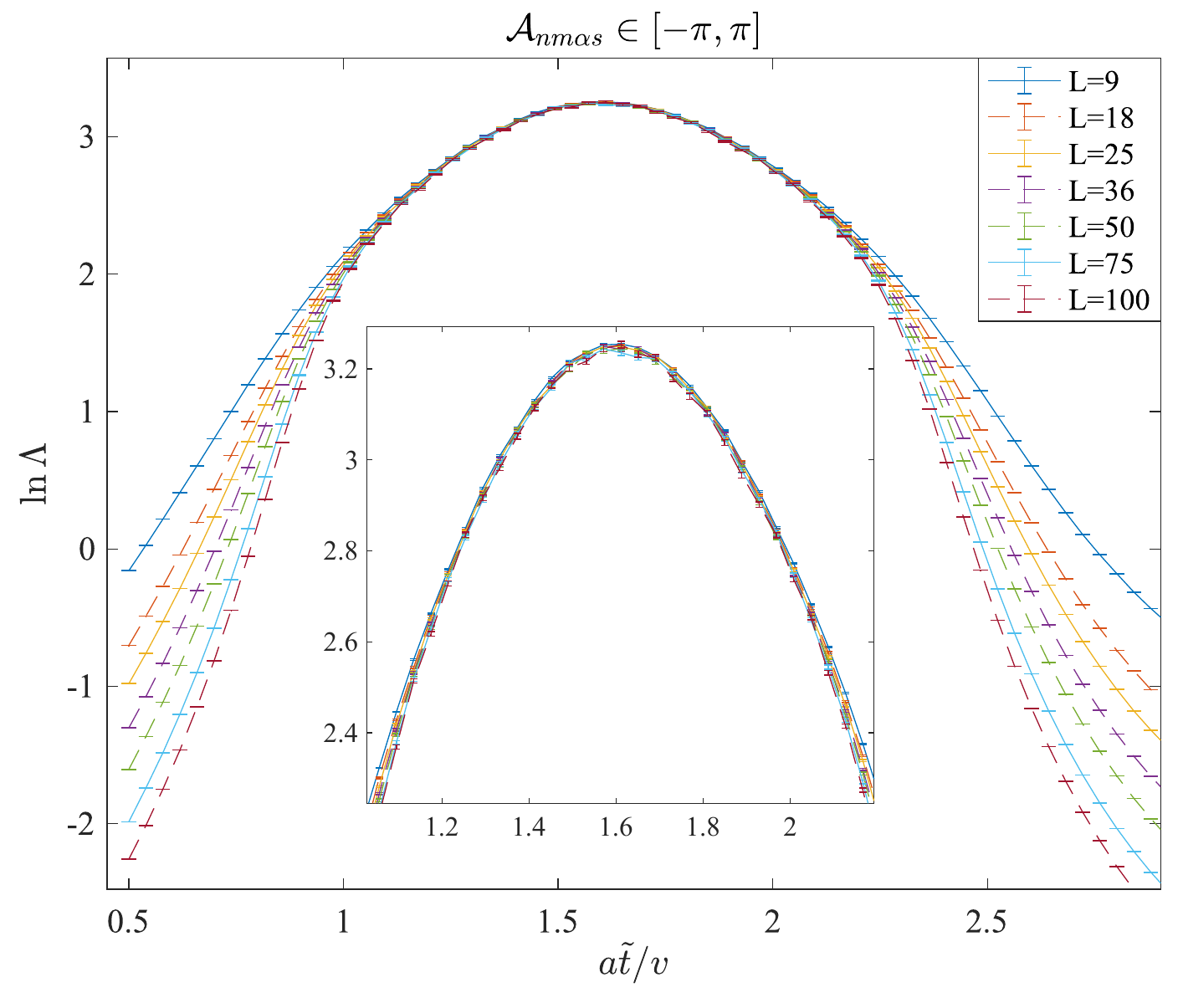}}
    \caption[]{{The normalized localization length $\Lambda$ versus $\tilde{t}$ in $H_{8B}$ (\cref{TB-H-real}) with all the random potentials in \cref{H-rand-corner-state}. The random vector potential $\mathcal{A}_{nm\alpha s}$ evenly distributes in $[-\pi,\pi]$ to consist with the network model. The longitudinal length is $M=5\times 10^7$ and the data precision ($\sigma_{\Lambda}/\Lambda$) has reached $0.7\%$. The inset is the zoomed critical region. }}
	\label{ll_full_fulldisorder_Wpi}
\end{figure}

This is a quite complicated random potential ($\sum_{s}\mathcal{A}_{nm\alpha s}D_{\alpha s}$) because all its matrix elements are nonzero. 
Nevertheless, we should point out that the most prominent terms, whose norm is at least as $\frac{8+3\pi}{7\sqrt{2}}\approx 2$ times large as other terms, locate in the diagonal line. 
This inspired us to only keep the on-site (chemical potential) disorder for ${H}_{8B}$ \& $H_{8B}'$. For simplicity and efficiency, we would like to ignore the correlations among on-site random potentials at different corners further.
It turns out that uncorrelated on-site disorder is sufficient to reproduce the (de)localization behaviors of the network model. Benchmarking and test calculations (see \cref{ll_full_fulldisorder_Wpi} for an example) show no qualitative difference between using uncorrelated on-site disorder only and using the full disorders in \cref{H-rand-corner-state}.
Therefore, we take the evenly distributed, uncorrelated on-site disorder (in the range $[-W/2,W/2]$) for $H_{8B}$ and $H_{8B}^{\prime}$.

\clearpage 
\section{Quasi-1D localization length and Transfer matrix method }
\label{sec:analytical-localization-length}
\subsection{General introduction}
\label{subsec:basic-localization-length}
\begin{figure}[h]
	\centering
	\subfigure{
    \includegraphics[width=.5\linewidth]{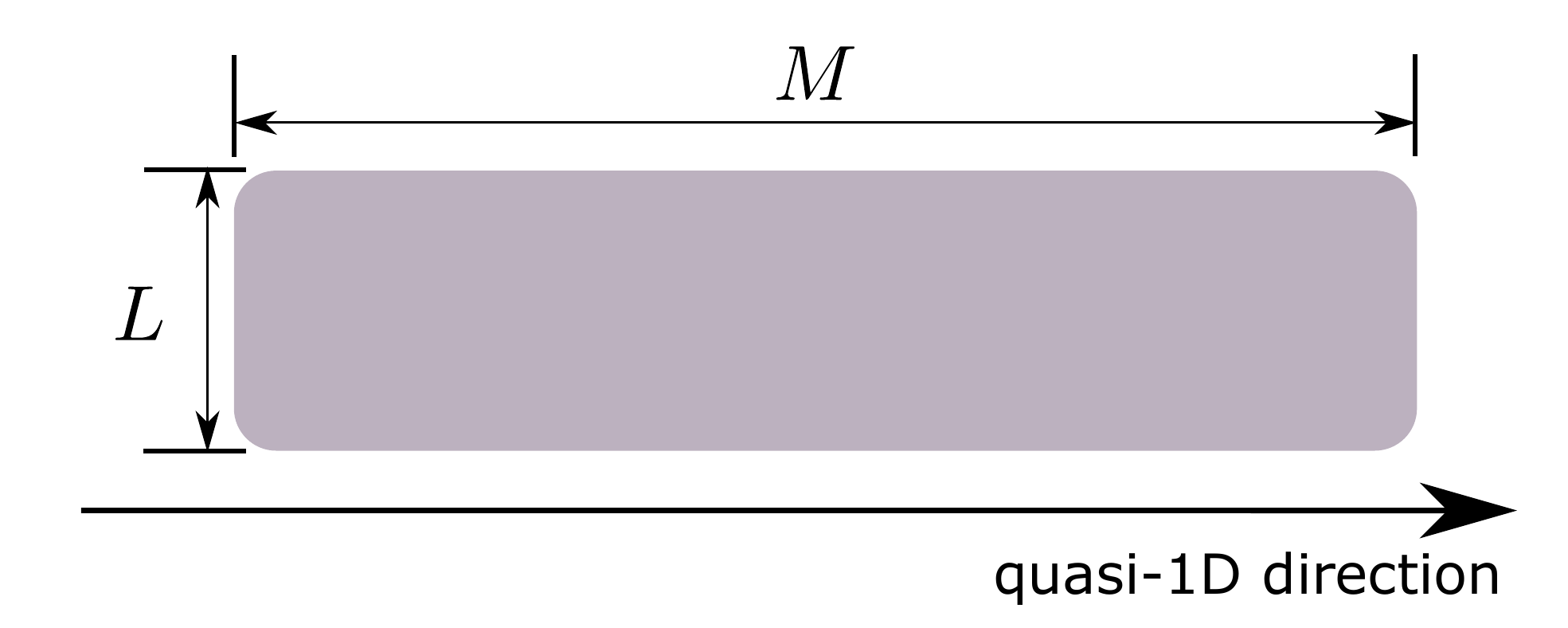}}
    \caption[]{Illustration of a quasi-1D sample with transversal size $L$ and longitudinal size $M\gg L$. The quasi-1D localization length $\rho_{\mathrm{q-1D}}$ based on the Lyapunov exponents is calculated on such a sample.}
	\label{quasi-1D-sample}
\end{figure}

A commonly used physical quantity in researches of localization is the \emph{quasi-1D localization length} $\rho_{\mathrm{q-1D}}$. It is defined on a 2D/3D sample prepared in quasi-1D shape, \eg a long thin cylinder with $L_{\rm{axial}}\gg L_{\rm{radius}}$. 
$\rho_{\mathrm{q-1D}}$ reflects the decaying rate of eigenstates in the quasi-1D direction, \eg the axial direction of a long thin cylinder. Since any 1D system will be localized under nonzero disorder strength, $\rho_{\mathrm{q-1D}}$ will always be finite except for a perfectly clean sample. 
Localization of the original 2D/3D system (in 2D/3D shape) can be determined by a scaling analysis of the dimensionless quasi-1D localization length $\Lambda=\rho_{\mathrm{q-1D}}/L$, where $L$ is the transversal size of the quasi-1D sample (see \cref{quasi-1D-sample}). 
We denote the localization length of a normal shaped (scales of different directions are similar) and sufficiently large sample as $\rho$.
For a metallic system, $\rho$ in a (normal shaped and sufficiently large) sample is much larger than the sample size. Thus, $\rho_{\mathrm{q-1D}}(L)$ increases faster than $L$, \ie $\Lambda(L)\rightarrow\infty$ in the limit $L\rightarrow\infty$. 
For an insulating system, $\rho$ is finite in a (normal shaped and sufficiently large) sample, so that $\rho_{\mathrm{q-1D}}$ will converge to $\rho$ when $L\gg\rho$, \ie $\Lambda(L)\rightarrow 0$ as $L\rightarrow\infty$. 
In practice, we identify the region where $\Lambda(L)$ monotonically increases as the metallic phase and the region where $\Lambda(L)$ monotonically decreases as the localized phase. If a system contains both localized and extended phases in some parameter space, there will be a critical region (usually a point) in the parameter space where $\Lambda(L)$ is independent on (sufficiently large) $L$. Note that such an analysis can only demonstrate the localization behavior (of a normal shaped sample) along one direction. For instance, in a 2D system, if we take the quasi-1D direction along the x axis (and the transversal direction will be the y direction), the scaling analysis of $\rho_{\mathrm{q-1D}}(L_y)$ will tell us the localization behavior (of a normal shaped sample) along the x direction. In principle, a normal shaped 2D system can be localized in the x direction while extended in the y direction. If one cannot rule out this possibility, it is necessary to do the afore-mentioned analyses of $\rho_{\mathrm{q-1D}}$ both along the x and y directions. Nevertheless, if one can rule out this possibility, \ie the normal shaped system is localized or extended in the x and y directions simultaneously, then analysing $\rho_{\mathrm{q-1D}}$ along one direction is enough. To see this, we can denote the angle between two directions $\hat{o}_1,\hat{o}_2$ as $\langle\hat{o}_1,\hat{o}_2\rangle$. For a state on a normal shaped 2D sample, if it is extended in direction $\hat{o}_1$, it should also be extended in $\hat{o}_2$ as long as $\langle \hat{o}_1,\hat{o}_2\rangle<\pi/2$. Notice that $\min\{\langle\hat{o}_1,\hat{x}\rangle,\langle\hat{o}_1,\hat{y}\rangle\}<\pi/2$ for any direction $\hat{o}_1$. Hence, a normal shaped 2D system extended (localized) in both the x and y directions should be extended (localized) in any direction. In such a case, scaling analyses of $\rho_{\mathrm{q-1D}}$ along different quasi-1D directions are similar, since the localization behaviors (of a normal shaped sample) in different directions have no qualitative difference. Our models all have $C_4$ symmetries. Hence, the systems are extended (localized) in the x and y directions simultaneously, and we can choose the quasi-1D directions for numerical convenience without resulting in qualitative difference. 

 However, it is questionable to apply the above strategy on a localized system with $\rho\gg$ \emph{practical} $L$ since the finite size effect is prominent there. For such a system, a localized phase can behave as a critical phase for practical $L$. This is not the case for the network model since its critical phase begins with $\Lambda\!<\!e^2\approx7$ (\cref{Lambda-theta--L-network}). However, it seems to be the case of $H_{8B}$ and $H_{8B}'$. Because $\Lambda\approx e^3\approx20$ 
 in their critical metal phases (\cref{ll-full} \& \ref{ll-simplify}). To rule out this possibility, we calculate $\Lambda$'s in the critical phases for a wide range of transversal sizes where $L_{\max}\approx 50 L_{\min}$ (see \cref{ll-test}) and no significant decline of $\Lambda(L)$ has been observed. Second, if the critical phases we observed in \cref{ll-full} \& \ref{ll-simplify} are actually localized phases with large localization lengths, they will eventually become localized as $L\rightarrow\infty$. For finite $L$, these fake critical phases should shrink as $L$ increases. However, according to \cref{ll-full} \& \ref{ll-simplify}, the boundary between the localized phase and critical phase does not change as $L$ increases. In addition, we calculate the local Chern marker in \cref{subsec:LCM}, which also demonstrate the delocalization of $H_{8B}$ and $H_{8B}'$ in the critical phases. (We leave the detailed explanation in \cref{subsec:LCM}.) Based on these observations, we claim that the critical phase in $H_{8B}$ and $H_{8B}'$ is not due to the finite size effects and will persist in the thermodynamic limit.

The transfer matrix method \cite{pichard_finite_1981} is a widely used numerical approach in calculating $\rho_{\mathrm{q-1D}}$. 
Although it has different formulae for (general) network and lattice models, the basic ideas are the same. 
A quasi-1D sample is divided into layers with normals along the quasi-1D direction. 
The amplitudes of an energy-eigenstate on different layers are related by the Schrodinger equation. Note that the layer division on a concrete system should be specially designed for numerical convenience (we will see this in \cref{subsec:transfer-network} \& \ref{subsec:transfer-lattice}). 
A ($2s\times2s$ shaped) transfer matrix $T_n$ in general is a transformation from the amplitudes on the $(\rm{l\!-\!r_1)th\sim (l\!+\!r_2\!-\!1)th}$ layers to those on the $\rm{(l\!- \!r_1\!+\!1)th \sim ( l\!+\!r_2) th}$ layers. Here $s\in \mathbb{N}^+$ is proportional to the number of degrees of freedom in one layer. And $r_1,\,r_2\in \mathbb{N}^+$ represent that only the $\rm{(l\!-\!r_1)th\sim(l\!+\!r_2)th}$ layers can hop/propagate (in one step) to the $\rm{l\thinspace th}$ layer. The values of $r_1,\,r_2$ depend on the hopping of concrete models. In our models, $r_1=r_2=1$. Nevertheless, for generality, we keep $r_1,\,r_2$ in general discussions about the transfer matrix method. From this point of view, network model is a special case. Since the transfer matrix of a (general) network model is determined by the transmission matrix, the amplitudes in the $\mathrm{(l+1)th}$ layer depend only on the $\mathrm{lth}$ layer. Hence, $r_1=r_2=1$ for general network models (not only ours). We will see this more clearly in \cref{subsec:transfer-network}. For general lattice models, that is not the case, \ie $r_1$ and $r_2$ can take arbitrary finite non-negative integer values.

For a quasi-1D system containing $M$ layers (we also use $M$ to represent the longitudinal size), we can define a consecutive product $O_M=\prod_{n=1}^M T_n$ that transforms the amplitudes on the first $r_1+r_2+1$ layers to the last $r_1+r_2+1$ layers. 
$\rho_{\mathrm{q-1D}}$ can be extracted from $O_M$. 
Because of the disorder, some elements in $T_n$ are random variables. 
According to the Oseledec’s theorem, the limit $P=\lim_{M\rightarrow\infty} (O_M^{\dagger}O_M)^{1/2M}$ exists and has eigenvalues $\{\exp(\nu_1),\,\exp(-\nu_1),...\exp(\nu_{s}),\,\exp(-\nu_{s})\}$ where $\nu_{i}\geq\nu_{i+1}\geq0,\,i=1,2...s$. These (positive) exponents are so-called \emph{Lyapunov exponents} (LEs). 
The definition of $P$ indicates that an eigenvector $\vec{\eta}_i$ of $P$ with eigenvalue $\exp(-\nu_i)$ satisfies $\Vert O_M\vec{\eta}_i
\Vert^2=\vec{\eta}_i^{\dagger}(O_M^{\dagger}O_M)\,\vec{\eta}_i=\vec{\eta}_i^{\dagger}[(O_M^{\dagger}O_M)^{1/2M}]^{2M}\,\vec{\eta}_i\approx\vec{\eta}_i^{\dagger}P^{2M}\,\vec{\eta}_i=\Vert\exp(-M\nu_i)\vec{\eta}_i\Vert^2$ for sufficiently large M. 
Therefore, the smallest LE $\nu_s$ determines the decaying rate of energy-eigenstates (along the quasi-1Dc direction) since any energy-eigenstate is a superposition of eigenvectors of $P$ with eigenvalues $\exp(-\nu_i)$ (the amplitudes cannot grow exponentially hence $\exp(\nu_i)$ is excluded). 
In this paper, we define $\rho_{\mathrm{q-1D}}=1/\nu_s$.

It is worth introducing some numerical details briefly. 
The definition of $P$ requires us to first calculate $O_M^{\dagger}O_M$ whose eigenvalues are $\exp(\pm 2M\nu_i)$ when $M\rightarrow\infty$. (For our calculations, we usually take $M=10^6\!\sim\!10^8$.) 
When $M$ is sufficiently large, $\exp( M\nu_i) \gg \exp( M\nu_j)$ if $\nu_i>\nu_j$, and most of the computational resources are consumed by the large LEs. The small LEs will acquire vast round-off errors since they are stored in the low digits. Unfortunately, we are concerned with the smallest LE $\nu_s$. To circumvent such errors, the practical numerical method should extract information after every several $T$'s are multiplied to $O$. 
We point out without proof \cite{pichard_finite_1981} that
\begin{equation}
\label{LEs}
    \begin{aligned}
    & O_M =UR\\
    & \nu_i=\lim_{M\rightarrow\infty}\frac{\ln(R)_{i,i}}{M}\ .
    \end{aligned}
\end{equation}
The first line is the QR decomposition of $O_M$, and the second line means that the LEs are determined by diagonal elements in the upper triangular matrix $R$. 
Suppose there are two upper triangular matrices $R_1,\,R_2$, it is direct to prove that $(R_1R_2)_{i,i}=(R_1)_{i,i}(R_2)_{i,i}$. Because of this property, we do not have to decompose the entire $O_M$ after all $T$'s are multiplied. 
Instead, QR decomposition can be executed after every $q$ ($2\sim10$) $T$'s are multiplied to $O$. 
After each decomposition, diagonal elements in $R$ are stored, and the next $q$ $T$'s will be multiplied to $U$. 
Repeat this treatment one will obtain $M/q$ arrays of diagonal elements $(R_j)_{i,i},\,(j=1,2...M/q)$ in the end. 
We can rewrite the second line of Eq.~(\ref{LEs}) as 
\begin{equation}
\label{true-mean-LE}
    \nu_i=\lim_{M/q\rightarrow\infty} \frac{1}{M/q - n_0}\sum_{j=n_0+1}^{M/q}\frac{\ln(R_{j})_{i,i}}{q} 
\end{equation}
In this formula, the vast round-off error ($\sim\exp(M(\nu_1-\nu_s))$) is replaced by a marginal round-off error ($\sim\exp(q(\nu_1-\nu_s))$). 
We have excluded the first $n_0$ (set to 10) groups ($n_0 q$ layers) in practical calculations to avoid possible boundary effects. 

Another advantage of Eq.~(\ref{true-mean-LE}) is that we can estimate the numerical precision of LEs.
One can view Eq.~(\ref{true-mean-LE}) as an average of $\ln(R_j)_{i,i}/q$ over $M/q - n_0$ samples.
Therefore, we can estimate the error of the sample average by unbiased estimation and use it to control the precision. 
Prudent readers may suspect the independence between ``data points'' $\ln(R_j)_{i,i}/q$ for different $j$ when $q$ is small. 
In practice, we will group up $r\sim 10^1$ ``data points'' and view the ($(M/q-n_0)/r$) groups independent to each other.

\subsection{Transfer matrices of the network model on the Manhattan lattice}
\label{subsec:transfer-network}

\begin{figure}[h]
	\centering
	\subfigure[~Layer division of the network model Chiral edges between two adjacent green dashed lines belong to one layer. Purple indices $\{(\mathcal{Y},\nu)\vert \mathcal{Y}\in\mathbb{Z},\nu=0,1\}$ in the bottom are layer indices. Purple indices $\{(\mathcal{X},\chi)\vert \mathcal{X}\in\mathbb{Z},\chi=0,1\}$ along square edges denote different edges in one layer. The golden arrows indicate the directions of accumulating random phases $\vartheta(\mathcal{X},\chi,\mathcal{Y},\nu)$.]{\label{transfer-network-a}
    \includegraphics[width=.6\linewidth]{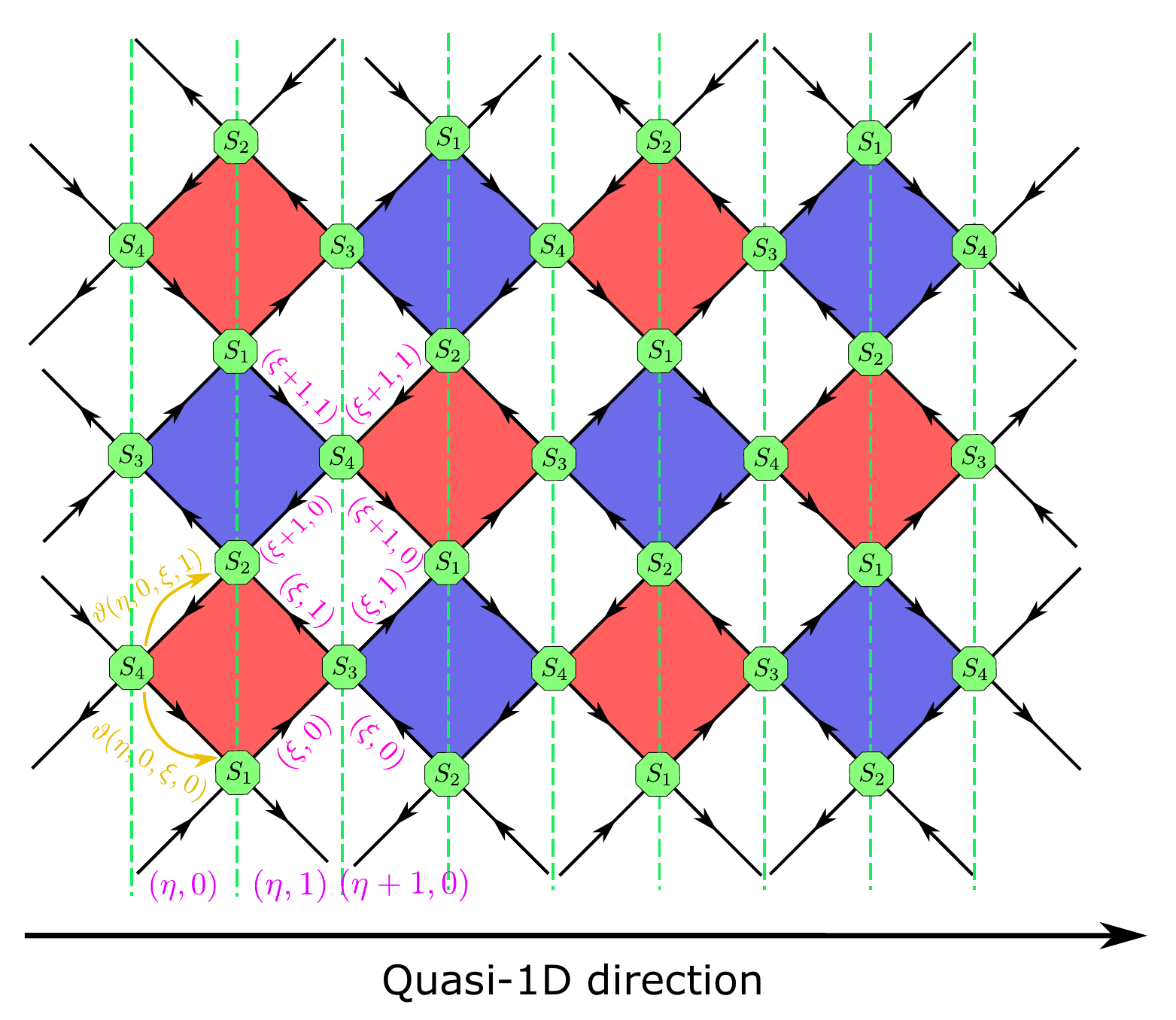}}
	\subfigure[~Scattering node $S_{i},\,(i=1,\,2,\,3,\,4)$ relating chiral edges in adjacent layers. $in/out_{1,2}$ indicate scattering channels. The scattering effect controlled by $\theta$ can be found in \cref{transfer-scattering-nodes}.]{\label{transfer-network-b}
    \includegraphics[width=.3\linewidth]{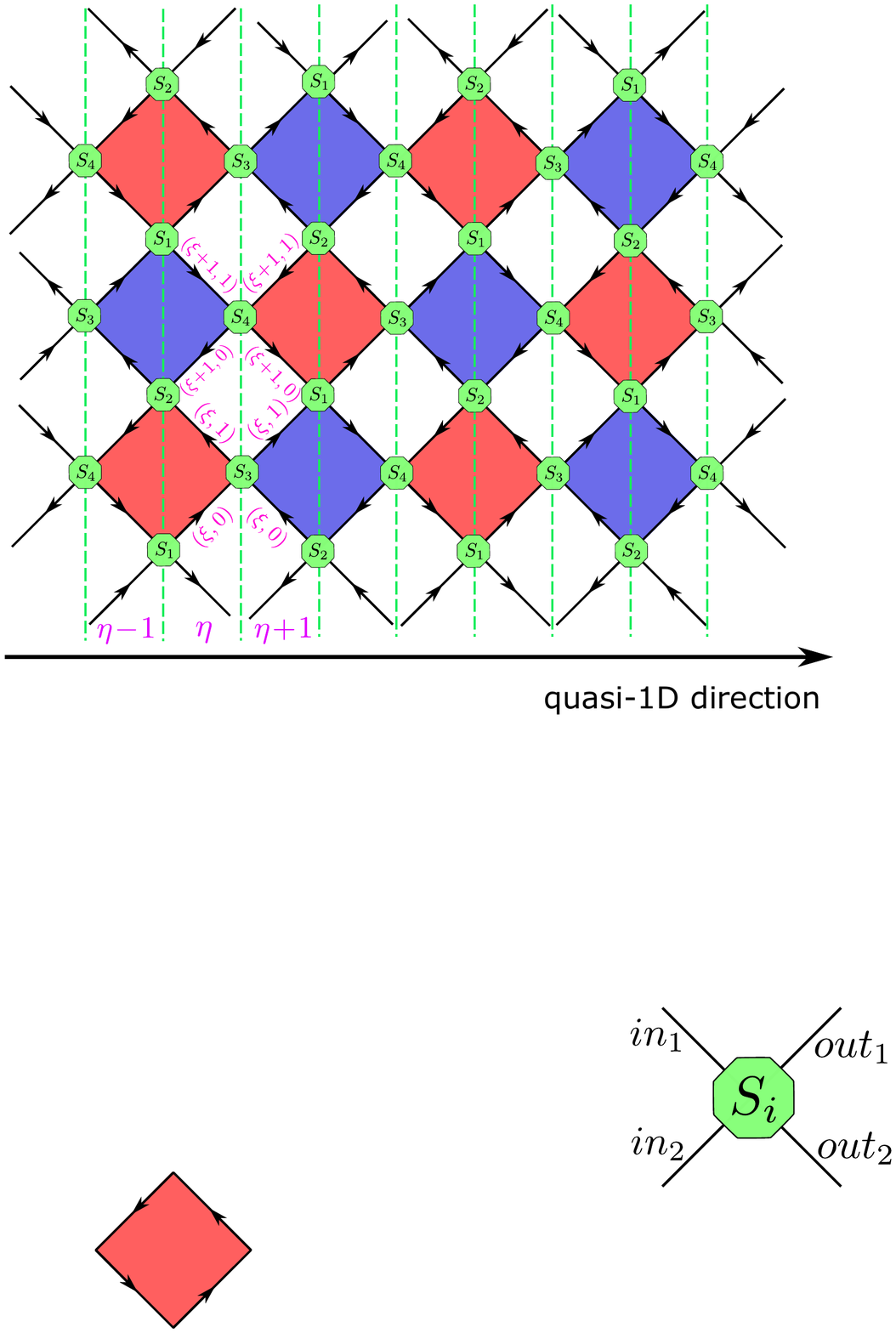}}
    \caption[]{Layer division and scattering nodes of the network model (\cref{network-decoupled-chiral-wires} \& (\ref{network-delta-scattering-potential}))}
	\label{transfer-network}
\end{figure}

In this subsection, we will introduce the concrete formulae of transfer matrices in the network model defined in \cref{network-decoupled-chiral-wires} \& (\ref{network-delta-scattering-potential}).
Fig.~\ref{transfer-network} shows some details of transfer matrices in the network model. We choose $x+y$ as the quasi-1D direction since only in this direction the layer division and inter-layer relation are simple.
The green octagons marked by $S_1\cdots S_4$ indicate four kinds of scattering nodes, respectively. The green dashed lines indicate the layer division along the quasi-1D direction, \ie chiral edges sitting inside two adjacent green dashed lines belong to one layer. 
Each square edge is represented by two composite indices $(\mathcal{X},\chi)$ \& $(\mathcal{Y},\nu)$, where $\mathcal{X}\in\mathbb{Z}$ is the square index along the quasi-1D direction, $\mathcal{Y}\in\mathbb{Z}$ is the square index along the transversal direction, and $\chi,\nu=0,1$ further distinguish four edges in one square. 
As shown in \cref{transfer-network-a}, three successive layers are labeled by $(\mathcal{X},\chi)=(\eta,0)$, $(\eta,1)$, $(\eta+1,0)$, respectively. Within each layer, four successive chiral edges are labeled by $(\mathcal{Y},\nu)=(\xi,0),\,(\xi,1),\,(\xi+1,0),\,(\xi+1,1)$, respectively. 
We take the convention that $(-1)^{\mathcal{X+Y}}=-1$ corresponds to the blue squares and $(-1)^{\mathcal{X+Y}}=1$ corresponds to the red ones. Hence, in \cref{transfer-network-a}, $\eta+\xi$ is even.
The disorder is the collection of random phases accumulated along the edges, which reflects the random local fluxes (\cref{subsec:disorder}).

There are two parts that contribute to the transfer matrix.
The first part is the random phase accumulated (when moving from left to right) on each chiral mode, denoted as $\exp[ i \vartheta(\mathcal{X},\chi,\mathcal{Y},\nu)]$. The golden arrows in \cref{transfer-network-a} indicate the directions along which an electron will acquire the random factor $\exp[\vartheta(\eta,0,\xi,0)]$ or $\exp[\vartheta(\eta,0,\xi,1)]$.
We take completely random phases $\vartheta$, which uniformly distribute in $[-\pi,\pi]$ and set $\vartheta$ on different edges independent to each other. In principle, we should include a third part that comes from free propagation $e^{\pm i Ea/v}$ along chiral edges. (Inside the chiral edges, we have $iv\partial\psi=E\psi$, \ie the eigenstate behaves like a plane wave with wave vector $E/v$. Also recall that the length of an edge is $a$). However, since we uniformly take values of random phases $e^{i\vartheta}$ in entire $U(1)$, the free propagation factor $e^{\pm iaE/v}$ becomes irrelevant and can be omitted. (This is also true for generic network models with uniform $U(1)$ random phases. Hence, localization behavior of a network model with uniform $U(1)$ random phases is independent on the Fermi level. In common sense, the Fermi level is important since the system will become a band insulator/metal if we put the Fermi level inside a gap/band. For a network model with uniform $U(1)$ random phases, the situation is modified: the model can be delocalized only if the spectrum has no gap everywhere.)

\begin{figure}[h]
	\centering
    \includegraphics[width=.6\linewidth]{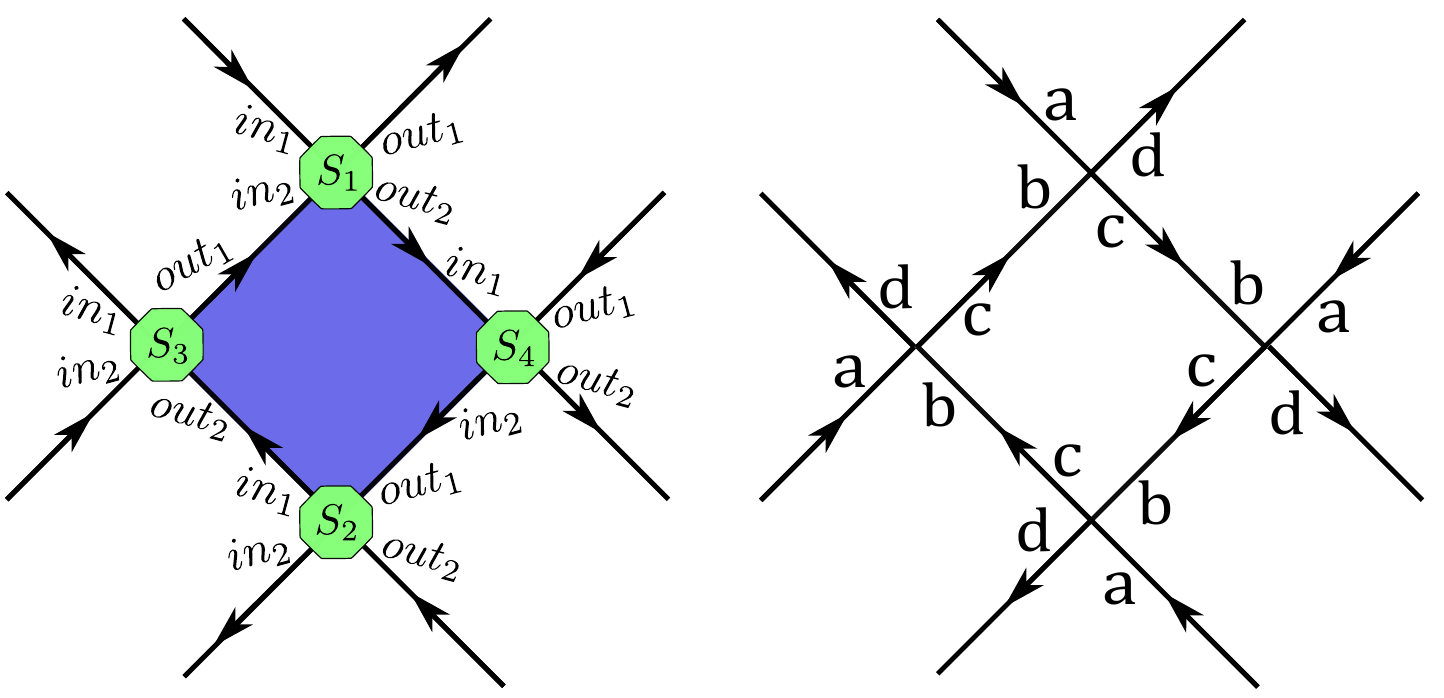}
    \caption[]{Scattering channel correspondence between \cref{transfer-network-b} and \cref{network-b}. The left side is a patch of \cref{transfer-network-a} containing four kinds of scattering nodes. Correspondence between chiral edges and the $in/out$ channels is illustrated on the left side. And the correspondence between chiral edges and $a\sim d$ channels in \cref{network-b} is illustrated on the right side. Hence we can see the correspondence between the $in/out$ channels and the $a\sim d$ channels for $S_1\cdots S_4$.}
	\label{channel-correspondence-network}
\end{figure}

The second part is the scattering nodes. Although there are four kinds of scattering nodes $S_1\cdots S_4$, all of them can be rotated into Fig.~\ref{network-b}. See \cref{channel-correspondence-network} for the correspondence between $in/out$ channels in \cref{transfer-network-b} and the $a\sim d$ channels in \cref{network-b}. For $S_1$, $in_{1,2}$ and $out_{1,2}$ correspond to $a,\,b$ and $d,\,c$ channels, respectively. 
For $S_2$, $in_{1,2}$ and $out_{1,2}$ correspond to $c,\,d$ and $b,\,a$ channels, respectively. 
For $S_3$, $in_{1,2}$ and $out_{1,2}$ correspond to $d,\,a$ and $c,\,b$ channels, respectively. 
For $S_4$, $in_{1,2}$ and $out_{1,2}$ correspond to $b,\,c$ and $a,\,d$ channels, respectively. Then, after some simple calculations, we can write the scattering effects on the basis of $in/out$ channels:
\begin{equation}
\label{transfer-scattering-nodes}
    \begin{aligned}
        &S_1:\quad 
        \begin{pmatrix}
            out_1\\ out_2
        \end{pmatrix}
        =\begin{pmatrix}
            -i\sin\theta & \cos\theta\\
            \cos\theta & -i\sin\theta
        \end{pmatrix}
        \begin{pmatrix}
            in_1\\ in_2
        \end{pmatrix}
        \\
        &S_2:\quad 
        \begin{pmatrix}
            out_1\\ out_2
        \end{pmatrix}
        =\begin{pmatrix}
            i\sin\theta & \cos\theta\\
            \cos\theta & i\sin\theta
        \end{pmatrix}
        \begin{pmatrix}
            in_1\\ in_2
        \end{pmatrix}
        \\
        &S_3:\quad 
        \begin{pmatrix}
            out_1\\ out_2
        \end{pmatrix}
        =\begin{pmatrix}
            -i\tan\theta & \sec\theta\\
            \sec\theta & i\tan\theta
        \end{pmatrix}
        \begin{pmatrix}
            in_1\\ in_2
        \end{pmatrix}
        \\        
        &S_4:\quad 
        \begin{pmatrix}
            out_1\\ out_2
        \end{pmatrix}
        =\begin{pmatrix}
            i\tan\theta & \sec\theta\\
            \sec\theta & -i\tan\theta
        \end{pmatrix}
        \begin{pmatrix}
            in_1\\ in_2
        \end{pmatrix}
    \end{aligned}
\end{equation}
For instance, we have the relation $a-in_2,\,b-out_2,\,c-out_1,\,d-in_1$ in $S_3$. Then \cref{scatter matrix} can be written as %
\[\begin{aligned}
    S_3:\quad 
    \begin{pmatrix}
        out_1\\ in_1
    \end{pmatrix}
    &=\begin{pmatrix}
        \cos\theta & -i\sin\theta\\
        -i\sin\theta & \cos\theta
    \end{pmatrix}
    \begin{pmatrix}
        in_2\\ out_2
    \end{pmatrix}
    \\
    \\
     \Rightarrow out_2&=\sec\theta \,in_1+i\tan\theta\, in_2
    \\    
     \Rightarrow out_1&=\cos\theta\,in_2-i\sin\theta (\sec\theta\, in_1+i\tan\theta \,in_2)
     \\
     &= -i\tan\theta\,in_1+\sec\theta\, in_2,
\end{aligned}\]
which is the third equation of \cref{transfer-scattering-nodes}. Other equations in \cref{transfer-scattering-nodes} can be derived similarly. 

Combining the scattering and the random propagation phases, we obtain the transfer matrices between different layers. For instance, the transfer matrix from $(\eta,0)$ to $(\eta,1)$ in \cref{transfer-network-a} is
\begin{equation}
\label{concrete-transfer-matrix-network-1}
T_{(\eta,1),(\eta,0)}=
    \begin{pNiceArray}{cccccccc}[first-row,first-col,nullify-dots]
    &(1,0) &\Ldots&(\xi\!-\!1,1)&(\xi,0)&(\xi,1)&(\xi\!+\!1,0)&\Ldots & (L,1) \\
    (1,0)& -i\sin\theta & & & & & & &\cos\theta\\
    \vdots & &\ddots & & & & &\\
    (\xi\!-\! 1,1)& & & -i\sin\theta & \cos\theta & & & \\
    (\xi,0)& & & \cos\theta & -i\sin\theta & & &\\
    (\xi,1)& & & & & i\sin\theta &  \cos\theta &\\
    (\xi\!+\!1,0)& & & & &\cos\theta & i\sin\theta &\\
    \vdots & & & & & & & \ddots\\
    (M,1)& \cos\theta & & & & & & &-i\sin\theta\\
    \end{pNiceArray}
    \times\mathrm{diag}\left[e^{i\vartheta(\eta,0,\xi,\mu)}\right].
\end{equation}
and the transfer matrix from $(\eta,1)$ to $(\eta+1,0)$ is
\begin{equation}
\label{concrete-transfer-matrix-network-2}
T_{(\eta+1,0),(\eta,1)}=
    \begin{pNiceArray}{cccccc}[first-row,first-col,nullify-dots]
    &\Ldots&(\xi,0)&(\xi,1)&(\xi\!+\!1,0)&(\xi\!+\!1,1)&\Ldots \\
    \vdots & \ddots &  &  &  & & \\
    (\xi,0)& & i\tan\theta & \sec\theta & & & \\
    (\xi,1)& & \sec\theta & -i\tan\theta & & & \\
    (\xi\!+\!1,0)& & & & -i\tan\theta &  \sec\theta & \\
    (\xi\!+\!1,1)& & & & \sec\theta & i\tan\theta & \\
    \vdots & & & & & &\ddots
    \end{pNiceArray}
    \times\mathrm{diag}\left[e^{i\vartheta(\eta,1,\xi,\mu)}\right],
\end{equation}

The diagonal matrix ``$\mathrm{diag}\,[\cdot]$" represents the random propagation phases. 
The transversal size $L$ in Eq.~(\ref{concrete-transfer-matrix-network-1}) is defined as the total number of squares per layer. In addition, from the scattering part of $T_{(\eta,1),(\eta,0)}$, one can see that we take the periodic boundary condition (PBC). Hence, $L$ has to be even to make the PBC well-defined. The longitudinal length $M$ is defined as the number of layers, \ie $\eta=1,2,\cdots,M/2$.

Transfer matrices $T_{(\eta+1,1),(\eta+1,0)},T_{(\eta+2,0),(\eta+1,1)}$ can be derived in a similar way, one just needs to exchange the scattering effects in \cref{transfer-scattering-nodes} by $S_1\!\leftrightarrow\!S_2$ \& $S_{3}\!\leftrightarrow\!S_{4}$ and use the corresponding random phases. 
The transfer matrices for other layers, apart from the random phase part, must be the same to one of $T_{(\eta,1),(\eta,0)},T_{(\eta+1,0),(\eta,1)},T_{(\eta+1,1),(\eta+1,0)},T_{(\eta+2,0),(\eta+1,1)}$. 
By successively multiplying these transfer matrices and taking the statistical procedure described in Sec.~\ref{subsec:basic-localization-length}, we can obtain the $\rho_{\mathrm{q-1D}}$ of the network model. The numerical results are shown in Sec.~\ref{subsec:numerical-network}.

\begin{figure}[h]
	\centering
    \includegraphics[width=.4\linewidth]{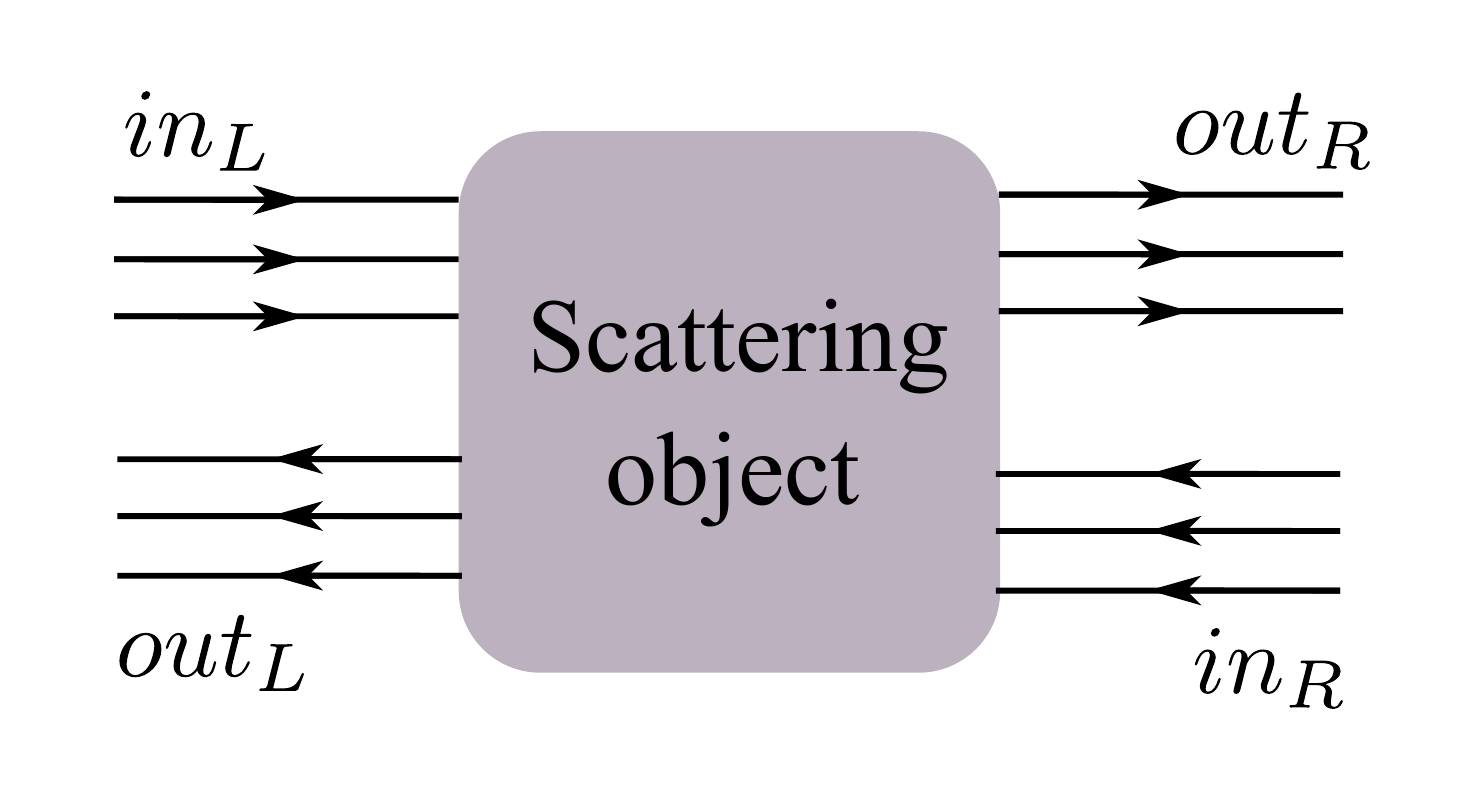}
    \caption[]{1D scattering process on a sample. $in_L\,\&\,out_L$ ($in_R\,\&\,out_R$) represent the incoming \& outgoing channels on the left (right) side, respectively. The arrows indicate the propagation directions of these channels. }
	\label{sample-scatter}
\end{figure}

With the help of transfer matrix, we can also calculate the conductivity $\sigma$ of the network model \cite{PhysRevB.22.3519}. Consider a general 1D scattering process on a sample, the amplitudes of incoming \& outgoing channels on the left \& right sides (see \cref{sample-scatter}) satisfies
\begin{equation}
\begin{aligned}
    &\begin{pmatrix}
         out_R \\
         out_L
    \end{pmatrix}
    =
    \begin{pmatrix}
         t & r'\\
         r & t'
    \end{pmatrix}
    \begin{pmatrix}
         in_L\\
         in_R
    \end{pmatrix}
\end{aligned}
\label{Smatrix}
\end{equation}

According to the Landauer formula, the conductance of the sample is determined by the transmission coefficients
\begin{equation}
    G=\frac{e^2}{h}\mathrm{Tr}[t^{\dagger}t]
    =\frac{e^2}{h}\mathrm{Tr}[t^{\prime\dagger}t'].
    \label{Landauer formula}
\end{equation}

Thus, we need $t$ or $t'$ to calculate the conductance. \cref{Smatrix} transforms the incoming channels to outgoing channels, while the transfer matrix transform the amplitudes from left to right. Hence, in order to obtain $t$ or $t'$, we should reshape \cref{Smatrix} into

\begin{equation}
    \begin{pmatrix}
         out_R \\
         in_R
    \end{pmatrix}
    =
    \begin{pmatrix}
         t-r't^{\prime -1} r & r't^{\prime -1} \\
         -t^{\prime -1} r & t^{\prime -1}
    \end{pmatrix}
    \begin{pmatrix}
         in_L\\
         out_L
    \end{pmatrix}
    =V_R^{\dagger}O_M V_L
    \begin{pmatrix}
         in_L\\
         out_L
    \end{pmatrix}
    =V_{R}^{\dagger}\prod_{\eta=1}^{M/2}T_{(\eta+1,0),(\eta,1)}T_{(\eta,1),(\eta,0)} V_L
    \begin{pmatrix}
         in_L\\
         out_L
    \end{pmatrix},
\label{Tmatrix}
\end{equation}
where $O_M=\prod_{n=1}^M T_n$ (defined in \cref{subsec:basic-localization-length}) is a product of $M$ consecutive transfer matrices. Since the bases of transfer matrices may not coincide with the $in/out_{L,R}$ channels, unitary matrices $V_{L}\,\&\,V_R$ are introduced to transform $(in_L,\,out_L)^T$ \& $(out_R,\,in_R)^T$ into the bases of $T_1$ \& $T_M$, respectively. (We will see a concrete construction on the network model later.) In the third equation of \cref{Tmatrix}, we apply the general formula to our network model. 

The remaining part of this subsection could be too technical for most readers. Thus, in this paragraph, we summarize the main idea for readers that is not interested in the details. In \cref{concrete-transfer-matrix-network-1} \& (\ref{concrete-transfer-matrix-network-2}), we have already chosen a basis (an index of chiral edges in one layer) for the transfer matrix. We denote this basis as $\mathcal{I}_1$. However, $\mathcal{I}_1$ is designed by the spatial positions of chiral edges, not by their correspondence with $in$ \& $out$ channels. Hence, we need to identify the correspondence between chiral edges and $in$ \& $out$ channels, and then design a new basis $\mathcal{I}_2$ to index these channels. Finally, we use $V_L,\,V_R$ to translate $\mathcal{I}_2$ into $\mathcal{I}_1$ and show how to obtain $t'$. Now we turn to the explicit construction.

\begin{figure}[h]
	\centering
    \includegraphics[width=.7\linewidth]{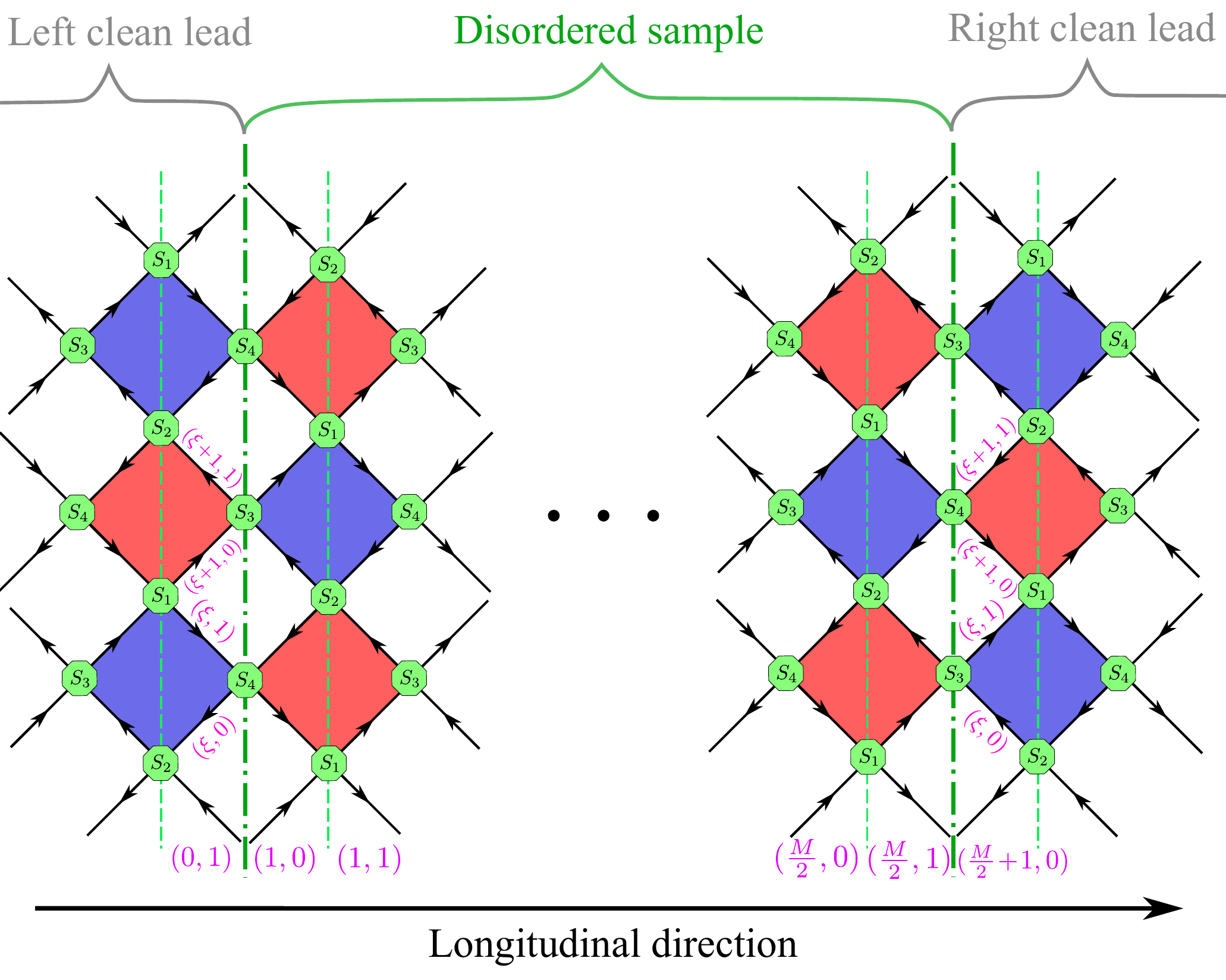}
    \caption[]{The setting of leads of the network model (\cref{network-decoupled-chiral-wires} \& (\ref{network-delta-scattering-potential})). The disordered sample contains layers $(1,0)\sim(M/2,1)$, the left lead contains $\{(\mathcal{X},\chi)\vert \mathcal{X}\leq 0, \chi=0,1\}$, and the right lead contains $\{(\mathcal{X},\chi)\vert \mathcal{X}\!>\! \frac{M}{2}, \chi=0,1\}$. In this figure, $(1,\chi,\xi,\nu)$ \& $(M/2,\chi,\xi,\nu)$ correspond to red squares. In order to conform the convention $(-1)^{\mathcal{X+Y}}=1\sim \mathrm{red\,square}$,  $M/2$ and $\xi$ are odd numbers. Therefore, in this figure, $in_L$ corresponds to edges $\{(0,1,\mathcal{Y},\nu)\vert (-1)^{\mathcal{Y}+\nu}=1\}$ and $out_R$ corresponds to $\{(\frac{M}{2}+1,0,\mathcal{Y},\nu)\vert (-1)^{\mathcal{Y}+\nu}=1\}$. Other edges in layers $(0,1)$ and $(\frac{M}{2}+1,0)$ correspond to $out_L$ and $in_R$, respectively. }
	\label{leads-network}
\end{figure}

The concrete form of channels depends on the setting of leads. For convenience, we take the clean Manhattan networks as leads, \ie we add several clean layers to two terminals of the disordered sample (see \cref{leads-network}). Now, we try to identify the incoming and outgoing channels in terms of the notation $(\mathcal{X},\chi,\mathcal{Y},\nu)$ defined at the beginning of this subsection. We take $\mathcal{X}=1 \,\&\,\chi=0$ as the first layer inside the sample and $\mathcal{X}=M/2\,\&\,\chi=1$ as the last layer. The left and right leads contain layers $\{(\mathcal{X},\chi)\vert \mathcal{X}\leq 0, \chi=0,1\}$ and $\{(\mathcal{X},\chi)\vert \mathcal{X}\!>\! \frac{M}{2}, \chi=0,1\}$, respectively. Recall that $\{(\mathcal{X},\mathcal{Y})\vert(-1)^{\mathcal{X+Y}}=1\}$ $\left(\{(\mathcal{X},\mathcal{Y})\vert(-1)^{\mathcal{X+Y}}=-1\}\right)$ corresponds to the red (blue) squares. Hence, $\xi$ and $\frac{M}{2}$ in \cref{leads-network} are odd numbers. In the left lead, the $in_L$ (right-going) channels end at edges $\{(0,1,\mathcal{Y},\nu)\vert (-1)^{\mathcal{Y}+\nu}=1\}$, and the $out_L$ (left-going) channels start at edges $\{(0,1,\mathcal{Y},\nu)\vert (-1)^{\mathcal{Y}+\nu}=-1\}$. Similarly, $out_R$ starts at $\{(\frac{M}{2}+1,0,\mathcal{Y},\nu)\vert (-1)^{\mathcal{Y}+\nu}=1\}$ and $in_R$ ends at $\{(\frac{M}{2}+1,0,\mathcal{Y},\nu)\vert (-1)^{\mathcal{Y}+\nu}=-1\}$. If the sample ends up with $\frac{M}{2}=$ even number, the correspondence between $in/out_R$ and the edges in layer $(\frac{M}{2}+1,0)$ is reversed.

If there are $L$ squares in the transversal direction, $in_{L},\,in_R,\,out_{L},\,out_R$ all contain $L$ channels. We define the $\mathrm{mth}$ $in_L$ channel as the channel ends at edge $(0,1,m,\mathrm{mod}(m,2))$. Similarly, the channel starts at $(0,1,m,\mathrm{mod}(m+1,2))$ is defined as the $\mathrm{mth}$ $out_L$ channel. When $\frac{M}{2}=$ odd number (as it is in \cref{leads-network}), the $\mathrm{m th}$ $out_R$ channel starts at $(\frac{M}{2},0,m,\mathrm{mod}(m,2))$ and the $\mathrm{m th}$ $in_R$ channel ends at $(\frac{M}{2},0,m,\mathrm{mod}(m+1,2))$. 

Since each lead contains both incoming and outgoing channels, we also need to index these channels together. See \cref{leads-network}, there are $2L$ channels in each lead, hence we need to index $in\,\&\,out$ channels in each lead by integers in $[1,2L]$. We first label the left channels: the $\mathrm{m th}$ $in_L$ \& $\mathrm{m' th}$ $out_L$  channels are labeled by $m$ \& $L+m'$, respectively. Since $in_L$ will transmit to $out_R$, we use the same integer to label the $\mathrm{m th}$ $in_L$ and the $\mathrm{m th}$ $out_R$ channels. Similarly, we label the $\mathrm{m' th}$ $out_L$ and the $\mathrm{m' th}$ $in_R$ by the same integer. Hence, we label $\mathrm{m th}$ $out_R$ and $\mathrm{m'th}$ $in_R$ channels by $m$ and $L+m'$, respectively. We can denote this label strategy by $\mathcal{I}_2$ and rewrite it in a vector form:
\begin{equation}
\begin{aligned}
\label{index-channels-network}
    L_{L+m}^T=(\overbrace{0\cdots\cdots0}^{in_L},\overbrace{0\cdots \underbrace{1}_{m th} \cdots 0}^{out_L})^T \quad m\in[1,L]\\
    R_{L+n}^T=(\overbrace{0\cdots\cdots0}^{out_R},\overbrace{0\cdots \underbrace{1}_{n th} \cdots 0}^{in_R})^T \quad n\in[1,L]
\end{aligned}
\end{equation}
The vectors of $in_L$ and $out_R$ are defined in a similar way.

On the other hand, as a basis of the transfer matrix, edges in one layer $\{(\mathcal{Y},\nu)\vert\mathcal{Y}\in[1,L],\,\nu=0,1\}$ have already been labeled by $\{2\mathcal{Y}+\nu\}$ (see \cref{concrete-transfer-matrix-network-1} \& (\ref{concrete-transfer-matrix-network-2})). We denote this label manner as $\mathcal{I}_1$. Therefore, we need $V_L$ \& $V_R$ to relate $\mathcal{I}_1$ and $\mathcal{I}_{2}$. $V_L$ is a $2L\times 2L$ orthogonal matrix that transforms $in_L\,\&\,out_L$ channels indexed by $\mathcal{I}_2$ to edges indexed by $\mathcal{I}_1$ on the left side, \ie layer $(0,1)$. Similarly, $V_R$ transforms $in_R\,\&\,out_R$ to the edges on right side, \ie layer $(\frac{M}{2}+1,0)$. We denote the entry of $V_{L,R}$ by $V_{L,R}[\,\mathrm{row\,index};~\mathrm{column \,index}]$. Hence, the nonzero entries of $V_{L,R}$ have the form: $V_{L,R}[\mathcal{I}_1 \mathrm{\,index\,of\,some\,channel},\mathcal{I}_2 \mathrm{\,index\,of\,the\,same\,channel}]=1$. It can be checked that the nonzero entries explicitly are
\begin{equation}
    \left\{V_{L,R}[2m+\mathrm{mod}(m,2);m]=V_{L,R}[2m+\mathrm{mod}(m+1,2);m+L]=1\,\vert\, m\in[1,L]\right\}.
\end{equation}

According to \cref{Tmatrix}, we can obtain $t'$ by calculating the amplitudes of $\{\mathrm{n th}\,in_R\rightarrow\mathrm{m th}\, out_L\vert n,m\in[1,L]\}$:
\begin{equation}
\begin{aligned}
    (t^{\prime-1})_{nm}=R_{L+n}^{\dagger} V_{R}^{\dagger}\prod_{\eta=1}^{M/2}T_{(\eta+1,0),(\eta,1)}T_{(\eta,1),(\eta,0)} V_L L_{L+m},
\end{aligned}
\label{t-prime-inverse}
\end{equation}
where $L_{L+m},\,R_{L+n}$ are defined in \cref{index-channels-network}. After taking inverse of \cref{t-prime-inverse}, we can calculate the conductance (and hence the conductivity) by \cref{Landauer formula} and the results are shown in \cref{subsec:numerical-network}. 

To end the discussion of conductivity, we should emphasize that the conductivity is not calculated on a quasi-1D sample. The calculation of $\rho_{\mathrm{q-1D}}$ needs a precise evaluation of the Lyapunov exponents, so we need to multiply a large number of random matrices together and utilize the self-average effect to do statistical analysis (the strategy described in \cref{subsec:basic-localization-length}). However, conductivity does not explicitly rely on the Lyapunov exponents. Therefore, such a statistical analysis (on segments of a quasi-1D sample) is not necessary. In addition, this strategy has two drawbacks that prevent us from using it. First, what we can directly calculate is the conductance $G=\sigma L/M$. In a quasi-1D sample ($M\!\gg\!L$), the error of conductance $\Delta G$ will be amplified by the large factor $\frac{M}{L}$ and result in a considerable error in conductivity $\Delta\sigma=\frac{M}{L}\Delta G$. Second, the $\sigma$ derived from a quasi-1D sample may intrinsically differ from that of a normal shaped 2D sample. Thus, we instead take the statistical analysis on a large number (usually $10^4\sim10^5$) of normal shaped samples with different disorder configurations.

\subsection{Transfer matrices of eight-band lattice models $H_{8B}$ and $H_{8B}'$}
\label{subsec:transfer-lattice}

Before moving to our lattice models $H_{8B}$ (\cref{TB-H-real}) and $H_{8B}'$ (\cref{simplified-H-reciprocal}), we derive the transfer matrices in a general lattice model whose Hamiltonian is
\begin{equation}
\label{general-lattice-H}
    H=\sum_{i j} t_{ij} c_i^{\dagger}c_j=\sum_{\substack{q\,=-r_1}}^{r_2} \sum_{l=1}^M\sum_{\alpha,\beta=1}^s t_{l\alpha,(l+q)\beta}\,c_{l\alpha}^{\dagger} c_{(l+q)\beta}
\end{equation}
where $i,\,j$ are site indices and $t_{i,j}$ is the hopping coefficient from $j$ to $i$. The communication relation is $\{c_i,c_j^{\dagger}\}=\delta_{i,j}$. In the second equation, we decompose site indices into layer indices ($l$ \& $l+q$) and indices inside one layer ($\alpha\,\&\, \beta$). $M$ is the total number of layers and $s$ is the total number of degrees of freedom per layer.
The restriction on the summation over $q$, \ie $q\in[-r_1,r_2]$ means that only the $(l-r_1)$th $\sim (l+r_2)$th layers can hop to the $l\thinspace$th layer (in one step). This restriction does not significantly damage the generality of \cref{general-lattice-H} since it merely forbids the hopping to be infinite-long range in the longitudinal direction. 

The eigenstates of $H$ should satisfy the stationary equation $E\vert\psi\rangle=H\vert\psi\rangle$ and we extract one component:
\[
    E\psi_{l\alpha}=\sum_{q=\,-r_1}^{r_2}\sum_{\beta}t_{l\alpha,(l+q)\beta}\psi_{(l+q)\beta}
\]
where $\psi_{l\alpha}$ is the amplitude on site $(l\alpha)$, \ie $\ket{\psi}=\sum_{l,\alpha} \psi_{l\alpha}c^{\dagger}_{l\alpha}\ket{0}$. 
We can combine amplitudes in one layer into a vector $\vec{\psi}_l=(\psi_{l1},\psi_{l2},...\psi_{ls})^T$ and define the  Hamiltonian block $(H_{l,l'})_{\alpha\beta}=\bra{0} c_{l,\alpha} H c^{\dagger}_{l',\beta}\ket{0}=t_{l\alpha,l'\beta}$. Therefore
\begin{equation}
\begin{aligned}
    &E\vec{\psi}_l= \sum_{q=\,-r_1}^{r_2} H_{l,l+q}\vec{\psi}_{l+q}\\
    \Longrightarrow\;
    &H_{l,l+r_2}\vec{\psi}_{l+r_2}=-\left(\sum_{q=1}^{r_2-1}+\sum_{q=-r_1}^{-1}\right)H_{l,l+q}\vec{\psi}_{l+q}-\left(H_{l,l}-E\right)\vec{\psi}_{l}
\end{aligned}
\end{equation}
If $H_{l,l+r_2}$ is invertible, we can reformulate the second equation as
{
\linespread{2} \selectfont
\begin{equation}
\label{TM-general-lattice}
   \begin{bNiceMatrix}
       \vec{\psi}_{l+r_2}\\
       \hline
       \vec{\psi}_{l+r_2-1}\\
       \vdots\\ \vdots \\
       \vec{\psi}_{l-r_1+1}
   \end{bNiceMatrix}                                         
   =\begin{bNiceArray}{ccccc|c}[margin]
       -H_{l,l+r_2}^{-1}H_{l,l+r_2-1}  &\cdots & -H_{l,l+r_2}^{-1}(H_{l,l}-E) & \cdots &-H_{l,l+r_2}^{-1}H_{l,l-r_1+1} & -H_{l,l+r_2}^{-1}H_{l,n-r_1}\\
       \hline
       \Block{4-5}<\Large>{I_{s(r_1+r_2-1)\times s(r_1+r_2-1)}} & & & & & \Block{4-1}<\Large>{0_{s(r_1+r_2-1)\times s}} \\
        & & & & & \\
        & & & & & \\
        & & & & & 
   \end{bNiceArray}
   \begin{bNiceMatrix}
       \vec{\psi}_{l+r_2-1}\\
       \vdots\\
       \vec{\psi}_{l}\\
       \vdots\\
       \hline
       \vec{\psi}_{l-r_1}
   \end{bNiceMatrix}
\end{equation}}
Eq.~(\ref{TM-general-lattice}) relates $\vec{\psi}_{l-r_1}\sim\vec{\psi}_{l+r_2-1}$ with $\vec{\psi}_{l-r_1+1}\sim\vec{\psi}_{l+r_2}$. And we can take the transformation matrix in Eq.~({\ref{TM-general-lattice}}) as the transfer matrix of $l$th layer ($T_{l}$). 

If different choices of the quasi-1D directions do not demonstrate qualitative different localization behaviors, we are free to choose the quasi-1D direction to optimize the numerical performance. First, we should take the quasi-1D direction that makes $(r_1\!+\!r_2)$ as small as possible since the transfer matrix in \cref{TM-general-lattice} is $s(r_1\!+\!r_2)$-dimensional, and we want the transversal size $L\!\propto\!s$ to be as large as possible.
Second, we should avoid a nearly singular $H_{l,l+r_2}$, \ie $\det (H_{l,l+r_2})\approx0$, since \cref{TM-general-lattice} explicitly depends on $H_{l,l+r_2}^{-1}$. If one, unfortunately, encounters a nearly singular $H_{l,l+r_2}$, he/she can try changing the quasi-1D direction or taking a finer layer division to resolve it.

\begin{figure}[h]
	\centering
    \includegraphics[width=.6\linewidth]{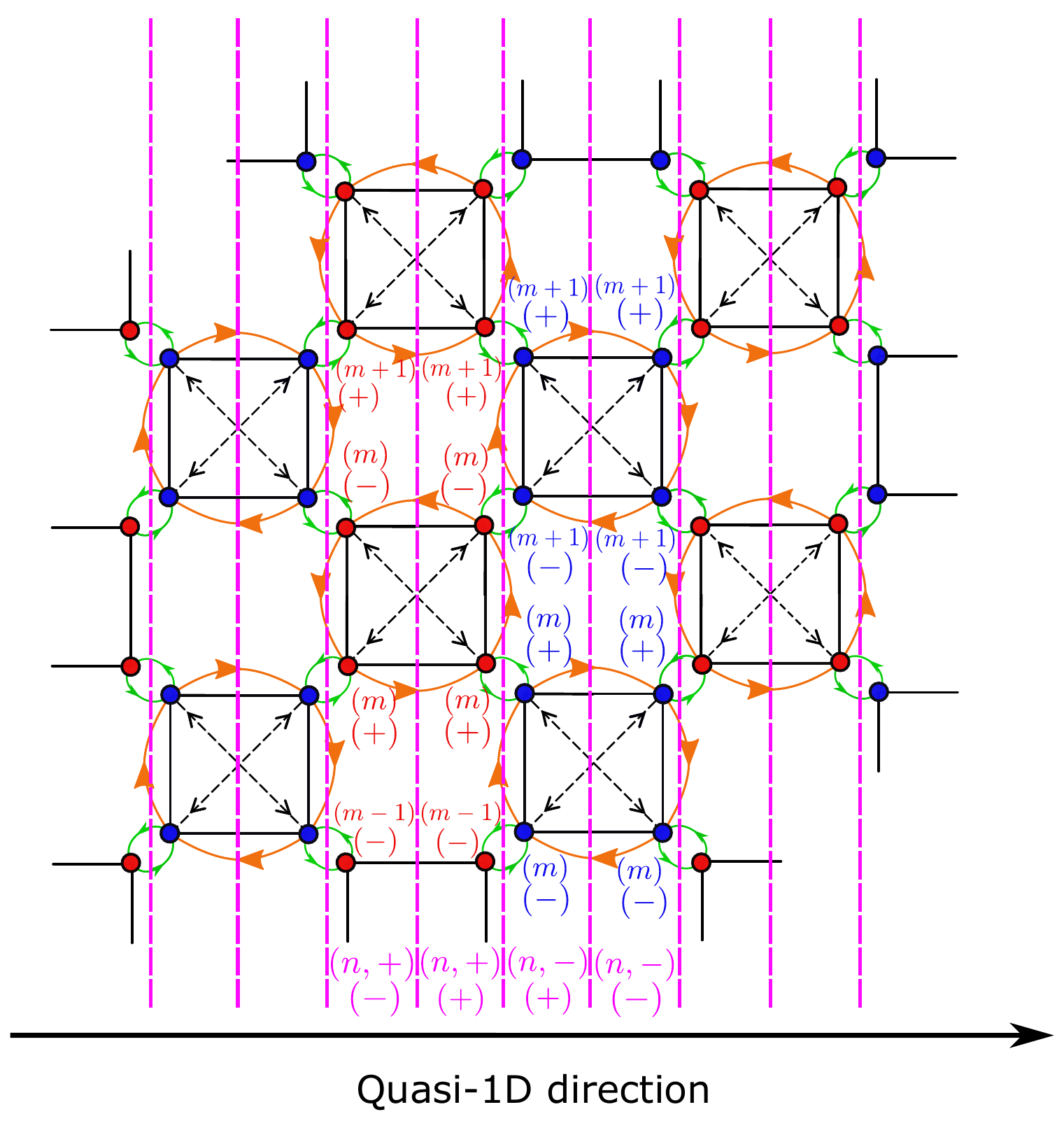}
    \caption[]{Layer division of $H_{8B}$ defined in \cref{TB-H-real}. For $H_{8B}'$ defined in \cref{simplified-H-reciprocal}, the division is the same but without square diagonal hopping. Corners inside two adjacent pink dashed lines belong to one layer. The pink indices $\{(n,\alpha,\beta)\vert n\in\mathbb{Z},\alpha,\beta=\pm\}$ in the bottom are layer indices. The red and blue indices $\{(m,\gamma)\vert m\in\mathbb{Z},\gamma=\pm\}$ further distinguish different corners in one layer. Among these indices, $(n,m)$ and $\alpha$ are previous cell index and chiral index, respectively. Two new indices $\beta\,\&\,\gamma$ together play the same role as the corner index $w$. The relation between them is $w=\frac{5}{2}-\beta(1+\frac{\gamma}{2})$. }
	\label{transfer-lattice}
\end{figure}

Now, we turn to our lattice models $H_{8B}$ and $H_{8B}'$. In \cref{subsec:basic-localization-length}, we have argued that different choices of the quasi-1D directions will not make qualitative difference in our models. Therefore, we are free to choose the quasi-1D directions of $H_{8B}$ and $H_{8B}'$ without losing the comparability with the network model. In order to minimize $(r_1\!+\!r_2)$, we take a different quasi-1D direction (x direction) in $H_{8B}$ and $H_{8B}'$ from the network model (x+y direction).
As shown in Fig.~\ref{transfer-lattice}, sites inside two adjacent pink dashed lines belong to one layer.
Such a choice gives $r_1=r_2=1$, no matter whether the square diagonal hopping is removed or not. In other words, we only have hoppings between the nearest neighbor layers for both $H_{8B}$ and $H_{8B}'$. The disorder, as we have explained in Sec.~\ref{subsec:disorder}, is restricted to the on-site form $U_{rand}(i)c_i^{\dagger}c_i$, where the random potentials $U_{rand}(i)$ uniformly distribute in $[-W/2,W/2]$ and are independent to each other.

The concrete formulae of transfer matrices in our lattice models is quite complicated, since every unit cell is divided into four layers.
As illustrated in Fig.~\ref{transfer-lattice}, the layer index has three components $(n,\alpha,\beta)$ and intra-layer index has two components $(m,\gamma)$, where $(n,m)$ and $\alpha$ are previous cell and chiral indices, respectively. Two new indices $\beta,\gamma=+,-$ together play the same role as the corner index $w$ defined in \cref{standing-wave-to-corner-dot}. The reason for splitting $w$ into $\beta$ and $\gamma$ is to fit the transfer matrix formula, \ie indicate the degrees of freedom by (layer indices)+(intra-layer indices).  The relation between them is $w=5/2-\beta(1+\gamma/2)$, which can be directly checked by comparing Fig.~\ref{transfer-lattice} with Fig.~\ref{TB-cell}.

Since we only have hoppings between nearest neighbor layers, we can denote the non-zero Hamiltonian blocks as $h_{l,\pm}=H_{l,l\pm1},\,h_{l,0}=H_{l,l}$. 
Explicit expressions of these blocks are 
\begin{equation}
\label{concrete-transfer-matrix-lattice}
    \begin{aligned}
        &\left(h_{(n,\alpha,\beta),0}\right)_{m\gamma,m\gamma}=U_{rand}(n,m,\alpha,\beta,\gamma)\qquad\quad \alpha,\beta,\gamma=\pm1\\
        &\left(h_{(n,\alpha,\beta),0}\right)_{m\gamma,m\bar{\gamma}}=(-\alpha\beta\gamma A i-1)\frac{\pi v}{4a}\\
        &\left(h_{(n,\alpha,\beta),\overline{\alpha\beta}}\right)_{m\gamma,m\gamma}=(\alpha\beta\gamma Ai-1)\frac{\pi v}{4a}\\
        &\left(h_{(n,\alpha,\beta),\overline{\alpha\beta}}\right)_{m\gamma,m\bar{\gamma}}=t'\\
        &\left(h_{(n,\alpha,\beta),\alpha\beta}\right)_{m+,m+}=\left(h_{(n,\alpha,\beta),\alpha\beta}\right)_{m-,(m+\alpha)-}=\tilde{t}\\
        &\mathrm{Other\;elements\;in\;} h_{(n,\alpha,\beta),0}\mathrm{\;and\;}h_{(n,\alpha,\beta),\pm} \mathrm{\;are\;zero.}
    \end{aligned}
\end{equation}
Here, $\overline{(\cdots)}$ means $(-1)\times(\cdots)$, $t'\,\&\,\tilde{t}$ are square diagonal hopping \& inter-square hopping, and $A$ is the coefficient controlling $DOS_F$ defined in Eq.~(\ref{simplified-H-reciprocal-coefficient}). 
For $H_{8B}$, $A=1,t'=-\frac{\pi v}{4a}$ (\cref{TB-H-real}) and for $H_{8B}'$, $A>0,t'=0$ (\cref{simplified-H-reciprocal}).  
The transversal size $L$ is defined by the number of cells involved in one layer, \ie $m=1,2,\cdots,L$. 
We use the periodic boundary condition in the transversal direction. 
The definition of longitudinal size $M$ is the number of layers, \ie $n=1,2,\cdots,M/4$.
We choose $A=1.2$ in our calculations and exhibit the results in \cref{subsec:numerical-lattice-full} \& \ref{subsec:numerical-lattice-simplified}.

If one finds Eq.~(\ref{concrete-transfer-matrix-lattice}) confusing, one can span it with explicit $\alpha,\beta$ indices. 
For example, for $\alpha=+,\beta=+$, the nonzero matrix elements are given by 
\[
\begin{aligned}
    &\left(h_{(n,+,+),0}\right)_{m+,m-}=\left(h_{(n,+,+),0}\right)_{m-,m+}^*=(-Ai-1)\frac{\pi v}{4a}\\
    &\left(h_{(n,+,+),-}\right)_{m+,m+}=\left(h_{(n,+,+),-}\right)_{m-,m-}^*=(Ai-1)\frac{\pi v}{4a}\\
    &\left(h_{(n,+,+),-}\right)_{m+,m-}=\left(h_{(n,+,+),-}\right)_{m-,m+}=t'\\
    &\left(h_{(n,+,+),+}\right)_{m+,m+}=\left(h_{(n,+,+),+}\right)_{m-,(m+1)-}=\tilde{t},
\end{aligned}\ .
\]

\clearpage
\bibliography{refs.bib}

\end{document}